\documentclass[a4paper,12pt,customfont,custommargin,numbered,index,oneside]{Classes/PhDThesisPSnPDF}

\usepackage{setspace}
\usepackage{amsfonts}
\usepackage{amsmath}
\usepackage{amssymb}
\usepackage{enumitem}
\usepackage{array}
\usepackage{relsize}
\usepackage{natbib}
\usepackage{graphicx}
\usepackage{import}
\usepackage{color,hyperref}
\usepackage{accents}
\usepackage{scalerel,stackengine}
\stackMath

\definecolor{darkblue}{rgb}{0.0,0.6,0.3}
\hypersetup{colorlinks,breaklinks,
            linkcolor=darkblue,urlcolor=darkblue,
            anchorcolor=darkblue,citecolor=darkblue}

\newcommand\doilink[1]{\href{http://dx.doi.org/#1}{#1}}
\newcommand\doi[1]{doi:\doilink{#1}}

\newcommand{\be}{\begin{equation}}
\newcommand{\ee}{\end{equation}}
\newcommand{\de}{\mbox{d}}
\newcommand{\e}{\mbox{e}}
\newcommand{\pa}{\partial}
\newcommand{\ba}{\begin{eqnarray}}
\newcommand{\ea}{\end{eqnarray}}
\newcommand{\ph}{\phantom{a}}
\newcommand{\pha}[1]{\hspace{-6pt}\ph^{#1}\hspace{-2pt}} 
\newcommand{\arccosh}{\mbox{arccosh}}
\newcommand\reallywidehat[1]{%
\savestack{\tmpbox}{\stretchto{%
  \scaleto{%
    \scalerel*[\widthof{\ensuremath{#1}}]{\kern-.6pt\bigwedge\kern-.6pt}%
    {\rule[-\textheight/2]{1ex}{\textheight}}
  }{\textheight}%
}{0.5ex}}%
\stackon[1pt]{#1}{\tmpbox}%
}
\input{Preamble/preamble}

\title{Cosmological consequences of Quantum Gravity proposals}

\author{Marco de Cesare}

\dept{Theoretical Particle Physics and Cosmology Group\\Department of Physics}

\university{King's College London}
\crest{\includegraphics[width=0.2\textwidth]{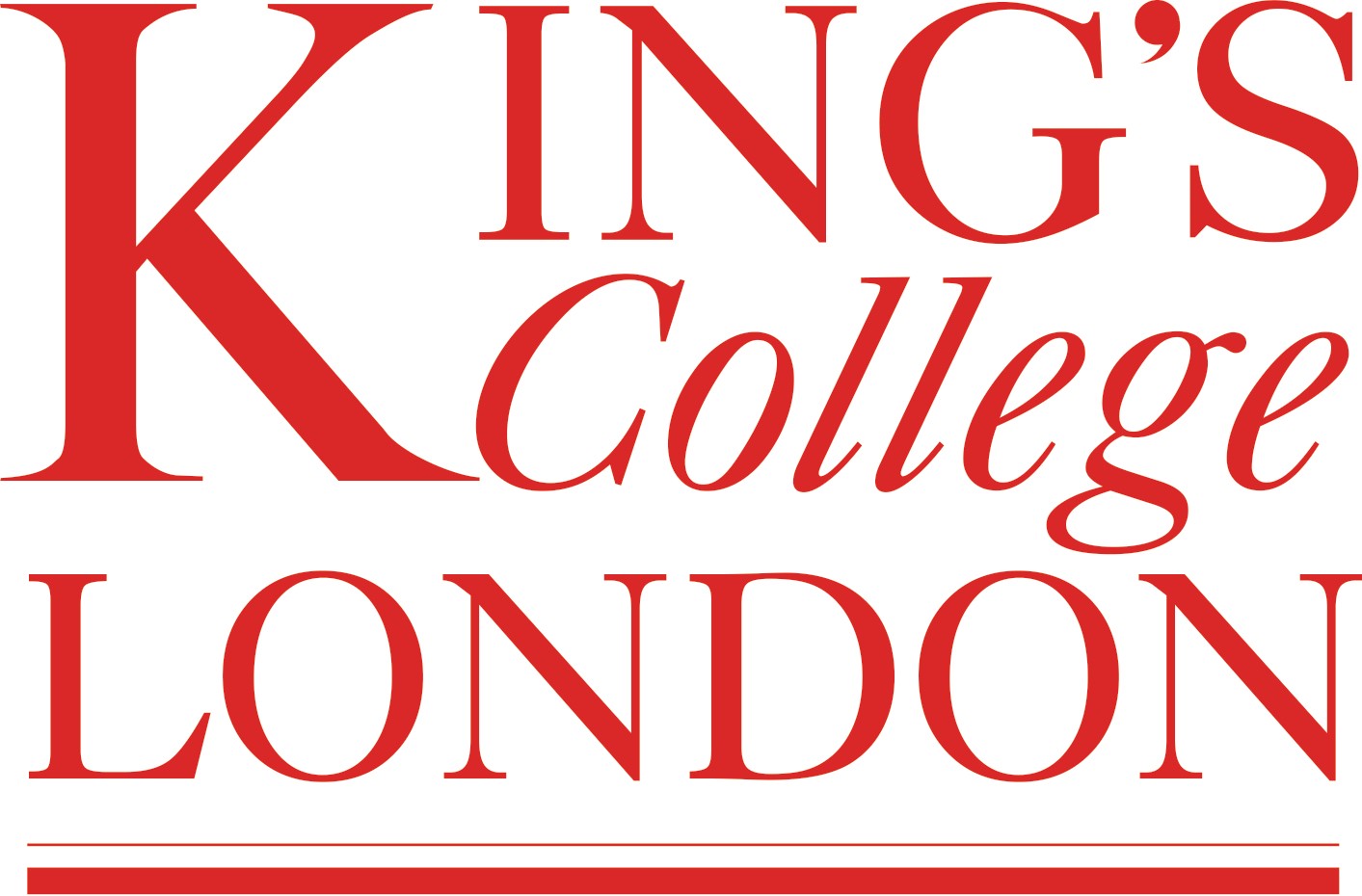}}

\supervisor{Professor Mairi Sakellariadou}

\supervisorlinewidth{0.6\textwidth}

\renewcommand{\submissiontext}{Submitted in partial fulfillment of the requirements for the degree of}

\degreetitle{Doctor of Philosophy}

\advisor{Professor G.A. Mena Marug\'an \newline
Professor J. Magueijo}
     
\advisorlinewidth{0.6\textwidth}

\degreedate{September 2017} 

\subject{LaTeX} 

\ifdefineAbstract
\fi

\ifdefineChapter
\fi

\begin{document}

\frontmatter

\maketitle


\begin{declaration}

The contents of this thesis are the result of the author's own work and of the scientific collaborations listed below, except where specific reference is made to the work of others. This thesis is based on the following research papers, published in peer reviewed journals. They are the outcome of research conducted at King's College London between October 2013 and September 2017.

\begin{enumerate}[label=$\diamond$]
\item  M.~de Cesare, D.~Oriti, A.~Pithis and M.~Sakellariadou, ``Dynamics of anisotropies close to a cosmological bounce in
quantum gravity'',Class. \ Quant.\ Grav.\  {\bf 35}, no. 1, 015014 (2018)\\
  \doi{10.1088/1361-6382/aa986a}
  [arXiv:1709.00994 [gr-qc]].
\item  M.~de Cesare, J.~W.~Moffat and M.~Sakellariadou,
  ``Local conformal symmetry in non-Riemannian geometry and the origin of physical scales'', Eur.Phys.J. C77 (2017) no.9, 605\\ \doi{10.1140/epjc/s10052-017-5183-0}
  [arXiv:1612.08066~[hep-th]]
	\item
  M.~de Cesare, A.~Pithis and M.~Sakellariadou
  ``Cosmological implications of interacting Group Field Theory models: cyclic Universe and accelerated expansion'', Phys.\ Rev.\ D {\bf 94}, no. 6, 064051 (2016)\\ \doi{10.1103/PhysRevD.94.064051}
  [arXiv:1606.00352 [gr-qc]]
    \item
  M.~de Cesare and M.~Sakellariadou,
  ``Accelerated expansion of the Universe without an inflaton and resolution of the initial singularity from GFT condensates'',
  Phys.\ Lett.\ B764, 49 (2017)\\ \doi{10.1016/j.physletb.2016.10.051}~[arXiv:1603.01764 [gr-qc]]
  \item
  M.~de Cesare, F.~Lizzi and M.~Sakellariadou,
  ``Effective cosmological constant induced by stochastic fluctuations of Newton's constant,''
  Phys.\ Lett.\ B {\bf 760}, 498 (2016)\\
  \doi{10.1016/j.physletb.2016.07.015}
  [arXiv:1603.04170 [gr-qc]]
  \item
  M.~de Cesare, M.~V.~Gargiulo and M.~Sakellariadou,
  ``Semiclassical solutions of generalized Wheeler-DeWitt cosmology,''
  Phys.\ Rev.\ D {\bf 93}, no. 2, 024046 (2016)\\
  \doi{10.1103/PhysRevD.93.024046}
  [arXiv:1509.05728 [gr-qc]]
\end{enumerate}
During the course of my doctoral studies, I also worked on the following papers. However, they are beyond the scope of this thesis and will not be discussed here.
\begin{enumerate}[label=$\diamond$]
\item M.~de Cesare, R.~Oliveri and J.~W.~van Holten,
  ``Field theoretical approach to gravitational waves'', Fortschritte der Physik - Progress of Physics {\bf 65}, no.5, 1700012 (2017)\\ \doi{10.1002/prop.201700012}
  [arXiv:1701.07794 [gr-qc]]
\item M.~de Cesare, N.~E.~Mavromatos and S.~Sarkar,
  ``On the possibility of tree-level leptogenesis from Kalb-Ramond torsion background'',
  Eur.\ Phys.\ J.\ C {\bf 75}, no. 10, 514 (2015)\\
  \doi{10.1140/epjc/s10052-015-3731-z}
  [arXiv:1412.7077 [hep-ph]]
\end{enumerate}



\end{declaration}

\begin{abstract}
In this thesis, we study the implications of Quantum Gravity models for the dynamics of spacetime and the ensuing departures from classical General Relativity. The main focus is on cosmological applications, particularly the impact of quantum gravitational effects on the dynamics of a homogenous and isotropic cosmological background. Our interest lies in the consequences for the evolution of the early universe and singularity resolution, as well as in the possibility of providing an alternative explanation for dark matter and dark energy in the late universe.

The thesis is divided into two main parts, dedicated to alternative (and complementary) ways of tackling the problem of Quantum Gravity. The first part is concerned with cosmological applications of background independent approaches to Quantum Gravity, as well as minisuperspace models in Quantum Cosmology. Particularly relevant in this work is the Group Field Theory approach, which we use to study the effective dynamics of the emergent universe from a full theory of Quantum Gravity (\emph{i.e.} without symmetry reduction). We consider both approaches based on loop quantisation and on quantum geometrodynamics. 

In the second part, modified gravity theories are introduced as tools to provide an effective description of quantum gravitational effects, and show how these may lead to the introduction of new degrees of freedom and symmetries. Particularly relevant in this respect is local conformal invariance, which finds a natural realisation in the framework of Weyl geometry. We build a modified theory of gravity based on such symmetry principle, and argue that new fields in the extended gravitational sector may play the role of dark matter.  New degrees of freedom are also natural in models entailing fundamental `constants' that vary over cosmic history, which we examine critically.

Finally, we discuss prospects for future work and point at directions for the derivation of realistic cosmological models from Quantum Gravity candidates.

\end{abstract}


\begin{acknowledgements}      
There are many people I need to thank for their help while I was working on this thesis. First of all, I want to thank my supervisor Mairi Sakellariadou for her guidance during my doctoral studies, and especially for teaching me to think like a researcher. I also would like to thank all of my collaborators, and in particular Giampiero Esposito, Maria Vittoria Gargiulo, Fedele Lizzi, Nick Mavromatos, John Moffat, Roberto Oliveri, Daniele Oriti, Andreas Pithis, Jan Wilhelm van Holten. I greatly benefitted from discussions with Steffen Gielen, Nick Houston and Edward Wilson-Ewing. I would like to thank my examiners, Jo\~ao Magueijo and Guillermo Mena Marug\'an, for their valuable feedback and helpful comments.  I am also grateful to many people in the department of Physics for their help and advice, particularly Jean Alexandre and Julia Kilpatrick.

During these years in London I met many extraordinary people, whom I can now count among my friends. In particular, I would like to thank Alix, Andreas, Apostolos, Claire, Dan, Krzysziek, Marcello, Marco, Pooya, Tom, Laura for making it such a great experience. Special thanks go to Juliette, and to Vittorio and Federica (Cuc\'u); I could write a book to explain why, but they already know it. I am thankful to my brother Lorenzo and to my aunt Lilla for being the persons they are and for supporting me. Perhaps I should thank Flavia too. Many thanks also to my old friends Casimiro, Mario and Salvatore. I am indebted to Andreas Pithis, Roberto Oliveri and Salvatore Castrignano for reading parts of this thesis and for their invaluable comments. I want to thank Pam from the North London Buddhist centre for many enjoyable conversations and for her endless curiosity. We would need more people like her in this world, and this work is also for them. Finally, I am deeply grateful to Federica for her love and constant encouragement, and for supporting me in difficult times.

\end{acknowledgements}


\begin{dedication}
\vspace{-4cm}
\emph{By means of the easy and the simple we grasp the laws of the whole world.}\par
\begin{flushright} Ta Chuan/The Great Treatise \end{flushright}\bigskip ~\\

\emph{I think the best viewpoint is to pretend that there are experiments and calculate.\\ In this field since we are not pushed by experiments we must be pulled by imagination.}\par
\begin{flushright} Richard P. Feynman at the 1957 Chapel Hill Conference \end{flushright}\medskip

\null\vspace{\stretch{1}}
\begin{center}
 This thesis is dedicated to the memory of my mother
\end{center}
\vspace{\stretch{4}}

\end{dedication}


\chapter{Preface}

\section*{Thesis Aim}
The purpose of this thesis is to study the impact of quantum gravitational effects in cosmology and the modifications they bring to the standard picture for the history of our Universe. This is necessary in order to bring Quantum Gravity closer to the point of being predictive and to bridge the gap between candidate fundamental theories and cosmological observations. Looking at the cosmological consequences is also an important way of comparing between different approaches and to gain a deeper understanding of their relative strengths and weaknesses.

This research represents a first step towards making the connection between Quantum Gravity and more conventional model building in cosmology. At the same time, it offers the opportunity to consider alternative scenarios, such as \emph{e.g.} emergent cosmologies with a bounce.
I have explored different possibilities for the study of quantum gravity effects in cosmology, which are complementary among them. Specifically, I worked using both a top-down approach and an effective field theory approach. The former aims at recovering cosmology from a given theory of Quantum Gravity, with possible departures from standard cosmology at early and late times. The latter aims at capturing quantum gravity effects by considering modifications of gravity as a classical effective theory, \emph{e.g.}~by the introduction of new symmetry principles or degrees of freedom.

A substantial effort went into making this thesis as self-contained as possible. Chapters can be read independently from one another to a very large extent, since they deal with different approaches. Nevertheless, serious effort was made to show their complementarity and, where applicable, the relations between them. Each chapter contains enough introductory material to make it suitable as a primer on the topic discussed. To this end, several appendices have also been included with introductory and review material as a complement to the discussions in the chapters. Technical appendices with detailed calculations are included for the benefit of the reader. An attempt was made to provide the reader with a full picture of the topics discussed, which goes beyond the particular applications that we considered in this thesis. A comprehensive list of bibliographical references is given.

\section*{Thesis Outline}
This thesis consists of two main parts. The first part deals with the study of the cosmological sector upon quantisation of the gravitational field.
This is done both in the context of a full theory of Quantum Gravity (specifically, Group Field Theory) and in reduced symmetry quantisation (Quantum Cosmology).
The focus of the second part is instead on classical models (effective field theories) of modified gravity, which aim at encoding quantum gravitational effects by introducing suitable modifications of classical General Relativity.

A brief outline is the following. In the Introduction, after giving a concise overview of the motivations for seeking a quantum theory of gravity, we discuss the distinguished role of cosmology as an arena for competing theories. In Chapter~\ref{Chapter:StandardCosmology}  we review the formulation of the Standard Cosmological Model. In Chapter~\ref{Chapter:WDW} we review the quantum geometrodynamics approach, then focusing on the evolution of the universe wave-function in a minisuperspace model which generalises Wheeler-DeWitt theory. In Chapter~\ref{Chapter:GFT}, after reviewing the fundamentals of the Group Field Theory formalism for Quantum Gravity, we consider its hydrodynamics approximation and study the dynamics of the background in the ensuing emergent cosmology scenario. The consequences for cosmology of the early and late universe are discussed in detail for different models.  Chapter~\ref{Chapter:VariableG} deals with the expansion of a homogeneous and isotropic Universe, in the case where the gravitational constant is dynamical and given by a stochastic process. Chapter~\ref{Chapter:Weyl} presents an extension of classical General Relativity based on the principle of local conformal invariance, entailing the shift from the framework of Riemannian geometry to that of Weyl geometry. Finally, in the Conclusion we review our results and discuss how they fit in the bigger picture of research in Quantum Gravity, hinting at directions for future work.


\tableofcontents


\chapter{Notation and Conventions}

We consider units in which $\hbar=c=1$, unless otherwise stated. Greek indices denote spacetime components of tensor field. Latin indices are used for their spatial components. When Latin indices appear, sometimes letters from the first part of the Greek alphabet are used to denote internal indices. However, whenever there is a risk of any ambiguity this is clearly spelled out. The metric has signature $(-+++)$, \emph{i.e.} mostly plus. Our conventions for the Riemann curvature tensor and its contractions are the same as in Wald's book \cite{Wald:1984rg}.

\

The gravitational constant will be denoted by $G$ in most chapters. In some chapters, a different notation was preferred to avoid the risk of confusion. In particular, in Chapter~\ref{Chapter:GFT} Newton's constant is denoted by $G_{\rm N}$, whereas $G$ denotes a generic Lie group. In Chapter~\ref{Chapter:VariableG}, the notation $G$ is used for the dynamical gravitational constant.

\

The reader must be aware that due to the heterogeneous nature of the subject, and in order to keep the notation as close as possible to the published literature, the notation used in any two distinct chapters is not necessarily consistent. However, the notation is certainly consistent within each chapter taken individually. The meaning of a symbol is always explained on its first occurrence in a given chapter, and often recalled when appropriate. A list of common symbols, having the same meaning in different parts of the thesis, is the following:

\newpage

{\bf \large List of common symbols}
\vskip 0.5cm
\begin{tabular}{ll}
$g_{\mu\nu}$ & spacetime metric\\
$\nabla_\mu$ & affine connection (non necessarily metric-compatible)\\
$R_{\mu\nu\rho}^{\ph\ph\ph\sigma}$ & Riemann curvature of the connection $\nabla_\mu$\\
$R_{\mu\nu}$ & Ricci tensor\\
$R$    & Ricci scalar\\
$\Lambda$ & cosmological constant\\
$T_{\mu\nu}$ & stress-energy tensor of matter\\
$h_{ij}$ & spatial metric\\
$p^{ij}$ & canonically conjugated momentum to $h_{ij}$ in ADM\\
$\pha{(3)}R$ & curvature of three-space\\
$G_{ijkl}$ & DeWitt supermetric\\
$G_{AB}$ & DeWitt supermetric on minisuperspace\\
$a$  &  scale factor\\
$H$ & Hubble rate\\
$\ell_{\rm Pl}$    & Planck length
\end{tabular}

\vskip 1cm

{\bf \large List of common abbreviations}
\vskip 0.5cm
\begin{tabular}{ll}
l.h.s.  &   left hand side (of an equation)\\
r.h.s.  &   right hand side (of an equation)\\
w.r.t.  &    with respect to\\
cf.     &     compare with\\
d.o.f. &     degree(s) of freedom\\
Eq.   &     an equation (followed by its number)\\
Ref.   &   a bibliographic reference (followed by its number)\\
QG    &   Quantum Gravity\\
QC    &   Quantum Cosmology\\
LQG  &    Loop Quantum Gravity\\
LQC  &     Loop Quantum Cosmology\\
GFT   &    Group Field Theory\\
QFT   &    Quantum Field Theory\\
WDW   &   Wheeler-DeWitt

\end{tabular}

\listoffigures




\printnomenclature

\mainmatter



\newpage

\chapter*{Introduction}\label{sec:Introduction}
\addcontentsline{toc}{chapter}{Introduction}
The problem of formulating a theory of Quantum Gravity, which would combine in a satisfactory way General Relativity and Quantum Field Theory, started nearly as far back as Quantum Mechanics was established as a universal theoretical framework for the description of microscopic phenomena \cite{Rovelli:2000aw}.
It can be argued that, given a quantum system, all physical systems interacting with it must also be quantum. Therefore, the quantization of the gravitational field appears as a necessary consequence of the universality of the gravitational interaction and that of Quantum Mechanics  \cite{DeWitt:1962cg,Woodard:2009ns}.

The standard quantization procedures, which have been extremely successful in the case of the electroweak and the strong interaction, fail in the case of gravity. The well-known perturbative non-renormalizability of Einstein's theory of General Relativity thus prompted the investigation of alternative paths.
Modern approaches to Quantum Gravity fall essentially into two classes: unified theories and background independent approaches. A candidate in the first class is a modern incarnation of Superstring Theory, known as M-theory \cite{becker2006string}; such theory would provide a unified description of all fundamental interactions, including gravity, in terms of more fundamental objects (strings and branes) living in a higher-dimensional spacetime. At the present stage, M-theory is only known in some limits, corresponding to the five known superstring theories or to eleven-dimensional supergravity, related to each other by dualities. Approaches belonging to the second class insist on a central property of General Relativity, namely its \emph{background independence} \cite{Smolin:2005mq}, which is elevated to the status of a fundamental principle and used as a guide in the quest for the fundamental theory of Quantum Gravity. 

A physical theory is said to be background independent if and only if it is diffeomorphism invariant and has no absolute structures (\emph{i.e.} non-dynamical ones) \cite{Giulini:2006yg}. It is precisely the absence of absolute structures, such as a background metric, which makes the quantization of General Relativity particularly challenging. This also has important consequences for the quantum theory, making its interpretation particularly problematic, due to the absence of a fixed causal structure \cite{Isham:1995wr}, or a preferred choice of a time parameter (problem of time) \cite{Isham:1992}.

Particularly relevant for this thesis will be the so-called \emph{canonical approaches}, based on a Hamiltonian formulation of General Relativity which appropriately takes into account the constraints stemming from background independence. The first such approach, known as \emph{quantum geometrodynamics}, is based on the reformulation of Einstein's theory by Arnowitt, Deser and Misner (ADM) \cite{Arnowitt:1962hi}, with the quantum theory obtained by applying the standard heuristic quantization rules. In this approach, the fundamental phase space variables are represented by the spatial metric and its canonical momentum. The quantization of General Relativity following this path was first proposed by Wheeler and DeWitt \cite{DeWitt:1967yk}. One of the main merits of this approach lies in the fact that it allows to recover General Relativity in the semiclassical limit \cite{gerlach1969derivation}, thus hinting that it may offer a valid description of Quantum Gravity at least at an effective level \cite{Kiefer:2008bs}. Despite mathematical ambiguities in the implementation of constraints in the full theory \cite{Kiefer:2007ria}, this approach has the clear advantage of allowing for analytic control of simple, yet physically relevant, systems, such as cosmological models and black holes \cite{Kiefer:2008bs}. In 
this thesis we will study some applications of the quantum geometrodynamics approach to minisuperspace models and consider the possibility of generalizing the framework to display the extant connections with other approaches.

An alternative approach to canonical quantum gravity is based on a different choice of variables, namely Ahstekar-Barbero variables \cite{Ashtekar:1986yd,Barbero:1994ap}, which enable one to recast the theory in a form displaying many similarities with non-Abelian Yang-Mills theories. The corresponding quantum theory is known as Loop Quantum Gravity and has the merit of providing rigorous mathematical foundations of the canonical approach \cite{Thiemann:2007zz}. The loop quantization programme has led to remarkable insights in the structures arising from the quantization of geometry such as, for instance, the discreteness of the spectra of geometric operators \cite{Rovelli:1994ge,Ashtekar:1996eg,Ashtekar:1997fb}. Such quantum discreteness of geometry also has an impact on the dynamics of spacetime on large scales, as shown in the symmetry reduced version of the theory, known as Loop Quantum Cosmology, which predicts a bounce resolving the initial `big bang' singularity of classical cosmological models \cite{Ashtekar:2011ni}. The major open problem remains the implementation of the Hamiltonian constraint in the full theory. Different programmes have been developed to this end, such as \emph{e.g.} Spin Foam models \cite{Perez:2012wv}, the master constraint programme \cite{Thiemann:2003zv}, the iterative coarse graining scheme \cite{Dittrich:2013xwa}.


The Group Field Theory approach is intimately related to Loop Quantum Gravity. In fact, it provides a non-perturbative completion of Spin Foam models, which allows to make sense of the sum over triangulations of spacetime in terms of a path-integral \cite{Oriti:2006se,Freidel:2005qe}. The picture of quantum geometry offered by Group Field Theory is that of a many body system, in which the fundamental degrees of freedom are open spin network vertices labelled by data of group theoretic nature \cite{Oriti:2006se}. These are sometimes referred to as `particles' or `quanta of geometry'. The generic state of the system contains combinatorial information about the way such quanta are linked to each other. The Group Field Theory formalism can thus be understood as a second quantized formulation of Loop Quantum Gravity, providing an interpretation of the spin network states of Loop Quantum Gravity as `many-particle states' \cite{Oriti:2013}. In this approach, spacetime is not a fundamental concept and has been argued to emerge dynamically from the collective dynamics of many such quanta \cite{Oriti:2013jga}. Particularly relevant for this thesis are the applications of the Group Field Theory approach to early universe cosmology.

Background independent approaches also include: Causal Dynamical Triangulations \cite{Ambjorn:2004qm,Ambjorn:2011cg}, Regge Calculus \cite{regge1961general,Williams:1991cd}, Causal Sets \cite{Bombelli:1987aa,Sorkin:2003bx} and Quantum Graphity \cite{Konopka:2006hu}. The realization of background independence in Asymptotic Safety is a delicate issue, due to the reliance of the set up on the splitting of the metric into a background and fluctuation. Such splitting generally introduces an artificial background dependence in the results, which has to be restored at the level of physical observables \cite{Eichhorn:2017egq}. There are also approaches that do not necessarily fall in the two categories defined above. 
This is for instance the case of Non-Commutative Geometry \emph{\`a la} Connes, that can be seen in more general terms as a bottom-up approach, in which our knowledge of low-energy physic is used to determine the geometric data defining the non-commutative space \cite{Connes:2017oxm}.
Some non-commutative geometric structures are also known to arise in String Theory \cite{Lizzi:1997yr,Seiberg:1999vs}. At the present stage of development it is not clear whether a fully background independent formulation of Non-Commutative Geometry exists.

All known approaches to Quantum Gravity have their own strengths and weaknesses, which may make them more suitable for some applications compared to others. Some of them display remarkable connections, as stressed above in the case of Group Field Theory, Spin Foam models and Loop Quantum Gravity. However, it is not known at present whether any of the theories that are currently available can represent a fundamental theory of Quantum Gravity, and the recovery of General Relativity in the continuum limit is perhaps the most pressing issue in many approaches. At any rate, it is important to improve our understanding of the connections between different proposals and unravel common mathematical structures, which may encourage progress in the field and lead to the convergence of different lines of investigation.

Background independence is a common feature of many different approaches; however, it is not clear whether it is realized in String Theory. In fact, the formulation of all known superstring theories is only known at a perturbative level. It is usually argued that it will only be clear whether background independence is realized in String Theory once M-theory is formulated. Nevertheless, some background independent features of String Theory are already well-known. A remarkable example is provided by the so-called holographic principle \cite{Maldacena:1997re,Witten:1998qj}. In fact, it has been argued that the holographic principle must actually represent one of the fundamental principles of a theory of Quantum Gravity \cite{Bousso:2002ju}. Recent results in Loop Quantum Gravity also hint at a possibility of realizing the holographic principle in this context \cite{Donnelly:2016auv,Livine:2017xww}. Thus, there may be chances that String Theory and Loop Quantum Gravity are closer than it has been anticipated so far \cite{Jackson:2014nla}. 

While making progress in examining the structure of quantum geometry at a fundamental level, it is also important to attempt at making contact with experiments. In fact, the primary reason for our limited understanding of Quantum Gravity, despite many decades of theoretical effort, can ultimately be traced back to the lack of experiments which can probe the fabric of spacetime at the smallest length scales (Planckian). Thus, it is crucial to bridge the gap which currently exists between Quantum Gravity theories and phenomenology of the gravitational interaction. The final aim must be that of being able to extract predictions from alternative candidate theories which can be potentially put to test, thus enabling us to compare them and rule out some alternatives.

 Given the difficulty of the task, it is appropriate to identify suitable systems and regimes in which we expect Quantum Gravity effects to play a role. Such effects are expected to become relevant at extremely high energy scales, of the order of the Planck scale. Although unaccessible to terrestrial experiments, such energy scales were typical soon after the big bang in the so-called Quantum Gravity era.
Therefore, cosmology of the very early universe represents the natural place to look for observable signatures of Quantum Gravity, which may for instance be lurking in the spectrum of primordial fluctuations \cite{Kiefer:2011cc}. This thesis represents a preliminary step going in this direction. We are not able yet to extract measurable quantities from a full theory of Quantum Gravity, although, as we will discuss, considerable progress has been made in making contact to more conventional model building in cosmology and in understanding the dynamics of the cosmological background near classical singularities.

Quantum gravitational effects are also expected to play an important role in cosmology for different reasons. In fact, the standard inflationary scenario assumes the occurrence of an era of exponential expansion taking place in the very early universe. The seeds for structure formation would be generated during inflation. However, the inflationary scenario does not provide a justification for the choice of initial conditions, nor it offers a resolution of the initial singularity. Moreover, a sufficient amount of homogeneity is necessary at the onset of inflation for the mechanism to take place \cite{Calzetta:1992bp,Calzetta:1992gv}.
Perhaps more importantly, inflation is affected by the so-called trans-Planckian problem: fluctuation modes corresponding to scales that are observable today originated as sub-Planckian during inflation. The mechanism is thus very sensitive to ultraviolet modifications, which could potentially undermine its success \cite{Martin:2001aa}. Hence, a successful realization of this scenario requires its emebedding in a theory of Quantum Gravity.
It is important to remark that Quantum Gravity proposals may also provide an alternative to the inflationary paradigm, as in the case of bouncing cosmologies \cite{WilsonEwing:2012pu,Brandenberger:2016vhg}, cyclic and ekpyrotic models \cite{Lehners:2008vx}, and in the emergent cosmology scenario in string gas cosmology \cite{Brandenberger:2008nx}. Work is under way to determine whether Group Field Theory can also represent such an alternative.

An alternative way to approach the problem of Quantum Gravity is to assume an effective field theory point of view and study modifications of Einstein's theory. Modified gravity theories can thus be motivated as `emerging' from some more fundamental theory of Quantum Gravity, and extend the Einstein-Hilbert action by the inclusion of suitable quantum corrections. The term emergence here may have different meanings, depending on the theory \cite{Butterfield:1998dd}. It can be understood, for instance, in the sense of a semiclassical limit, or as the appearance of novel properties in certain regimes of the underlying fundamental theory \cite{Oriti:2013jga}.
 As seen from the discussion above, we are not yet at the stage of deriving such an effective theory from a background-independent theory of Quantum Gravity. Hence, motivation for a modified gravity theory must also seek support in phenomenology \cite{Sotiriou:2007yd,Capozziello:2011et}. In fact, there is a hope that Quantum Gravity may lead to an alternative resolution of the tension between the predictions of General Relativity and observational data, usually resolved by the introduction of dark matter and dark energy. This approach is pursued in the second part of this thesis.
  
While the computation of the effective action from a background-independent theory is still an open problem, this is certainly possible in perturbative approaches. For instance, in String Theory the low-energy spacetime effective action is known and features two new fields in the gravitational multiplet: the dilaton and the Kalb-Ramond field \cite{Green:1987sp}. 
Generalizations also exist for M-theory \cite{Witten:1996md}. Double Field Theory offers a description of closed string field theory on a torus, which is able to capture T-duality symmetry \cite{Hull:2009mi}; the realization of background independence in this framework was considered in Ref.~\cite{Hohm:2010jy}.
 Additional terms in the action of gravity also arise in the perturbative quantization of General Relativity, where higher-order curvature terms naturally arise as radiative corrections. Moreover, the formal computation of the path-integral for Quantum General Relativity leads to a term proportional to the square of the Weyl tensor, hinting at a possible role played by conformal invariance in Quantum Gravity \cite{Hooft:2010ac,Hooft:2014daa}. These results show that quantum corrections arising from the behaviour of the gravitational field at high energies also induce significant departures from General Relativity on large scales, thus hinting at a deeper connection between the microscopic structure and dynamics of quantum spacetime and macroscopic gravitational phenomena, particularly on cosmological scales \cite{Sotiriou:2007yd}. 
     
Extra degrees of freedom also play an important role in this context. They may arise from a given Quantum Gravity proposal, as in the case of String Theory, which predicts the existence of many scalar fields such as moduli and ultra-light axions \cite{Arvanitaki:2009fg,Svrcek:2006yi}. Alternatively, new degrees of freedom can be phenomenologically motivated, as \emph{e.g.} in scalar-tensor theories, or scalar-tensor-vector theories. Yet another possibility is that they are required by the adoption of a more general geometric framework than Riemannian geometry, as in the case of metric-affine gravity \cite{Hehl:1994ue}, and particularly in Weyl geometry which is discussed in this thesis \cite{Smolin:1979uz,Cheng:1988zx,deCesare:2016mml}. Scalar fields play an important role in this respect, as they can be used to promote fundamental constants of Physics (such as the gravitational constant \cite{Brans:1961sx}, or the fine structure constant \cite{Sandvik:2001rv}) to dynamical variables.

Finally, in light of our discussion, it seems wise to pursue different directions in the quest for Quantum Gravity. Progress in this field will depend on many factors, not least the ability to relate different approaches and combine their insights. The fundamental theory may or may not be among the candidates that are available today; its formulation may require the revision of principles that have so far been regarded as fundamental, or the introduction of new principles altogether. This problem cannot be settled at the outset, as all physical principles, whether new or old, must be grounded in experimental results. In the absence of experiments that can probe the full Quantum Gravity regime, the best strategy is to proceed in two opposite directions. On the one hand, one must address the problem of recovering a continuum spacetime from a background independent quantum theory of the gravitational field, and obtain the corresponding effective dynamics at low energies in a top-down approach. This will be given by some modified gravity theory, which must reduce to General Relativity in the regimes where the latter has been tested. Departures from General Relativity are expected, which may explain the dark sector of our Universe and provide an alternative to the inflationary paradigm (or a completion thereof). On the other hand, while progress is done in the top-down approach, it is also possible to work at an effective level; thus classical General Relativity is appropriately modified in order to reach agreement with observational data. The effective approach and the top-down approach may sustain each other as they develop, motivating for instance the existence of new fields or symmetries (\emph{e.g.} conformal symmetry or dualities from String Theory). The hope is that such radically different approaches may converge at some point, shedding new light on the relation between quantum geometry and continuum spacetime, and at the same time leading to a better understanding of cosmic evolution. 
  


\chapter{Standard Cosmology}\label{Chapter:StandardCosmology} 
In this chapter we review the Friedmann-Lema\^itre-Robertson-Walker (FLRW) model for homogeneous and isotropic cosmology, which lies at the foundations of the standard $\Lambda$CDM model. We discuss the foundational aspects and the formulation of such model, highlighting the questions left unaddressed in this framework. We briefly review the standard cosmological puzzles and how they are addressed by the inflationary paradigm. Finally we will discuss the occurrence of the Big-Bang singularity in classical Cosmology and examine possible ways to prevent its occurrence. 

\section{The Cosmological Principle}\label{Sec:CosmologicalPrinciple}
The subject of Cosmology is the study of the dynamics of our Universe as a whole. It aims at understanding the history and the present state of our Universe by rooting it in fundamental physics. The development of modern Cosmology as a science is fairly recent compared to other branches of Physics. It was only made possible by the formulation of General Relativity (GR), which provided the necessary conceptual framework to treat spacetime as a dynamical entity. Cosmology deals with the dynamics of spacetime on large scales, \emph{i.e.} much larger than those of visible structures in the Universe. On such scales, the Universe has remarkably simple properties which are concealed on smaller scales.

\begin{figure}
\begin{center}
\includegraphics[width=\columnwidth]{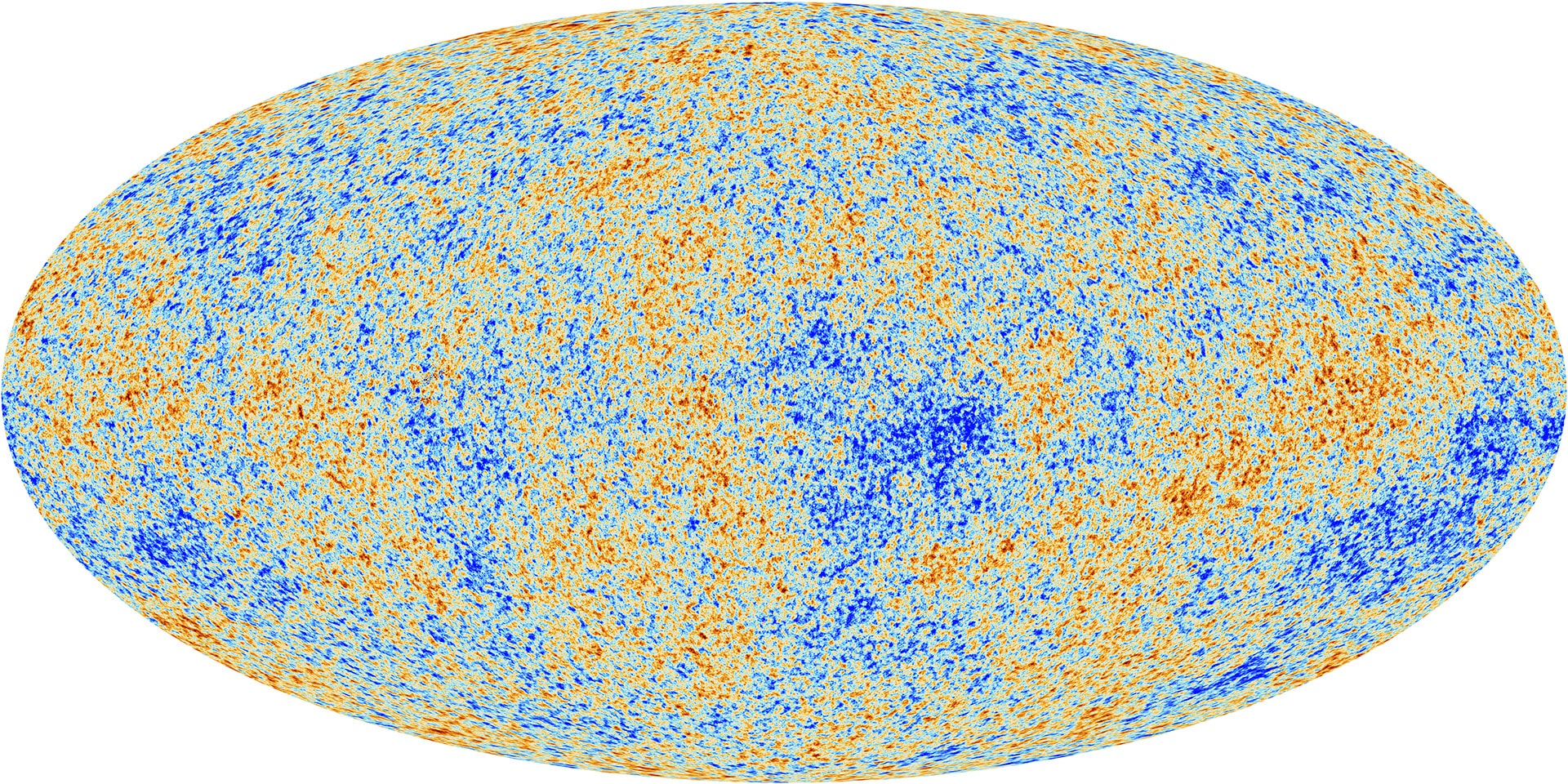}
\caption[CMB anisotropy spectrum]{Temperature anisotropies in the CMB, measured by the Planck collaboration. Each point is characterised by perfect blackbody radiation at the temperature 2.726 K. The tiny temperature differences observed correspond to differences in the matter density (inhomogeneities) at the time of photon decoupling. Such primordial inhomogeneities represent the seeds of structure formation. Copyright: ESA and the Planck Collaboration}\label{Fig:CMB}
\end{center}
\end{figure}

Despite the lumpiness exhibited by the distribution of matter which is apparent on smaller scales, the Universe looks very homogeneous when it is observed on scales that are of cosmological interest. Although direct tests of homogeneity are difficult, due to the uncertainties involved in the measurements of distant objects, there is very good evidence that this is indeed the case. The strongest direct evidence for the homogeneity of the Universe comes from the Sloan Digital Sky Survey, which showed that the galaxy distribution is homogeneous on scales larger than about 300 million light years \cite{Weinberg:2008zzc,Yadav:2005vv}. Furthermore, besides the absence of any privileged point, the Universe also lacks a preferred direction. In fact, the Planck mission showed that the Cosmic Microwave Background (CMB) is isotropic, \emph{i.e.} it has the same properties in every direction in the sky, within one part in $10^5$ (see Fig.~\ref{Fig:CMB}).
 These two basic facts provide strong support of what has been for decades only an assumption, which represented the starting point for much work in Cosmology. The so-called \emph{cosmological principle} states that the Universe is homogeneous and isotropic with respect to every point\footnote{For a suitably defined family of observers.}. As such, it is a generalisation of the Copernican principle \cite{Bondi:1960,Hawking:1973uf}. The cosmological principle allows us to single out one preferred frame (\emph{i.e.} a family of freely falling observers, one for each spatial point), namely the one in which the CMB looks isotropic. Deviations from perfect homogeneity and isotropy are thus treated as small perturbations on the expanding cosmological background.

The dynamics of a homogeneous and isotropic Universe was first derived independently by Friedmann \cite{Friedman:1922kd,Friedmann:1924bb} in order to exhibit a simple solution of Einstein's field equations. Lema{\^i}tre derived independently what later became known as the Friedmann equation and related it to the expansion of the Universe \cite{Lemaitre:1927zz}. Remarkably, these discoveries came years before the first evidence of the expansion of the Universe, provided by the observations made by Hubble \cite{Hubble:1929ig}.
The results of Friedmann and Lema{\^i}tre were later re-obtained in a series of papers by Robertson \cite{Robertson:1935zz,Robertson:1936zza,Robertson:1936zz} and in the work of Walker \cite{Walker:1937}, showing that the particularly simple form of the metric they assumed can be obtained on the basis of the cosmological principle alone, and is in fact independent of Einstein's equations. In particular, Robertson showed that the cosmological principle can equivalently be re-formulated as the requirement that the Universe is isotropic (\emph{i.e.} spherically symmetric) about every point \cite{Hawking:1973uf}. Hence, it is customary to refer to such cosmological models as FLRW models, using the initials of those authors.

In this thesis, unless otherwise stated, we will deal with a perfectly homogeneous and isotropic Universe and will not consider the role of perturbations. Thus, the problem of Cosmology is reduced to the study of the dynamics of the expansion of the cosmological FLRW background.

\section{FLRW models}
The cosmological principle implies that there is a frame (up to time reparametrisation) in which the metric takes the following form
\be\label{Eq:FLRW-Metric}
\de s^2=g_{\mu\nu}\de x^{\mu}\de x^{\nu}=-N(t)^2\de t^2 +a(t)^2\de l^2~,
\ee
where $N(t)$ is the \emph{lapse function} and is linked to time reparametrisation invariance. The choice $N(t)=1$ corresponds to the frame in which comoving observers are freely falling. In this case the time coordinate coincides with proper time as measured by such observers. 
$a(t)$ is the scale factor, which represents a conversion factor from comoving to physical distances. 
 $\de l^2=h_{ij}\de x^i\de x^j$ is the spatial metric on the three dimensional sheets and is time independent. The cosmological principle implies that sheets of the foliation are spaces of constant curvature. Hence, there are only three possibilities for the form of $\de l^2$, depending on whether the spatial curvature $\hspace{-6pt}\phantom{e}^{\scriptscriptstyle (3)}R$ is positive, negative or vanishing. It is possible to encode the three different possibilities in a single expression \cite{Weinberg:2008zzc}
 \be\label{Eq:Metric3Space}
 \de l^2=\left({\bf \de x}^2+K~\frac{({\bf x\cdot \de x})^2}{1-K {\bf x}^2}\right)~,
 \ee
 where ${\bf x}$ denotes quasi-Cartesian coordinates. Quantities such as ${\bf x}^2$ and ${\bf x\cdot \de x}$ in this expression are computed using the flat Euclidean metric $\delta_{ij}$. The parameter $K$ can take the values $1,~0,~-1$. If we consider the maximally extended spaces whose metric has the form given by Eq.~(\ref{Eq:Metric3Space}), those values correspond, respectively, to the following three dimensional geometries: a sphere, flat Euclidean space and a hyperboloid.
\begin{figure}
\begin{center}
\def\svgwidth{0.7\columnwidth}
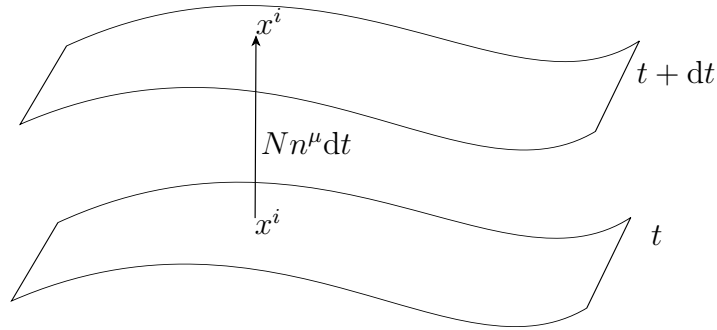
\end{center}
\caption[Homogeneous foliation]{Foliating the universe. The sheets of the foliation are assumed to be homogeneous and isotropic, so as to satisfy the \emph{cosmological principle}. Thus, the comoving frame is assumed to be the CMB frame. The vector $n^\mu$ is hypersurface orthogonal and the lapse $N(t)$ is an arbitrary function of time. In the homogeneous case there is no need to introduce a shift vector $N^i$. Points on different sheets having the same comoving coordinates give the trajectory of a comoving (freely falling) observer. }
\end{figure}

The metric of a FLRW universe given in Eq.~(\ref{Eq:FLRW-Metric}) is entirely specified once we know the two functions $N(t)$ and $a(t)$. However, it must be noted that the former does not represent any physical degree of freedom. In fact, given a second metric with lapse $N^{\prime}(t^{\prime})$ and scale factor $a^{\prime}(t^{\prime})$, it represents the same physical configuration as above provided that
\be
a^{\prime}(t^{\prime})=a(t)~,
\ee
where the two time variables are related by
\be
\frac{\de t^{\prime}}{\de t}=\frac{N(t)}{N^{\prime}\left(t^{\prime}(t)\right)}~.
\ee
Therefore, the scale factor $a(t)$ is the only physical degree of freedom of such a universe. $N(t)$ is a completely arbitrary function, which serves to identify a time coordinate. We expect the dynamics of the system to be compatible with such physical understanding.

The dynamics of a FLRW universe in classical GR can be obtained from the Einstein-Hilbert action by considering the particular ansatz Eq.~(\ref{Eq:FLRW-Metric}). The Einstein-Hilbert action reads as
\be\label{Eq:EH-Action}
S_{\rm EH}[g_{\mu\nu}]=\frac{1}{16\pi G}\int\de^4x\; \sqrt{-g}\, (R-2\Lambda)~,
\ee
where $R$ is the Riemann scalar of the metric $g_{\mu\nu}$ and we included the gravitational constant $\Lambda$. Integration is performed with respect to the invariant volume element $\de^4x\, \sqrt{-g}$. 
Variation of the action\footnote{A boundary term needs to be included in order to make the variational problem well posed \cite{Kiefer:2007ria}. The boundary term involves the extrinsic geometry of the boundary.} Eq.~(\ref{Eq:EH-Action}) with respect to the metric yields the Einstein equations for the dynamics of the gravitational field (see \emph{e.g.} \cite{Wald:1984rg})
\be\label{Eq:Einstein}
G_{\mu\nu}=8\pi G~T_{\mu\nu}~.
\ee
$T_{\mu\nu}$ is the stress-energy tensor of matter and represents the source of the gravitational field. Considering a matter action $S_{\rm m}$ the stress-energy tensor is given by
\be\label{Eq:DefStressEnergy}
T_{\mu\nu}=-\frac{2}{\sqrt{-g}}\frac{\delta S_{\rm m}}{\delta g^{\mu\nu}}~.
\ee

The dynamics of the expansion of the universe is usually obtained from the Einstein equations, considering the particular ansatz Eq.~(\ref{Eq:FLRW-Metric}). Here we wish to give an alternative derivation, taking the action rather than the field equations as a starting point. By plugging the ansatz Eq.~(\ref{Eq:FLRW-Metric}) into Eq.~(\ref{Eq:EH-Action}), we get the following action for the gravitational sector of the model \cite{Kiefer:2007ria}
\be\label{Eq:FRLWaction}
S_{\rm g}[a]=\frac{3}{8\pi G}\int\de t\; N\left(-\frac{a\dot{a}^2}{N^2}+K a-\frac{\Lambda a^3}{3}\right)~. 
\ee
Note that in order to get the action (\ref{Eq:FRLWaction}), integration over the comoving coordinates must be carried out in the Einstein-Hilbert action (\ref{Eq:EH-Action}). For infinite spatial slices this integral is clearly divergent. However, the problem is easily solved by restricting the spatial integration to a fiducial cell\footnote{In the case $K=1$ this is not necessary and the integration can be carried out over the whole 3-sphere as in Ref.~\cite{Kiefer:2007ria}. The same reasoning also applies to a flat compact space, such as a torus.} with finite comoving volume $V_0$, which can then be reabsorbed in the definition of the scale factor and lapse\footnote{Such a rescaling has an effect on the cosmological constant term in the action (\ref{Eq:FRLWaction}). In fact, one has $\Lambda\to \Lambda/V_0$.}.
Note the unusual minus sign in front of the kinetic term of the scale factor. 
Matter fields can be introduced by starting from their covariant action in four dimensions and imposing homogeneity and isotropy. For instance, the action of a minimally coupled scalar field in four dimensions with potential $V(\phi)$ reads as
\be
S_{\rm m}[\phi]=-\int\de^4x\; \sqrt{-g}\, \left(\frac{1}{2}\partial_\mu\phi\partial^\mu\phi +V(\phi)\right)~.
\ee
In a FLRW universe it reduces to
\be\label{Eq:ActionScalarFieldFLRW}
S_{\rm m}[\phi]=\int\de t\; N a^3 \left(\frac{1}{2}\frac{\dot{\phi}^2}{N^2}-V(\phi)\right)~.
\ee

The full action of the system, including both gravitational and matter contributions, can be rewritten more compactly as (see Ref.~\cite{Kiefer:2007ria})
\be\label{Eq:MinisuperspaceAction}
S=S_{\rm g}+S_{\rm m}=\int\de t\; N\left(\frac{1}{2}G_{AB}\frac{\dot{q}^A\dot{q}^B}{N^2}-U(q)\right)~,
\ee
where we have introduced the matrix
\be
G_{AB}=\begin{pmatrix} -\frac{3}{8\pi G}a &0\\0&a^3\end{pmatrix}~
\ee
and the minisuperspace potential
\be
U(q)=a^3V(\phi)-\frac{3}{8\pi G}K a+\frac{\Lambda }{8\pi G}a^3~.
\ee
In the specific example of the scalar field considered above, capital latin indices can take only two values $A=1,2$. We have $q=(a,\phi)$, which defines a point in a new configuration space, known as \emph{minisuperspace}. The term \emph{superspace}\footnote{It must be stressed that it has nothing to do with supersymmetry.} was coined by Wheeler to denote the configuration space of geometrodynamics \cite{Wheeler:1988zr}, \emph{i.e.} canonical gravity, which we discuss in Appendix~\ref{Appendix:ADM} and in Chapter~\ref{Chapter:WDW}. For systems having a finite number of degrees of freedom, such space is finite dimensional; whence the name minisuperspace.
$G_{AB}$ is known as the DeWitt supermetric, which defines a distance on minisuperspace\footnote{The expression of the DeWitt supermetric on the superspace of the full theory is given in Appendix~\ref{Appendix:ADM}.}. 
The indefiniteness of the DeWitt supermetric is responsible for the negative sign in the kinetic term of the scale factor.

We notice that the action does not contain the time derivative of the lapse $\dot{N}$. This is a direct consequence of time-reparametrisation invariance and implies that we are dealing with a constrained system. The Hamiltonian formulation of constrained systems developed by Dirac is discussed in Appendix~\ref{Appendix:Dirac}. The case of GR is reviewed in Appendix~\ref{Appendix:ADM}. For the time being we simply remark that the observation made above leads to the conclusion that the lapse $N$ does not represent a physical degree of freedom. The discussion of technical and conceptual subtleties related to the dynamics of GR as a constrained system is postponed to later chapters. Hence, we will proceed to derive brute-force the equations of motion for the scalar factor $a$ from the action (\ref{Eq:MinisuperspaceAction}).
Variation of the action with respect to the lapse yields
\be
\begin{split}
\delta S=&\delta S_{\rm FLRW}+\delta S_{\rm m}=\\
&\int\de t\; \left[\frac{3}{8\pi G}\left(\frac{a\dot{a}^2}{N^2}+K a-\frac{\Lambda a^3}{3}\right)-a^3\left(\frac{1}{2}\frac{\dot{\phi}^2}{N^2}+V(\phi)\right)\right]\delta N~.
\end{split}
\ee
When the first variation vanishes, and for lapse $N=1$ (comoving coordinates) we get the Friedmann equation
\be
\left(\frac{\dot{a}}{a}\right)^2=\frac{8\pi G}{3}\rho_\phi-\frac{K}{a^2}+\frac{\Lambda}{3}~,
\ee
where $\rho_\phi$ is the energy density of the scalar field
\be\label{Eq:ScalarFieldEnergyDensity}
\rho_\phi=\left(\frac{1}{2}\dot{\phi}^2+V(\phi)\right)~.
\ee
The Friedmann equation (\ref{Eq:AccelerationSecondFriedmannScalarField}) gives the expansion of a homogeneous and isotropic universe. It shows that the expansion is sourced by the energy density of matter in the universe (including the cosmological constant), plus a contribution coming from the three-dimensional geometry of space.

Variation of the action (\ref{Eq:MinisuperspaceAction}) with respect to $a$ yields instead, after using the Friedmann equation (\ref{Eq:AccelerationSecondFriedmannScalarField}) and requiring that we are in the comoving frame
\be\label{Eq:AccelerationSecondFriedmannScalarField}
\frac{\ddot{a}}{a}=-\frac{4\pi G}{3}(3p_\phi+\rho_\phi)+\frac{\Lambda}{3}~,
\ee
where we introduced the pressure of the scalar field
\be\label{Eq:ScalarFieldPressure}
p_\phi=\frac{1}{2}\dot{\phi}^2-V(\phi)~.
\ee
The second Friedmann equation, Eq.~(\ref{Eq:AccelerationSecondFriedmannScalarField}), relates the acceleration of the expansion of the universe, defined in the comoving frame, with some particular combination of the pressure and the energy density of the scalar field. Notice that there is no contribution coming from the curvature of the spatial slices. We will say more about the implications of the second Friedmann equation in the following, in particular for what concerns the sign of the acceleration $\ddot{a}$.

The equations of motion of the scalar field are also obtained from the action Eq.~(\ref{Eq:MinisuperspaceAction}), by requiring that its first variation with respect to $\phi$ be vanishing. Thus, we have (in the comoving frame $N=1$)
\be\label{Eq:EoMotionScalar}
\ddot{\phi}+3H\dot{\phi}+\frac{\partial V}{\partial\phi}=0~.
\ee
We observe that the curvature of the gravitational background\footnote{More precisely the extrinsic curvature, defined as $K_{ij}=\frac{1}{2N}\frac{\partial h_{ij}}{\partial t}=\frac{\dot{a}}{N a}h_{ij}$, where $h_{ij}$ is the spatial metric \cite{Kiefer:2007ria}. Tracing this tensor using the inverse spatial metric $h^{ij}$ we get $K_{i}^{~i}= K_{ij}h^{ij}=\frac{3\dot{a}}{N a}$. Finally, in the comoving gauge we have $K_{i}^{~i}=\frac{H}{3}$.} is responsible for the second term in Eq.~(\ref{Eq:EoMotionScalar}). This is a friction term\footnote{The Hubble rate is positive $H>0$. Hence the term $3H\dot{\phi}$ implies a loss of kinetic energy.} due to the expansion of the background. This is an important point, which will be taken into account when discussing the conditions for inflation to take place in Section~\ref{Sec:SlowRollInflation}.

The dynamics of a FLRW model filled with a generic fluid, characterised by its energy density $\rho$ and pressure $p$, can be obtained as a straightforward generalisation of the case studied above, with $\rho_\phi$ and $p_\phi$ replaced by the corresponding quantities for the fluid. Depending on the type of matter under consideration the functional relation between $\rho$ and $p$ (\emph{i.e.} the equation of state of the fluid) will be different. We rewrite the more general equations below, for convenience of the reader. The Friedmann equation reads as
\be\label{Eq:Friedmann}
H^2=\left(\frac{\dot{a}}{a}\right)^2=\frac{8\pi G}{3}\rho-\frac{K}{a^2}+\frac{\Lambda}{3}~,
\ee
where we introduced the Hubble rate, defined as the logarithmic derivative of the scale factor $H=\frac{\dot{a}}{a}$. The second Friedmann equation, or acceleration equation, reads as follows
\be
\frac{\ddot{a}}{a}=-\frac{4\pi G}{3}(3p+\rho)+\frac{\Lambda}{3}\label{Eq:AccelerationSecondFriedmann}~.
\ee
The presence of the minus sign in the r.h.s of Eq.~(\ref{Eq:AccelerationSecondFriedmann}) is crucial. In fact, most types of matter satisfy the \emph{strong energy condition} (see Appendix \ref{Appendix:ECs}), implying that
\be
\rho+3 p\geq 0~.
\ee
As a consequence, the acceleration is always negative. In order to accommodate for a positive acceleration, the strong energy condition (SEC) must be violated. In Appendix \ref{Appendix:ECs} we discuss the formulation of SEC, as well as the weaker null energy condition (NEC) that is usually assumed in singularity theorems. Violation of SEC is achieved in inflationary models and guarantees that the universe expands with a positive acceleration. This is crucial in order to solve some standard problems in the Hot Big Bang cosmology, as we will discuss later in this chapter, in Section~\ref{Sec:CosmologicalPuzzles}.

\section{The dynamics of matter}
This section and the next one are more general than the rest of this chapter for their purpose. In fact, the domain of applications of many of the results obtained here is not restricted to the cosmological case or to GR and will be used in Chapter~\ref{Chapter:Weyl}. We will study the problem of deriving the dynamics of matter fields within the framework of a classical theory of gravity. In particular, we will show that the conservation law of the stress-energy tensor of matter can be obtained under very general conditions, namely diffeomorphism invariance. Such conservation law is particularly important in the case of a perfect fluid (studied in the next section) since it turns out to be equivalent to the equations of motion of the fluid.

Let us start from the definition of the stress-energy tensor, Eq.~(\ref{Eq:DefStressEnergy})
\be
T_{\mu\nu}=-\frac{2}{\sqrt{-g}}\frac{\delta S_{\rm m}}{\delta g^{\mu\nu}}~.
\ee
$S_{\rm m}$ is the action of the matter field under consideration, which we consider to be minimally coupled to the gravitational field and uncoupled to other matter species that may be present.
In order to prove the conservation of $T_{\mu\nu}$, we consider an infinitesimal diffeomorphism parametrised by a vector field $\xi^{\mu}$. Under such a diffeomorphism, the metric transforms as follows\footnote{In Eq.~(\ref{Eq:DiffeoMetric}) $\mathcal{L}_{\xi}$ denotes the Lie derivative along the flow generated by the vector field $\xi^{\mu}$, $\nabla$ is the Levi-Civita connection (\emph{i.e.} the unique symmetric metric-compatible affine connection). The notation with round brackets around tensor indices is used to mean symmetrisation over those indices, \emph{e.g.} considering a rank-two tensor $S_{\mu\nu}$ its symmetric part reads as $S_{(\mu\nu)}=\frac{1}{2}\left(S_{\mu\nu}+S_{\nu\mu}\right)$.} \cite{Wald:1984rg}
\be\label{Eq:DiffeoMetric}
\delta g_{\mu\nu}=\mathcal{L}_{\xi}g_{\mu\nu}=2\nabla_{(\mu}\xi_{\nu)}~.
\ee
We observe that the action $S_{\rm m}$ is invariant under diffeomorphisms\footnote{For an example of a theory where diffeomorphism invariance is explicitly broken, see Refs.~\cite{Charmousis:2009tc,Kimpton:2010xi}. In fact, in Ho\v{r}ava gravity the symmetry group of the action is broken to a subgroup of the group of diffeomorphisms, namely the group of foliation-preserving diffeomorphisms. As a consequence, the stress-energy tensor of matter is in general not conserved.}. Therefore, under such a transformation $\delta S_2=0$ by definition. Thus, we have
\be
0=\delta S_2=-\frac{1}{2}\int\de^4x\sqrt{-g}\; T_{\mu\nu}\delta g^{\mu\nu}=-\int\de^4x\sqrt{-g}\; T_{\mu\nu}\nabla^{(\mu}\xi^{\nu)}~.
\ee
Recalling that $T_{\mu\nu}$ is a symmetric tensor and integrating by parts, we find
\be\label{Eq:ConservationStressEnergyLastStep}
\int\de^4x\sqrt{-g}\; \left(\nabla^{\mu}T_{\mu\nu}\right)\xi^{\nu}=0.
\ee
Since the infinitesimal generator of the diffeomorphism, \emph{i.e.} the vector field $\xi^{\mu}$, is arbitrary, it follows that the stress-energy tensor is covariantly conserved\footnote{When there is more than one matter species coupled to the gravitational field, interactions between them are possible. In this case, the stress-energy tensors of the different species are \emph{not} conserved separately, \emph{e.g.} $\nabla^{\mu}T^1_{\mu\nu}\neq0$, $\nabla^{\mu}T^2_{\mu\nu}\neq0$, since exchange of energy and momentum between species is allowed. However, the \emph{total} stress-energy tensor of matter is always conserved $\nabla^{\mu}\left(T^1_{\mu\nu}+T^2_{\mu\nu}\right)=0$, following a straightforward generalisation of the argument given in this section. In the limit in which non-gravitational interactions are negligible, Eq.~(\ref{Eq:ConservationStressEnergy}) is a good approximation to the dynamics of a single matter species.}
 with respect to the Levi-Civita connection
\be\label{Eq:ConservationStressEnergy}
\nabla^{\mu}T_{\mu\nu}=0.
\ee
The last equation (\ref{Eq:ConservationStressEnergy}) gives the dynamics of matter. It is worth stressing that its derivation did not require using the specific form of the stress-energy tensor for a perfect fluid. Hence, it is completely general and holds for all matter fields minimally coupled to gravity\footnote{When a modified gravity theory is considered (\emph{e.g.} $f(R)$) or when non-minimal couplings are included, a conservation law like Eq.~(\ref{Eq:ConservationStressEnergy}) still holds if $T_{\mu\nu}$ is replaced by a suitably defined effective quantity $T^{\rm eff}_{\mu\nu}$, see \emph{e.g.} Refs.~\cite{Duruisseau:1986ga,Visser:1999de,Deser:1970hs}. The exact form of the correction terms in $T^{\rm eff}_{\mu\nu}$ is model-dependent and may involve second derivatives of the matter fields or higher.} . Moreover, since no use has been made of Einstein's equations, the result is valid also in modified gravity theories\footnote{For a review of modified gravity (or extended theories of gravity) see \emph{e.g.} Refs.~\cite{Capozziello:2011et,Sotiriou:2008rp,Sotiriou:2007yd}.} \cite{Sotiriou:2007yd}.

Before closing this section we would like to reconsider the conservation of $T_{\mu\nu}$ from an alternative point of view, which will be useful for the applications of Chapters~\ref{Chapter:VariableG},~\ref{Chapter:Weyl}. 
In fact, the covariant conservation law (\ref{Eq:ConservationStressEnergy}), which we obtained in this section from the sole requirement of diffeomorphism invariance of the matter action, can also be obtained from Einstein's equations. In fact, taking the covariant divergence of both sides of Eq.~(\ref{Eq:Einstein}) we find
\be\label{Eq:StressEnergyConsFromBianchi}
0=\nabla^\mu G_{\mu\nu}=8\pi G~\nabla^\mu T_{\mu\nu}~.
\ee
In fact, the l.h.s. of Eq.~(\ref{Eq:StressEnergyConsFromBianchi}) is always vanishing thanks to the contracted Bianchi identities. On the r.h.s. we also used the fact that the gravitational coupling $G$ is a constant in GR. This observation will be important in the following chapters. The Bianchi identities represent a kinematical constraint which is always valid in Riemannian geometry, since it follows only from the definition of the Riemannian curvature tensor \cite{Wald:1984rg}. 
Although the contracted Bianchi identities have a much more general status than then specific dynamical law of the gravitational field in GR, it is remarkable that they can also be obtained as a conservation law following from diffeomorphism invariance of the Einstein-Hilbert action. In fact, its variation is computed as (see Ref.~\cite{Wald:1984rg})
\be\label{Eq:VariationEH}
\delta S_{\rm EH}=\frac{1}{16\pi G}\int\de^4x\sqrt{-g}\; \left(G_{\mu\nu}\delta g^{\mu\nu}+\nabla^\mu v_\mu\right)~,
\ee
where
\be
v_{\mu}=\nabla^{\nu}\delta g_{\mu\nu}-g^{\rho\sigma}\nabla_{\mu}\delta g_{\rho\sigma}~.
\ee
Disregarding the surface term\footnote{This is clearly not possible when $G$ is promoted to a dynamical variable, as in scalar-tensor theories and as in the theory considered in Chapter~\ref{Chapter:Weyl}.} in Eq.~(\ref{Eq:VariationEH}) and considering an infinitesimal diffeomorphism as in Eq.~(\ref{Eq:DiffeoMetric}), we have
\be
0=\delta S_{\rm EH}=-\frac{1}{16\pi G}\int\de^4x\sqrt{-g}\;(\nabla^{\mu} G_{\mu\nu})\xi^\nu~.
\ee
Hence, as for the stress-energy tensor in Eq.~(\ref{Eq:ConservationStressEnergyLastStep}) we conclude
\be
\nabla^{\mu} G_{\mu\nu}=0~.
\ee

\section{The action of a relativistic fluid}\label{Sec:Fluid}
Whereas the discussion, as well as the conclusions, of the previous section are fully general and apply to any (minimally coupled) matter field, our purpose here is to study in more detail the dynamics of a fluid in a generally covariant framework. More specifically, we will deal with perfect relativistic fluids, \emph{i.e.} fluids with vanishing viscosity and anistropic stress. Although fluids do not offer a fundamental description of the dynamics of matter, they can be used as a good approximation in many situations of cosmological and astrophysical interest.

In order to find the dynamics we must construct a suitable action functional. A hint comes from the action of a scalar field in Eq.~(\ref{Eq:ActionScalarFieldFLRW}) which, using Eq.~(\ref{Eq:ScalarFieldPressure}), can be rewritten as
\be\label{Eq:ActionScalarFluid}
S_{\rm m}[\phi]=\int\de t\; Na^3 p_{\phi}~.
\ee
Hence, we can generalise Eq.~(\ref{Eq:ActionScalarFluid}) and make the following ansatz for the action of a fluid
\be\label{Eq:FluidAction1}
S_1=\int\de^4x\sqrt{-g}\; p~.
\ee
In the particular case of a FLRW background, with metric given by Eq.~(\ref{Eq:FLRW-Metric}) and considering a scalar field, we trivially recover our starting point Eq.~(\ref{Eq:ActionScalarFieldFLRW}). The action (\ref{Eq:FluidAction1}) was considered in Ref.~\cite{Schutz:1970aa} and it can be used to derive, by extremisation, the conservation law of the stress-energy tensor of a perfect fluid. Moreover, the author of Ref.~\cite{Schutz:1970aa} showed that by including suitable terms depending on the fluid potentials in the action, the Euler equations for a perfect fluid can be recovered from an action principle.

For our purposes it will be more convenient to use a different action, which reads as
\be\label{Eq:FluidAction2}
S_2=-\int\de^4x\sqrt{-g}\; \rho~.
\ee
This will be our starting point to derive the equations of motion of the fluid. The action $S_2$ and $S_1$ can be shown to be equivalent, \emph{i.e.} they are the same up to boundary terms\footnote{The proof is in Ref.~\cite{Brown:1992kc} and references therein. It is necessary to introduce suitable extra terms in $S_1$ and $S_2$ involving the fluid potentials. This in order to recover the Euler equations by extremising the action. The two actions are then seen to be equivalent when the equations of motion are satisfied.}\textsuperscript{,}\footnote{It must be pointed out that the equivalence is lost if non-minimal couplings to curvature are allowed, see Ref.~\cite{Faraoni:2009aa}}. 
In order to derive the dynamics of the fluid we must understand how the energy density, which appears in the action $S_2$, depends on the other dynamical variables which characterise the fluid, namely the particle number density $n$ and the four-velocity $U^\mu$. We start by making some considerations of themordynamical nature, following Ref.~\cite{Misner:1974qy}. Let us consider a comoving cell of volume $v\propto a^3$, and consider a homogenous distribution of particles at thermal equilibrium characterised by their energy density $\rho$ and their number density $n$. We also introduce the pressure of the fluid $p$. Since the fluid is in thermal equilibrium its entropy is conserved\footnote{The assumption of perfect homogeneity also excludes entropy flow from neighbouring fluid elements. For a more detailed discussion see Ref.~\cite{Misner:1974qy}.}, \emph{i.e.} the fluid is isentropic. The total number of particles in the volume is $A=n v$. Hence, from the first principle of thermodynamics we have
\be
\de \left(\rho v\right)=-p\,\de v~.
\ee
Dividing both sides by $A$, this equation can be rewritten as 
\be
\de \left(\frac{\rho}{n}\right)=-p\,\de\left(\frac{1}{n}\right)~.
\ee
After some trivial manipulations we obtain
\be\label{Eq:1stThermoChemicalPot}
\de \rho=\mu\,\de n~,
\ee
where we defined the chemical potential as
\be\label{Eq:DefChemicalPot}
\mu=\frac{\rho+p}{n}~.
\ee
Using Eqs.~(\ref{Eq:1stThermoChemicalPot}) and (\ref{Eq:DefChemicalPot}) we find the following relation between $p$, $\rho$ and $n$, which holds regardless of our assumption on the fluid being isentropic \cite{Misner:1974qy}
\be\label{Eq:PressureMTW}
p=\frac{\pa\rho}{\pa n}n-\rho~.
\ee

At this stage it is convenient to introduce a covariant quantity which can describe the flow of fluid particles in spacetime. The fluid in a given region of spacetime is characterised by its four-velocity $U^{\mu}$ (normalised as $U^{\mu}U_{\mu}=-1$) and the particle number density $n$. Therefore, it is natural to define the densitised particle number flux vector as in Ref.~\cite{Brown:1992kc}
\be\label{Eq:DefinitionFluidFlow}
J^{\mu}=n\sqrt{-g}\,U^{\mu}.
\ee
We can then express $n$ as
\be\label{Eq:NumberDensity}
n=\frac{\sqrt{-J_{\mu}J^{\mu}}}{\sqrt{-g}}=\frac{|J|}{\sqrt{-g}}~,
\ee
where in the last step we defined $|J|=\sqrt{-J_{\mu}J^{\mu}}$. We will assume that $\rho$ is a function only of the particle number density\footnote{In the general case it will also depend on the entropy density, see Refs.~\cite{Brown:1992kc,Misner:1974qy}.} given by Eq.~(\ref{Eq:NumberDensity})
\be
\rho=\rho\left(\frac{|J|}{\sqrt{-g}}\right)~.
\ee

The machinery is now all set and we can work out the stress-energy tensor by varying the action $S_2$ with respect to the inverse metric. The variation of the energy density is computed as
\be\label{Eq:VariationRho}
\delta\rho=\frac{\pa\rho}{\pa n}\delta n~.
\ee
In order to compute the variation $\delta n$ we recall
\be\label{Eq:VariationDetMetric}
\delta\sqrt{-g}=-\frac{\sqrt{-g}}{2}g_{\mu\nu}\delta g^{\mu\nu}~.
\ee
Thus, using Eq.~(\ref{Eq:NumberDensity}) and Eqs.~(\ref{Eq:DefinitionFluidFlow}),~(\ref{Eq:VariationDetMetric}) we find
\be
\delta n=\frac{n}{2}\left(g_{\mu\nu}+U_{\mu}U_{\nu}\right)\delta g^{\mu\nu}~.
\ee
Hence, plugging this result into Eq.~(\ref{Eq:VariationRho}) we have
\be
\delta\rho=\frac{n}{2}\frac{\pa\rho}{\pa n}\left(g_{\mu\nu}+U_{\mu}U_{\nu}\right)\delta g^{\mu\nu}~.
\ee
We can then compute the variation of the action $S_2$ given in Eq.~(\ref{Eq:FluidAction2})
\be
\delta S_2=\frac{1}{2}\int\de^4x\sqrt{-g}\;\Big[\left(-\frac{\pa\rho}{\pa n}n\right)U_{\mu}U_{\nu}+\left(\rho-\frac{\pa\rho}{\pa n}n\right)g_{\mu\nu}\Big]\delta g^{\mu\nu}~.
\ee
The stress-energy tensor, defined in Eq.~(\ref{Eq:DefStressEnergy}), has in this case the following form
\be
T_{\mu\nu}=\left(\frac{\pa\rho}{\pa n}n\right)U_{\mu}U_{\nu}+\left(\frac{\pa\rho}{\pa n}n-\rho\right)g_{\mu\nu}~.
\ee
Using Eq.~(\ref{Eq:PressureMTW}), this expression can be rewritten in the familiar form 
\be\label{Eq:StressEnergyPefectFluid}
T_{\mu\nu}=(\rho+p)U_{\mu}U_{\nu}+p\,g_{\mu\nu}~.
\ee
Eq.~(\ref{Eq:StressEnergyPefectFluid}) gives the stress-energy tensor for a perfect fluid.

For a perfect fluid, Eq.~(\ref{Eq:ConservationStressEnergy}) implies (using the normalisation $U^\mu U_\mu=-1$)
\begin{align}
&U^{\mu}\nabla_{\mu}U^{\nu}=0~,\label{Eq:GeodesicsEquation}\\
&U^{\mu}\nabla_{\mu}\rho+(\rho+p)\nabla_{\mu}U^{\mu}=0~\label{Eq:ContinuityEquationGeneral}.
\end{align}
Eq.~(\ref{Eq:GeodesicsEquation}) implies that fluid particles follow geodesics. Eq.~(\ref{Eq:ContinuityEquationGeneral}) is a continuity equation, expressing local energy conservation. In the case of a FLRW universe, fluid particles are freely falling. Hence, their four-velocity is given by
\be
U^{\mu}\frac{\pa}{\pa x^{\mu}}=\frac{1}{N}\frac{\pa}{\pa t}~.
\ee
Thus, the four-velocity is the unit normal to the spacelike hypersurfaces of the foliation and can be identified (up to a gauge-dependent factor) with the generator of `time flow' (see Ref.~\cite{Kiefer:2007ria}). 
In comoving gauge, Eq.~(\ref{Eq:ContinuityEquationGeneral}) reads as\footnote{We used $N=1$, $\nabla_{\mu}U^{\mu}=\Gamma^i_{0i}=3\frac{\dot{a}}{a}$.}
\be\label{Eq:ContinuityFLRW}
\dot{\rho}+3H(\rho+p)=0~.
\ee
Once the equation of state of the fluid is given, the equation (\ref{Eq:ContinuityFLRW}) gives the evolution of the fluid in a FLRW universe. It must be noted that the dynamics of the fluid is not at all independent from that of the gravitational field. In fact, Eq.~(\ref{Eq:ContinuityFLRW}) can also be derived using the Friedmann equations\footnote{ Such derivation can be found in all standard textbooks, see \emph{e.g.} Refs.~\cite{Wald:1984rg},~\cite{Weinberg:2008zzc}.} (\ref{Eq:Friedmann}),~(\ref{Eq:AccelerationSecondFriedmann}). This can be seen as a particular case of the more general result represented by Eq.~(\ref{Eq:StressEnergyConsFromBianchi}). In fact, it is a direct consequence of the contracted Bianchi identities.

\section{The Standard Cosmological Model $\Lambda$CDM}\label{Sec:StandardCosmology:LCDM}
The Standard Cosmological Model rests on the assumption that the expanding cosmological background is well described by a FLRW model and that the dynamics of the gravitational field and matter is given by classical GR. It has been immensely successful in fitting the Planck data and, so far, provides the best physical description of our Universe. However, this success comes at a price. In fact, as it is apparent from its name, the $\Lambda$CDM model introduces two dark components in the energy budget of the Universe: a positive cosmological constant $\Lambda$ and cold dark matter.

It must be stressed that, although there are several (some may add, compelling) reasons to include a dark sector in our picture of the Cosmos, so far there is no evidence (direct or indirect) for the existence of dark matter 
from particle physics experiments. 
Rather, dark components were introduced \emph{ad hoc} in order to fit the cosmological and astrophysical data and solve specific drawbacks of GR\footnote{In particular, dark matter explains the flatness of galaxy rotation curves and the stability of spiral galaxies. Dark matter is also needed in the standard cosmological model to account for a substantial part of the energy budget of the Universe, as well as to explain structure formation. See Ref.~\cite{Bertone:2004pz} for a review of particle dark matter and a discussion of different candidates.}.  Although this is no argument to discard \emph{a priori} the dark matter hypothesis, it is worth exploring alternative scenarios which may provide a more natural explanation of observational data. In the rest of this section we will discuss the standard $\Lambda$CDM model. Some alternatives will be studied in the next chapters.

Let us start by recalling the Friedmann equation (\ref{Eq:Friedmann})
\be\label{Eq:FriedmannRiscritta}
H^2=\frac{8\pi G}{3}\rho-\frac{K}{a^2}+\frac{\Lambda}{3}~.
\ee
For uniformity of notation, we can introduce an energy density corresponding to the observed positive cosmological constant (dark energy)
\be\label{Eq:VacuumEnergyDensity}
\Lambda=8\pi G\, \rho_{\Lambda}~.
\ee
We assume that the total energy density is the sum of the contributions of radiation $\rho_{\rm R}$, baryonic matter $\rho_{\rm B}$, dark matter $\rho_{\rm DM}$, and dark energy $\rho_{\Lambda}$. Hence, Eq.~(\ref{Eq:FriedmannRiscritta}) can be generalised to include all of these contributions
\be\label{Eq:FriedmannLambdaCDM}
H^2=\frac{8\pi G}{3}\left(\rho_{\rm R}+\rho_{\rm B}+\rho_{\rm DM}+\rho_{\Lambda}\right)-\frac{K}{a^2}~.
\ee
We introduce the density parameters
\be\label{Eq:DefDensityParameter}
\Omega_{\rm R}=\frac{8\pi G}{3}\frac{\rho_{\rm R}}{H^2}~,\hspace{1em}\Omega_{\rm B}=\frac{8\pi G}{3}\frac{\rho_{\rm B}}{H^2}~,\hspace{1em}\Omega_{\rm DM}=\frac{8\pi G}{3}\frac{\rho_{\rm DM}}{H^2}~, \hspace{1em}\Omega_{\Lambda}=\frac{8\pi G}{3}\frac{\rho_{\Lambda}}{H^2}~.
\ee
Using Eqs.~(\ref{Eq:FriedmannLambdaCDM}),~(\ref{Eq:DefDensityParameter}), we can rewrite the Friedmann equation in the form of an `energy balance' equation
\be\label{Eq:FriedmannBalance}
\Omega_{\rm R}+\Omega_{\rm B}+\Omega_{\rm DM}+\Omega_{\Lambda}+\Omega_{\rm K}=1~,
\ee
where we defined
\be
\Omega_{\rm K}=-\frac{K}{a^2 H^2}~.
\ee

Accurate measurements of the cosmological parameters of the $\Lambda$CDM model were performed by the Planck collaboration \cite{Ade:2013zuv,Ade:2015xua}. The value of the Hubble parameter today is
\be
H_0=100\times h~ {\rm Km}~ {\rm s}^{-1}{\rm Mpc}^{-1}=
67.74\pm0.46~ {\rm Km}~ {\rm s}^{-1}{\rm Mpc}^{-1}~,
\ee
at 68\% confidence level. Here we introduced the dimensionless Hubble parameter $h$, as it is customary. A lower `0' index refers to quantities measured today. The other cosmological parameters in Eq.~(\ref{Eq:FriedmannBalance}) are found to be
\be
\Omega_{\rm B,0}h^2=0.02230\pm0.00014~,\hspace{1em} \Omega_{\rm DM,0}h^2=0.1188\pm0.0010~,\hspace{1em}\Omega_{\Lambda,0}=0.6911\pm0.0062~.
\ee
Spatial curvature is found to be tightly constrained
\be
|\Omega_{\rm K,0}|<0.005~.
\ee
The fact that $\Omega_{\rm K,0}$ is extremely small is used to indicate the flatness of the Universe\footnote{Notice that this definition of flatness, used in Cosmology, does not immediately correspond to flatness of the spatial geometry, which only occurs when $K$ is vanishing \emph{exactly}. It must be understood as the requirement that spatial curvature gives a negligible contribution to the total energy budget of the Universe. However, if one uses the Friedmann equation and the definition of distances conventionally used in Cosmology (\emph{e.g.} angular and luminosity distances), it is immediate to realise that in the limit $\Omega_{\rm K}\to0$ the Euclidean expressions are recovered (see \emph{e.g.} Ref.~\cite{Weinberg:2008zzc}).}.
Therefore, the analysis of the CMB data by Planck gives us a picture of a flat Universe, dominated by the cosmological constant and with matter being predominantly non-luminous ($\Omega_{\rm DM,0}\approx0.259$, $\Omega_{\rm B,0}\approx0.049$).

Notice that $\rho_{\Lambda}$ introduced in Eq.~(\ref{Eq:VacuumEnergyDensity}) can be interpreted as vacuum energy density. Hence, it is natural to compare its value with the natural value given by the Planck density $\rho_{\rm\scriptscriptstyle Pl}=\frac{M_{\rm\scriptscriptstyle Pl}}{\ell_{\rm\scriptscriptstyle Pl}^3}$. The ratio of the two numbers gives
\be
\frac{\rho_{\Lambda}}{\rho_{\rm\scriptscriptstyle Pl}}=\frac{3}{8\pi}\Omega_{\Lambda}(H t_{\rm\scriptscriptstyle Pl})^2\simeq 1.15\times10^{-123}~.
\ee
The extreme smallness of the cosmological constant with respect to the Planck scale is known as the \emph{cosmological constant problem} \cite{Weinberg:1988cp}.

\section{Classic Cosmological Puzzles}\label{Sec:CosmologicalPuzzles}
Despite its success in predicting the expansion of the Universe and the fact that they represent the underpinning of all the successful predictions of theoretical cosmology, such as \emph{e.g.} primordial nucleosynthesis, standard cosmology based on FLRW models has some serious drawbacks. These are related to the cosmological dynamics at early times. In particular, there are three major problems, commonly known as `cosmological puzzles'. In this section we will discuss the origin of these problems in the framework of the so-called `hot big bang cosmology'. Our discussion will be based on Ref.~\cite{Weinberg:2008zzc}. The underlying assumption in the hot big bang cosmology scenario is that only standard matter species are present and the dynamics of the cosmic expansion is given by the Friedmann equation. The discussion of possible solutions of the cosmological puzzles will be postponed until later sections.

\subsection{Flatness Problem}
As we saw at the beginning of this section, the Planck data favours very small values of the spatial curvature parameter $\Omega_{\rm K}$. In particular, such measurements are compatible with $\Omega_{\rm K}=0$. This is simply achieved if we assume $K=0$. However, this would mean that our Universe is very special. It would be much more satisfactory if we were able to provide a physical mechanism that could explain such a small value. In fact, $\Omega_{\rm K}=0$ is not only a very special case, but it is also unnatural if we assume the hot big bang cosmology. The reason for this is readily seen by means of a simple argument.

As we go back in time, the contribution of non-relativistic matter to the energy density becomes subdominant with respect to that of radiation. Hence, in the radiation dominated era $a\propto t^{1/2}$ and $H=\frac{1}{2t}$, implying $|\Omega_{\rm K}|=\frac{|K|}{\dot{a}^2}\propto t$. During matter domination, this value keeps increasing as $t^{2/3}$.
It follows that, in order to have a very small value today, there must have been a remarkable amount of fine tuning in the initial conditions. The precise amount of fine-tuning needed can be estimated by making reference to the specifics of the thermal history of the Universe. The reader interested in the details of the calculation is referred to Ref.~\cite{Weinberg:2008zzc}. As it turns out, having $|\Omega_{\rm K}|<1$ at the present time implies that its value at the time of electron-positron annihilation 
had to be smaller than $10^{-16}$ , and even smaller at earlier times \cite{Weinberg:2008zzc,Dicke:1979}.

\subsection{Horizon Problem}
Particle horizons are a peculiar feature of FLRW spacetimes. A particle horizon is defined as the largest distance at which an observer $O$ at time $t$ will be able to receive light signals from other observers \cite{Hawking:1973uf}. This is given by the following expression
\be\label{Eq:HorizonDefinition}
d_{\rm max}(t)=a(t)\int_0^t\de t^{\prime}\;\frac{1}{a(t^{\prime})}~.
\ee
The observer sees a particle horizon at $t$ if and only if the integral on the r.h.s. of Eq.~(\ref{Eq:HorizonDefinition}) converges. Observers at a distance larger than $d_{\rm max}(t)$ have not been in causal contact with  $O$. Hence, we can think of $d_{\rm max}(t)$ as the spatial extent of a causally connected region at time $t$.

The occurrence of particle horizons is indeed typical in the hot big bang cosmology. In fact, if radiation dominated at early times, the size of a causally connected region during radiation domination would be equal to the Hubble radius $d_{\rm max}(t)=1/H$. Therefore, we would expect to be able to detect the imprints of this in the CMB, where it should manifest itself in the form of patches in the last scattering surface. Each patch would be characterised by distinct properties and uncorrelated to others, giving rise to substantial anisotropies in the CMB. Remarkably, this turns out not to be the case. Patches with the size of the Hubble radius at the time of last scattering now subtend an angle of the order of one degree \cite{Weinberg:2008zzc}. Thus, the isotropy of the CMB at large angular scales is in stark contrast with the implications of the hot big bang scenario. Here lies the essence of the horizon problem: how can the Universe be so homogeneous? If we only assume causal physics, the hot big bang cosmology provides no explanation for the isotropy of the CMB radiation, which would thus seem to be the result of a very unreasonable fine tuning over a large number of causally disconnected patches.

\subsection{Cosmic Relics}\label{Sec:CosmicRelics}
Cosmic relics are exotic particles or structures (topological defects) whose existence is predicted by a large class of theoretical models in high energy physics. An example is offered by monopoles in Grand Unified Theories (GUT).
Monopoles are field configurations of the Higgs and the gauge field characterised by a non-trivial topological structure; they are produced as a result of the spontaneous breakdown of the gauge symmetry of GUT to the Standard Model gauge group $\mbox{SU(3)}\times\mbox{SU(2)}\times\mbox{U(1)}$ \cite{Weinberg:2008zzc}. They can be regarded as `particles' with a magnetic charge whose nature is topological. Their mass is typically very large, of the order of the GUT scale ($M\approx10^{16}~{\rm GeV}$). The main problem with monopoles in Cosmology is that they are overproduced during symmetry breaking. Given their extremely large mass, they would represent the dominant contribution to the energy density. Hence, if there is no other mechanism taking place that may dilute them, they would overclose the Universe. 
 
 \section{A Solution to the Cosmological Puzzles}\label{Sec:InflationSolvesPuzzles}
 The mechanism of inflation was introduced in order to provide a solution to the cosmological puzzles, described in the previous sections. This is in fact possible if, before entering the radiation dominated era, the Universe went through an era of accelerated expansion\footnote{This was first shown by Guth \cite{Guth:1981aa}, who introduced the first example of inflationary mechanism, now known as `old inflation'.}. By looking at the second Friedmann equation (\ref{Eq:AccelerationSecondFriedmann}), and knowing that the cosmological constant $\Lambda$ observed today was negligible in the Early Universe compared to other forms of energy, this is possible only if $\rho+3p<0$. In order to achieve this, a violation of the strong energy condition is required. In fact, as it was first realised by Guth this is possible if one considers a fluid characterised by the equation of state $p=-\rho$, \emph{i.e.} some form of energy that behaves like a cosmological constant. The stress-energy tensor would then be given by\footnote{Notice that, with this definition, the units of measurement of $\lambda$ are not the same of $\Lambda$, defined before. In fact, in our units $[\lambda]=[\Lambda][G]^{-1}$.}
\be\label{Eq:StressEnergyInflation}
T_{\mu\nu}=\lambda g_{\mu\nu}~.
\ee
 However, this cannot be a cosmological constant in a strict sense. In fact, if that was the case, it would be dominant at all times, since no mechanism can dilute a cosmological constant, by definition. If it is dominant at early times, it stays dominant throughout the whole history of the Universe. Therefore, although this would solve the three cosmological puzzles, it would not allow for structure formation and is for this reason incompatible with our own existence. The conclusion we draw from this argument is that, if we want to pursue this path, we must look for some form of energy which behaves as a cosmological constant \emph{only approximately}. In the following, we will briefly discuss how such an assumption is a solution to the classic cosmological puzzles discussed in the previous section.
 
 \subsection{Flatness Problem}
 During the inflationary era the energy density of the Universe is dominated by a cosmological constant-like term, with stress-energy tensor  given by Eq.~(\ref{Eq:StressEnergyInflation}). It follows that the Universe undergoes exponential expansion. Therefore, we have $a(t)\sim \mbox{e}^{Ht}$, with the Hubble rate $H$ being slowly varying. In this case, the curvature parameter is given by
 \be
 |\Omega_{\rm K}|=\frac{|K|}{\dot{a}^2}\propto \mbox{e}^{-2Ht}~.
\ee 
Hence, a flat Universe characterised by $\Omega_{\rm K}=0$, which is unstable under small changes in the initial conditions in the hot big bang cosmology, is turned into an attractor by inflation.

 \subsection{Horizon Problem}
 The horizon problem is also solved in the inflationary scenario. We define $t_i$ as the time at the beginning of inflation, while $t_e$ marks its end. Computing the horizon distance using Eq.~(\ref{Eq:HorizonDefinition}) we have
 \be\label{Eq:HorizonInflation}
 d_{\rm max}=a(t_e)\int_{t_i}^{t_e}\de t^{\prime}\;\frac{1}{a(t^{\prime})}=H^{-1}(\mbox{e}^{\mathcal{N}}-1)~,
\ee
having defined the number of e-folds of expansion during inflation as
\be
\mathcal{N}=\log\left(\frac{a(t_e)}{a(t_i)}\right)=H(t_e-t_i)~.
\ee
We can assume the value of $d_{\rm max}$ computed in Eq.~(\ref{Eq:HorizonInflation}) as giving the main contribution to the size of the particle horizon today, since the radiation and the matter dominated eras do not give significant contributions to the integral in Eq.~(\ref{Eq:HorizonDefinition}). 
 
 \subsection{Cosmic Relics}
 The problem of the abundance of cosmic relics discussed in Section~\ref{Sec:CosmicRelics} can also be resolved similarly by introducing a source of energy-momentum that behaves approximately like a cosmological constant. In fact, the original motivation for inflation was the overproduction of magnetic monopoles in $\mbox{SU(5)}$ GUT. Although the $\mbox{SU(5)}$ model was abandoned due to its prediction of proton instability, the problem persists with other GUT theories based on simple Lie groups. Moduli fields represent yet a different kind of cosmic relics predicted by many superstring-inspired models of particle physics \cite{deCarlos:1993wie,Peloso:2002rx}. They would typically have masses of the order of the scale of supersymmetry (SUSY) breaking. Hence, similar considerations would apply as in the case of monopoles discussed in Section~\ref{Sec:CosmicRelics}.
 
 It is important to remark that the most serious among the classic cosmological puzzles is represented by the horizon problem. In fact, in inflationary models any solution to the horizon problem also solves the flatness problem and the cosmic relics problem, although the converse is not true in general \cite{Weinberg:2008zzc}. This must be compared with the case of bouncing cosmologies, which generally solve the horizon problem, whereas the solution of other cosmological problems (\emph{e.g.} the flatness problem) depends on the particular model considered \cite{Battefeld:2014uga,Brandenberger:2016vhg}.

 \section{The Inflationary Mechanism}\label{Sec:SlowRollInflation}
 The simplest possible realisation of the idea underlying inflation is obtained by considering a scalar field with self-interactions. In fact, as we saw in Eqs.~(\ref{Eq:ScalarFieldEnergyDensity}),~(\ref{Eq:ScalarFieldPressure}), we have in that case
 \begin{align}
 \rho_\phi=\left(\frac{1}{2}\dot{\phi}^2+V(\phi)\right)~,\\
  p_\phi=\left(\frac{1}{2}\dot{\phi}^2-V(\phi)\right)~.
 \end{align}
 We immediately realise that, if in some regime the kinetic energy term is negligible, we have
 \be\label{Eq:SlowRoll}
 \rho_\phi\approx-p_\phi\approx V(\phi)~.
 \ee
 Therefore, the problem is to study under which conditions Eq.~(\ref{Eq:SlowRoll}) can be consistently imposed when studying the dynamics of a FLRW universe. This will lead us to the study of \emph{slow-roll inflation}. In particular, we observe that, when Eq.~(\ref{Eq:SlowRoll}) holds and $\phi$ can be considered as nearly constant, the Friedmann equation implies
 \be
 a(t)\sim \mbox{e}^{H t}~,\hspace{1em} \mbox{with}\hspace{1em} H=\sqrt{\frac{8\pi G}{3}V(\phi)}~.
 \ee
 $V(\phi)$ is taken to be positive in order to have an exponential expansion.
 
During the inflationary era, \emph{the dynamics} of spatial geometry is close to that of de Sitter spacetime, which is a maximally symmetric spacetime and an exact solution of the Einstein equations in vacuo with a positive cosmological constant $\Lambda>0$. In fact, in comoving gauge the scale factor of de Sitter spacetime is an exponential of proper time. 
Therefore, it is common practice to dub the inflationary era as \emph{quasi-de Sitter}.

The first slow-roll condition is
\be\label{Eq:FirstSlowRollCondition}
\frac{1}{2}\dot{\phi}^2\ll V(\phi)~,
\ee
which justifies neglecting the kinetic term of the scalar in the first Friedmann equation. We must also require that this condition is preserved under time evolution. Thus, we obtain
\be\label{Eq:SecondSlowRollCondition}
\ddot{\phi}\ll V_{,\phi}~.
\ee
This motivates the additional requirement that the inertial term be negligible compared to the friction term in the equation of motion for the scalar field (\ref{Eq:EoMotionScalar}). Thus, we impose the second slow-roll condition
\be
|\ddot{\phi}|\ll H |\dot{\phi}|~.
\ee
At this stage, it is convenient to introduce the so-called \emph{slow-roll parameters} \cite{Baumann:2009ds}
\begin{align}
\varepsilon&=\frac{1}{16\pi G}\left(\frac{V^{\prime}}{V}\right)^2~,\\
\eta&=\frac{1}{8\pi G}\frac{V^{\prime\prime}}{V}~,
\end{align}
where a prime denotes differentiation w.r.t. $\phi$.
Using the Friedmann equations (\ref{Eq:Friedmann}),~(\ref{Eq:AccelerationSecondFriedmann}) and the equation of motion of the scalar field (\ref{Eq:EoMotionScalar}), the two slow-roll conditions, Eqs.~(\ref{Eq:FirstSlowRollCondition}),~(\ref{Eq:SecondSlowRollCondition}) can equivalently be expressed as a condition on the smallness of the slow-roll parameters
\be
\varepsilon,|\eta|\ll1~.
\ee
This is also not enough, since inflation has to end after the Universe has expanded for a sufficiently large number of e-folds, marking the transition (\emph{reheating}) to a radiation dominated era. Therefore, slow-roll inflation requires by its very construction a particular choice of the profile of the potential function and a suitable tuning of the parameters which enter in its definition.

\section{Structure Formation}
The major strength of the inflationary mechanism lies in the fact that it provides a simple and elegant explanation for the observed anisotropies in the CMB\footnote{We will not attempt at a systematic discussion of the theory of cosmological perturbations and its applications. The reader is referred to Refs.~\cite{Mukhanov:1990me,Brandenberger:2003vk,Mukhanov:2005sc,Baumann:2009ds} for comprehensive reviews.}. Moreover, it provides the seeds for structure formation, which does not have an explanation in the old hot big bang cosmology scenario. The main tool is the theory of cosmological perturbations, which serves to link models of the very early Universe to the observational data obtained in Precision Cosmology \cite{Brandenberger:2003vk}; it is so powerful and general that it can be rightly regarded as a cornerstone of modern Cosmology. In order to discuss this topic, we have to go beyond the case of a perfectly homogeneous and isotropic Universe discussed in previous sections of this chapter. A FLRW spacetime is considered as a \emph{background} on which cosmological perturbations, \emph{i.e.} metric and matter fields perturbations, dynamically evolve. During the inflationary era, the cosmological background is characterized by a quasi-DeSitter geometry.

Perturbations of the inflaton field start as quantum fluctuations in the Bunch-Davies vacuum. As the Universe expands, modes characterized by a given wave vector ${\bf k}$ are redshifted until they exit the (comoving) Hubble radius, which happens when $|{\bf k}|\geq (aH)^{-1}$. When a given mode exits the Hubble radius the corresponding quantum state becomes highly squeezed, so that it is highly sensitive to decoherence induced by interactions with the environment \cite{Kiefer:1998qe,Kiefer:2006je}. This allows
inflaton perturbations to be treated as classical stochastic fluctuations after horizon exit.

Curvature perturbations are frozen out for modes outside the horizon, implying that they can be computed at horizon exit. When modes eventually re-enter the horizon, after the end of inflation, the corresponding curvature perturbations are the ones characterising primordial inhomogeneities in the observable Universe. This gives rise to the anisotropies observed in the CMB, and also provides the seeds for structure formation. It is worth stressing that, due to the conservation law that holds for curvature perturbation modes outside the horizon, the mechanism is insensitive to the details of the re-heating stage at the end of inflation.

It must be pointed out that the success of the inflationary mechanism heavily relies on the initial conditions assumed at the onset of inflation. In fact, homogeneity of the inflaton field over several Hubble lengths must be assumed in order for inflation to take place \cite{Calzetta:1992gv}. Such initial conditions are not generic and can hardly be seen as natural, and it has been argued that they must find an explanation in events occurring during the QG era before the onset of inflation \cite{Calzetta:1992bp}. Then, the problem is shifted to that of finding suitable boundary conditions for the universe wave function in QC, which must at the same time realize the necessary homogeneity for the onset of inflation, and explain Bunch-Davies initial conditions for cosmological perturbations\footnote{The latter point has been studied recently in Ref.~\cite{Feldbrugge:2017fcc} in the context of Lorentzian QC, where it is claimed that the celebrated Hartle-Hawking no-boundary proposal does not lead to Bunch-Davies initial conditions for the perturbations.}.

\section{The Initial Singularity}
The occurrence of singularities is generic in GR, as shown by the singularity theorems of Hawking and Penrose \cite{Hawking:1973uf}. They hold provided that Einstein's equations are satisfied and matter fields satisfy suitable energy conditions\footnote{A certain number of additional technical assumptions is also assumed, see Ref.~\cite{Hawking:1973uf}.}. In particular, in the cosmological case the singularity takes place at zero scale factor. Singularities are problematic in GR since they imply a breakdown of determinism, unless they are protected by an event horizon  as in the case of black holes\footnote{Black holes are nevertheless problematic for different reasons since they lead to the information loss problem, encountered in the context of QFT on curved spacetimes.}. This is clearly not the case in a cosmological setting. In particular, it follows that we cannot set the initial conditions for our Universe at the origin of the cosmic expansion, but we have to arbitrarily single out a past space-like hypersurface (provided it is a Cauchy hypersurface, see Ref.~\cite{Hawking:1973uf}), where to assign the initial data for the gravitational field and other physical fields.

At this stage, we would like to study in more detail what are the relevant properties of matter that ultimately lead to the occurrence of singularities in GR. From the definition of the Hubble rate $H=\frac{\dot{a}}{a}$, taking its first derivative with respect to proper time, we have
\be\label{Eq:DerivativeHubbleRate}
\dot{H}=\frac{\ddot{a}}{a}-\left(\frac{\dot{a}}{a}\right)^2=-4\pi G (\rho +p)~.
\ee
In the last step we used the two Friedmann equations (\ref{Eq:Friedmann}),~(\ref{Eq:AccelerationSecondFriedmann}). For matter satisfying the null energy condition (NEC, see Appendix \ref{Appendix:ECs}) we have
\be
\rho+p>0~,
\ee
implying that the Hubble rate is always decreasing $\dot{H}<0$. Moreover, the continuity equation (\ref{Eq:ContinuityFLRW}) for matter subject to NEC implies that $\dot{\rho}<0$. As we go back in time these quantities are both diverging, leading to the initial singularity and the breakdown of GR. Hence, we have identified NEC as the culprit for such drawback of the classical theory of GR. In the literature there are many examples of NEC violating theories which solve the initial singularity problem, with applications to Cosmology, wormholes and the creation of a `universe in a lab' \cite{Rubakov:2014jja}. In a cosmological setting, in these classical models the would-be singular region is replaced by a regular region of spacetime and there is a minimal volume of the Universe, \emph{i.e.} a bounce. This is achieved by introducing some exotic type of matter which violates NEC.

Bouncing cosmologies are also interesting since they offer an alternative to the inflationary paradigm. The way in which the classic cosmological puzzles are solved in such scenarios is reviewed in Ref.~\cite{Brandenberger:2016vhg,Battefeld:2014uga}. In particular, we observe that the horizon problem is automatically solved in all bouncing models.
 However, there are also new problems that may arise. In fact, as a generic feature of all bouncing models, inhomogeneities and anisotropies tend to increase during the contracting phase, as a result of the Belinsky--Lifshits--Khalatnikov phenomenon \cite{Belinsky:1970ew} (also see Refs.~\cite{Rubakov:2014jja,Brandenberger:2016vhg} and references therein). There may also be further model-specific problems such as \emph{e.g.} curvature and vector modes instabilities. For a review of all these different aspects of bouncing cosmologies the reader is referred to Ref.~\cite{Brandenberger:2016vhg}.
 
Before closing this section we find it worth remarking that inflation does not lead to singularity resolution. In fact, suitable initial conditions must be imposed at the onset of inflation in order to recover compatibility with observations. Thus, inflation fails to give a complete picture of the history of the universe. We will not review the open issues in inflation here, since it would be beyond the purpose of this thesis. We refer the interested reader to the published literature, and in particular to Refs.~\cite{Brandenberger2000,Martin:2001aa}. It will be enough for us to observe that the inflationary paradigm relies on the existence of a hypothetical new field, the inflaton, with a potential that does not seem to follow from any fundamental physical principles, but rather its profile is chosen so as to be compatible with observations.
It would be surprising if there was not a more fundamental theory to describe physics at
high energy scales and classical cosmology was valid up to the Planck scale.

In later chapters we will study singularity resolution in frameworks that are different from classical GR. To be specific, we will see that a bounce can also be achieved by modifying the laws of gravity, \emph{e.g.} by means of quantum gravity corrections as in case of Group Field Theory (GFT) models considered in Chapter~\ref{Chapter:GFT}.

\part{Cosmology from Quantum Gravity}


\chapter{Quantum Cosmology and Minisuperspace Models}\label{Chapter:WDW} 
In this chapter we present an extension of Wheeler-DeWitt (WDW) minisuperpace cosmology with additional interaction terms that preserve the linear structure of the theory, which was considered by the author in Ref.~\cite{deCesare:2015vca}. 
Particular attention will be paid to the link between Quantum Cosmology \emph{\`a la} WDW and other approaches, such as Loop Quantum Cosmology (LQC) and Group Field Theory (GFT). In fact, a motivation for our model comes from a toy model inspired by GFT, which was proposed in Ref.~\cite{Calcagni:2012vb}. In that model, the Hamiltonian constraint in LQC defines the kinetic term of a third quantised theory in minisuperspace.
In the large volume limit a generalisation of WDW is obtained, characterised by the presence of extra terms representing interactions in minisuperspace. 

Section~\ref{sec:intro} is a general introduction to the geometrodynamics (WDW) approach to Quantum Gravity and Quantum Cosmology, where we discuss the ideas underlying it, its strong points and weaknesses. This will also give us the opportunity to further contextualise our model within the framework of canonical approaches. In Section~\ref{Sec:Geometrodynamics} we review the formulation of the geometrodynamics approach to Quantum Gravity. In Section~\ref{Sec:QCWDW} we focus on its symmetry reduced version, \emph{i.e.} the WDW approach to Quantum Cosmology. We discuss its standard formulation, the connection with path-integrals, and review some recent results obtained in this approach. In Section~\ref{Sec:InterpretWDW} we will review some of the issues related to the physical interpretation  of the theory and some alternatives that have been proposed in the literature.
 In particular, we will discuss the ideas underlying the possibility of a `third quantisation' that may solve those issues. This is also relevant in light of the third quantised perspective which is also adopted in the formulation of GFT, discussed in Chapter~\ref{Chapter:GFT} of this thesis. In Section~\ref{section 2} we review the GFT inspired model of Ref.~\cite{Calcagni:2012vb}  and show how our model can be recovered from it in the large volume limit. This will allow us to discuss the relation between geometrodynamics and other canonical approaches to Quantum Cosmology, such as LQC.

General perturbative methods are then developed and applied to our model. In particular, we study the effect of the new interaction terms on
known semiclassical solutions for a closed ($K=1$) FLRW universe filled with a
massless scalar, which plays the role of an internal clock. The unperturbed case is studied in Section~\ref{Section unperturbed}, where we review the construction of wave-packets solutions done in Ref.~\cite{Kiefer:1988}.
 In Section~\ref{Section:TimeIndependent} we obtain a Helmoltz-like equation in the
case of interactions that do not depend on the internal time. The
corresponding Green kernel is obtained exactly and
turns out to depend on a real parameter, which is linked to the choice of
boundary conditions at the singularity.  In Section~\ref{Section:TimeDependent} the exact Feynman propagator of the WDW operator is
computed non-perturbatively by means of a conformal transformation in minisuperspace. The problem is thus reduced to
that of finding the Klein-Gordon propagator in a planar region with a
boundary. The solution is found by applying techniques that are similar to those used in analogous problems in electrostatics.
In Section~\ref{section noise}, as an application of the `time-dependent' perturbation theory developed in the previous section, we consider the case of an additional interaction given by white noise. It is argued that such term can be used to give an effective description of the interaction of the cosmological background with the microscopic d.o.f. of the quantised gravitational field.  The corrections to the wave-packet solutions to first perturbative order are then computed numerically. We discuss the implications of our results for semiclassicality of the perturbed states and for the arrow of time.
Finally, we review our results and their physical consequences in Section~\ref{Section:WDW:Conclusions}.

\section{Introduction}\label{sec:intro}
The framework adopted will be that of quantum geometrodynamics, also known in the literature as Wheeler-DeWitt (WDW) theory.  It was
originally derived by applying Dirac's quantisation scheme (Appendix \ref{Appendix:Dirac}) to
Einstein's theory of gravity, formulated in Arnowitt-Deser-Misner
(ADM) variables (Appendix \ref{Appendix:ADM}). In this approach, the dynamics of the gravitational field is entirely determined by \emph{constraint equations}, known as the diffeomorphism constraint and the Hamiltonian constraint. The latter is frequently referred to as WDW equation. Such constraints express the invariance of the dynamics under time reparametrisations and spatial diffeomorphisms, respectively. They close an algebra, known as the hypersurface deformation algebra\footnote{Also known as the Bergmann-Komar algebra, see Appendix \ref{Appendix:ADM} for a discussion and references.}, which describes deformations of three-dimensional spatial hypersurfaces embedded in a four-dimensional spacetime. The specific form of the constraints depends on the particular gravitational theory at hand. However, it reduces to the algebra of spacetime diffeomorphisms on-shell, \emph{i.e.} when the constraints are satisfied. The fact that the dynamics of the gravitational field in the canonical formulation of GR turns out to be totally constrained\footnote{This is actually true in any relativistic theory of gravity, not necessarily GR. However, the particular form of the constraints and their algebra structure depend on the theory, see Ref.~\cite{Thiemann:2007zz}.} has important consequences for the interpretation of the quantum theory, as we will discuss below.

Geometrodynamics was the first canonical approach\footnote{There are several other approaches to Quantum Gravity. It would be well beyond the purpose of this thesis to review all of them and explore their mutual relations. The interested reader is referred to Ref.~\cite{Kiefer:2007ria} for an accessible introduction to different approaches.}  to be historically developed. There are essentially two reasons for our choice of this approach. In first place, besides representing a path to Quantum Gravity in its own respect, it is also expected that it can be recovered approximately as the low energy limit of some other (canonical) approaches, such as \emph{e.g.} Loop Quantum Gravity (LQG) \cite{Kiefer:2011cc,Kiefer:2008bs}. This is essentially due to the fact that the Einstein equations can be recovered from it in the eikonal approximation, in analogy with the way classical mechanics is recovered from ordinary quantum mechanics \cite{gerlach1969derivation,Kiefer:2008bs}.
In second place, this approach allows for a rigorous study of the quantum dynamics of simple models, more specifically cosmological models and black holes \cite{Kiefer:2008bs} . The former class of models turns out to be also the most relevant one for the determination of potentially observable effects in Quantum Gravity \cite{Bini:2013fea,Brizuela:2015tzl,Brizuela:2016gnz,Kiefer:2012sy,Kiefer:2014qpa}.

It must be stressed that geometrodynamics has several well-known limitations as a
\emph{full theory} of Quantum Gravity\footnote{For a general account of the WDW theory and the
  problems it encounters see \emph{e.g.} Ref.~\cite{Rovelli:2015}.}, which prompted the development of alternative canonical approaches. 
  The difficulties in geometrodynamics are essentially of a mathematical character.
   Some well-known examples are given by the factor ordering
problem in the Hamiltonian constraint, the anomaly problem\footnote{See discussion in Section~\ref{Sec:Geometrodynamics}.}, and the construction of a
physical Hilbert space, \emph{i.e.} the implementation of the constraints\footnote{Also in LQG the implementation of the Hamiltonian constraint represents a major open problem. Different proposals exists to implement the dynamics in LQG, \emph{e.g.} the Spin Foam approach \cite{Perez:2012wv}, the master constraint programme \cite{Thiemann:2003zv}, or the GFT approach \cite{Oriti:2006se}.}.
  Nevertheless, it can be used as a powerful tool in simple systems, as discussed above.
Particularly relevant are its applications to Cosmology, in cases where the number of degrees of freedom is restricted \emph{a
  priori}, \emph{i.e.} in minisuperspace and midisuperspace\footnote{In midisuperspace models one applies a less drastic symmetry reduction than in the case of minisuperspace, \emph{i.e.} such that the space of Riemannian metrics on spacetime (modulo diffeomorphisms) is still a functional space rather than a finite dimensional one. For a review and references to the literature see Ref.~\cite{FernandoBarbero:2010qy}.}
   models.
  In Cosmology, geometrodynamics has also been
extremely useful to gain insight into some of the deep conceptual issues
raised by a quantum theory of the gravitational field, which are shared by all
nonperturbative approaches to quantum gravity. Among those we mention the
\emph{problem of time} (\emph{i.e.} the lack of a universal
parameter used to describe the evolution of the gravitational field) \cite{Isham:1992,Kuchar:1991qf,Bergmann:1961zz,Halliwell}, the closely related problem of the \emph{arrow of time} \cite{Hawking:1985,Hawking:1993tu,Kiefer:1994gp,Kiefer:2009xr,Moffat:1992bf},
and the occurrence of \emph{stable macroscopic branches} for the state of the universe (as in the
consistent histories approach~\cite{Hartle:1992as,Hartle:1997dc,Halliwell:2002th,Anastopoulos:2004gk,Craig:2010ai}). All of those problems are particularly relevant for Cosmology. Their resolution would have a great impact on our understanding of Quantum Gravity and of the history of our Universe.
 
The Hamiltonian constraint, commonly known as the WDW equation after its inventors, deserves particular attention.
It has the peculiar form of a timeless
Schr{\"o}dinger equation, in which the unknown is a function defined on the
space of three-dimensional geometries\footnote{Those are defined on each sheet $\Sigma_t$ of a foliation of spacetime, which is assumed to be globally hyperbolic $\mathcal{M}\cong \mathbb{R}\times\Sigma$ (where $\Sigma$ is a Cauchy surface). See Appendix \ref{Appendix:ADM}.}

\be
\hat{H}\psi[h_{ij}]=0~,
\label{wdw}
\ee
where $h_{ij}$ is the spatial metric\footnote{We must stress that the wave-function $\psi$ only depends on the class of three-geometries corresponding to $h_{ij}$, \emph{i.e.} $\psi[h^{\prime}_{ij}]=\psi[h_{ij}]$ if $h^{\prime}_{ij}$ and $h_{ij}$ are related by a diffeomorphism. This is a consequence of the momentum constraint $D_i \frac{\delta}{\delta h_{ij}} \psi[h_{ij}]=0$.} and $\hat{H}$ is the quantum Hamiltonian constraint (see Appendix \ref{Appendix:ADM} and Eq.~(\ref{Eq:WDW:QuantumHamiltonianConstraint})).
The solution of Eq.~(\ref{wdw}) is referred to as the wave-function of the universe.
The WDW equation is a constraint imposed on physical states, stemming from the
invariance of the classical theory under reparametrisations of the time
variable. In order to talk about physical evolution one has to introduce the notion of \emph{relational observables},
expressing how one physical variable evolves with respect to another \cite{Rovelli:2001bz,Tambornino:2011vg,Rovelli:1990ph}.

One of the fundamental properties of WDW is its linearity. This is indeed
a consequence of the Dirac quantisation procedure and a property of
any first-quantised theory. One can nonetheless find in the literature
several proposals for non-linear extensions of the geometrodynamics
equation. They are in general motivated by an interpretation of $\psi$
in Eq.~(\ref{wdw}) as a field operator (rather than a function)
defined on the space of geometries. This idea has been applied to
Cosmology and found concrete realisation in the old baby-universes
approach~\cite{Giddings:1988}. More recently, a model with a similar spirit and motivated by GFT (see Chapter~\ref{Chapter:GFT}) has been
proposed in Ref.~\cite{Calcagni:2012}, which assumes instead
LQC~\cite{Ashtekar:2003hd} as a starting point (\emph{i.e.} to define the free theory).

It is worth stressing that WDW theory can be recovered in the
continuum limit of LQC (see Ref.~\cite{Ashtekar:2006} for details). In this sense it can be interpreted as
an effective theory of Quantum Cosmology, valid at scales such that the
fundamentally discrete structure of spacetime which arises from the loop quantisation cannot be
probed. Therefore WDW, far from being of mere historical
relevance, is rather to be considered as an important tool to
extract predictions from cosmological models when a continuum, semiclassical behaviour is to be expected.

In Section~\ref{section 2} we consider as a starting point the Hamiltonian
constraint of LQC, leading to the free evolution of the universe, as
discussed in the GFT-inspired model of Ref.~\cite{Calcagni:2012}.
Non-linear terms are allowed, which can be interpreted as interactions between
disconnected homogeneous components of an isotropic universe
(scattering processes describe topology change). Instead of resorting
to a third-quantised formalism built on minisuperspace models, we
follow a rather conservative approach which does not postulate the
existence of universes disconnected from ours. Thus, we take
into account only linear modifications of the theory, so as to guarantee
that the superposition principle remains satisfied. These extra
terms can be interpreted as self-interactions of the universe, as well
as violations of the Hamiltonian constraint.

We aim at studying the dynamics in a regime such that the free
dynamics (given by the finite difference equation of LQC in improved dynamics) can be approximately
described in terms of differential equations. In this regime, the
dynamics is given by a modified WDW equation. Since WDW is obtained
in the limit of large volumes, it is clear that this effective theory
cannot be used to study the dynamics of the universe near the
big bang or big crunch singularities. The additional
interactions should be such that deviations from the Friedmann
equation in the semiclassical limit are small; therefore, they will be treated as small
perturbations.

We consider a closed FLRW
 universe, for which
semiclassical solutions are known explicitly. 
In fact, it was shown in Ref.~\cite{Kiefer:1988} that it is
possible to construct a quantum state whose evolution mimics a solution of the
classical Friedmann equation, and is given by a quantum superposition
of two Gaussian wave packets (one for each of the two phases of the
universe, expanding and contracting) centered on the classical
trajectory. 
We build on the results of Ref.~\cite{Kiefer:1988} and, considering the new interaction term as a perturbation, we
determine the corrections to evolution of the wave packets
arising from the extra linear interactions in our model.

Particularly interesting is the case in which the
perturbation is represented by white noise. In fact, this could be a
way to model the effect of the underlying  discrete structure of spacetime
on the evolution of the macroscopically relevant degrees of
freedom. This is in analogy with standard applications of the Langevin equation to study the dynamics of a subset of the degrees of freedom of a physical system, where the interaction with the other (microscopic) degrees of freedom is modelled as a stochastic driving term.
Thus, it should be possible to make contact (at least
qualitatively) with GFT
cosmology, according to which the dynamics of the universe is that of
a condensate of elementary spacetime constituents. In particular,
describing the interaction term as stochastic white noise, we aim at capturing
from an effective point of view the interaction with extra degrees of freedom
(such as \emph{e.g.} those arising from a possible third quantisation of gravity)
which are not included in the formalism and whose exact nature we ignore. This will be done is
Section~\ref{section noise}.
 A comparison with the case of Stochastic Electrodynamics (SED)~\cite{Lamb-shift-SED, de1983stochastic, Lamb-shift-spontaneous-emission-SED, QM-SED} may be useful, which treats vacuum fluctuations as a classical stochastic process and can provide an alternative expanation of the Lamb shift.  The analogy with SED has of course its limitations,
since in the case at hand there is no knowledge about the dynamics of
the vacuum, which should instead come from a full theory of Quantum
Gravity. In spite of this limitation, the methods developed here are
fully general, so as to allow for a perturbative analysis of the
solutions of the linearly modified WDW equation for any possible form
of the additional interactions.

\section{Geometrodynamics}\label{Sec:Geometrodynamics}
Geometrodynamics is a canonical approach to Quantum Gravity. The starting point is the Hamiltonian formulation of GR in ADM variables (see Appendix \ref{Appendix:ADM}). The system is fully constrained, \emph{i.e.} the Hamiltonian is a linear combination of constraints, and the dynamics is given by the two equations
\be\label{Eq:WDW:ADMconstraints}
H[h_{ij},p^{ij}]\approx0~,\hspace{1em} H_a[h_{ij},p^{ij}]\approx0~,
\ee
called the Hamiltonian constraint and the (spatial) diffeomorphism constraint, respectively. Here $h_{ij}$ represents the metric of three-space and $p^{ij}$ its canonically conjugated momentum. In Appendix~\ref{Appendix:ADM} we also give a review of the ADM formalism and the general form of the classical constraints is given in Eq.~(\ref{Eq:WDW:ADMconstraints}).
The constraints are understood as weak equations, \emph{i.e.} they do not hold everywhere in phase space but only on a region called the \emph{constraint hypersurface}. Since they span the algebra of first class constraints, they are interpreted as the generators of gauge transformations, \emph{i.e.} they do not change the physical state (see Appendix~\ref{Appendix:Dirac}). Thus, the canonical formulation of GR in ADM variables makes it clear that time, as an evolution parameter, has no actual physical interpretation already at the classical level.

Following the Dirac procedure, first class constraint are imposed as restrictions on the space of wave functionals $\psi[h_{ij}]$ (spanning the \emph{kinematical} Hilbert space). In this way physical states are selected as the solutions of the following equations
\be
\hat{H}\psi[h_{ij}]=0,\hspace{1em} \hat{H}_a\psi[h_{ij}]=0~.
\ee
In other words, the \emph{physical} Hilbert space is a subspace of the kinematical Hilbert space, whose elements satisfy the constraints.

In quantum geometrodynamics, the canonically conjugated variables $h_{ij}$ and $p^{ij}$ are assumed as the fundamental ones, and the system is quantised by applying the standard heuristic quantisation rules
\be
h_{ij}\to \hat{h}_{ij}~,\hspace{1em} \hat{p}^{ij}\to \frac{\hbar}{i}\frac{\delta}{\delta h_{ij}}~.
\ee
In this way, one has
\be
[h_{ij}({\bf x}),p^{hk}({\bf y})]=i\delta^h_{(i}\delta^k_{j)}\delta({\bf x},{\bf y})~.
\ee
The diffeomorphism constraint reads as
\be
\hat{H}_a\psi=-2D_j h_{ik}\frac{\hbar}{i}\frac{\delta}{\delta h_{jk}}\psi=0~,
\ee
while the Hamiltonian constraint is given by
\be\label{Eq:WDW:QuantumHamiltonianConstraint}
\hat{H}\psi=\Bigg(-16\pi G\hbar^2 G_{ijhk}\frac{\delta^2}{\delta h_{ij}\delta h_{hk}}-\frac{\sqrt{h}}{16\pi G}\left(\hspace{-6pt}\phantom{e}^{\scriptscriptstyle (3)}R-2\Lambda\right) \Bigg)\psi=0~,
\ee
where $G_{ijhk}$ is the DeWitt supermetric (see Appendix \ref{Appendix:ADM} for definitions). Equation~(\ref{Eq:WDW:QuantumHamiltonianConstraint}) is also known as the WDW equation. Classical GR is recovered in the eikonal approximation of Eq.~(\ref{Eq:WDW:QuantumHamiltonianConstraint}) (see Ref.~\cite{gerlach1969derivation}), similarly to the way classical mechanics can be recovered from quantum mechanics. This suggests that geometrodynamics should be at least approximately valid below the Planck scale, regardless of what the fundamental theory of QG is \cite{Kiefer:2008bs}.

Our considerations above were limited to the pure gravity case. When extra non-gravitational fields are allowed, they contribute additional terms to the constraints. We observe that the kinetic term in the Hamiltonian constraint (\ref{Eq:WDW:QuantumHamiltonianConstraint}) is ill-defined due to the presence of a second order functional derivative and to the non-linearity of the DeWitt supermetric. In fact, there is a factor ordering ambiguity in the definition of such term. Besides that, there is a possibility that quantum anomalies (Schwinger terms) may arise in the constraint algebra. Such terms are potentially dangerous since they can spoil the consistency of the Dirac scheme.  The factor ordering problem and the anomaly problem are intimately related. In fact, it is possible to fix the factor ordering by requiring anomaly cancellation \cite{Barvinsky:1985jt}, although a generally consistent regularisation scheme that can preserve such results is not known \cite{Kiefer:2007ria}. For a more detailed review of these issues, see Ref.~\cite{Kiefer:2007ria} and references therein. 

Before closing this section, we would like to stress the relation between the canonical approach to quantum geometrodynamics and the path-integral approach. In fact, it turns out that they are (at least formally) equivalent. This was discussed from a heuristic point of view in Ref.~\cite{Hawking:1983hj} considering the path-integral over Euclidean metrics. Similar considerations also apply to the Lorentzian case\footnote{For the Lorentzian case in full quantum geometrodynamics the reader is referred to Refs.~\cite{Barvinsky:1986qn,Barvinsky:1986bt,Teitelboim:1981ua}.}. We will discuss in Section~\ref{Sec:QCWDW} the Lorentzian path-integral formulation of quantum cosmological models, which was derived rigorously in Ref.~\cite{Halliwell:1988wc}.

\section{Quantum Cosmology \emph{\`a la} Wheeler-DeWitt}\label{Sec:QCWDW}
The underlying assumption in Quantum Cosmology (QC\footnote{The acronym QC will be used in a generic sense to refer to \emph{any} approach to Quantum Cosmology. When a particular approach is implied, this will be specified by a different acronym, such as \emph{e.g.} WDW or LQC.}) is that Quantum Mechanics is really universal and, as such, its rules must also apply to the universe as a whole. In particular, the fundamental laws determining the large scale structure of spacetime must be quantum. From a formal point of view, QC is based on a reduced symmetry quantisation of the gravitational field, \emph{i.e.} suitable symmetry properties are imposed on field configurations \emph{before} quantisation. Particularly relevant in this respect are homogeneous and isotropic configurations, \emph{i.e.} those that satisfy the Cosmological Principle (discussed in Section~\ref{Sec:CosmologicalPrinciple}).

Symmetry reduction is a crucial aspect of QC. On the one hand, it allows analytic treatment of simple models which are relevant for cosmological applications. On the other hand, its relation to the full theory is in general not clear; more precisely, a given QG\footnote{Similarly to Footnote~10.} proposal motivates the particular quantisation rules used in its symmetry reduced version, although it is not known whether QC can be recovered in some regime from the full quantum theory. These features are shared by all of the known different proposals for (canonical) QG, such as \emph{e.g.} quantum geometrodynamics and LQG.
Another drawback of symmetry reduction lies in the fact that neglecting some degrees of freedom \emph{and} their conjugate momenta is inconsistent with the Heisenberg uncertainty relations from the point of view of the full theory \cite{Kiefer:2007ria}. Nevertheless, it can be a good approximation as long as such degrees of freedom can be regarded as small perturbations. For instance, this assumption is justified when studying cosmological perturbations.

In this section we will show the derivation of the QC dynamical equations in the geometrodynamics approach, \emph{i.e.} quantising the cosmological background \emph{\`a la} WDW. We will start from the Einstein-Hilbert action of GR, symmetry reduced so as to satisfy the cosmological principle. The action of a FLRW universe filled with a scalar field $\phi$ is given in Eq.~(\ref{Eq:MinisuperspaceAction})
\be\label{Eq:WDW:MinisuperspaceAction}
S=S_{\rm g}+S_{\rm m}=\int\de t\; N\left(\frac{1}{2}G_{AB}\frac{\dot{q}^A\dot{q}^B}{N^2}-V(q)\right)~,
\ee
with $q^A=(a,\phi)$ and the DeWitt supermetric being (with $a$ being the scale factor and in units such that $\frac{8\pi G}{3}=1$)
\be
G_{AB}=\left(\begin{array}{cc} -a & 0\\ 0 & a^3\end{array}\right)~.
\ee
The indefiniteness of $G_{AB}$, already observed in the general case, has important consequences for the interpretation of the quantum theory.

At this stage, we would like to rewrite the action principle in Hamiltonian form, which is suitable for the Dirac quantisation scheme of constrained systems. Obviously the Hamiltonian constraint for the cosmological background can be obtained as a particular case of the general ADM action, given in Appendix \ref{Appendix:ADM}. 
We wish to give an alternative derivation, by symmetry reducing the Einstein-Hilbert action and applying the Dirac procedure to the symmetry reduced Lagrangian theory. This is known as the principle of \emph{symmetric criticality}, which we will simply assume to hold in our case\footnote{The reader must be aware that symmetry reduction at the level of the action and at the level of the equations of motion are in general not equivalent. In our case their equivalence is guaranteed by the spherical symmetry of the field configuration. For the general criteria for the validity of the \emph{symmetric criticality principle}, see Ref.~\cite{Palais:1979rca}. Further references and a discussion which is particularly relevant for the application of this principle to cosmology are given in Ref.~\cite{Kiefer:2007ria}.}.

We will follow Ref.~\cite{Kiefer:2007ria} in our derivation of the WDW equation for a FLRW universe. To get started, we need to compute the momenta canonically conjugated to $a$ and $\phi$ from the action (\ref{Eq:WDW:MinisuperspaceAction}). We get
\be
p_a=-\frac{a\dot{a}}{N}~,\hspace{1em}p_\phi=\frac{a^3\dot{\phi}}{N}
\ee
The canonically conjugated momentum to $N$ is found to be identically vanishing, which should be interpreted according to the Dirac scheme (Appendix~\ref{Appendix:Dirac}) as a constraint equation in phase space
\be
p_N\approx 0~.
\ee
Note that the system is fully constrained, as one would expect on general grounds from the Dirac analysis of the constraints of GR. In particular, since homogeneity has been imposed at the outset, the diffeomorphism constraint is automatically satisfied. The Hamiltonian is proportional to the Hamiltonian constraint, stemming from time reparametrisation invariance of the classical theory; it reads as
\be
H=N\left(\frac{1}{2}G^{AB}p_Ap_B+U(q)\right)\approx0~.
\ee
The kinetic term is given by
\be\label{Eq:WDW:KineticTerm}
\frac{1}{2}G^{AB}p_Ap_B=\frac{1}{2}\left(-\frac{p_a^2}{a}+\frac{p_\phi^2}{a^3}\right)~,
\ee
while the interaction term depends both on the form of the scalar field potential and on the spatial geometry (\emph{i.e.} the parameter $K$)
\be\label{Eq:WDW:Potential}
U(q)=\frac{1}{2}\left(V(\phi)-Ka+\frac{\Lambda a^3}{3}\right)~.
\ee

The quantum theory is obtained by applying the standard heuristic quantisation rules, as in the derivation of the complete WDW theory discussed in the previous section\footnote{The reader may appreciate that the minisuperspace model does not feature functional derivatives, but only ordinary ones. This is a major advantage in comparison with the full theory, thus allowing full analytic control.}. In this case we have two degrees of freedom, namely the scale factor $a$ and the scalar field $\phi$. The rules for quantisation are then
\be
\hspace{1em}p_a\to\hat{p}_a=\frac{\hbar}{i}\frac{\pa}{\pa a}~,\hspace{1em}\hspace{1em}p_\phi\to\hat{p}_\phi=\frac{\hbar}{i}\frac{\pa}{\pa \phi}~.
\ee
The operators corresponding to $a$ and $\phi$ act instead multiplicatively.
We observe that factor ordering ambiguities inevitably arise in the quantisation of the kinetic term (\ref{Eq:WDW:KineticTerm}), due to the non-linearity in the DeWitt supermetric $G_{AB}$. A possible solution corresponds to the so-called \emph{covariant factor-ordering}, \emph{i.e.} a choice is made such that the kinetic term is given by the Laplacian corresponding to $G_{AB}$. Thus, we require
\be\label{Eq:WDW:KineticCovariant}
G^{AB}p_A p_B\rightarrow-\frac{\hbar^2}{\sqrt{-\det G_{AB} }}\pa_A\left(\sqrt{-\det G_{AB} }~G^{AB}\pa_B\right)=\frac{\hbar^2}{a^2}\frac{\pa}{\pa a}\left(a\frac{\pa}{\pa a}\right)-\frac{\hbar^2}{a^3}\frac{\pa^2}{\pa \phi^2}~.
\ee
Having fixed the factor ordering, we can finally write down the WDW equation for a FLRW universe. From the Hamiltonian constraint in the classical theory and using Eqs.~(\ref{Eq:WDW:KineticCovariant}),~(\ref{Eq:WDW:Potential}) we obtain (setting $\hbar=1$)
\be\label{Eq:WDW:StillScaleFactor}
\frac{1}{2}\Bigg( \frac{1}{a^2}\frac{\pa}{\pa a}\left(a\frac{\pa}{\pa a}\right)-\frac{1}{a^3}\frac{\pa^2}{\pa \phi^2}+V(\phi)-Ka+\frac{\Lambda a^3}{3}\Bigg)\psi=0~.
\ee
In the rest of this chapter we will consider only the case of a closed universe ($K=1$).
Making the variable transformation $a\to \exp\alpha$, we can bring Eq.~(\ref{Eq:WDW:StillScaleFactor}) to a simpler form. 
\be
\e^{-3\alpha} \Bigg(\frac{\pa^2}{\pa \alpha^2}-\frac{\pa^2}{\pa \phi^2}-\e^{4\alpha}+\e^{6\alpha}\left(V(\phi)+\frac{\Lambda}{3}\right)\Bigg)\psi=0~.
\ee
This is a wave equation with a potential, which in general depends on both variables $\alpha$ and $\phi$. For our purposes we will consider the relatively simple case of a massless scalar field and vanishing cosmological constant. In this case, the equation above reduces to
\be\label{Eq:WDW:NoInteractions}
\Bigg( \frac{\pa^2}{\pa \alpha^2}-\frac{\pa^2}{\pa \phi^2}-\e^{4\alpha}\Bigg)\psi=0~.
\ee
In Section~\ref{Section unperturbed} we will review the construction of wave packet solutions for Eq.~(\ref{Eq:WDW:NoInteractions}), which were obtained in Ref.~\cite{Kiefer:1988}. Such solutions will then be perturbed using techniques developed by the author in Ref.~\cite{deCesare:2015vca}.

As remarked in the final part of Section~\ref{Sec:Geometrodynamics}, the canonical formulation of geometrodynamics and the path-integral one are formally equivalent. For minisuperspace models this equivalence actually holds beyond the merely formal level and can be established on solid grounds \cite{Halliwell:1988wc}. The path-integral of a minisuperspace model is given by
\be\label{Eq:WDW:MinisuperspacePathIntegral}
\mathcal{Z}=\int\mathcal{D}p\mathcal{D}q\mathcal{D}N\;\exp\left[i\int\de t\;(p_h\dot{q}^h-NH)\right]~.
\ee
Due to time-reparametrisation invariance, physical configurations in the naively formulated path-integral (\ref{Eq:WDW:MinisuperspacePathIntegral}) are counted more than once, which makes it badly divergent. This situation is entirely analogous to the one which is familiar from Yang-Mills theories. In order to provide a path-integral quantisation of the system one introduces a gauge-fixing term
and ghost fields, such that the resulting action is invariant under BRST\footnote{After Becchi, Rouet, Stora \cite{Becchi:1975nq}, and Tyutin \cite{Tyutin:1975qk}.} symmetry. This is done following the general BFV (Batalin-Fradkin-Vilkovisky) method, which provides general prescriptions for the path-integral quantisation of a constrained Hamiltonian system \cite{Batalin:1977pb,Fradkin:1975cq}. With a judicious choice of gauge-fixing and by means of a suitable discretisation of the path-integral (skeletonisation), integration over the ghost fields can then be performed \cite{Halliwell:1988wc}.  The result is the following
\be
\mathcal{Z}=\int\de N(t^{\prime\prime}-t^{\prime})\int\mathcal{D}p\mathcal{D}q\;\exp\left[i\int_{t^{\prime}}^{t^{\prime\prime}} \de t\;(p_h\dot{q}^h-NH)\right]~.
\ee
Hence, the path-integral of a parametrised system involves an additional \emph{ordinary} integration over the Lagrange multiplier $N$ (lapse). It turns out that the range of integration of $N$ is directly related to the physical interpretation of the result. In fact, the wave-function $\psi$ (\emph{i.e.} a solution of the WDW equation) is obtained if one integrates the lapse over the whole real line. Instead, by restricting the integration range to the positive half-line, one obtains the Green function of the WDW operator. However, the Green function cannot be interpreted as a propagator, as in the case of quantum mechanics or in QFT \cite{Kiefer:1990ms}. The reason for this difference lies in the fact that only zero modes of the WDW operator define physical states.

The path-integral formulation also allows to show the relation between factor-ordering in the WDW equation and the integration measure in the path-integral. It was suggested in Ref.~\cite{Halliwell:1988wc} that the choice of factor-ordering must be such that the WDW equation includes an extra term $\hbar \xi \mathcal{R}$ which makes it conformally invariant in minisuperspace. In fact, such choice would correspond to the invariance of the path-integral measure under field redefinitions ($\mathcal{R}$ being the scalar curvature of minisuperspace).

A different motivation for considering the path-integral (\ref{Eq:WDW:MinisuperspacePathIntegral}) lies in the fact that boundary conditions for solutions of the WDW equation correspond in this alternative formulation to the choice of an integration contour \cite{Kiefer:1990ms}.
In fact, recent applications have shown that the Lorentzian path-integral can be computed analytically by means of an appropriate deformation of the integration contour, which is fixed uniquely by \emph{Picard-Lefschetz} theory \cite{Feldbrugge:2017kzv,Feldbrugge:2017fcc}. These results are not compatible with those obtained in previous works based on the Euclidean path-integral and the ensuing \emph{no-boundary} proposal (see Ref.~\cite{Hartle:1983}). This shows that the Lorentzian and the Euclidean path-integral are actually two distinct theories\footnote{In particular, the results obtained in those works show that perturbations cannot be kept under control in their approach. To overcome such a problem, one may consider radically different choices of boundary conditions. In fact, one of the renowned features of the Euclidean path-integral approach, based on no-boundary initial conditions, is the fact that it offers a natural explanation for Bunch-Davies initial conditions for inflation. This position does not seem tenable in the more rigorous Lorentzian path-integral approach \cite{Feldbrugge:2017fcc}.}.
 At the background level, the Lorentzian theory leads to a result which is qualitatively similar to the \emph{tunneling proposal} introduced in Ref.~\cite{Vilenkin:1986cy}.  For an alternative position on the results of Refs.~\cite{Feldbrugge:2017kzv,Feldbrugge:2017fcc}, see Ref.~\cite{DiazDorronsoro:2017hti}. In Refs.~\cite{Gielen:2015uaa,Gielen:2016fdb} a model with conformally coupled matter was considered; using a suitable deformation of the lapse integration contour, the Lorentzian path-integral was shown to lead to a cosmological bounce\footnote{Delicate aspects related to the propagation of cosmological perturbations through the quantum bounce are systematically studied in Ref.~\cite{Gielen:2016fdb}, which represents an important viability check for such scenario.}.

\section{Interpretation of the WDW equation}\label{Sec:InterpretWDW}
As remarked above, due to the absence of an absolute time parameter, the WDW equation has the form of a `timeless' Schr\"odinger equation. 
In fact, due to the indefinite nature of the DeWitt supermetric\footnote{This is a general property, which holds for a general field configuration as well as for minisuperspace models, see Appendix \ref{Appendix:ADM}.} $G_{AB}$ and the fact that only second-order derivatives are involved, the WDW equation for a minisuperspace model has the form of a Klein-Gordon (KG) equation with a potential. The KG equation in a $d$-dimensional spacetime reads as
\be
\Box_d \phi=0~,
\ee
where $\phi$ is a scalar field and $\Box_d$ denotes the $d$-dimensional d'Alembertian.
As in the KG case, there is no direct probabilistic interpretation of the solutions of the WDW equation.

For definitenes, let us consider the WDW equation for a closed universe ($K=1$) with vanishing cosmological constant and filled with a massless scalar field with vanishing potential. This is given in Eq.~(\ref{Eq:WDW:NoInteractions}), which can also be obtained from the following Lagrangian
\be\label{Eq:WDW:LagrangianWDW}
L={\partial\psi^{*}\over\partial\alpha}{\partial\psi\over\partial\alpha}-{\partial\psi^{*}\over\partial\phi}{\partial\psi\over\partial\phi}+e^{4\alpha}\psi^{*}\psi~.
\ee
Since $L$ in Eq.~(\ref{Eq:WDW:LagrangianWDW}) is invariant under global $\rm U(1)$ transformations, the following conservation law holds as a consequence of Noether's theorem
\begin{align}
\partial_{\mu}j^{\mu}=0,
\end{align}
where the Noether current $j^{\mu}$ is the standard KG current\footnote{We introduced a compact notation to denote the
  antisymmetric combination of right and left derivatives, namely
$$
   {\overset{\leftrightarrow}{\partial}\over\partial\phi}
\equiv{\overset{\rightarrow}{\partial}\over\partial\phi}-
{\overset{\leftarrow}{\partial}\over\partial\phi}
   $$
   }
 \be\label{Noether current}
j_{\mu}=-i\left(\psi^{*}{\overset{\leftrightarrow}{\partial}
\psi\over\partial\phi},\;\psi^{*}{\overset{\leftrightarrow}
{\partial}\psi\over\partial\alpha}\right).
\ee

In the cosmological case, the absence of a probabilistic interpretation along the lines of the Copenaghen interpretation of QM is closely related to the fact the universe, when considered as a whole, is by definition a closed system. Hence, there is a priori no distinction between system and observer.
Nevertheless, a number of different alternative interpretations of the WDW equation is possible. Some of them may also offer a solution to the problem of time. In particular, the solutions we will discuss below all deal with the problem of identifying time \emph{after} quantisation (following the classification of Ref.~\cite{Isham:1992}). Alternatively, it is also possible to approach the problem of time by fixing time \emph{before} quantisation. We will not deal with such approaches here, for which the reader is referred to Refs.~\cite{Isham:1992,Kiefer:2007ria} and references therein. We will only mention that it is in principle possible to solve the constraints before quantisation (local deparametrisation), although it may be very difficult in practice. In this case, the system can be brought in canonical form with respect to a locally defined \emph{bubble-time} on the classical phase space\footnote{See also Ref.~\cite{Bojowald:2010xp} for a bubble-time approach to the problem of time using effective equations. However, one may argue that such approach would be better qualified as solving the problem of time \emph{after} quantisation, since the effective equations, though being classical, already encode quantum effects.}  \cite{kuchavr1972bubble}. Yet another possibility is offered by the \emph{consistent histories} interpretation, proposed in the seminal paper \cite{Griffiths:1984rx} (see also Ref.~\cite{Isham:1992} and references therein). This is a very general \emph{timeless} interpretation that can be used for all quantum mechanical systems, in which the concept of measurement is secondary. An important role in this approach is played by the decoherence functional, which defines the consistency of a family of histories of the quantum system \cite{Isham:1992}.

\subsection{Third Quantisation}
In standard QFT, the problems related to the lack of a probabilistic interpretation of the KG equation for a relativistic scalar particle are solved by introducing a second quantisation of the KG field. The probabilistic interpretation is then guaranteed by unitarity of the $S$-matrix in the second quantised theory. In the case of the gravitational field, quantum geometrodynamics is already, strictly speaking, a second quantised theory. In order to be consistent with the terminology, one then talks about a third quantisation of the gravitational field.

The concept of a third quantisation was introduced in order to study topology change in Quantum Gravity. In fact, in the standard second quantised formulation of geometrodynamics topology is fixed at the outset, which would seem to exclude a range of phenomena that could potentially play an important role in Quantum Gravity. This possibility is reintroduced in the third quantised theory. In this formalism, the universe wave function is turned into a field operator, which acts on the vacuum of a Fock space to create `particles', \emph{i.e.} universes with a given topology. In Ref.~\cite{Giddings:1988}, a general third quantised formalism for a system of interacting universes is described (see also Refs.~\cite{Kuchar:1991qf,Isham:1992} for further references). In particular, non-linear interaction terms are introduced. The interaction between two such universes is represented by pants diagrams, similar to those used in the description of closed strings scattering\footnote{Another interesting possibility is the nucleation of \emph{baby universes} from a parent universe, to which they are connected by means of a wormhole \cite{Coleman:1988cy}. The latter is interpreted in this formalism as a gravitational instanton (\emph{i.e.} a minimum of the \emph{Euclidean action}, interpreted as a quantum tunneling effect).} in String Theory. The non-linear interaction terms make the vacuum parameters of the theory dynamical \cite{Giddings:1988}. This has important consequences for the effective low energy description of the theory, notably leading to a dynamical cosmological constant \cite{Giddings:1988,Coleman:1988tj}.

Some new conceptual problems arise in this formalism. In fact, although the mathematical formalism parallels the one normally employed in conventional field theories, some of the concepts that have a clear interpretation in that context do not necessarily have one in the case of gravity. In particular, it is not clear in general what the Hilbert space of a `single particle' is \cite{Isham:1992}. Moreover, a field theory formalism requires fixing the statistics obeyed by the field operator. Since statistics is intimately related to microcausality in conventional QFTs, which does not have an obvious counterpart in the scattering of separate universes, it is not clear what its meaning in this framework  is \cite{Isham:1992}. It is also not clear how the problem of time is addressed in such formulation.

A third quantised formalism is also adopted in modern background independent approaches to QG, such as Group Field Theory \cite{Oriti:2006se}. However, the interpretation of the theory is completely different from that of the WDW theory in third quantisation. GFT is reviewed in Chapter~\ref{Chapter:GFT}, where we also discuss its connections with other approaches. In that framework, some of the issues we just listed, such as the construction and the interpretation of the Fock space, have a solution.

\subsection{Construction of the Hilbert Space Inner Product from the KG Inner Product} 
The existence of a probabilistic interpretation in QG is linked to the notion of time evolution \cite{Isham:1992}, which makes the physical interpretation of the theory particularly difficult and quite different from other quantum theories. Since there is no absolute time structure, one must single out a suitable clock (internal time) variable among the other physical variables. 
In the cosmological case, there is a striking formal analogy between the WDW equation and the KG equation in curved spacetime. The WDW equation resembles the KG equation also in the general case, due to the hyperbolic character of superspace \cite{DeWitt:1967yk}. One could then attempt the construction of the physical Hilbert space of solutions of the WDW equation in a similar fashion to the KG equation. In the KG case, if we assume the existence of a timelike Killing vector field\footnote{Its existence depends on the background spacetime.} and provided the potential is positive, we can split the space of solutions in two and construct separately the Hilbert space of positive and that of negative frequency solutions. In fact, the KG inner product turns out to be positive-definite on such solutions and directly leads to the definition of a Hilbert space inner product.
We can then try to apply the same strategy to the WDW equation \cite{Isham:1992}. In fact, the indefiniteness of the DeWitt supermetric also allows to define a notion of `spacelike', `null', and `timelike' vectors in superspace, although these notions are not necessarily related to the notion of a physical time. However, it was shown that in general no such Killing vector exists in superspace \cite{kuchar1991problem,1981qugr.conf..329K}. Moreover, the potential in the WDW equation contains a curvature term which is not necessarily positive definite; thus positive and negative frequencies would generally be mixed \cite{Isham:1992}. In the following, we will discuss how this problem can be solved in the specific cosmological example we gave in the previous section, in which many of the difficulties of the general case do not occur.

In the cosmological model considered above, a possible choice for a (globally defined) internal time is represented by the scalar field $\phi$. We remark that, given the structure of Eq.~(\ref{Eq:WDW:NoInteractions}), the parameter $\alpha$ would also be a valid choice for a \emph{local} time variable\footnote{A consistent way of using local bubble-times from the point of view of effective equations was studied
  in Ref.~\cite{Bojowald:2010xp}.}. In fact, this is the interpretation of $\alpha$ that was given in Ref.~\cite{Kiefer:1988}.
  However, such a choice leads to a potential that is not
  bounded from below\footnote{Even from a classical perspective $\alpha$ is
  not a good choice for a \emph{global} time since it is not a
  monotonic function of the proper time.}. We remark that in the general case the existence of a globally defined internal time variable depends both on the form of the DeWitt supermetric and on the potential.
  
  With the interpretation given above of the massless scalar $\phi$ as an internal clock, the Noether current (\ref{Noether current}) leads to the following Noether charge %
\be\label{Noether charge}
Q=\int_{-\infty}^{\infty}\mbox{d}\alpha\; j_{\phi}~, \ee
which is obviously conserved under `time' evolution 
\be\label{Charge conservation}
{\mbox{d}Q\over\mbox{d}\phi}=0~.  \ee
Clearly, the conservation of $Q$ is in general incompatible with
 that of the $L^2$ norm of $\psi$, defined as
\be
\lVert\psi\rVert_{L^2(\mathbb{R})}=\int_{-\infty}^{\infty}\mbox{d}\alpha\;
|\psi|^2~.  \ee
Furthermore, the Noether charge $Q$ has no definite sign unless one
considers either only positive frequency solutions or only those with negative frequency,
since a generic solution of the WDW equation admits a decomposition in terms of both.
Hence, Eq.~(\ref{Noether charge}) does not define in general a positive-definite norm on the space of solutions.

We observe that positive frequency solutions and negative frequency solutions
span superselection sectors that are preserved by the algebra of Dirac
observables\footnote{Dirac observables in the classical theory are functions on phase space with vanishing Poisson brackets with the generators of the first class constraints algebra (see Appendix~\ref{Appendix:Dirac}). In the quantum theory they are represented as Hermitian operators. } identified \emph{e.g.} in Ref.~\cite{Ashtekar:2006D73}.
States
in a given superselection sector satisfy a Schr{\"o}dinger equation
with Hamiltonian $\pm\sqrt{\hat{\Theta}}$, with
\be
\hat{\Theta}=-\frac{\pa}{\pa\alpha^2}+\e^{4\alpha}~.
\ee
Thus, one has within each sector
\be\label{Eq:WDW:Schroedinger}
i\frac{\pa}{\pa\phi}\psi_{\pm}=\pm\sqrt{\hat{\Theta}}~\psi_{\pm}~,
\ee
where $\psi_{\pm}$ denotes a positive (negative) frequency solution of the WDW equation.
Within each sector, the
conservation of the Hilbert norm holds as in ordinary QM. Thus, Eq.~(\ref{Charge conservation}) becomes in this case a statement about conservation of the expectation value of the operator\footnote{The conservation of $p_\phi$ holds in the classical theory, but also \emph{by definition} in the Schr\"odinger reformulation. In fact, the generator of $\phi$-translations is identified with the Hamiltonian $\pm\sqrt{\hat{\Theta}}$ in Eq.~(\ref{Eq:WDW:Schroedinger}).}
 $\hat{p}_{\phi}=-i\pa_{\phi}$.

  As
far as the free dynamics is concerned (\emph{i.e.} no third quantisation is introduced), it is perfectly legitimate to
work in a given superselection sector and use it to construct the
physical Hilbert space.
However, when the
theory includes either non-linearities or potentials that are not positive-definite,
superselection no longer holds and both
sectors have to be taken into account at the same time. In that case, a possible way to remedy the lack of a natural Hilbert space structure could be represented by a
third quantisation (see above)~\cite{Isham:1992}.

The non-equivalence of $Q$ and
$\lVert\psi\rVert_{L^2(\mathbb{R})}$ can be easily seen for a wave
packet
\be
\psi(\alpha,\phi)=\int_{-\infty}^{\infty}\mbox{d}k\;\psi_{k}(\alpha,\phi)A(k)~,
\ee
where $\psi_{k}$ is a fundamental solution to the WDW equation and $A(k)$ its amplitude.
In particular, if the $\psi_{k}$ are the ones for the model considered in Section~\ref{Section unperturbed}, one finds (see Eq.~(\ref{Eq:WDW:FundamentalSolutions})\,)
\be\label{Eq:WDW:NoetherChargeQ}
Q=\int_{-\infty}^{\infty}\mbox{d}k\; 2k|A(k)|^2~.  \ee 
Note that when $A(k)$ is supported only on one side of the real line (consistently with superselection),
Eq.~(\ref{Charge conservation}) can be interpreted as a conservation
law for the average momentum\footnote{In the QFT context the same
  relation is interpreted as a relativistic normalisation, which
  amounts to having a density of $2k$ particles per unit volume, cf. Ref.~\cite{Peskin:1995ev}.}, \emph{i.e.} the expectation value $\langle \hat{p}_{\phi}\rangle$. In agreement with the discussion above, a
similar interpretation is not possible when the support of the wave packet is not contained within a half-line (in momentum space $k$). 

\section{A generalisation of the WDW equation}\label{section 2}
In this section we introduce a generalisation of the WDW equation, characterised by new terms representing interactions in the minisuperspace of a FLRW universe. The model is quite general and different interpretations are possible.
In fact, by assuming a third quantised perspective, the model represents the most general field theory on minisuperspace, obtained by introducing new terms which could be in principle non-linear and non-local. Alternatively, it can be obtained in the large volume limit of the model in Ref.~\cite{Calcagni:2012}. The latter can be understood as a third quantised field theory based on LQC, or equivalently as a toy model for GFT based on the Lie group $\mbox{U}(1)$. Given the relevance of the GFT approach for our work, we will follow this alternative path to motivate our model. However, the reader must be aware that this is not the only possibility and the present model can also be considered in its own respect.
 
 \subsection{Motivation and the relation to LQC and GFT}
 The model introduced in Ref.~\cite{Calcagni:2012} is a field theory on the minisuperspace of a FLRW universe, filled with a minimally coupled massless scalar field, whose kinetic term is given by the Hamiltonian constraint of LQC, see Eq.~(\ref{Eq:WDW:HamiltonianConstraintLQC}). The kinetic term in our model will be obtained from the one in Ref.~\cite{Calcagni:2012} in the large volume limit.
For this reason, it will be useful to review how the WDW equation is recovered from the
Hamiltonian constraint of LQC\footnote{The actual way in which WDW
  represents a large volume limit of LQC is put in clear mathematical
  terms in Refs.~\cite{Ashtekar:2007, Corichi:2007}, where the
  analysis is based on a special class of solvable models (sLQC). See also Appendix~\ref{Appendix:LQC} and references therein.}  following
Ref.~\cite{Ashtekar:2006}.
  Essential elements of LQC are reviewed in Appendix~\ref{Appendix:LQC}.

For our purposes it is convenient to consider a massless scalar field
$\phi$, minimally coupled to the gravitational field. As explained above, this choice has
the advantage of allowing for a straightforward deparametrisation of
the theory, thus defining a clock. The Hamiltonian constraint has the
general structure~\cite{Ashtekar:2006, Calcagni:2012}
\be\label{Eq:WDW:HamiltonianConstraintLQC}
\hat{\mathcal{K}}\psi(\nu,\phi)\equiv-B(\nu)
\left(\Theta+\partial_{\phi}^2\right)\psi(\nu,\phi)=0~,
\ee
where $\psi$ is the wave function of the universe, defined in this model on a discrete minisuperspace, $\Theta$ is a
finite difference operator acting on the gravitational sector in the
kinetimatical Hilbert space of the theory $\mathcal{H}^{\rm g}_{\rm kin}$,
and $\nu$ labels eigenvectors of the volume operator\footnote{\label{Footnote:LQCdifference1}In LQC one usually introduces the dimensionless variable $\upsilon$ and correspondingly the states $|\upsilon\rangle$ on which the holonomies act as shift operators, see Appendix \ref{Appendix:LQC}. It turns out that $|\upsilon\rangle$ is also an eigenvector of the volume operator $\hat{V}$, with eigenvalue $|\nu|=\left(\frac{8\pi\beta}{6}\right)^{3/2} \frac{|\upsilon|}{K}\ell_{Pl}^3$ \cite{Ashtekar:2006improved}. The constant $K$ is given below Eq.~(\ref{kinetic operator}). The absolute value $\nu$ indicates that both orientations of the volume of the cell are possible.}.

The action on minisuperspace that was considered in Ref.~\cite{Calcagni:2012} reads\footnote{The model was originally formulated for $K=0$, but it admits a straightforward generalisation to include the case $K=1$.}
\be\label{action}
S[\psi]=S_{\rm free}[\psi]+\sum_{n}\frac{\lambda_{n}}{n!}\sum_{\nu_1\dots\nu_{n}}\int\mbox{d}\phi_1\dots\mbox{d}\phi_{n}\; f^{(n)}(\{\nu_i\},\{\phi_i\})\prod_{j=1}^n\psi(\nu_{j},\phi_{j})~,
\ee
where the first term gives the dynamics of the free theory, namely a homogeneous and isotropic gravitational background coupled to a massless scalar field 
\be
S_{\rm free}[\psi]=\sum_{\nu}\int\mbox{d}\phi\;\psi(\nu,\phi)\hat{\mathcal{K}}\psi(\nu,\phi)~.
\ee
The operator $\hat{\mathcal{K}}$ is defined in Eq.~(\ref{Eq:WDW:HamiltonianConstraintLQC}) and gives the Hamiltonian constraint in LQC. The terms containing the functions $f^{(n)}$ represent additional interactions, that can be interpreted as violating the constraint.

The action (\ref{action}) defines a toy model for Group Field Cosmology~\citep{Calcagni:2012}, given by a GFT with Lie group $G=\mbox{U}(1)$  and $\nu$ a Lie algebra element\footnote{Up to a constant \emph{dimensionful} factor. In fact, Lie algebra elements are dimensionless, whereas for us $\nu$ has the physical dimensions of a volume.}. The free dynamics, \emph{i.e.} the precise form of the operator $\hat{\mathcal{K}}$, depends on the specific LQC model adopted. However, the continuum limit should be the same regardless of the model considered and must give the WDW equation for  the corresponding three-space topology\footnote{The WDW equation obtained in the continuum limit from LQC for a generic lattice refinement model displays factor ordering ambiguities that are inherited from those in the Hamiltonian constraint of LQC. The ambiguities are solved and the factor ordering in the WDW equation is \emph{uniquely determined} for a particular choice of lattice refinement, which corresponds to the \emph{improved dynamics} scheme of LQC \cite{Nelson:2008vz}. The unique factor ordering thus determined turns out to correspond to the \emph{covariant} one (see below).}.
From this point of view, the WDW approach to QC should be interpreted as an \emph{effective theory}, valid at scales such that the discreteness introduced by the polymer quantisation cannot be probed. We will show that, even from this more limited perspective, the theory defined by the action~(\ref{action}) leads to a novel effective theory of QC that introduces significant modifications to the standard WDW equation. If one holds the loop quantisation as more fundamental, one must conclude that the validity of our approach is limited to large volumes; therefore, it cannot be used to study the dynamics in regimes that would correspond to \emph{classical} spacetime singularities such as, \emph{e.g.}, big bang and big crunch singularities in cosmology.

It was hinted in Ref.~\cite{Calcagni:2012} that the additional interactions could also be interpreted as interactions occurring between homogeneous patches of an inhomogeneous universe\footnote{However, this may pose problems related to the absence of a natural notion of contiguity of such patches.}. Another possible interpretation is that they actually represent interactions among different, separate, universes. The latter turns out to be a natural option in the framework of third quantisation (see Ref.~\cite{Isham:1992} and references therein), which naturally allows for topology change\footnote{For examples of topology change in the `baby universes' literature see \emph{e.g.} Refs.~\cite{Giddings:1988,Giddings:1988cx,Coleman:1988cy}.}. The interactions are in principle completely general and can be non-local in minisuperspace. However, this case will not be considered in our work.

\subsection{The large volume limit}
The model described in the previous subsection is based on a discrete minisuperspace. Here we will show how a continuum minisuperspace is recovered in the limit of large volumes, thus providing a generalisation of WDW cosmology. It turns out that only the gravitational sector is discretised.
In fact, the
discreteness introduced by the LQC formulation, and by its extension represented by the action (\ref{action}), does not affect the
matter sector, which is still the same as in the continuum WDW quantum
theory\footnote{However, it is important to remark that it is also possible to introduce a \emph{polymer quantisation} for the matter degrees of freedom. This is similar to the one used for the quantisation of the homogeneous gravitational field, and inequivalent to the standard Fock quantisation (Schr\"odinger quantisation in the homogeneous case) \cite{Varadarajan:1999it,Varadarajan:2001nm,Ashtekar:2002vh,Ashtekar:2001xp}. See also Ref.~\cite{Corichi:2007tf} for a polymer version of quantum mechanics.}
. The gravitational sector of the Hamiltonian constraint
operator of LQC in `improved dynamics'~\footnote{In the framework of
  LQC, the Hamiltonian constraint contains the gravitational
  connection $c$. However, only the holonomies of the connections are
  well defined operators, hence to quantise the theory we replace $c$
  by $\sin\bar\mu c/\bar \mu$, where $\bar\mu$ represents the
  `length' of the line segment along which the holonomy is
  evaluated.  Originally $\bar\mu$ was set to a constant $\mu_0$,
  related to the area-gap. To cure severe issues in the ultraviolet
  and infrared regimes which plague the $\mu_0$ quantisation, a new
  scheme called `improved dynamics' was
  proposed~\cite{Ashtekar:2006improved}. In the latter, the dimensionless
  length of the smallest plaquette is $\bar\mu$. See Appendix \ref{Appendix:LQC} for further discussion and references.}, reads (cf. Eq.~(\ref{Eq:WDW:HamiltonianConstraintLQC}))
\be\label{Eq:WDW:HamiltonianConstraintLQCFiniteDifferences}
-B(\nu)\Theta\,\psi(\nu,\phi)\equiv
A(\nu)\psi(\nu+\nu_0,\phi)+C(\nu)\psi(\nu,\phi)+D(\nu)\psi(\nu-\nu_0,\phi)~,
\ee 
where the finite increment $\nu_0$ represents an elementary volume unit\footnote{\label{Footnote:LQCdifference2}Assuming for instance the Hamiltonian constraint given in Ref.~\cite{Ashtekar:2006improved} $\hat{\mathcal{C}}_{\rm grav}\psi(\upsilon)=f_{+}(\upsilon)\psi(\upsilon+4)+f_{0}\psi(\upsilon)+\psi_{-}(\upsilon)\psi(\upsilon-4)$, with $\upsilon$ dimensionless. In our notation, this corresponds to Eq.~(\ref{Eq:WDW:HamiltonianConstraintLQCFiniteDifferences}) with elementary step $\nu_0=\left(\frac{8\pi\beta}{6}\right)^{3/2} \frac{4}{k}\ell_{Pl}^3$. Also cf. Appendix~\ref{Appendix:LQC} and Footnote~\ref{Footnote:LQCdifference1} for the definition of $\upsilon$.}
 and $A, B, C, D$ are
functions which depend on the chosen quantisation scheme.  In order to
guarantee that $\Theta$ is symmetric\footnote{In the large volume
  limit it will be formally self-adjoint w.r.t. the measure
  $B(\nu)\mbox{d}\nu$.} in $\nu$, the coefficients must satisfy the
$D(\nu)=A(\nu-\nu_0)$ condition~\cite{Calcagni:2012}; it holds in both
the $K=0$ and the $K=1$ case.

It is a general result in LQC that the WDW equation can be recovered in the
continuum (\emph{i.e.} large volume) limit~\cite{Ashtekar:2007}.
 In particular, it was shown in Ref.~\cite{Ashtekar:2006} that for
 $K=1$ one recovers the Hamiltonian constraint of
 Ref.~\cite{Kiefer:1988}. In fact, it turns out that $\Theta$ can be
 expressed as the sum of the operator $\Theta_0$ relative to the
 $K=0$ case, and a $\phi$-independent potential term (\emph{i.e.}
 diagonal in the $\nu$ basis) as
\be\label{kinetic operator}
\Theta=\Theta_0+\frac{\pi G l_0^2\beta^2}{3 k^{4/3}}|\nu|^{4/3}~.
\ee
In the above expression $k=\frac{2\sqrt{2}}{3\sqrt{3\sqrt{3}}}$,
$\beta$ is the Barbero-Immirzi parameter of Loop Quantum Gravity
(LQG), $G$ the gravitational constant and $l_0$ is the (comoving) size of the fiducial cell on the spatial manifold in
the $K=1$ model. The latter can be formally sent to zero in order to
recover the $K=0$ case.

Restricting to wave functions $\psi(\nu)$ which are smooth and slowly
varying in $\nu$, we obtain the WDW limit of the Hamiltonian
constraint
\be
\Theta_0\psi(\nu,\phi)=-12\pi G (\nu\partial_{\nu})^2\psi(\nu,\phi)~,
\ee
which is exactly the same constraint one has in WDW theory.  Thus,
LQC naturally recovers the factor ordering (also called covariant
factor ordering, in the sense that the quantum
constraint operator is of the form $G^{AB}\nabla_A\nabla_B$, where
$G^{AB}$ is the inverse WDW metric and $\nabla_A$ denotes the covariant
derivative associated with $G_{AB}$) which was obtained in
Ref.~\cite{Halliwell} under the requirement of field reparametrisation
invariance of the minisuperspace path-integral. 

Since $\nu$ represents a
proper volume, it is proportional to the physical\footnote{As opposed to comoving.} volume of a cell
with linear dimensions equal to the scale factor $a$
\be
\nu\propto a^3~.
\ee
With the change of variable $\alpha=\log a$, we can rewrite the constraint
operator $\hat{\mathcal{K}}$ as
\be\label{Eq:WDW:FreeTheory}
\hat{\mathcal{K}}
=e^{-3\alpha}\left(\frac{\partial^2}{\partial\alpha^2}-\frac{\partial^2}{\partial\phi^2}-e^{4\alpha}\right)~.
\ee
The operator in Eq.~(\ref{Eq:WDW:FreeTheory}) will be used as the kinetic kernel in our model. Thus, we have the general extension of WDW theory, with action
\be\label{Eq:WDW:MinisuperspaceAction}
S[\psi]=\int\mbox{d}\phi\mbox{d}\alpha\;\psi(\alpha,\phi)\hat{\mathcal{K}}\psi(\alpha,\phi)+\sum_{n}\frac{\lambda_{n}}{n!}\int\prod_{i=1}^n\mbox{d}\phi_i\,\mbox{d}\alpha_i \; f^{(n)}(\{\alpha_i\},\{\phi_i\})\prod_{j=1}^n\psi(\alpha_{j},\phi_{j})~,
\ee
where $\hat{\mathcal{K}}$ is the one defined in Eq.~(\ref{Eq:WDW:FreeTheory}).
\subsection{Simplified model}
The model specified by the minisuperspace action (\ref{Eq:WDW:MinisuperspaceAction}) is the most general third quantised field theory\footnote{We implicitly assumed above that the universe wave function (now promoted to a field) is real. This is actually for mere simplicity of notation, and the model can be formulated even more generally for a complex wave function. Nevertheless, these considerations do not affect our discussion below, since we will only consider the linear dynamics.} that extends WDW cosmology without introducing new degrees of freedom\footnote{Such as new fields or higher dimensions in minisuperspace. The latter may correspond to other degrees of freedom of the gravitational field or extra matter fields that were neglected in the two-dimensional minisuperspace truncation.}.
If we consider only the quadratic term in the action (\ref{Eq:WDW:MinisuperspaceAction}), we get a modified dynamical equation of the form\footnote{\label{Footnote:ViolationHamiltonianConstraint}This equation can be interpreted as a violation of the Hamiltonian constraint, or a modification achieved by the inclusion of extra degrees of freedom. As long as only the cosmological background is considered and we work at an an effective level, both interpretations are possible and the mathematical methods used to solve it are the same.}
\be\label{general equation of motion}
\hat{\mathcal{K}}\psi(\alpha,\phi)+\int\de{\alpha^{\prime}}\mbox{d}\phi^{\prime}
\; g(\alpha,{\alpha^{\prime}};\phi,\phi^{\prime})
\psi(\alpha^{\prime},\phi^{\prime})=0~, \ee
with
\be
g(\alpha,{\alpha^{\prime}};\phi,\phi^{\prime})=\lambda_2 f^{(2)}(\alpha,{\alpha^{\prime}};\phi,\phi^{\prime})~.
\ee
We will assume for simplicity that the interaction $g$ is local in minisuperspace,
\emph{i.e.}
\be\label{Eq:WDW:LocalInteractions}
g(\alpha,\alpha^{\prime};\phi,\phi^{\prime})=g(\alpha,\phi)\delta(\alpha-\alpha^{\prime})\delta(\phi-\phi^{\prime})~.
\ee
Using Eqs.~(\ref{Eq:WDW:FreeTheory}) and (\ref{Eq:WDW:LocalInteractions}) we find that Eq.~(\ref{general equation of motion}) reduces to a Klein-Gordon equation
with a `space' and `time' dependent potential 
\be\label{Eq:WDW:equation of motion general potential}
\left(\frac{\partial^2}{\partial\alpha^2}-\frac{\partial^2}{\partial\phi^2}-e^{4\alpha}\right)\psi(\alpha,\phi)=-e^{3\alpha}g(\alpha,\phi)\psi(\alpha,\phi)~.
\ee
Note that the functional form of $g(\alpha,\phi)$ is kept completely general.

	It should be pointed out that more general interaction terms than the ones considered in (\ref{general equation of motion}) can in principle be conceived. In fact, from the point of view of a third quantised theory of gravity, the most general equation of motion would involve non-linear terms, as in the action (\ref{Eq:WDW:MinisuperspaceAction}).  However, the methods developed in this chapter are completely general and their application to the study of non-linearities in perturbation theory is straightforward, following the analogous well-known procedure in perturbative relativistic Quantum Field Theory (QFT). The main ingredient for such calculations is given by the Green function, together with the Feynman rules for the vertices associated with the various interactions. A perturbative study of the WDW equation with an additional linear term can thus be seen as the starting point for a more general study. Note that, even though such a generalisation may seem more natural from a merely formal point of view, it makes the physical interpretation of the theory even more problematic. A linear term instead, as we will show explicitly with the example given in Section~\ref{section noise}, can be used to model the interaction of the degree of freedom represented by the scale factor with other microscopic degrees of freedom of the gravitational field without having to change the interpretation of the wave function $\psi$.

\subsection{Perturbation theory}
Since the interaction term represented by the r.h.s. of Eq.(\ref{general equation of motion}) is unknown, we cannot determine an exact
solution of the equation without resorting to a case by case
analysis. However, since the solutions of the WDW equation in the
absence of a potential are known explicitly, it is convenient to adopt a
perturbative approach. The methods we develop are fully general and can
thus be applied for any possible choice of the function $g$.

We formally expand the wave function and the WDW operator in terms of a dimensionless  parameter $\lambda$. Such parameter only serves book-keeping purposes and will be eventually set equal to unity. Thus, we have
\begin{align}
\psi&=\psi^{(0)}+\lambda\,\psi^{(1)}+\dots~,\\
\hat{T}&=\hat{T}_{0}+\lambda\, e^{3\alpha} g~,
\end{align}
with
\be\label{constraint operator}
\hat{T}_{0}\equiv\left(\frac{\partial^2}{\partial\alpha^2}-\frac{\partial^2}{\partial\phi^2}-e^{4\alpha}\right)~.
\ee
Therefore, the dynamics up to first order in perturbation theory is given by
\begin{align}
\hat{T}_{0}\psi^{(0)}&=0~,\\
\hat{T}_{0}\psi^{(1)}&=-e^{3\alpha}g\,\psi^{(0)}\label{first order perturbation}~.
\end{align}
The zero-th order contribution to the wave function $\psi^{(0)}$ is a solution of the wave equation with an exponential potential; this was obtained in Ref.~\citep{Kiefer:1988} and will be reviewed in the next section. If we were able to invert $\hat{T}_{0}$, we would get the wave function corrected to first order from Eq.~(\ref{first order perturbation}). However, finding the Green function of $\hat{T}_{0}$ is not as straightforward in this case as it would be in the $K=0$ case\footnote{For $K=0$, the kinetic operator is simply $\hat{T}_{0}$ the d'Alembertian, whose Green kernels are well-known for all possible choices of boundary conditions.}. Moreover, as for the d'Alembertian, the Green kernel will depend on the boundary conditions. The problem of determining which set of boundary conditions is more appropriate depends on the physical situation we have in mind and will be dealt with in the next sections.


\section{Analysis of the unperturbed case}\label{Section unperturbed}

In this section, we review the construction of wave packet solutions for the cosmological background, which were obtained in Ref.~\cite{Kiefer:1988}. The WDW equation for a homogeneous and isotropic universe (positive curvature, $K=1$) with a massless scalar field is 
\be\label{WDW equation}
\left(\frac{\partial^2}{\partial\alpha^2}-\frac{\partial^2}{\partial\phi^2}-e^{4\alpha}\right)\psi(\alpha,\phi)=0~.
\ee
We impose the boundary condition
\be\label{boundary condition}
\lim_{\alpha\to\infty} \psi(\alpha,\phi)=0~,
\ee
which is necessary in order to reconstruct semiclassical states describing the dynamics of a closed universe. In fact, regions of minisuperspace corresponding to arbitrary large scale factors are not accessible due to the potential barrier in Eq.~(\ref{WDW equation}).

Equation~(\ref{WDW equation}) can be solved by separation of variables
\be\label{sol-psi}
\psi_{k}(\alpha,\phi)=N_k\, C_{k}(\alpha)\varphi_{k}(\phi)~,
\ee
leading to
\begin{align}
&\frac{\partial^2}{\partial\phi^2}\varphi_{k}+k^2\varphi_{k}=0~,\label{EQUATION
  FOR
  PHI}\\
  &\frac{\partial^2}{\partial\alpha^2}C_{k}-(e^{4\alpha}-k^2) C_{k}=0~.\label{EQUATION
  FOR ALPHA}
\end{align}
Note that $\psi_{k}(\alpha,\phi)$ stands for the elementary solution of
Eq.~(\ref{WDW equation}) and should not be confused with the Fourier
transform of $\psi$, for which we will use instead the notation
$\tilde{\psi}(\alpha,k)$.  In Eq.~(\ref{sol-psi}) $N_k$ denotes a
normalisation factor depending on $k$ and whose value will be fixed
later.

Let us proceed with the solutions of Eqs.~(\ref{EQUATION FOR PHI}),
(\ref{EQUATION FOR ALPHA}).  Equation~(\ref{EQUATION FOR PHI}) yields
complex exponentials as solutions
\be
\varphi_{k}=e^{ik\phi}~.
\ee
Eq.~(\ref{EQUATION FOR ALPHA}) has the same form as the
stationary Schr\"odinger equation for a non-relativistic particle in
one dimension, with potential $V(\alpha)= e^{4\alpha}-k^2$ and zero
energy. We observe that the particle is free for
$\alpha\to -\infty$, whereas the potential barrier becomes
infinitely steep as $\alpha$ takes increasingly large positive
values. Given the boundary condition (\ref{boundary condition}), Eq.~(\ref{EQUATION FOR
  ALPHA}) admits as an exact solution the modified Bessel function of
the second kind (also known as MacDonald function)
\be\label{MacDonald
  function} C_{k}(\alpha)=K_{ik/2}\left(\frac{e^{2\alpha}}{2}\right)~.
\ee
Wave packets are then constructed as linear superpositions of the
(appropriately normalised) solutions
\be\label{Wave Packet}
\psi(\alpha,\phi)=\int_{-\infty}^{\infty}\mbox{d}k\;\psi_{k}(\alpha,\phi)A(k)~,
\ee
with a suitable amplitude $A(k)$.
Since the $C_{k}(\alpha)$ are improper eigenfunctions of the
(one-parameter family of) Hamitonian operator(s) in Eq.~(\ref{EQUATION
  FOR ALPHA}), they do not belong to the space of square integrable
functions on the real line $L^{2}(\mathbb{R})$. However, it is still
possible to define some sort of normalisation by fixing the
oscillation amplitude of the improper eigenfunctions for
$\alpha\to-\infty$. For this purpose, we recall the WKB expansion of the
MacDonald function
\be
K_{ik/2}\left(\frac{e^{2\alpha}}{2}\right)
\simeq\frac{\sqrt{\pi}}{2}e^{-k\frac{\pi}{4}}(k^2-e^{4\alpha})^{-1/4}
\cos\left(\frac{k}{2}\arccosh\frac{k}{e^{2\alpha}}-\frac{1}{2}
\sqrt{k^2-e^{4\alpha}}
-\frac{\pi}{4}\right)~,
\ee
which provides a very accurate approximation for large values of $k$. Therefore, we define
\be\label{Eq:WDW:FundamentalSolutions}
\psi_{k}(\alpha,\phi)=e^{k\frac{\pi}{4}}\sqrt{k}\;K_{ik/2}
\left(\frac{e^{2\alpha}}{2}\right)e^{ik\phi}~,
\ee
which for large enough values of $k$ gives elementary waves all having
same amplitude to the left of the potential barrier. Note that for
small $k$ Eq.~(\ref{Eq:WDW:FundamentalSolutions}) is insufficient to normalise the amplitudes to the same value, and the elementary waves (\ref{Eq:WDW:FundamentalSolutions}) will still exhibit their dependence on $k$.

\begin{figure}
\centering
\begin{minipage}{0.45\textwidth}
\centering
 \includegraphics[width=\columnwidth]{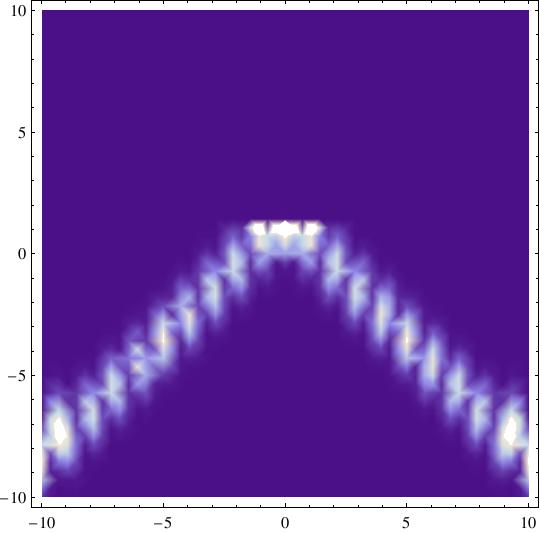}
    \caption[Semiclassical solution of the Wheeler-DeWitt equation]{Absolute square $|\psi^{(0)}|^2$ of the `wave function' of the universe,
      corresponding to the choice of parameters $b=1$,
      $\overline{k}=10$. Lighter shades correspond to larger values of
      the wave function. During expansion and recollapse the evolution
      of the universe can be seen as a freely propagating wave
      packet. From the plot it is also evident the reflection against
      the potential barrier at $\alpha_{\overline{k}}=\frac{1}{2}\log
      \overline{k}$, where the wave function exhibits a sharp peak.}\label{Semiclassical}
      \end{minipage}\hfill
\begin{minipage}{0.45\textwidth}
\centering
\includegraphics[width=\columnwidth]{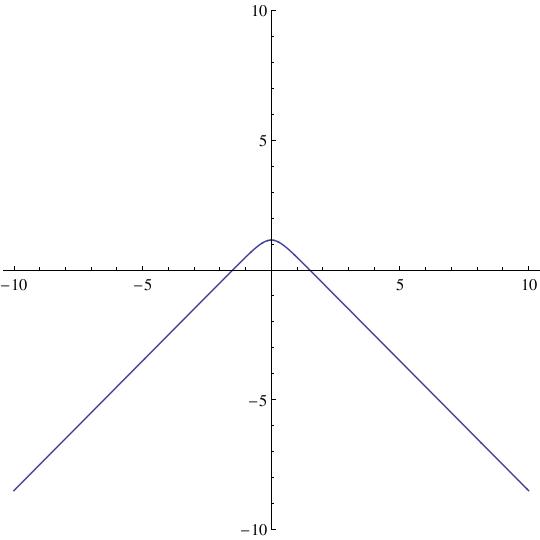}
    \caption[Classical trajectory of a closed universe]{Classical trajectory of the universe in
      minisuperspace. It is closely followed by the peak of
      semiclassical states, as we can see by comparison with
      Fig.~\ref{Semiclassical}. Here $\overline{k}=10$ as in the previous figure.}
   \label{ClassicalTrajectory}
   \end{minipage}
\end{figure}

Let us assume a Gaussian profile for the amplitude $A(k)$ in
Eq.~(\ref{Wave Packet})
\be
A(k)=\frac{1}{\pi^{1/4}\sqrt{b}}e^{-\frac{(k-\overline{k})^2}{2b^2}}~,
\ee
where $\overline{k}$ should be taken large enough so as to guarantee
the normalisation of the function $\psi_{k}(\alpha,\phi)$ according to Eq.~(\ref{Eq:WDW:FundamentalSolutions}). One then
finds that the solution has the profile shown in
Fig.~\ref{Semiclassical}.  The solution represents a wave packet that
starts propagating from a region where the potential vanishes
(\emph{i.e.} at the initial singularity, where $\alpha$ and $\phi$ are both large and negative) towards the potential barrier
located approximately at $\alpha_{\overline{k}}=\frac{1}{2}\log
\overline{k}$, whence it is reflected back. The region where $\alpha$ and $\phi$ are both large and positive corresponds to the classical big crunch singularity. For large values of
$\overline{k}$ the wave packet is practically completely reflected
back from the barrier.  The parameter $b$ gives a measure of the
semiclassicality of the state, \emph{i.e.} it accounts for how much it
peaks on the classical trajectory.  We remark that the peak of the
wave packet follows closely the classical trajectory (see Fig.~\ref{ClassicalTrajectory})
\be\label{Classical Solution}
e^{2\alpha}= {\bar{k}\over \cosh (2\phi)}~.
\ee
At a classical level, the conserved quantity $\bar{k}$ is identified with the canonical momentum $p_\phi$.

It is worth observing that quantum gravitational effects are significant also in regions where classically there is no singularity, \emph{i.e.} at the turning point corresponding to the maximum expansion and where the universe enters the phase of recollapse. In the corresponding quantum mechanical model, the point of recollapse corresponds to a potential barrier. Hence, quantum tunneling takes place, which significantly alters the profile of the wave function at recollapse.

\subsection{Noether charge}
The Noether charge of the semiclassical universe considered in
this section 
can actually be computed
analytically. In fact, since it is a conserved quantity, one can evaluate it in the regime where the computation is as easy as possible.
This turns out to be the minisuperspace region corresponding to the classical singularity, given that the potential in the WDW equation vanishes for $\alpha\to
-\infty$. 
We note that when both $\alpha$ and $\phi$ are in a neighborhood
of $-\infty$, \emph{i.e.} close to the initial singularity, one can
approximate the exact solution with a Gaussian wave packet (given by
the WKB approximation, see Appendix \ref{WKB solution} and Ref.~\cite{Kiefer:1988}), which
describes the evolution of the universe in the expanding phase\footnote{A similar expansion holds in the region of the big crunch singularity, with the wavefunction propagating in the opposite $\alpha$ direction after reflection against the potential barrier \cite{Kiefer:1988}.}. One has
\be\label{WDW:WKBapproximation}
\begin{split}
\psi\overset{\mathsmaller{\alpha\to-\infty}}{\mathlarger{\approx}}
&c_{\overline{k}}\,b\sqrt{\frac{\pi}{2}}\exp\left\{\frac{i}{2}\left[\overline{k}\left(2\phi +\arccosh\left(\frac{\overline{k}}{e^{2\alpha}}\right)+\sqrt{1-\frac{e^{4\alpha}}{\overline{k}^2}}\right)-\frac{\pi}{2}
\right]\right\}      \nonumber\\
&\times\exp\left[-\frac{b^2}{2}\left(\phi+\frac{1}{2}\arccosh\left({\overline{k}\over
    e^{2\alpha}}\right)\right)^2\right]~,
    \end{split}
    \ee
Using Eq.~(\ref{WDW:WKBapproximation}) and the definition of the Noether current (\ref{Noether current}), the charge density reads 
\be
j_{\phi}=|c_{\overline{k}}|^2 \pi b^2 \overline{k}\exp\left[-{b^2\over
    4}\left(2\phi+\arccosh\left({\overline{k}\over
    e^{2\alpha}}\right)\right)^2\right]~.  \ee 
For large negative values of $\alpha$ one can approximate the inverse
hyperbolic function in the argument of the exponential as
\be \arccosh\left({\overline{k}\over
  e^{2\alpha}}\right)\overset{\mathsmaller{\alpha\to-\infty}}
{\mathlarger{\approx}}\log(2\overline{k})-2\alpha~.
\ee 
Thus, the computation of the Noether charge using Eq.~(\ref{Eq:WDW:NoetherChargeQ}) is reduced to a Gaussian integral, leading to the result
\be
Q=|c_{\overline{k}}|^2 b\,\pi^{3/2}\; \overline{k}~.  \ee 

From the calculations above we conclude that for such wave packet solutions the conservation of the KG current
leads to momentum conservation in the classical limit. We recall
that classically the momentum is related to the velocity of the scalar field
\be
\overline{k}=p_\phi=a^3\dot{\phi}~.
\ee
%

\section{Time-independent perturbation potential}\label{Section:TimeIndependent}

In the case in which the potential in the r.h.s of Eq.~(\ref{Eq:WDW:equation of motion general potential}) does not depend on the internal time $\phi$, 
  we can resort to time-independent
perturbation theory to calculate the corrections to the wave
function. More precisely, we consider the representation of the operator $\hat
T_0$ defined in Eq.~(\ref{constraint operator}) in Fourier space\footnote{Or
equivalently its action on a monochromatic wave.}, which leads
to the following Helmoltz equation with a potential
\be\label{homogeneous Helmoltz}
\left(\frac{\partial^2}{\partial\alpha^2}
+k^2-e^{4\alpha}\right)\tilde{\psi}^{(0)}(\alpha,k)=0~.
\ee
Note that the potential cannot be considered as a small perturbation with
respect to the standard Helmoltz equation. In fact, reflection from an
infinite potential barrier requires boundary conditions that are
incompatible with those adopted in the free particle case.
Therefore, we have to solve Eq.~(\ref{homogeneous Helmoltz}) exactly.
The solution of the unperturbed equation are known and were given above in Eq.~(\ref{MacDonald
  function}).

In order to compute the first perturbative corrections, we need to
solve Eq.~(\ref{first order perturbation}), which in Fourier space
leads to
\be\label{perturbation in Fourier space}
\left(\frac{\partial^2}{\partial\alpha^2}
+k^2-e^{4\alpha}\right)\tilde{\psi}^{(1)}(\alpha,k)
=U(\alpha)\tilde{\psi}^{(0)}(\alpha,k)~,
\ee
upon defining $U(\alpha)=-e^{3\alpha}g(\alpha)$.
Equation~(\ref{perturbation in Fourier space}) is easily solved once
the Green function of the Helmoltz operator on the l.h.s is
known. The equation for the Green function is
\be
\left(\frac{\partial^2}{\partial\alpha^2}
+k^2-e^{4\alpha}\right)G^{k}(\alpha,\alpha^{\prime})
=\delta(\alpha-\alpha^{\prime})~.
\ee
The homogeneous equation admits two linearly independent solutions,
which we can use to form two distinct linear combinations that satisfy
the boundary conditions at the two extrema of the interval of the real
axis we are considering. The remaining free parameters are then fixed
by requiring continuity of the function at $\alpha^{\prime}$ and the
condition on the discontinuity of the first derivative at the same
point.  

Equation~(\ref{homogeneous Helmoltz}) has two linearly independent
solutions $p(\alpha)$ and $h(\alpha)$ given by
\be
p(\alpha)=K_{i\frac{k}{2}}\left(\frac{e^{2\alpha}}{2}\right)
\ee
and
\be
h(\alpha)=I_{-i\frac{k}{2}}\left(\frac{e^{2\alpha}}{2}\right)
+I_{i\frac{k}{2}}\left(\frac{e^{2\alpha}}{2}\right)~.
\ee
Thus, we make the following ansatz
\be\label{ansatz}
G^{k}(\alpha,\alpha^{\prime})=\begin{cases} 
      \gamma\; p(\alpha) ~ & \alpha\geq\alpha^{\prime}~, \\
      \delta\; h(\alpha)+\eta\; p(\alpha) ~ & \alpha<\alpha^{\prime}~.
   \end{cases}
\ee
Note that $G^{k}(\alpha,\alpha^{\prime})$ satisfies the boundary
condition Eq.~($\ref{boundary condition}$) by construction.
Since we do not know what is the boundary condition for
$\alpha\to-\infty$, \emph{i.e.} near the classical
singularity\footnote{In fact, even popular choices like the
  Hartle-Hawking \emph{no-boundary} proposal~\cite{Hawking:1981,
    Hartle:1983} or the \emph{tunneling condition} proposed by
  Vilenkin~\cite{Vilenkin:1987}, only apply to massive scalar
  fields.}, we will not be able to fix the values of all constants
$\gamma,\delta,\eta$.
 Therefore, we will end up with a one-parameter family of Green functions.

The Green function must be continuous at the point
$\alpha=\alpha^{\prime}$, which implies
\be\label{eq1above} \gamma\; p(\alpha^{\prime})=\delta\;
h(\alpha^{\prime})+\eta\; p(\alpha^{\prime})~.  \ee
Moreover, in order for its second derivative to be a Dirac delta
function, the following condition on the discontinuity of the first
derivative must be satisfied
\be\label{eq2above} \gamma\; p^{\prime}(\alpha^{\prime})-\delta\;
h^{\prime}(\alpha^{\prime})-\eta\; p^{\prime}(\alpha^{\prime})=1~.
\ee
Upon introducing a new constant $\Omega=\gamma-\eta$, we can rewrite
Eqs.~(\ref{eq1above}), (\ref{eq2above}) in the form of a Kramer's
system
\[
\begin{cases}
\Omega\; p(\alpha^{\prime})-\delta\; h(\alpha^{\prime})&=0~,\\
\Omega\; p^{\prime}(\alpha^{\prime})-\delta\; h^{\prime}(\alpha^{\prime})&=1~,
\end{cases}
\]
which admits a unique solution, given by
\begin{align}
\Omega&=-\frac{h(\alpha^{\prime})}{W(\alpha^{\prime})}~,\\
\delta&=-\frac{p(\alpha^{\prime})}{W(\alpha^{\prime})}~,
\end{align}
where
$W(\alpha^{\prime})=p(\alpha^{\prime})h^{\prime}(\alpha^{\prime})
-h(\alpha^{\prime})p^{\prime}(\alpha^{\prime})$
is the Wronskian. Therefore, substituting back in the ansatz (\ref{ansatz}), we obtain the desired expression for the Green function
\[
G^{k,\eta}(\alpha,\alpha^{\prime})=\begin{cases}
p(\alpha)\left(\eta-\frac{h(\alpha^{\prime})}{W(\alpha^{\prime})}\right)
& \alpha\geq\alpha^{\prime}~,
\\ -\frac{p(\alpha^{\prime})}{W(\alpha^{\prime})}\; h(\alpha)+\eta\;
p(\alpha) & \alpha\leq\alpha^{\prime}~.
\end{cases}
\]
Here we have explicitly introduced the parameter $\eta$ in our
notation for the Green function in order to stress its
non-uniqueness. Note that the phase shift at $-\infty$ varies with
$\eta$.

Finally, the solution of Eq.~(\ref{perturbation in Fourier space}) reads
\be\label{corrections tilde}
\tilde{\psi}^{(1)}(\alpha,k)=\int\mbox{d}\alpha^{\prime}\;
G^{k,\eta}(\alpha,\alpha^{\prime})U(\alpha^{\prime})
\tilde{\psi}^{(0)}(\alpha,k)~.
\ee
The possibility of studying the perturbative corrections using a
decomposition in monochromatic components is viable because of the
validity of the superposition principle. This is in fact preserved by
additional interactions of the type considered here, which violate the Hamiltonian
constraint while preserving the linearity of the wave equation. Note
that in general modifications of the scalar constraint in the WDW
theory would be non-linear and non-local, as for instance within the
proposal of Ref.~\cite{Oriti:2010}, where classical geometrodynamics
arises as the hydrodynamics limit of GFT. However, such non-linearities
would spoil the superposition principle, hence making the analysis of
the solutions much more involved.

The time dependence of the perturbative corrections is recovered by
means of the inverse Fourier transform of Eq.~(\ref{corrections
  tilde})
\be
\psi^{(1)}(\alpha,\phi)=\int\frac{\mbox{d}k}{2\pi}\; 
\tilde{\psi}^{(1)}(\alpha,k)e^{-ik\phi}~.
\ee

Expectation values of observables can be defined using the measure
determined by the `time' component of the KG current $j_{\phi}$ as
\be \langle f\rangle=\int\mbox{d}\alpha\; f j_{\phi}~.  \ee 
The KG current is defined as in Eq.~(\ref{Noether current})
\be
j_{\mu}=-i\left(\psi^{*}{\overset{\leftrightarrow}{\partial}
\psi\over\partial\phi},\;\psi^{*}{\overset{\leftrightarrow}
{\partial}\psi\over\partial\alpha}\right).
\ee
Using the conservation law of the KG current we can then derive an
analogue of Ehrenfest theorem, namely
\be\label{Ehrenfest}
\frac{\mbox{d}}{\mbox{d}\phi}\langle f\rangle=\int\mbox{d}\alpha\;
\left( j_{\phi}\partial_{\phi} f +f
\partial_{\phi}j_{\phi}\right)=\int\mbox{d}\alpha\; \left(
j_{\phi}\partial_{\phi} f+f \partial_{\alpha}j_{\alpha}\right)~.  \ee
When the observable $f$ does not depend explicitly on the internal time
$\phi$, the first term in the integrand vanishes. Considering for
instance the scale factor $a=e^{\alpha}$ then, after integrating by
parts, we get 
\be \frac{\mbox{d}}{\mbox{d}\phi}\langle
a\rangle=-\int\mbox{d}\alpha\; e^{\alpha}j_{\alpha}~.  \ee 
In an analogous fashion, one can show that 
\be\label{Ehrenfest2}
\frac{\mbox{d}}{\mbox{d}\phi}\langle a^2\rangle=-2\int\mbox{d}\alpha\;
e^{2\alpha}j_{\alpha}~.  \ee 
Formulae (\ref{Ehrenfest}) and (\ref{Ehrenfest2}), besides their
simplicity, turn out to be quite handy for numerical computations,
especially when dealing with `time'-independent observables. In fact
they can be used to compute time derivatives of the averaged
observables without the need for a high resolution on the $\phi$ axis,
\emph{i.e.} they can be calculated using data on a single
`time'-slice. Expectation values can thus be propagated forwards or
backwards in `time' by solving first order ordinary differential
equations.

\section{First perturbative corrections with time-dependent potentials}\label{Section:TimeDependent}

In the previous section we considered a time-independent perturbation,
which can be dealt with using the Helmoltz equation. This is in
general not possible when the perturbation depends on the internal
time. In order to study the more general case we have to resort to
different techniques to find the exact Green function of the
operator $\hat{T}_{0}$.  Since the potential $e^{4\alpha}$ breaks
translational symmetry, Fourier analysis, which makes the
determination of the propagator so straightforward in the $K=0$ case\footnote{As remarked above, in the $K=0$ case the WDW equation is equivalent to the wave equation.}, is of
no help.

Let us perform the following change of variables
\be X=\frac{1}{2}e^{2\alpha}\cosh(2\phi)~~,\hspace{1em}
Y=\frac{1}{2}e^{2\alpha}\sinh(2\phi)~, \ee
which represents a mapping of minisuperspace into the wedge $X>|Y|$.
The minisuperspace interval (corresponding to DeWitt's supermetric)
can be expressed in the new coordinates as
\be \mbox{d}\phi^{2}-\mbox{d}\alpha^2
=\frac{1}{X^2-Y^2}(\mbox{d}Y^{2}-\mbox{d}X^2)~, \ee with
$(X^2-Y^2)^{-1}$ the conformal factor\footnote{Recall that any two
  metrics on a two-dimensional manifold are related by a conformal
  transformation.}. As an immediate consequence of conformal
invariance, uniformly expanding (contracting) universes are given by
straight lines parallel to $X=Y$ ($X=-Y$) within the wedge. The
two lines $X=-Y$ and $X=Y$ represent the initial and final
singularity, respectively.  It is worth pointing out that the
classical trajectory Eq.~(\ref{Classical Solution}) now takes the much
simpler expression
\be \overline{k}=2X~, \ee 
\emph{i.e.} classical trajectories are
represented by straight lines parallel to the $Y$ axis and with the
extrema on the two singularities.

In the new coordinates $X, Y$ the operator $\hat{T}_{0}$ in
Eq.~(\ref{constraint operator}) reads
\be\label{Conformal to KG}
\hat{T}_{0}=(X^2-Y^2)(\partial_X^2-\partial_Y^2-4)~,
\ee
which is, up to the inverse of the conformal factor, a Klein-Gordon
operator with $m^2=4$. This is a first step towards a perturbative
solution of Eq.~(\ref{first order perturbation}), which we rewrite
below for convenience of the reader in the form
\be\label{equation for the perturbation with U}
\hat{T}_{0}\psi^{(1)}=U(\alpha,\phi)\psi^{(0)}~,
\ee
where
\be
U(\alpha,\phi)=-e^{3\alpha}g(\alpha,\phi)~.
\ee
Note that the above is the same equation as the one considered in the
previous section, but we are now allowing for the interaction
potential to depend also on $\phi$. Given Eq.~(\ref{Conformal to KG}),
we recast Eq.~(\ref{equation for the perturbation with U}) in the form
that will be used for the applications of the next section, namely
\be\label{eq-above}
(\partial_X^2-\partial_Y^2-4)\psi^{(1)}=\frac{U(X,Y)}{(X^2-Y^2)}\psi^{(0)}~.
\ee
The formal solution to Eq.~(\ref{eq-above}) is given by a convolution
of the r.h.s with the Green function satisfying suitably chosen
boundary conditions
\be\label{Convolution solution}
\psi^{(1)}(X,Y)=\int\mbox{d}X^{\prime}\mbox{d}Y^{\prime}\;
G(X,Y;X^{\prime},Y^{\prime})\frac{U(X^{\prime},Y^{\prime})}{(X^{\prime
    2}-Y^{\prime2})}\psi^{(0)}(X^{\prime},Y^{\prime})~.  \ee
The Green  function of the Klein-Gordon operator \emph{in free space} is
well-known for any dimension $D$ (see \emph{e.g.}
Ref.~\cite{Zhang}). In $D=2$ it is formally given by\footnote{Notice
  that here $Y$ plays the role of time.}
\be\label{Green function} G(X,Y;X^{\prime},Y^{\prime})=\int
\frac{\mbox{d}^2 k}{(2\pi)^2}\frac{e^{-i
    \left(k_Y(Y-Y^{\prime})-k_X(X-X^{\prime})\right)}}
{\left(k_{Y}^{2}-k_{X}^{2}\right)-4}~,
\ee
and satisfies the equation
\be\label{Green equation}
(\partial_X^2-\partial_Y^2-4)G(X,Y;X^{\prime},Y^{\prime})
=\delta(X-X^{\prime})\delta(Y-Y^{\prime})~.
\ee
Evaluating (\ref{Green function}) explicitly using Feynman's
integration contour, which is a preferred choice in the context of a
third quantisation, we get~\cite{disessa, Zhang}
\be
G(X,Y;X^{\prime},Y^{\prime})=-\frac{1}{4}\theta(s)H_{0}^{(2)}(2\sqrt{s})-\frac{i}{2\pi}\theta(-s)K_{0}(2\sqrt{-s})~,
\ee
where we introduced the notation $s=(Y-Y^{\prime})^2-(X-X^{\prime})^2$
for the interval.  However, the present situation is quite different
from that of a free particle in the plane, since there is a physical boundary represented by
the edges of the wedge $X>|Y|$. Therefore, boundary conditions must be 
appropriately discussed and implemented. A preferred choice is the one that leads to
the Feynman boundary conditions in the physical coordinates
$(\alpha,\phi)$. In the following, we will see the form that such boundary
conditions take in the new coordinate system, finding the
transformation laws of operators annihilating progressive and
regressive waves.

We start with the observation that the generator of dilations in the $(X,Y)$
plane acts as a tangential derivative along the edges
\be\label{dilation}
X\partial_{X}+Y\partial_{Y}=\partial_{t}~.
\ee
Moreover on the upper edge $X=Y$ (corresponding to the big crunch) we have
\be
\partial_{t}|_{X=Y}=\frac{1}{4}\left(\partial_{\alpha}+\partial_{\phi}\right)~,
\ee
while on the lower edge $X=-Y$ (corresponding to the initial singularity) we have
\be
\partial_{t}|_{X=-Y}=\frac{1}{4}\left(\partial_{\alpha}-\partial_{\phi}\right)~.
\ee
Therefore the boundary conditions
\be\label{boundary conditions}
\partial_{t}|_{X=Y}G=\partial_{t}|_{X=-Y}G=0
\ee
are equivalent to the statement that the Green function is a
positive (negative) frequency solution of the wave equation at the
final (resp. initial) singularity. This is in agreement with the Feynman
prescription and with the fact that the potential $e^{\alpha}$
vanishes at the singularity $\alpha\to-\infty$.

There is a striking analogy with the classical electrostatics problem
of determining the potential generated by a point charge inside a
wedge formed by two conducting plates. In fact, it is well-known that
the electric field is normal to the surface of a conductor, so that
the tangential derivative of the potential vanishes. The similarity
goes beyond the boundary conditions and holds also at the
level of the dynamical equation. In fact, after performing a Wick rotation $Y\to -i\,Y$
Eq.~(\ref{Green equation}) becomes the Laplace equation with a
constant mass term, while the operator $\partial_t$ defined in
Eq.~(\ref{dilation}) keeps its form.

\begin{figure}
\centering
\begin{minipage}{0.45\textwidth}
\centering
 \includegraphics[width=\columnwidth]{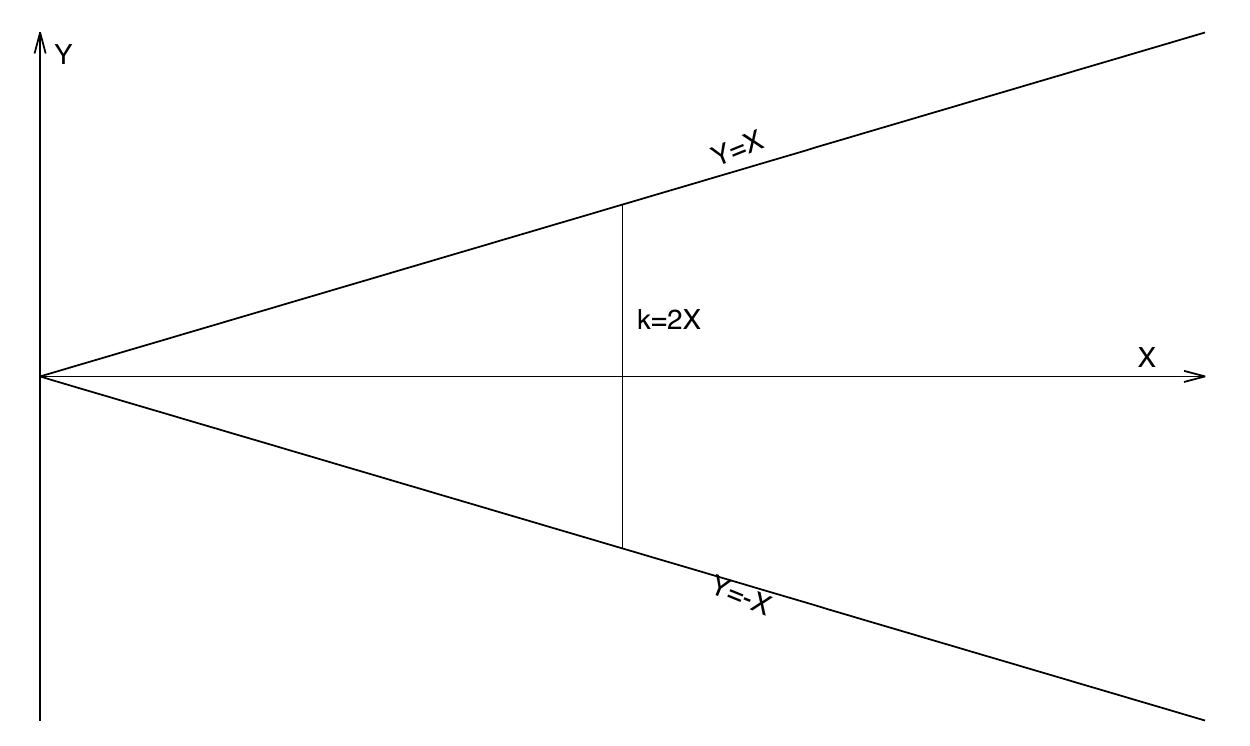}
\caption[Conformal map in minisuperspace]{The whole of minisuperspace $(\alpha,\phi)$ is conformally mapped into the wedge $X>|Y|$. The straight lines $X=\pm Y$ correspond to the two classical singularities, while the vertical line $\bar{k}=2X$ is the classical trajectory of a closed universe.}
\end{minipage}\hfill
\begin{minipage}{0.45\textwidth}
\centering
\hspace*{-1cm}\includegraphics[width=1.3\columnwidth]{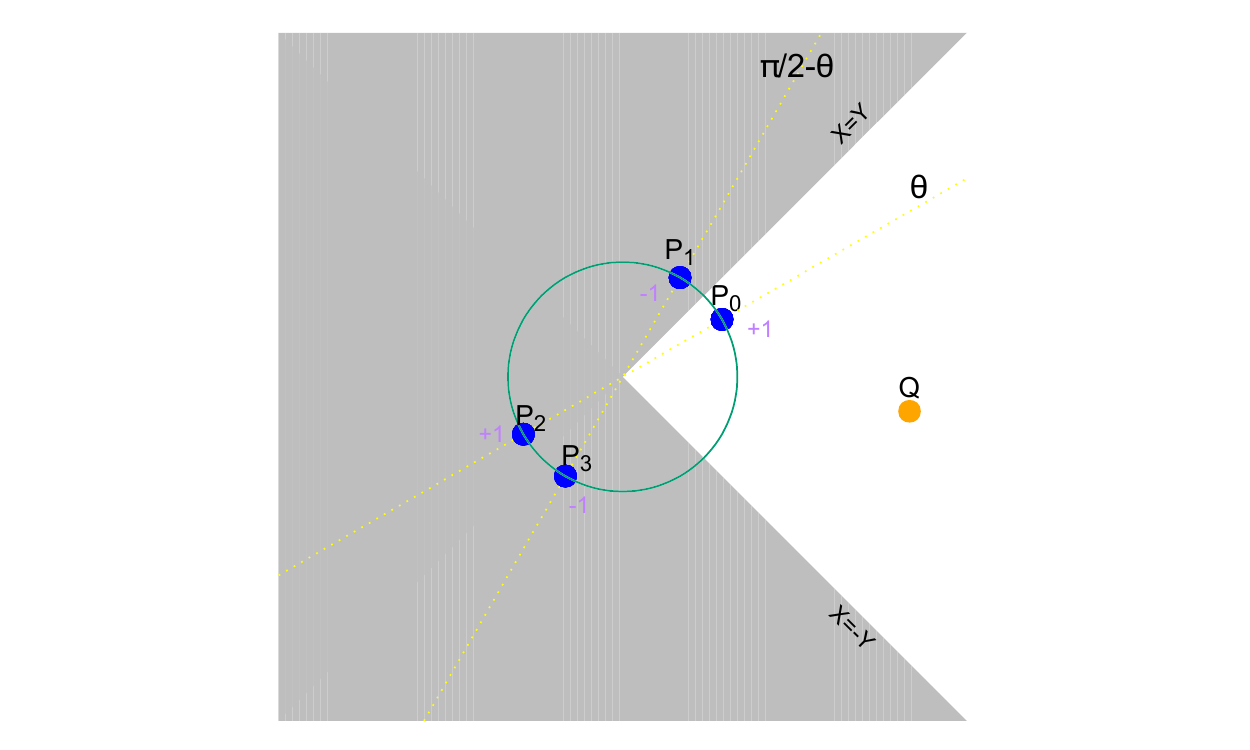}
\caption[Method of image charges]{The Green function is a function of the point $Q$ and the source $P_0$. The positions and the charges of the images $P_j$'s are determined so as to satisfy the boundary conditions.}
\label{fig:wedge}
\end{minipage}
\end{figure}

After performing the Wick rotation, the problem can be solved with the
method of image charges. Given a source (charge) at point
$P_0=(r^{\prime},\theta^{\prime})$, three image charges as in Fig. \ref{fig:wedge} are
needed to guarantee that the boundary conditions are met. The
Euclidean Green function, as a function of the point $Q=(r,\theta )$
and source $P_0$ in the $(X,Y)$ plane, is given by
\be\label{Feynman's Green function}
G(r,\theta ,r^{\prime},\theta^{\prime})=\frac{1}{2\pi}\Big(K_0(m
\,\overline{P_0 Q})-K_0(m \,\overline{P_1 Q})+K_0(m \,\overline{P_2
  Q})-K_0(m \,\overline{P_3 Q})\Big)~,
\ee
where $K_0$ is a modified Bessel function of the second kind and $m=2$ (cf. Eqs.~(\ref{Green equation}),~(\ref{Conformal to KG})).
The quantities $\overline{P_j Q}$ represent the Euclidean distances
between the charges $P_j$ and the point $Q$. The Lorentzian Green function
is then recovered by Wick rotating all the $Y$ time coordinates,
\emph{i.e.} those of $Q$ and of the $P_j$'s. We would like to remark that we obtained this
solution by treating the two edges symmetrically, thus maintaining time
reversal symmetry. The Green function with a source $P_0$ within the
wedge vanishes when $Q$ is on either of the two edges. In fact, this
can be seen as a more satisfactory
way of realising DeWitt's boundary
condition, regarding it as a property of correlators rather than of states\footnote{DeWitt originally proposed the vanishing of the
  wave function of the universe at singular metrics on superspace, suggesting in this way that the singularity problem would be solved \emph{a priori} with an appropriate choice of the boundary conditions.
 However, there are cases (see discussion in Ref.~\cite{Bojowald:2002, Bojowald:2003}) where DeWitt's proposal does not lead to a well posed boundary value problem and actually overconstrains the dynamics. For instance, the solution given in Ref.~\citep{Kiefer:1988} that we discussed in Section~\ref{Section unperturbed} satisfies it for $\alpha\to\infty$ but not at the initial singularity, since all the elementary solutions $C_{k}(\alpha)$ are indefinitely oscillating in that region. Imposing the same
  condition on the Green function does not seem to lead to such
  difficulties.}.
  
\section{Treating the interaction as white noise}\label{section noise}

The methods developed in the previous sections are completely general
and can be applied to any choice of the extra interaction terms using
Eq.~(\ref{Convolution solution}).  In this section we will consider,
as a particularly simple example, the case in which the additional
interaction is given by white noise. Besides the mathematical
simplicity, there are also physical motivations for doing so.  In fact,
we can make the assumption that the additional terms responsible for the violation of the
Hamiltonian constraint\footnote{See Footnote~\ref{Footnote:ViolationHamiltonianConstraint}.} can be used to model the effect of other degrees of freedom in QG, arising, \emph{e.g.}, from the
underlying discreteness of spacetime .

\subsection{Motivation}
Some approaches to Quantum Gravity (such as, \emph{e.g.}, GFT) suggest that an
appropriate description of the gravitational interaction at a
fundamental level must be given in the language of third
quantisation~\cite{Oriti:2013}. At this stage, one might make an
analogy with the derivation of the Lamb shift in Quantum
Electro-Dynamics (QED) and in effective stochastic approaches. As it is well-known, the effect stems from
the second quantised nature of the electromagnetic field. However, the
same prediction can also be obtained if one holds the electromagnetic
field as classical, but impose \emph{ad hoc} conditions on the
statistical distribution of its modes, which necessarily has to be the same as
the one in the vacuum state of QED. In this way, the
instability of excited energy levels in atoms and the Lamb shift are
seen as a result of the interaction of the electron with a
\emph{stochastic} background electromagnetic field~\cite{Lamb-shift-SED, Lamb-shift-spontaneous-emission-SED}. However, in the
present context we will not resort to a third quantisation of the
gravitational field, but instead put in \emph{ad hoc} stochastic terms
arising from interactions with degrees of freedom other than the scale
factor. Certainly, our position here is weaker than that of Stochastic
Electro-Dynamics (SED) (see Refs.~\cite{de1983stochastic, QM-SED} and the works
cited above), since a fundamental third quantised theory of gravity
has not yet been developed to such an extent so as to make observable
predictions in Cosmology. Therefore we are unable to give details
about the statistical distribution of the gravitational degrees of
freedom in what would correspond to the vacuum state of QG. Our model should
henceforth be considered as purely phenomenological; its link to
the full theory will be clarified only when the construction of the latter will eventually
be completed.

To be more specific, we treat the function $g$ in the perturbation as
white noise. Stochastic noise is used to describe the interaction of a
system with other degrees of freedom regarded as an
\emph{environment}\footnote{For a derivation of the
  Schr{\"o}dinger-Langevin equation using the methods of stochastic
  quantisation we refer the reader to Ref.~\cite{Yasue1977}.  We are
  not aware of other existing works suggesting the application of
  stochastic methods to Quantum Cosmology. However, there are
  applications to the fields of classical cosmology and inflation and
  the interested reader is referred to the published literature.}.
Hence we write
\begin{align}
\langle g(\alpha, \phi)\rangle&=0~,\\ \langle g(\alpha, \phi)
g(\alpha^{\prime}, \phi^{\prime}) \rangle &
=\varepsilon\;\delta(\alpha-\alpha^{\prime})\delta(\phi-
\phi^{\prime})~,\label{seconda proprieta' noise}
\end{align}
where $\langle \underline{\hspace{1em}}\rangle$ denotes an ensemble average and $\varepsilon$ is a parameter which can be regarded as the magnitude of the noise.
It is straightforward to see that
\be
\langle\psi^{(1)}\rangle=0~,
\ee
which means that the ensemble average of the corrections to the wave
function vanishes.

A more interesting quantity is represented by the second moment
\be\label{definition variance}
\mathcal{F}(X_1,Y_1;X_2,Y_2)=
\langle (\psi^{(1)}(X_1,Y_1))^{*}\psi^{(1)}(X_2,Y_2)\rangle~.
\ee
In fact, when evaluated at the same two points,
$\tilde{\mathcal{F}}(X,Y)\equiv\mathcal{F}(X,Y;X,Y)$ represents
the variance of the \emph{statistical} fluctuations of the wave
function at the point $(X,Y)$.  Using Eqs.~(\ref{first order
  perturbation}), (\ref{seconda proprieta' noise}) and
(\ref{definition variance}) we get
\be\label{expression for the variance}
\begin{split}
 &\left\langle\left|\psi^{(1)}\right|^2 (X,Y)\right\rangle=\tilde{\mathcal{F}}(X,Y)\\
 &~~~=16\,\varepsilon\int_{X\geq |Y|}
\mbox{d}X^{\prime}\,\mbox{d}Y^{\prime}\;(X^{\prime 2}-Y^{\prime
  2})^{\frac{1}{2}}|\psi^{(0)}(X^{\prime},Y^{\prime})|^2 \,
|G(X-X^{\prime},Y-Y^{\prime})|^2~.
\end{split}
\ee
In order to obtain the correct expression of the integrand, one needs
to use the transformation properties of the Dirac distribution, which
yield
\be
\langle g(\alpha, \phi) g(\alpha^{\prime}, \phi^{\prime})\rangle
=2\varepsilon(X^2-Y^2)\delta(X-X^{\prime})\delta(Y-Y^{\prime})~.
\ee
Note the resemblance of Eq.~(\ref{expression for the variance}) with the
two-point function evaluated to first order (using Feynman rules) for a
scalar field in two dimensions and interacting with a potential
$(X^2-Y^2)^{\frac{1}{2}} |\psi^{(0)}|^2$. Following this analogy, the
variance $\tilde{\mathcal{F}}$ can be seen as a vacuum bubble.

Equation~(\ref{expression for the variance}) implies that a white noise
interaction is such that the different contributions to the modulus
square of the perturbations add up incoherently. 
Moreover, the perturbative corrections receive contributions from all regions of minisuperspace where the unperturbed wave function is supported.

\subsection{Results}
A discussion about the implications of the results of this section is now in order. The boundary conditions (\ref{boundary conditions}) imposed on the Green function (\ref{Feynman's Green function}) treat the two directions of the internal time $\phi$ symmetrically. Therefore, the amplitude for a universe expanding from $X=-Y$ to the point with coordinates $(X_0,Y_0)$ is the same as that for having a collapsing universe\footnote{Or anti-universe, note the close analogy with CPT symmerty in ordinary relativistic QFT.} going from $(X_0,Y_0)$ to $X=-Y$. It is also clear from (\ref{expression for the variance}) and the property $G(X-X^{\prime},Y-Y^{\prime})=G(X-X^{\prime},Y^{\prime}-Y)$ that
\be\label{Eq:WDW:TimeInversion}
\tilde{\mathcal{F}}(X,Y)=\tilde{\mathcal{F}}(X,-Y).
\ee
Therefore, fluctuations in the perturbed wave function are symmetric under time inversion. The result expressed by Eq.~(\ref{Eq:WDW:TimeInversion}) is actually more general and holds for any interaction function $g(\alpha,\phi)$ which respects this symmetry property. In this case, the arrow of time is thus determined by the unperturbed solution for the background and points towards the direction in which the universe expands, in agreement with the interpretation of Ref.~\cite{Kiefer:1988}. Whether this cosmological arrow of time agrees with the thermodynamic one, defined by the direction of growth of inhomogeneities, remains an open problem\footnote{The problem remains open also in LQC. Progress on the inclusion of inhomogeneities in the LQC framework using lattice models can be found in Ref.~\cite{WilsonEwing:2012}. For a study of inhomogeneities using effective equations, see instead Ref.~\cite{Fernandez-Mendez:2014} }. The suggestive idea of such an identification between the two fundamental arrows of time is old and was first proposed by Hawking~\cite{Hawking:1985} in the context of the sum-over-geometries approach to quantum cosmology, but later disproved by Page~\cite{Page:1985}. Their arguments are based on formal properties of the wave function of the universe, defined by a path-integral over compact Euclidean metrics (\emph{i.e.} with no boundary). However, it is not clear whether they have a counterpart in other approaches to quantum cosmology, such as the one we considered in this work. In order to be able to provide a satisfactory answer to this question, one must not neglect the role played by inhomogeneities. In fact, their dynamics might display interesting features especially where quantum effects become more relevant, \emph{i.e.} close to the singularities or at the turning point (recollapse). In particular, the symmetry (or the lack thereof) of the dynamics around the latter will be crucial, since that is the point where the cosmological arrow of time defined by the background undergoes an inversion~\cite{Hawking:1985}.

\begin{figure}
  \centering
 \includegraphics[width=0.6\textwidth]{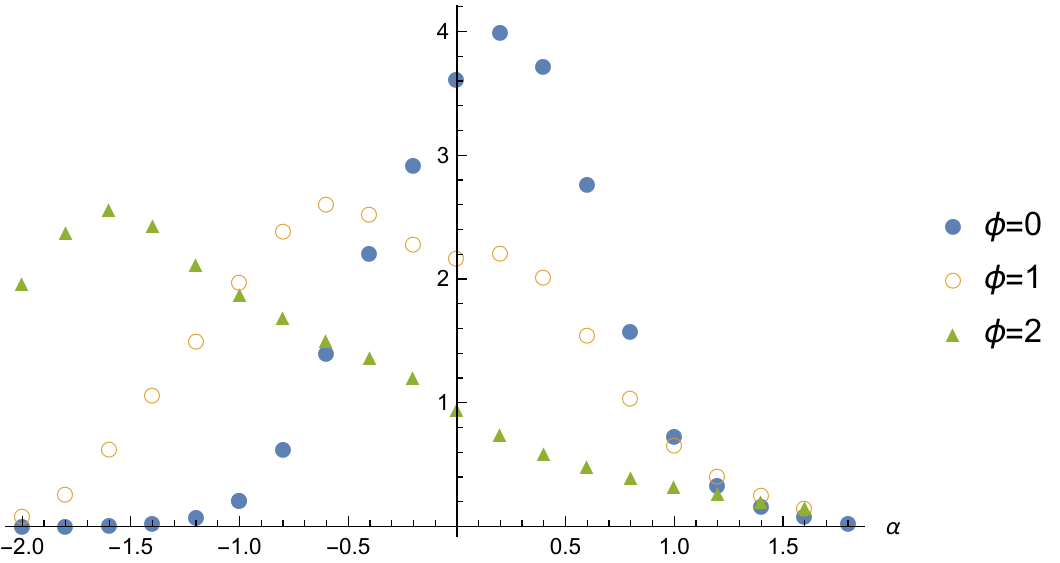}
\caption[Evolution of the perturbed wave packet]{Plot of $\langle|\psi^{(1)}|^2\rangle$ as a function of $\alpha$ for different values of the internal time $\phi$ and for $\bar{k}=b=1$. The perturbation is maximally focused at the time of recollapse, whereas it spreads  and develops a secondary maximum when approaching the singularities.}
\label{fig:curves_corrections}
    \end{figure}

Another important aspect concerns the semiclassical properties of the perturbed states that we constructed. From Figure \ref{fig:curves_corrections} one sees that the perturbation at a fixed $\phi$ is a rapidly decreasing function of $\alpha$ with Gaussian tails. Therefore, the result of a white noise perturbation on a wave packet solution of the WDW equation is still a wave packet. The position of the peak of the perturbation, like that of the unperturbed state, is a monotonically increasing function of $\phi$ for negative values of $\phi$, while it is monotonically decreasing for positive $\phi$. In this sense we can regard the perturbed state as retaining the property of semiclassicality of the unperturbed state. It is therefore possible to perform the classical limit, which can be obtained when $\hbar\to0$, or equivalently by considering infinitesimally narrow unperturbed wavepackets $\psi^{(0)}$, \emph{i.e.} in the limit $b\to 0$. More precisely, one should compute expectation values of physical observables (\emph{e.g.} the scale factor) on the perturbed state $\psi^{(0)}+\psi^{(1)}$ and perform an asymptotic expansion near $\hbar=0$. One can also simultaneously expand around $\varepsilon=0$, and the classical dynamics would then be seen to get corrections in the form of additional terms involving powers of $\varepsilon$ and $\hbar$, to be interpreted as stochastic and quantum effects respectively (or a combination of the two).

\section{Conclusion}\label{Section:WDW:Conclusions}

Motivated by LQC and GFT we considered an extension of WDW
 cosmology with additional interaction terms in minisuperspace. Such terms can be seen as a particular case of the model considered in Ref.~\cite{Calcagni:2012}, which was proposed as an approach to quantum dynamics of inhomogeneous cosmology. In general, in that model inhomogeneities would lead to non-linear differential equations for the quantum field. This is indeed the case when the `wave function' of the universe is interpreted as a quantum field or even as a classical field describing the hydrodynamics limit of GFT.

In the framework of a first quantised cosmology, we only considered linear modifications of the theory in order to ensure the validity of the superposition principle. Our work \cite{deCesare:2015vca} represents indeed a first step towards a more general study, that should take into account non-linear and possibly non-local interactions in minisuperspace. However, the way such terms arise, their exact form, and even the precise way in which minisuperspace dynamics is derived from a given candidate fundamental theory of Quantum Gravity, should be dictated by the full theory itself.

Assuming that the additional interactions are such that deviations from the standard WDW equation (and from the Friedmann equation in the classical limit) are small, we developed general perturbative methods which allowed us to solve the modified WDW equation. We considered a closed FLRW universe filled with a massless scalar field to define an internal time, for which wave packet solutions are known explicitly and propagate with no dispersion.
A modified WDW equation is then obtained in the large volume limit of a particular GFT inspired extension of LQC for a closed FLRW universe. Perturbative methods are then used to find the corrections given by self-interactions of the universe to the exact solution given in Ref.~\cite{Kiefer:1988}. To this end, the Feynman propagator of the WDW equation is evaluated exactly (\emph{i.e.} non-perturbatively) by means of a conformal map in minisuperspace and (after a Wick rotation) using the method of the image charges that is familiar from electrostatics. This is potentially interesting as a basic tool for any future perturbative analysis of non-linear minisuperspace dynamics for a closed universe. Our choice of the boundary conditions satisfied by the Green function is compatible with  a cosmological arrow of time given by the expansion of the universe.

A Helmoltz-like equation was obtained from WDW when the extra interaction does not depend on the internal time. Its Green kernel was evaluated exactly and turns out to depend on a free parameter $\eta$ related to the choice of boundary conditions. Further work is needed to clarify the possible link between the value of the parameter $\eta$ and different boundary proposals in Quantum Cosmology.

We illustrated our perturbative approach in the
simple and physically motivated case in which the
perturbation is given by stochastic white noise. In this phenomenological
model the stochastic interaction term can be seen as describing the
interaction of the cosmological background with other degrees of freedom
of the gravitational field, which are not included in the minisuperspace formulation of the model. Calculating the variance of the
statistical fluctuations of the wave function, we found that a white
noise interaction is such that the different contributions to the
modulus square of the perturbations add up incoherently. Moreover, the perturbed wavefunction retains the property of semiclassicality. Furthermore, the width of the perturbation of the wave function reaches a minimum at recollapse. We discussed the possible implications of the model for the cosmological arrow of time.



\newpage

\chapter{Group Field Theory approach and emergent Cosmology}\label{Chapter:GFT}
In this chapter we discuss the Group Field Theory (GFT) formalism and its applications to Cosmology. GFT is a non-perturbative and background independent approach to Quantum Gravity formulated as a field theory on a Lie group. It has several remarkable connections with other approaches, in particular with LQG and Spin Foam models. The fundamental degrees of freedom are open spin-network vertices, interpreted as the basic building blocks, or quanta, of geometry. The formalism allows one to study the dynamics of a large (and \emph{a priori} variable) number of microscopic degrees of freedom in quantum geometry by using a field theory formalism. In GFT the dynamics of the cosmological background is emergent and can be obtained from the full theory, \emph{i.e.} without symmetry reduction before quantisation. This is achieved by considering a particular class of states, namely `condensate' states. In particular, we will focus on the implications of the emergent cosmology scenario for early universe cosmology, studied by the author in Refs.~\cite{deCesare:2016axk,deCesare:2016rsf,deCesare:2017ynn}.

The plan of the chapter is as follows. In Section~\ref{Sec:GFT:Motivation} we give an introduction to the ideas underlying the GFT programme. We discuss the motivation for our work and relate it to the main results obtained in GFT, as well as other approaches to Quantum Gravity and early universe cosmology. In Section~\ref{sec:1} we give a brief introduction to the formalism. In Section~\ref{Sec:GFT:MeanField} we study the dynamics of the GFT field in a mean field approximation. We then apply the mean field method to the GFT formulation of the EPRL Lorentzian model. The additional conditions imposed on the mean field in order to recover the dynamics of the background are discussed in detail.

In Section~\ref{Sec:NoInteractions} we study the evolution of the emergent cosmological background in a model with no interactions between the geometry quanta. The dynamics of expansion is governed by an effective Friedmann equation, which includes quantum gravitational correction terms. Those can be understood as effective fluids coupled to the emergent classical background. The occurrence of a bounce which resolves the initial spacetime singularity is shown to be a general property of the model, \emph{i.e.} it holds regardless of the specific choice of initial conditions. An important feature of this model is the occurrence of an early era of
accelerated expansion, without the need to introduce an inflaton field with an appropriately chosen potential. However, the number of e-folds that can be obtained during this era, which we may call \emph{geometric inflation}, is very small compared to standard inflationary models.

In Section~\ref{Sec:GFT:Interactions} we assume a phenomenological perspective and include interactions between
\emph{spacetime quanta} in the model. In
particular, we show how GFT interactions lead to a recollapse of the
universe while preserving the bounce replacing the initial
singularity. It is remarkable that cyclic cosmologies are thus obtained in
this framework without any \emph{a priori} assumption on the geometry
of spatial sections of the emergent spacetime. Furthermore, we show
how interactions make it possible to have an early epoch of
accelerated expansion, which can be made to last for an arbitrarily
large number of e-folds.

Section~\ref{sec:Perturbations} deals with the dynamics of perturbations representing deviations from perfect isotropy. Working in the mean field approximation of the GFT formulation of the Lorentzian EPRL model, we derive the equations of motion for such perturbations to first order. We then study these equations around a specific simple isotropic background, characterised by the fundamental representation of SU(2), and in the regime of the effective cosmological dynamics corresponding to the bouncing region replacing the classical singularity, well approximated by the free GFT dynamics. In this particular example, we identify a region in the parameter space of the model such that perturbations can be large at the bounce but become negligible away from it, \emph{i.e.} when the background enters the non-linear regime.
We also study the departures from perfect isotropy by introducing specific quantities, such as the \emph{surface-area-to-volume} ratio and the \emph{effective volume per quantum}, which makes them quantitative.

\section{Motivation}\label{Sec:GFT:Motivation}
Our current understanding of the early universe based on Standard Cosmology is limited by the initial singularity, and more generally by the lack of control over the deep Quantum Gravity regime (around the Planck scale). The occurrence of spacetime singularities in General Relativity is generic for matter satisfying suitable energy conditions, as shown by Hawking and Penrose \cite{Penrose:1964wq,Hawking:1966sx,Hawking:1969sw,Hawking:1973uf}. It is believed that quantum gravitational effects taking place at the Planck scale could lead to a resolution of the singularities, as first suggested in Ref.~\cite{DeWitt:1967yk}. Moreover, since the onset of inflation is supposed to take place at Planckian times, the dynamics of the
universe at this stage should find a more suitable formulation so as
to take quantum gravitational effects into account. In fact, it is
conceivable that the quantum dynamics of the gravitational field
itself could effectively give rise to dynamical features similar to
those of inflationary models, without the need to introduce a new
hypothetical field (the inflaton) with an \emph{ad hoc} potential. 

The idea of singularity resolution in Quantum Gravity has been very fruitful in theoretical frameworks based on a fundamentally discrete quantum geometry, such as background independent approaches to Quantum Gravity. The first realisation was made within the context of Loop Quantum Cosmology (LQC), a quantisation inspired by Loop Quantum Gravity (LQG) of the symmetry-reduced, cosmological sector of GR. The results obtained in such framework have shown that the initial singularity is replaced by a quantum bounce, which is a robust feature \cite{Bojowald:2001xe,Ashtekar:2011ni,Banerjee:2011qu,Bojowald:2008zzb,Bojowald:2012we}.
Other models featuring a cosmological bounce, inspired by different approaches to Quantum Gravity, are reviewed in Ref.~\cite{Brandenberger:2016vhg} and references therein.

A particular background independent approach to Quantum Gravity in which bouncing cosmological solutions were found is Group Field Theory (GFT). Remarkably, this result was obtained within the complete theory, \emph{i.e.} without symmetry reducing before quantising, as it is done instead in LQC\footnote{Reviewed in Appendix~\ref{Appendix:LQC}.}. GFT represents a higher dimensional generalisation of matrix models \cite{Freidel:2005qe,Oriti:2006se,Oriti:2011jm,Baratin:2011aa,Krajewski:2012aw,Kegeles:2015oua,Kegeles:2016wfg}, like random tensor models, but further enriched by the group-theoretic data characterising the quantum states of geometry also in LQG. It can be understood indeed also as a second quantisation of LQG \cite{Oriti:2013,Thiemann:2007zz,Rovelli:2004tv}. In this formulation, the elementary degrees of freedom (single particle states) are open spin network vertices, with their edges labelled by irreducible representations of a Lie group $G$ (typically $G=\mbox{SU(2)}$). They are dual to quantum tetrahedra\footnote{See Appendix~\ref{sec:Intertwiner} for a review of the kinematics of a quantum tetrahedron.}, which can be understood as the basic building blocks, or quanta, of a spatial geometry.

GFTs for $d$-dimensional Quantum Gravity are field theories on a Lie group $G^d$, quantised \emph{e.g.} by path integral methods. In the models discussed in this chapter, we will be interested in the case $d=4$. For a given GFT, the perturbative expansion of the path integral can be expressed as a sum of transition amplitudes of a given Spin Foam (SF) model, a covariant definition of the LQG dynamics \cite{Ooguri:1992eb,Reisenberger:2000zc,Freidel:2007py,Krajewski:2010yq,Baratin:2011tx,Kaminski:2009fm,Han:2011rf,Baratin:2010wi,Rovelli:2011eq,Perez:2012wv}.

At a fundamental level, spacetime is absent in the formulation of GFT, and rather an emergent concept ~\cite{Oriti:2013jga}, reminiscent of the way we understand collective phenomena in condensed matter physics. This general perspective has been advocated and outlined in Ref.~\cite{Oriti:2013jga}, and in related contexts in Ref.~\cite{Konopka:2006hu}. The programme of recovering cosmology from GFT along this perspective was started in Refs.~\cite{Gielen:2013kla,Gielen:2013naa,Sindoni:2014wya}. In following studies, Refs.~\cite{Oriti:2016qtz,Oriti:2016ueo}, it was shown that GFT allows to derive an effective Friedmann equation from the evolution of the \emph{mean field}, within a generalised GFT formulation of the Lorentzian EPRL spin foam model (Ref.~\cite{Engle:2007wy}).

The mean field describes the collective dynamics of quanta of space/geometry. Solutions to the effective Friedmann equation describe then an emergent classical background obtained from a full theory of quantum gravity. As a general feature, the solutions of the model exhibit a quantum bounce which resolves the initial singularity, provided that the hydrodynamics approximation of GFT (where the mean field analysis is confined) holds \cite{Oriti:2016qtz,Oriti:2016ueo,deCesare:2016axk,deCesare:2016rsf}. This is discussed in Section~\ref{Sec:NoInteractions} for the non-interacting GFT model and it is shown in Section~\ref{Sec:GFT:Interactions} to be a robust feature under the inclusion of interaction terms. In Section~\ref{Sec:GFT:Interactions} we will discuss a number of other interesting results concerning interacting GFT models and their effective cosmological dynamics studied by the author, namely: the acceleration phase after the bounce can be long lasting without the need to introduce an inflaton field, but purely driven by the GFT interactions; the same interactions can produce a cyclic evolution for the universe, with infinite expansion and contraction phases (and no singularities). In Section~\ref{sec:Perturbations}, microscopic anisotropies of the fundamental building blocks are studied perturbatively and a region of parameter space is determined such that perturbations stay bounded away from the bounce; hence, they can be neglected compared to the expanding background in this regime.

\section{Brief review of the GFT formalism and its emergent cosmology}\label{sec:1}
In this Section, we review the basics of the GFT formalism for Quantum Gravity\footnote{The reader interested in a more comprehensive review of the mathematical and foundational aspects of GFT is referred to Refs.~\cite{Freidel:2005qe,Oriti:2006se,Krajewski:2012aw}.}. In particular, we will discuss in detail those aspects which will be needed for the cosmological applications considered in our work\footnote{Reviews of the recent developments of the cosmological applications of GFT can be found in Refs.~\cite{Gielen:2016dss,Oriti:2016acw}.}.

\subsection{The GFT formalism. Kinematics}\label{Sec:GFT:BoundaryStates}
The GFT field for pure gravity is defined on four copies of a Lie group $G$, to be extended when introducing matter degrees of freedom\footnote{For instance, when a scalar field is considered the domain is $G^4\times\mathbb{R}$, see below. Geometry and matter degrees of freedom are treated on the same level, \emph{i.e.} as arguments of the GFT field rather than fields defined on a differentiable manifold, as in classical GR.}. Therefore, its argument is an array of four group elements $(g_1,g_2,g_3,g_4)$ which we will denote as $g_\nu$, for brevity. We use instead the notation $g_{i}$ for specific components of $g_\nu$. Analogously, we introduce the notations $j_\nu$, $m_\nu$, etc. for the spins and their counterparts with Latin indices. Sometimes more than four spins are considered. In that case, only lower Latin indices appear in the equations and refer to the individual spins. Upper case Latin indices denote quadruples of holonomies $g^A_\nu$ attached to the edges emanating from a vertex. Throughout this chapter, Newton's constant will be denoted by $G_{\rm N}$ to avoid confusion with the notation used for the Lie group $G$. We also assume $\hbar=G_{\rm N}=c=1$, unless otherwise stated.

We introduce the field operator $\hat{\varphi}(g_\nu)$, which is a scalar on $G^4$, and its adjoint $\hat{\varphi}^{\dagger}(g_\nu)$. These operators act on a Fock space, whose vacuum state $|0\rangle$ is defined as the state which is annihilated by the field operator\footnote{Moreover, we expect that such state must be invariant under the automorphism group of $G^4$, \emph{i.e.} the group of isometries of the Cartan-Killing form on $G^4$. This is in analogy with standard QFT, in which the vacuum is invariant under the group of isometries of the Minkowski metric, \emph{i.e.} the Lorentz group (the Euclidean group in Euclidean field theories). We are not aware of any proof of the uniqueness of such vacuum for GFTs. In principle, different inequivalent vacua may exist that are invariant under global isometries on a curved manifold, as in the case \emph{e.g.} of the $\alpha$-vacua in DeSitter space.}
\be
\hat{\varphi}(g_\nu)|0\rangle=0~, \hspace{1em} \forall g_\nu~.
\ee
The vacuum is interpreted as a \emph{no geometry} state. This is to be contrasted with the perturbative quantisation of linearised GR on a fixed background (\emph{e.g.} Minkowski spacetime) in which the vacuum corresponds to a state with no gravitons. The field operators satisfy the following (bosonic) commutation relations
\begin{align}
[\varphi(g_\nu),\hat{\varphi}^{\dagger}(g^{\prime}_\nu)]=\mathbb{I}_{\rm G}(g_\nu,g^{\prime}_\nu)~,\label{Eq:GFT:CommutationRelations1}\\
\hspace{1em} [\varphi(g_\nu),\varphi(g^{\prime}_\nu)]=[\varphi(g^{\prime}_\nu),\varphi(g^{\prime}_\nu)]=0~\label{Eq:GFT:CommutationRelations2},
\end{align}
with\footnote{The Dirac delta function on a Lie group can be defined by means of the exponential map. Considering $g\in G$ we have $\delta(g)=\delta(\exp X)=\prod_{\alpha=1}^3 \delta(X^\alpha)$, where $X$ is an element of the Lie algebra $\mathfrak{g}$ and $\delta(X^\alpha)$ is the ordinary Dirac delta defined on $\mathfrak{g}$. 
Note that, while the ordinary Dirac delta is peaked at zero, its counterpart on a Lie group is peaked at the identity.}
\be\label{Eq:GFT:DeltaGaugeInvariant}
\mathbb{I}_{\rm G}(g_\nu,g^{\prime}_\nu)=\int_{\rm G}\de h\;  \prod_{i=1}^4\delta\left(g_i h (g^{\prime}_i)^{-1}\right)~.
\ee
Integration is performed with respect to the Haar measure on $G$, which is assumed to be normalised to unity, \emph{i.e.} $\int_G \de g=1$. The GFT field must satisfy the right-invariance property under the diagonal action of $G$ on $G^4$, \emph{i.e.}
\be
\varphi(g_\nu)=\varphi(g_\nu~h)~,\forall h\in G.
\ee
This implies that the field is actually a function defined on $G^4/G$.
We will come back later to the interpretation of the commutation relations (\ref{Eq:GFT:CommutationRelations1}),~(\ref{Eq:GFT:CommutationRelations2}) from a geometrical point of view. We remark that, in case a non-compact group $G$ is considered, the expression (\ref{Eq:GFT:DeltaGaugeInvariant}) needs appropriate regularisation \cite{Gielen:2016dss}. For our applications this will not represent an issue, since we will only be interested in the compact case, more specifically $G=\mbox{SU(2)}$.

A single excitation of the vacuum is given by the `one-particle state'
\be\label{Eq:GFT:SingleParticle}
|g_\nu\rangle\equiv\hat{\varphi}^{\dagger}(g_\nu)|0\rangle~.
\ee
This is interpreted as a single four-valent vertex with holonomies $g_\nu$ on its edges. 
\begin{figure}
\begin{center}
\includegraphics[width=0.5\columnwidth]{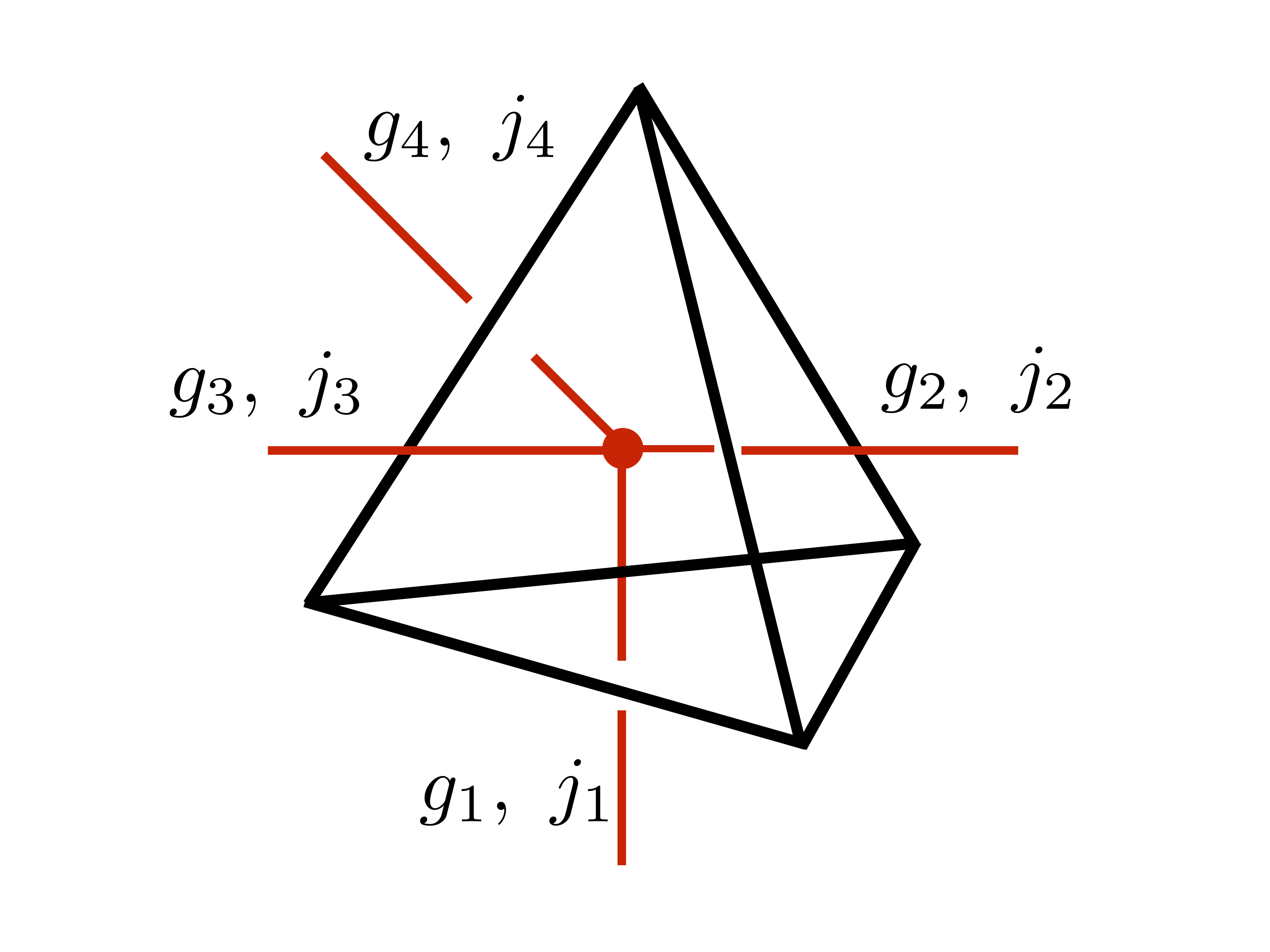}
\caption[Quantum tetrahedron]{An open spin network vertex (in red), corresponding to an elementary excitation of the GFT vacuum. To each link is attached a group element $g_i$ (\emph{holonomy}). The four-valent vertex is dual to a tetrahedron (in black), whose faces are labelled by $\mathfrak{su}(2)$ representations with spin $j_i$. The four spins must satisfy the closure condition ${\bf J}_1+{\bf J}_2+{\bf J}_3+{\bf J}_4=0$, implementing gauge invariance at the vertex.}\label{Fig:Tetrahedron}
\end{center}
\end{figure}
A basis of multiparticle states with $N$ vertices is represented by the states
\be\label{Eq:GFT:DisconnectedGraphs}
|g^1_\nu,\dots, g^N_\nu\rangle=\prod_{A=1}^N \hat{\varphi}^{\dagger}(g^A_\nu)|0\rangle~.
\ee
The states (\ref{Eq:GFT:DisconnectedGraphs}) correspond to a totally disconnected graph. More interesting graph structure can be obtained by considering particular superpositions of such states, as we are going to explain.

The analogue of a `wave packet' in this framework is given by the state
\be\label{Eq:GFT:OneParticle}
|f\rangle\equiv\int\de^4g\; f(g_\nu)\hat{\varphi}^{\dagger}(g_\nu)|0\rangle~,\hspace{1em} f(g_\nu)=\langle g_\nu|f\rangle~,
\ee
where we introduced the notational shorthand $\de^4g$ for $\prod_{i=1}^4\de g_i$, where $\de g_i$ denotes the Haar measure on the $i$-th copy of G. The most general multiparticle state is defined as
\be\label{Eq:GFT:MultiParticle}
|f_N\rangle\equiv\frac{1}{\sqrt{N!}}\int\prod_{A=1}^N\de^4g^A\;f(g_\nu^1,\dots,g_\nu^N)\prod_{A=1}^N \hat{\varphi}^{\dagger}(g^A_\nu)|0\rangle~,
\ee
where $g_\nu^A$ are quadruples of holonomies labelled by $A=1,\dots,N$. The wavefunction 
\be
f(g_\nu^1,\dots,g_\nu^N)=\frac{1}{\sqrt{N!}}\langle g^1_\nu,\dots, g^N_\nu| f_N\rangle~
\ee
 is completely symmetric under permutations of the quadruples of holonomies, as a consequence of the bosonic commutation relations (\ref{Eq:GFT:CommutationRelations2}). $|f_N\rangle$ can be interpreted as the state of a graph with $N$ vertices, with topological and geometrical properties depending on the properties of the wave-function $f(g_\nu^1,\dots,g_\nu^N)$. This will be explained in more detail below.

The commutation relation (\ref{Eq:GFT:CommutationRelations1}) implies further restrictions on the functional dependence of the single-particle and multi-particle wave functions. In fact, one finds for the one-particle state (\ref{Eq:GFT:OneParticle})
\be
f(g_\nu)=\langle0|\varphi(g_\nu)|f\rangle=\int\de k\;f(g_\nu k)~,
\ee
which implies that the wave function must be invariant under the right action of $G$ on the quadruple $g_\nu$, \emph{i.e.}
\be
f(g_\nu)=f(g_\nu k)~.
\ee
Even though in this formalism it is clearly a global symmetry, this property is usually referred to as gauge invariance since it represents the counterpart of $\rm SU(2)$ gauge transformations at each vertex of a spin network in LQG. 
Next we consider a two-particles state
\be
|f_2\rangle\equiv\frac{1}{\sqrt{2}}\int\prod_{A=1}^2\de^4g^A\;f(g_\nu^1,g_\nu^2) \hat{\varphi}^{\dagger}(g^1_\nu)\hat{\varphi}^{\dagger}(g^2_\nu)|0\rangle~,
\ee
for which we find
\be\label{Eq:GFT:Consistency2ParticleState}
f(g_\nu^1,g_\nu^2)=\frac{1}{\sqrt{2!}}\langle g^1_\nu, g^2_\nu| f_2\rangle=\frac{1}{2!}\int\de k~\de h\; \Bigg(f\left(g^1_\nu~ k,g^2_\nu~ h\right)+f\left(g^2_\nu ~k,g^1_\nu~ h\right)\Bigg)~.
\ee
The last equation is satisfied provided that: the multi-particle wave-function is right-invariant with respect to each of its arguments $g^A_\nu$
\be
f\left(g^1_\nu~ k,g^2_\nu\right)=f\left(g^1_\nu,g^2_\nu~k\right)=f\left(g^1_\nu,g^2_\nu\right)~,\hspace{1em} \forall k\in G
\ee
and it is invariant under permutations of its arguments
\be
f\left(g^1_\nu,g^2_\nu\right)=f\left(g^2_\nu,g^1_\nu\right)~.
\ee
A straightforward generalisation of the above shows that similar properties must also hold for the wave function of a generic multi-particle state with $N$ vertices.

It is clear at this stage that the non-standard bosonic commutation relations (\ref{Eq:GFT:CommutationRelations1}), (\ref{Eq:GFT:CommutationRelations2}) enforce gauge invariance (right-invariance) at each vertex. States $|f_N\rangle$ of the type (\ref{Eq:GFT:MultiParticle}) can be interpreted geometrically as graphs consisting of a collection of identical vertices, that are gauge invariant and symmetric under permutations, with wave-function $f(g_\nu^1,\dots,g_\nu^N)$ satisfying the properties discussed above. Such graphs are in general disconnected. A more interesting class of graphs, which is closer to the states in LQG (spin networks) can be obtained as follows. Let us consider a generic multiparticle state $|f_N\rangle$ with wavefunction $f(g_\nu^1,\dots,g_\nu^N)$ and pick two distinct vertices $A$, $B$ and edges emanating from them, which we label as $i$ and $j$, respectively. We define a modified wavefunction as (no summation over repeated indices) \cite{Gielen:2016dss}
\be\label{Eq:GFT:Gluing}
\tilde{f}(\dots,g_i^A,\dots, g_j^B,\dots)=\int_{G}\de h^{AB}_{ij}\; f(\dots,h^{AB}_{ij}g_i^A,\dots,h^{AB}_{ij}g_j^B,\dots)~.
\ee
It follows from the definition (\ref{Eq:GFT:Gluing}) that $\tilde{f}$ depends on the two holonomies $g_i^A$ and $g_j^B$ only through the combination $G^{AB}_{ij}=(g_j^B)^{-1}g_i^A $. The $i$-th edge emanating from the vertex $A$ (called the source) is incident on the vertex $B$ (the target), where it is identified with the $j$-th edge emanating from it. $G^{AB}_{ij}$ then represents the holonomy over the glued edges, in going from $A$ to $B$. Clearly, $G^{BA}_{ji}=(G^{AB}_{ij})^{-1}$ by definition. 
It is straightforward to generalise the construction to the case in which more than one pair of edges are glued together. We refer the reader to Ref.~\cite{Oriti:2013} for the general case.

As a simple example, we consider a two-particle state in which all edges with the same label $i$ are glued together. In this case, the gluing map is very simple and is given by $h^{12}_{ij}=\delta_{ij}h$, where $h\in G$. We have, using Eq.~(\ref{Eq:GFT:Gluing}) and the right-invariance of the Haar measure\footnote{\label{Footnote:LieGroupBiInvMetric}Every Lie group admits a right-invariant Haar measure and a left-invariant one. However, in general they do not coincide. It is possible to show that for compact Lie groups there is a Haar measure which is both right- and left-invariant \cite{taylorlectures:lie00}. In this case the Haar measure is said to be bi-invariant. Lie groups for which such a bi-invariant Haar measure exists are called unimodular. Some non-compact groups, such as $SL(2,\mathbb{R})$, also admit a bi-invariant metric.}
\be
\tilde{f}\left(g^1_\nu,g^2_\nu\right)=\int_G\de h\; f(h g^1_{\nu},hg^2_{\nu})=\int_G\de h\; f(h (g^2_{\nu})^{-1}g^1_{\nu},h)\equiv F\left( (g^2_{\nu})^{-1}g^1_{\nu}\right)~.
\ee
The holonomies for this graph are given by $G^{12}_{ij}=(G^{21}_{ji})^{-1}=(g^2_{j})^{-1}g^1_{i}\delta_{ij}$ (no summation).

Delicate aspects of the relation between the Fock space of GFTs and the kinematical Hilbert space of LQG are discussed in Ref.~\cite{Oriti:2013}. Here we would like to remark some differences in the kinematical properties of states in the two approaches, which are also highlighted in Ref.~\cite{Gielen:2016dss}. In particular, the Fock space inner product is not equivalent to the inner product in the kinematical Hilbert space of LQG. This is due to the fact that GFT states corresponding to different graphs are not necessarily orthogonal if they have the same occupation numbers. Another difference is the symmetry of the wave function in GFT under permutations of the vertices, which is not satisfied by all LQG states. The limitation to simplicial graphs is instead not an essential one, and can be overcome by a suitable generalisation of the formalism to allow for more general combinatorial complexes \cite{Oriti:2014yla}. The states discussed in this Section are interpreted as the \emph{boundary states}, \emph{i.e.} they encode information about the quantum geometry of the spatial sheets of a foliation. The implementation of the dynamics in GFT gives rise to a sum over triangulations of a four dimensional spacetime with assigned boundary data. This is discussed in Section~\ref{Sec:GFT:Dynamics}.

\subsection{Kinematical Operators}\label{Sec:GFT:KinematicalOperators}
\subsubsection{One-body Operators}
A complete specification of the kinematics of the theory requires the introduction of kinematical operators. In analogy with condensed matter physics, we can introduce one-body and two-body operators\footnote{Their generalisation to $n$-body operators, \emph{albeit} possible, is not needed for our purposes.}. The general expression of a one-body operator is the following
\be\label{Eq:GFT:OneBodyOperator}
\hat{T}_1=\int \de^4g \de^4g^{\prime}\; t(g_\nu,g^{\prime}_\nu)\hat{\varphi}^{\dagger}(g_\nu)\hat{\varphi}(g^\prime_\nu)~.
\ee
If $\hat{T}_1$ has a first-quantised counterpart $\hat{t}$, one has
\be\label{Eq:GFT:OneBodyOperatorFrom1stQuantized}
t(g_\nu,g^{\prime}_\nu)=\langle \psi_{g_\nu}|\hat{t}|\psi_{g^{\prime}_\nu}\rangle~,
\ee
where $\psi_{g_\nu}$ is the (improper) eigenfunction of the position operator\footnote{Note that, in order to avoid confusion with the one-particle states $|g_\nu\rangle$ introduced in Eq.~(\ref{Eq:GFT:SingleParticle}) we use a different notation from the one in Ref.~\cite{Gielen:2016dss}.} 
 on $G^4$ \emph{in the first quantised theory}. If Hermiticity of $\hat{T}_1$ is assumed, the following condition must be satisfied
 \be
 t(g_\nu,g^{\prime}_\nu)=\overline{t(g^{\prime}_\nu,g_\nu)}~.
 \ee
In case the operator is obtained as the second quantised counterpart of a Hermitian operator in the first quantised theory, Hermiticity is automatically satisfied, as one can easily verify using Eq.~(\ref{Eq:GFT:OneBodyOperatorFrom1stQuantized}).
 
 Equations~(\ref{Eq:GFT:OneBodyOperator}),~(\ref{Eq:GFT:OneBodyOperatorFrom1stQuantized}) allow for a straightforward implementation of a vocabulary between observables in LQG and in GFT. In LQG, observables are represented by geometric operators. Perhaps the most important examples are given by the volume and area operators. In GFT one has for the volume
 \be\label{Eq:GFT:Volume2ndQuantized}
 \hat{V}= \int \de^4g \de^4g^{\prime}\; \langle \psi_{g_\nu}|\hat{V}^{\scriptscriptstyle \rm LQG}|\psi_{g^{\prime}_\nu}\rangle \hat{\varphi}^{\dagger}(g_\nu)\hat{\varphi}(g^\prime_\nu)~.
\ee
The volume operator is the main observable which is relevant for cosmological applications of GFT.
The area of a plaquette dual to the $i$-th link is given by\footnote{\label{Footnote:SteffenGNewton}The definition of the area operator involves Newton's constant $G_{\rm N}$. In the formulation of the microscopic theory this constant does not appear explicitly. Rather, as we will show in the following, it is identified with some combinations of its parameters by comparing the dynamics of the emergent cosmological background with the standard Friedmann equation. In Ref.~\cite{Gielen:2016dss} it is argued that the gravitational constant is subject to renormalisation, and the constant $G_{\rm N}$ appearing in Eq.~(\ref{Eq:GFT:Area}) at a purely kinematical level should be interpreted as the low energy effective gravitational constant.}
 (see Ref.~\cite{Gielen:2016dss})
\be\label{Eq:GFT:Area}
\hat{A}_i=8\pi \beta \hbar G_{\rm N} \int \de^4g \de^4g^{\prime}\; \hat{\varphi}^{\dagger}(g_\nu)\sqrt{-\Delta_i}~\hat{\varphi}(g^\prime_\nu)~.
\ee
We remark that, although $\hat{A}_i$ is a perfectly well-defined operator, its geometrical interpretation is not clear when we go beyond the case of a single disconnected vertex. This is essentially due to the properties that we expect to be satisfied by the area observable in the classical limit. Namely, the area defined in Ref.~(\ref{Eq:GFT:Area}) is an extensive operator, whereas from the point of view of differential geometry it should not. Additional difficulties in the interpretation of $\hat{A}_i$ as an observable for generic states is due to the bosonic statistics of GFT states.

A very important one-body operator, which is a specific feature of the second quantised formalism and does not have an analogue in LQG is given by the number operator $\hat{N}$. It is defined as
\be\label{Eq:GFT:NumberOperator}
\hat{N}=\int \de^4g \; \hat{\varphi}^{\dagger}(g_\nu)\hat{\varphi}(g_\nu)~.
\ee
States of the type (\ref{Eq:GFT:MultiParticle}) are eigenstates of the number operator, with eigenvalue equal to the number of vertices
\be
\hat{N}|f_N\rangle=N |f_N\rangle~.
\ee
An arbitrary state in Fock space is expressed as a superposition of $|f_N\rangle$ states, which may as well have a different number of vertices $N$. In this case, it is still possible to talk about the expectation value of the number of vertices $\langle \hat{N}\rangle$ or higher order moments of its distribution $\langle \hat{N}^r\rangle$, due to the definition of $\hat{N}$ as an operator in second quantisation.

\subsubsection{Basis Change}
It can be convenient to introduce a change of basis, which may be motivated by simplifications in the evaluation of the matrix elements of one-body operators (\emph{e.g.} the volume) or to simplify the dynamics (see Section~\ref{sec:2}). We will consider the expansion of the GFT field in a basis corresponding to a given set of guantum numbers $\vec{\chi}$. We introduce creation and annihilation operators $a_{\vec{\chi}}$, $a_{\vec{\chi}}^{\dagger}$ satisfying the standard harmonic oscillator algebra
\be\label{Eq:GFT:HOalgebra}
[a_{\vec{\chi}},a_{\vec{\chi}^{\prime}}^{\dagger}]=\delta_{\vec{\chi},\vec{\chi}^{\prime}}~,\hspace{1em}[a_{\vec{\chi}},a_{\vec{\chi}^{\prime}}]=[a_{\vec{\chi}}^{\dagger},a_{\vec{\chi}^{\prime}}^{\dagger}]=0~.
\ee
The single-particle wave functions with quantum numbers $\vec{\chi}$ are denoted by $\psi_{\vec{\chi}}$. They satisfy the normalisation condition
\be\label{Eq:GFT:1PNormalization}
\int\de^4g\;\overline{\psi_{\vec{\chi}}}(g)\psi_{\vec{\chi}^{\prime}}(g)=\delta_{\vec{\chi}{\vec{\chi}^{\prime}}}~.
\ee
The expansion of the GFT field operators in the new basis is then given by
\be\label{Eq:GFT:ChangeBasis}
\hat\varphi(g_\nu)=\sum_{\vec{\chi}}\psi_{\vec{\chi}}(g_\nu)\hat{a}_{\vec{\chi}}~,\hspace{1em}\hat\varphi^{\dagger}(g_\nu)=\sum_{\vec{\chi}}\overline{\psi}_{\vec{\chi}}(g_\nu)\hat{a}^{\dagger}_{\vec{\chi}}~.
\ee
The number operator (\ref{Eq:GFT:NumberOperator}) has the following expression in the new basis
\be
\hat{N}=\sum_{\vec{\chi}}\hat{a}^{\dagger}_{\vec{\chi}}\hat{a}_{\vec{\chi}}~.
\ee
Additional constraints must be imposed on the quantum numbers $\vec{\chi}$ for each term in the sums (\ref{Eq:GFT:ChangeBasis}) in order to satisfy the commutation relation (\ref{Eq:GFT:CommutationRelations1}). In other words the quantum numbers representing the components of the array $\vec{\chi}$ are not all independent. The origin of the constraints lies in the right-invariance property of the GFT field. In fact, it can be shown that the commutation relations are satisfied provided that Eq.~(\ref{Eq:GFT:HOalgebra}) holds and that $\psi_{\vec{\chi}}(g)$ be right-invariant. The last requirement in turn implies constraints on the quantum numbers, as we will illustrate below by means of a particularly relevant example. Such constraints represent a significant difference with standard QFT and stem from the geometric interpretation of the theory.

As an example, we consider a GFT on the Lie group $G=\mbox{SU(2)}$ and take $\vec{\chi}$ to represent spin quantum numbers of the complete set of observables (for a single particle) $(({\bf J}^{i})^2,J_3^{~i})$, \emph{i.e.} $\vec{\chi}=\{(j_i,m_i)\}$ with $i=1,\dots,4$. We consider the generic single particle wave-function that is an eigenstate of the set of four spin operators defined on $G^4$. It can be expressed as
\be\label{Eq:GFT:GenericState}
F^{j_\nu}_{m_\nu}(g_\nu)=\sum_{n_\nu} \mathcal{C}^{j_1 j_2 j_3 j_4}_{m_1 m_2 m_3 m_4 n_1 n_2 n_3 n_4}~  D^{j_1}_{m_1 n_1}(g_1)D^{j_2}_{m_2 n_2}(g_2) D^{j_3}_{m_3 n_3}(g_3)D^{j_4}_{m_4 n_4}(g_4)~.
\ee
The Wigner matrices $D^{j_i}_{m_i n_i}(g_i)$ form a complete basis of eigenvectors\footnote{This is essentially the content of the Peter-Weyl theorem, see Ref.~\cite{faraut2008analysis}.} of $({\bf J}^{i})^2$ and $J_3^{~i}$ for functions defined on the $i$-th copy of SU(2). The index $n_i$ is linked to the degeneracy of the eigenvalue $m_i$ for fixed $j_i$, and has the same range of values of $m_i$. For a review of the properties of Wigner matrices see Appendix~\ref{Appendix:HarmonicAnalysis}. However, the state (\ref{Eq:GFT:GenericState}) is not right-invariant in general. In fact, right-invariance leads to a constraint on the coefficients in Eq.~(\ref{Eq:GFT:GenericState}), which we will from now on denote more compactly as $C^{j_\nu}_{m_\nu n_\nu}$. In fact, under the right diagonal action of $h\in {\rm SU(2)}$ we have
\be
\begin{split}
F^{j_\nu}_{m_\nu}(g_\nu h)=\sum_{n_\nu} \mathcal{C}^{j_\nu}_{m_\nu n_\nu}~  \prod_{i=1}^4 D^{j_i}_{m_i n_i}(g_i h)
&=\sum_{l_\nu}\left(\sum_{n_\nu}\mathcal{C}^{j_\nu}_{m_\nu n_\nu} \prod_{r=1}^4 D^{j_r}_{l_r n_r}(h)\right)\prod_{i=1}^4 D^{j_i}_{m_i l_i}(g_i)\\
&=\sum_{n_\nu}\left(\sum_{l_\nu}\mathcal{C}^{j_\nu}_{m_\nu l_\nu} \prod_{r=1}^4 D^{j_r}_{n_r l_r}(h)\right)\prod_{i=1}^4 D^{j_i}_{m_i n_i}(g_i)~,
\end{split}
\ee
where we used the properties of the Wigner matrices and relabelled indices in the last step. Right-invariance $F^{j_\nu}_{m_\nu}(g_\nu h)=F^{j_\nu}_{m_\nu}(g_\nu)$ demands
\be\label{Eq:GFT:Algebraic constraint}
\mathcal{C}^{j_\nu}_{m_\nu n_\nu}=\sum_{l_\nu}\mathcal{C}^{j_\nu}_{m_\nu l_\nu} \prod_{r=1}^4 D^{j_r}_{n_r l_r}(h)~, \hspace{1em} \forall h\in {\rm SU(2)}~.
\ee
Note that the l.h.s in Eq.~(\ref{Eq:GFT:Algebraic constraint}) does not depend on $h$. Given that the rows of the Wigner matrices (as well as their columns) span irreducible representations of SU(2), the problem of determining the coefficients $C^{j_\nu}_{m_\nu n_\nu}$ is isomorphic to that of constructing a singlet state with four spins $(j_i,n_i)$, with $i=1,\dots,4$. This is discussed in more detail in Appendix~\ref{sec:Intertwiner}, where the solutions of Eq.~(\ref{Eq:GFT:Algebraic constraint}) are constructed explicitly. Solutions of Eq.~(\ref{Eq:GFT:Algebraic constraint}) are elements of the \emph{intertwiner space} of a four-valent open spin network vertex, defined as
\be\label{eq:HilbertSpace4Vertex}
\accentset{\circ}{\mathcal{H}}_{j_\nu}=\mbox{Inv}_{\mbox{\scriptsize SU(2)}}\left[\mathcal{H}_{j_1}\otimes\mathcal{H}_{j_2}\otimes\mathcal{H}_{j_3}\otimes\mathcal{H}_{j_4}\right]~.
\ee
Elements of the vector space $\accentset{\circ}{\mathcal{H}}_{j_\nu}$ are called intertwiners. A basis in $\accentset{\circ}{\mathcal{H}}_{j_\nu}$ is denoted by $\mathcal{I}^{j_\nu\iota}_{m_\nu n_\nu}$, where $\iota$ is an additional quantum number used to label basis vectors, \emph{e.g.} corresponding to eigenspaces of an Hermitian operator that commutes with all of the operators in the set $(({\bf J}^{i})^2,J_3^{~i})$. Finally, we can write the right-invariant single-particle wave function in the spin basis as\footnote{The intertwiner $\mathcal{I}^{j_\nu\iota}_{m_\nu n_\nu}$ must be appropriately normalised so as to satisfy (\ref{Eq:GFT:1PNormalization}).}
\be
\psi_{\vec{\chi}}(g_\nu)=\sum_{n_\nu}\mathcal{I}^{j_\nu\iota}_{m_\nu n_\nu}\prod_{i=1}^4 D^{j_i}_{m_i n_i}(g_i)~.
\ee
Using Eq.~(\ref{Eq:GFT:ChangeBasis}), the field operator in the spin basis reads as
\be
\hat\varphi(g)=\sum_{\vec{\chi}}\psi_{\vec{\chi}}(g)\hat{a}_{\vec{\chi}}=\sum_{j_\nu m_\nu n_\nu}\mathcal{I}^{j_\nu\iota}_{m_\nu n_\nu}\prod_{i=1}^4 D^{j_i}_{m_i n_i}(g_i) \hat{a}_{j_\nu m_\nu}~.
\ee
The expansion of $\hat\varphi^{\dagger}(g)$ can be worked out similarly.

In the spin basis the second quantised volume operator (\ref{Eq:GFT:Volume2ndQuantized}) takes a simpler form. In fact, by inverting Eqs.~(\ref{Eq:GFT:ChangeBasis}), we can re-express it as
\be
\hat{V}=\sum_{\vec{\chi},\vec{\chi}^{\prime}}\langle \psi_{\vec{\chi}}|\hat{V}^{\scriptscriptstyle \rm LQG}|\psi_{\vec{\chi}^{\prime}}\rangle\hat{a}^{\dagger}_{\vec{\chi}}\hat{a}_{\vec{\chi}^{\prime}}~,
\ee
where $\langle \psi_{\vec{\chi}}|\hat{V}^{\scriptscriptstyle \rm LQG}|\psi_{\vec{\chi}^{\prime}}\rangle$ is the matrix element of the LQG volume operator on gauge-invariant four-vertices. Note that in the first quantised theory intertwiner spaces corresponding to given set of spins $j_\nu$ are invariant under the action of $\hat{V}^{\scriptscriptstyle \rm LQG}$, \emph{i.e.} one has
\be
\langle \psi_{\vec{\chi}}|\hat{V}^{\scriptscriptstyle \rm LQG}|\psi_{\vec{\chi}^{\prime}}\rangle=\delta_{j_\nu j_\nu^{\prime}}\langle \psi_{\vec{\chi}}|\hat{V}^{\scriptscriptstyle \rm LQG}|\psi_{\vec{\chi}^{\prime}}\rangle~.
\ee
This implies that the LQG volume operator can be studied separately in each intertwiner space $\accentset{\circ}{\mathcal{H}}_{j_\nu}$ corresponding to a set of four spins $j_\nu$.
The choice of an appropriate recoupling scheme can simplify the computation of the matrix elements considerably, see Ref.~\cite{Brunnemann:2004xi}. We will only be dealing with four-valent vertices, for which the recoupling scheme is straightforward and is reviewed in Appendix~\ref{sec:Intertwiner}.

\subsubsection{Many-body operators}\label{Sec:Many-BodyOperators}
Many-body operators can be defined in parallel with the standard definition in condensed matter physics. A two-body operator representing the interaction of two `particles' reads as (cf.~\emph{e.g.} Ref.~\cite{Schwabl:1997gf})
\be\label{Eq:GFT:TwoBodyOperator}
\hat{T}_2=\frac{1}{2}\int \de^4g \de^4g^{\prime}\; u(g_\nu,g^{\prime}_\nu)\hat{\varphi}^{\dagger}(g_\nu)\hat{\varphi}^{\dagger}(g^\prime_\nu)\hat{\varphi}(g^\prime_\nu)\hat{\varphi}(g_\nu)~,
\ee
with $u(g_\nu,g^{\prime}_\nu)$ representing an interaction kernel. $N$-body operators can be constructed as a straightforward generalisation of (\ref{Eq:GFT:TwoBodyOperator}).

Note that both the one-body operator (\ref{Eq:GFT:OneBodyOperator}) and the two-body operator (\ref{Eq:GFT:TwoBodyOperator}) have an equal number of creation and annihilation operators. Moreover, they respect the same ordering of the arguments for each of the field operators. However, the most interesting operators in GFT, which give the dynamics of GFT models, do not fall in this class. In fact, for the operators that we will consider, the arguments of the field operators are arranged following peculiar combinatorial patterns, which make such operators highly non-local. As we will explain below, such non-locality is necessary for the geometric interpretation of the theory.

We aim to give a few examples of such non-local operators, by exhibiting the interaction terms of some GFT models in four dimensions.
The most general interaction term of the \emph{simplicial}  type\footnote{Simplicial GFT interactions are such that vertices in the corresponding spin foam model  are associated to 4-simplices. For the GFT-Spin Foam correspondence see the following section.}
is five-valent and has the form (see Ref.~\cite{Oriti:2016qtz})
\be
\hat{V}_{\rm simpl}=\frac{\lambda}{5}\int\Bigg(\prod_{A=1}^5\de^4g^A\; \hat{\varphi}(g_\nu^A)\Bigg) \mathcal{V}_5(g^1_\nu,\dots,g^5_\nu)~.
\ee
By specifying the functional form of the kernel $\mathcal{V}_5$ it is possible to recover some well-known GFT models. For instance, the Ooguri model corresponds to the interaction term
\be
\hat{V}_{\rm Oo}=\frac{\lambda}{5}\int\de^{10}g\; \hat{\varphi}_{1234}\hat{\varphi}_{4567}\hat{\varphi}_{7389}\hat{\varphi}_{962~10}\hat{\varphi}_{10~851}~,
\ee
where we used the shorthand notation $\hat{\varphi}_{1234}=\hat{\varphi}(g_1,g_2,g_3, g_4)$.
The EPRL model, on which we will focus on for the cosmological applications of GFT, has the same combinatorial structure of the Ooguri model but a non-constant kernel
\be\label{Eq:GFT:EPRLgroupRep}
\hat{V}_{\rm EPRL}=\frac{\lambda}{5}\int\de^{10}g\; \mathcal{V}^{\rm \scriptscriptstyle EPRL}_5(g_1,\dots,g_{10})\hat{\varphi}_{1234}\hat{\varphi}_{4567}\hat{\varphi}_{7389}\hat{\varphi}_{962~10}\hat{\varphi}_{10~851}~.
\ee
The precise form of the function $\mathcal{V}_5^{\rm \scriptscriptstyle EPRL}$ is currently not known. It must be pointed out that it involves some ambiguities that arise in the formulation of the GFT model, as well as in the corresponding spin foam model. For a discussion of these issues we refer the reader to Ref.~\cite{Oriti:2016qtz}. Nevertheless, the precise expression of the interaction kernel will not be relevant for our applications, since only its combinatorial structure is needed. In Section~\ref{Sec:GFT:MeanField} we will present the EPRL interaction potential (\ref{Eq:GFT:EPRLgroupRep}) in the spin representation, which is the one that has been used in cosmological applications.

\subsection{GFT dynamics}\label{Sec:GFT:Dynamics}
The dynamics in GFT follows from a path-integral. Considering a real GFT field for simplicity, this is given by
\be\label{eq:PathIntegralReal}
\mathcal{Z}=\int\mathcal{D}\varphi\; \e^{-S[\varphi]}~,
\ee
where the \emph{classical} GFT action has the form
\be
S[\varphi]=K[\varphi]+V[\varphi]~.
\ee
The kinetic term corresponds to one-body operators\footnote{Note that both $K[\varphi]$ and $V[\varphi]$ are \emph{classical} functionals in this section, since quantisation is achieved via the path-integral.}
 of the type considered in Eq.~(\ref{Eq:GFT:OneBodyOperator}). More specifically, it reads as
\be\label{Eq:GFT:KineticReal}
K[\varphi]=\int \de^4g \de^4g^{\prime}\; \varphi(g_\nu)\hat{k}(g_\nu,g^{\prime}_\nu)\varphi(g^\prime_\nu)~,
\ee
where we made use of the assumption that the GFT field is real, \emph{i.e.} $\overline{\varphi}=\varphi$.
The kernel $\hat{k}(g_\nu,g^{\prime}_\nu)$ in the kinetic term (\ref{Eq:GFT:KineticReal}) can be in principle non-local. However, for our applications in the next sections we will restrict our attention to a local kinetic kernel, \emph{i.e.} we will assume
\be
\hat{k}(g_\nu,g^{\prime}_\nu)=\prod_{i=1}^4\delta\left(g_i(g^{\prime}_i)^{-1}\right)~\hat{\mathcal{K}}_{g^\prime_\nu},
\ee
where $\hat{\mathcal{K}}_{g_\nu}$ is a differential operator affecting only the $g_\nu$ dependence. In this particular case one has 
\be
 K[\varphi]=\int \de^4g\; \varphi(g_\nu)\hat{\mathcal{K}}_{g_\nu}\varphi(g_\nu)~.
\ee

It is natural to assume that the dynamics is invariant under global isometries of $G$\footnote{\label{Footnote:BiInvariant} More specifically, isometries of the bi-invariant metric 
defined on G, which is naturally induced by the (negative of the) Killing-Cartan form on the Lie algebra via the pull-back of the left-translation. A bi-invariant metric exists for all compact Lie groups. Moreover, if the Lie algebra of the group is assumed to be simple the bi-invariant metric is unique up to a constant factor. See Ref.~\cite{Alexandrino2015} and references therein for further details. For our applications we will consider $G=\mbox{SU(2)}$, for which the hypotheses of the uniqueness theorem are satisfied. Thus, the bi-invariant metric can be obtained as the metric induced by the natural embedding of $\mbox{SU(2)}\cong S^3$ in $\mathbb{R}^4$, as it is done in Appendix~\ref{Appendix:HarmonicAnalysis}.}.
By further requiring that the equations of motion are second order, the simplest ansatz is
\be
\hat{\mathcal{K}}_{g_\nu}=-\triangle_{ G}+\mu~,
\ee
where $\mu$ is a mass term. The Laplace-Beltrami operator on $G$ reads as
\be
\triangle_{\rm G}=\frac{1}{\sqrt{g}}\pa^{i}\left(\sqrt{g}g^{ij}\pa_j\right)~,
\ee
where $g_{ij}$ denotes the Riemannian metric on $G$. Under suitable conditions this is uniquely determined, see footnote~\ref{Footnote:BiInvariant}.
The potential will be given by one of the corresponding many-body operators discussed in Section~(\ref{Sec:Many-BodyOperators}). In the case of the EPRL model, the interaction term is thus given by
\be
V_{\rm EPRL}[\varphi]=\frac{\lambda}{5}\int\de^{10}g\; \mathcal{V}^{\rm \scriptscriptstyle EPRL}_5(g_1,\dots,g_{10})\varphi_{1234}\varphi_{4567}\varphi_{7389}\varphi_{962~10}\varphi_{10~851}~.
\ee
The dynamics of GFT models is quite different from that of standard  field theories already at the classical level, due to their high degree of non-locality which leads to integro-differential equations of motion. However, the existence of a continuous group of symmetries of the action (given \emph{e.g.} by invariance under the right-action of G) leads to conservation laws which generalise the standard Noether theorem, which holds in local field theories \cite{Kegeles:2015oua,Kegeles:2016wfg}.

At the quantum level, the non-locality of GFT is intimately related to its interpretation as a theory of QG. In fact, the edges of the Feynman diagrams of a GFT model are stranded\footnote{This is similar to the case of matrix models \cite{DiFrancesco:1993cyw}, of which GFTs represent a higher dimensional generalisation. GFT is also not affected by some problems encountered in a different generalisation of matrix models, \emph{i.e.}~the so-called tensor models \cite{Ambjorn:1990ge,Godfrey:1990dt,Sasakura:1990fs}, namely the absence of a parameter to control the topological expansion \cite{Boulatov:1992vp,Ooguri:1992eb}.} (fat diagrams)
 with one strand per each argument of the GFT field. The form of the interaction term gives the prescription for joining the strands of different edges at a vertex. The Feynman diagrams are thus interpreted as cellular complexes $\mathcal{F}$ dual to a triangulated topological spacetime\footnote{Although we will only be concerned with the case $d=4$, it is worth remarking that GFTs can be formulated as theories of QG in any dimensions $d$. The interpretation of the Feynman diagrams in GFT in terms of discrete geometry holds true in general and for any dimensions (see Refs.~\cite{Freidel:2005qe,DePietri:1999bx,DePietri:2000ii,DePietri:2000ke}).} \cite{Freidel:2005qe,Oriti:2005mb,Oriti:2006se}.
 
By means of an appropriate choice of the GFT action, it is possible to establish a precise correspondence between GFT and Spin Foam models. The general derivation of the GFT action for a wide class of Spin Foam models is given in Refs.~\cite{Reisenberger:2000zc,Reisenberger:2000fy}, which generalises the results obtained in Refs.~\cite{Boulatov:1992vp,Ooguri:1992eb,DePietri:1999bx} for specific models of simplicial quantum gravity. It must be pointed out that such connection between the dynamics in the two approaches holds regardless of the interpretation of the boundary data, \emph{i.e.} whether they are the ones described in Section~\ref{Sec:GFT:BoundaryStates} or in the case of LQG spin-network states. The perturbative expansion of a GFT model is given by a sum over cellular complexes
\be\label{Eq:GFTpertExpansion}
\mathcal{Z}=\int\mathcal{D}\varphi\; \e^{-S[\varphi]}=\sum_{\mathcal{F}}\frac{\lambda^{|V|}}{\mbox{sym}(\mathcal{F})}A_{\scriptscriptstyle\mathcal{F}}~,
\ee
where $\lambda$ is the coupling constant\footnote{We are assuming only one type of interactions. If different interaction terms are considered, Eq.~(\ref{Eq:GFTpertExpansion}) must be generalised accordingly.}, $|V|$ is the number of vertices in the Feynman diagram (\emph{i.e.} perturbative order), $\mbox{sym}(\mathcal{F})$ the symmetry factor of $\mathcal{F}$, and $A_{\scriptscriptstyle\mathcal{F}}$ the Feynman amplitude obtained from the GFT \cite{Freidel:2005qe}. In this sense, the GFT formalism provides a completion of a given Spin Foam model \cite{Oriti:2013,Oriti:2006se}. In fact, the non-perturbative information contained in the GFT path-integral allows for the possibility of making sense of the sum over triangulations of spacetime. This has been shown explicitly for 3d quantum gravity, in which case the perturbative expansion (\ref{Eq:GFTpertExpansion}) is uniquely Borel resummable \cite{Freidel:2002tg}.

\section{Mean field approximation}\label{Sec:GFT:MeanField}
The main advantage of the third quantised formalism of GFT lies in the fact that it provides a convenient way to handle arbitrary superpositions of graphs, also allowing to study the dynamics of states with an arbitrary number of vertices $N$. In particular, in GFT the number $N$ does not have to be fixed at the outset and is in principle a dynamical variable, computed as the expectation value $\langle \hat{N}\rangle$ of the number operator (\ref{Eq:GFT:NumberOperator}) on a given state. This is a feature that in LQC must instead be imposed by hand, with the adoption of some lattice refinement model \cite{Bojowald2006}. However, the relation between LQC and the full theory is still not clear\footnote{See footnote~\ref{Footnote:QRLG} in Appendix~\ref{Appendix:LQC} for a discussion of recent developments in Quantum Reduced Loop Gravity (QRLG).}. In LQG the Hamiltonian constraint acts on spin network states by changing the underlying graph \cite{Thiemann:1996aw,Thiemann:1996ay,Thiemann:1997rv}, although the interpretation of the graph changing Hamiltonian is still an open issue\footnote{We would like to mention a possible interpretation of the graph changing Hamiltonian in LQG as the generator of graph refinements, regarded as time evolution \cite{Dittrich:2013xwa}. The notion of a graph changing Hamiltonian also inspired the development of spin foams, viewed as the time evolution of spin networks \cite{Reisenberger:1996pu}.}.

As we have seen above in Section~\ref{Sec:GFT:BoundaryStates}, in GFT the open spin network vertices represent the fundamental degrees of freedom of the theory, which are interpreted as the basic building blocks of quantum geometry. It was suggested in Ref.~\cite{Oriti:2013jga} that the continuum spacetime \emph{emerges} from QG when considering the collective dynamics of a large number $N$ of such elementary constituents. This is analogous to the way continuum mechanics is recovered from condensed matter physics. Following this analogy, the next step is to identify a regime of the theory that allows us to perform a `hydrodynamics approximation', which in turns leads to the recovery of a continuum spacetime. Suitable observables must then be identified in the quantum theory to match the macroscopic ones. Such a dynamical \emph{emergence} of classical spacetime from QG in the GFT approach has been studied in the cosmological case, considering a particular class of states, \emph{i.e.} coherent states \cite{Gielen:2013kla,Gielen:2013naa}. Such states $|\sigma\rangle$ are defined as eigenstates of the GFT field operator (which is assumed to be complex)
\be\label{Eq:GFT:CoherentState}
\hat{\varphi}(g_\nu)|\sigma\rangle=\sigma(g_\nu)|\sigma\rangle~,\hspace{1em} |\sigma\rangle=\mathcal{N}\exp\left(\int d^4g\; \sigma(g_\nu)\hat{\varphi}^{\dagger}(g_\nu)\right)|0\rangle
\ee
where $\sigma(g_\nu)$ is a function on $G^4$ and $\mathcal{N}$ is a normalisation factor given by
\be
\mathcal{N}=\e^{-\frac{||\sigma||^2}{2}}~,\hspace{1em}  \mbox{with} \hspace{1em}||\sigma||^2=\int d^4g\; |\sigma(g_\nu)|^2 =\langle\sigma|\hat{N}|\sigma\rangle~.
\ee
In Refs.~\cite{Oriti:2016qtz,Oriti:2016ueo} a modified Friedmann equation for the emergent universe, including quantum gravitational corrections, was thus obtained. It has been conjectured that such field coherent states would correspond to a new phase in QG, characterised by substantially different properties from the non-geometric phase corresponding to the Fock (\emph{i.e.} perturbative) vacuum. According to this point of view, such field coherent states should be interpreted as non-perturbative vacua of the GFT, arising from a process similar to Bose-Einstein condensation. For this reason the coherent state $|\sigma\rangle$ is often referred to in the literature as a GFT condensate\footnote{Condensate representations based on coherent states have been considered in the algebraic formulation of GFT in Ref.~\cite{Kegeles:2017ems}. Such states are characterised by an infinite number of particles and are inequivalent to the standard Fock representation.}. The idea of the emergence of spacetime geometry as the result of a phase transition is known as \emph{geometrogenesis}, and was introduced in the context of the \emph{quantum graphity} approach \cite{Konopka:2006hu}.

At the present stage, geometrogenesis remains an unproven conjecture, whose validity will only be assumed for our purposes.  
In the following, we will work in the mean field approximation of the GFT dynamics, which is suitable for the description of a system with a large number of quanta (\emph{i.e.} graphs with many nodes). This is equivalent to study the quantum dynamics of the state $|\sigma\rangle$ provided that we neglect field-field correlations, \emph{i.e.} only the evolution of the mean value $\langle \sigma|\hat{\varphi}(g_\nu)|\sigma\rangle=\sigma(g_\nu)$ in Eq.~(\ref{Eq:GFT:CoherentState}) is considered and higher order momenta are disregarded. Tests of the validity of the mean field approximation include the study of the stability of the `condensate' under perturbations, \emph{i.e.} the study of the solutions of the analogue of the Bogoliubov-DeGennes equation in the GFT context, and a calculation of the back-reaction of the perturbations on the condensate itself\footnote{For a detailed study of a Bose-Einstein condensates and the Bogoliubov-DeGennes equation in the context of analogue gravity systems, see Refs.~\cite{Girelli:2008gc,Sindoni:2009fc}.}. In fact, this is needed in order to quantify the fraction of `particles' that are expelled from the condensate.

In the rest of this section, we introduce the mean field approximation of GFT, in view of the cosmological applications considered in this chapter. We will consider a GFT model with a complex scalar field $\varphi$, whose partition function reads as
\be\label{eq:PathIntegral}
\mathcal{Z}=\int\mathcal{D}\varphi\mathcal{D}\overline{\varphi}\; \e^{-S[\varphi,\overline{\varphi}]}~,
\ee
where $S[\varphi,\overline{\varphi}]$ is the classical GFT action and $\varphi$ is defined on $\mbox{SU(2)}^4 \times \mathbb{R}$
\be\label{eq:GFTfieldDef}
\varphi=\varphi(g_1,g_2,g_3,g_4;\phi) = \varphi(g_\nu;\phi)~.
\ee
$\phi$ is a matter field, which for simplicity can be taken so as to represent a massless scalar, used as a relational clock \cite{Oriti:2016qtz,Oriti:2016ueo,Li:2017uao}. Just as for the discrete geometric data encoded in the group elements (and their conjugate variables), the interpretation of the real variable $\phi$ as a discretised matter field is grounded in the expression of the (Feynman) amplitudes of the model corresponding to the action $S$, which take the form of lattice gravity path integrals for gravity coupled to a massless scalar field. 
The field is invariant under the diagonal right action of $\mbox{SU(2)}$
\be\label{eq:right invariance}
\varphi(g_\nu h;\phi)=\varphi(g_\nu;\phi), \hspace{1em} \forall h\in \mbox{SU(2)}.
\ee
This \emph{right-invariance} property of the GFT field, and of the corresponding quantum states, can be understood as the usual gauge invariance on spin network states that characterises any lattice gauge theory, and more geometrically as the closure of the tetrahedra dual to the vertices of the same spin network states (this becomes transparent in the formulation of the theory using non-commutative Lie algebra-valued flux variables \cite{Baratin:2011hp}). Some mathematical details are given in Appendices~\ref{Appendix:HarmonicAnalysis},~\ref{sec:Intertwiner}.

The classical GFT field and the wavefunctions associated to its quantum states are $L^2$ functions with respect to the Haar measure on the ${\rm SU(2)}$ group manifold. Therefore they can be expanded, according to the Peter-Weyl theorem, in a basis of functions labeled by the irreducible representations of the same group (see Appendix~\ref{Appendix:HarmonicAnalysis}). For quantum states, this is the decomposition in terms of spin network states. For the classical GFT field, this decomposition looks: 
\be\label{eq:GFTfield}
\varphi(g_\nu;\phi)=\sum_{j_\nu,m_\nu,n_\nu,\iota}\varphi^{j_\nu,\iota}_{m_\nu}(\phi)\;\mathcal{I}^{j_\nu,\iota}_{n_\nu}\prod_{i=1}^4 \sqrt{d_{j_{i}}}D^{j_{i}}_{m_{i},n_{i}}(g_{i})~,
\ee
where $D^{j}_{m,n}(g)$ are the Wigner functions, $d_{j}$ is the dimension of the corresponding irreducible representation, \emph{i.e.} $d_j=2j+1$. The representation label $j$ takes integer and half-integer values, \emph{i.e.} $j\in\{0,\frac{1}{2},1,\frac{3}{2},\dots\}$, the indices $m$, $n$ take the values $-j\leq m,n\leq j$. Furthermore, the right-invariance leads more precisely to the Hilbert space $\mathcal{H}=L^2\left(\mbox{SU(2)}^4/\mbox{SU(2)},\de\mu_{Haar}\right)$. This is the \emph{intertwiner space} of a four-valent open spin network vertex, and also the Hilbert space of states for a single tetrahedron, a basis for which is given by the intertwiners $\mathcal{I}^{j_\nu,\iota}_{n_\nu}$, which are elements in $\accentset{\circ}{\mathcal{H}}_{j_\nu}$ (defined in Eq.~(\ref{eq:HilbertSpace4Vertex})).

The index $\iota$ labels elements in a basis in $\accentset{\circ}{\mathcal{H}}_{j_\nu}$, and represents an additional degree of freedom in the kinematical description of the GFT field and of its quantum states.

For example, $\iota$ can be chosen so as to label eigenstates of the volume operator for a single tetrahedron. With this choice, the volume operator acts diagonally on a wavefunction for a single tetrahedron $\varphi$ (which we indicate with the same symbol as the classical GFT field, since they are functionally analogous) decomposed as in Eq.~(\ref{eq:GFTfield})
\be\label{eq:VolumeActionGFT}
\hat{V}\varphi (g_\nu)=\sum_{j_\nu,m_\nu,n_\nu,\iota}V^{j_\nu,\iota}\varphi^{j_\nu,\iota}_{m_\nu}\;\mathcal{I}^{j_\nu,\iota}_{n_\nu}\prod_{i=1}^4 \sqrt{d_{j_{i}}}D^{j_{i}}_{m_{i},n_{i}}(g_{i}).
\ee
This action of the volume operator is the same as in LQG, where the volume eigenvalues for four-valent vertices have been studied extensively, see Ref.~\cite{Brunnemann:2004xi}. 
More details on the quantum geometry of such GFT states are in Appendix~\ref{sec:Volume}. 

The number of quanta (at a given clock time $\phi$) can be defined as a second quantised operator following Ref.~\cite{Oriti:2013}. Its mean value in a coherent state of the field operator (the simplest type of condensate state) is the squared norm of the mean field $\varphi$
\be\label{eq:NumberOfQuanta}
N(\phi)=\int_{\rm SU(2)^4}\de\mu_{\textrm{Haar}}\;\overline{\varphi}(g_{\nu};\phi)\varphi(g_{\nu};\phi).
\ee
It is worth stressing that $N$ is in general not conserved by the dynamics in GFT. In fact, this is one of the main advantages of this approach, since it allows us to study efficiently the dynamics of quantum gravity states with a variable number of degrees of freedom.

\subsection{Group Field Theory for the Lorentzian EPRL model}\label{sec:2}
In the following, we work with the GFT formulation of the Lorentzian EPRL model for quantum gravity, developed first in the context of spin foam models. This was also used in Refs.~\cite{Oriti:2016qtz,Oriti:2016ueo}, in a slightly generalised form, to account for some of the ambiguities in the definition of the model, including quantisation ambiguities, other choices at the level of model building, and possible quantum corrections due to renormalisation flow. It was presented using the spin representation of the GFT field $\varphi$ (thus of the action), and using the $\rm SU(2)$ projection of the Lorentz structures in terms of which the model is originally defined. Like any GFT model, the action is decomposed into the sum of a kinetic and an interaction term
\be\label{eq:Action}
S=K+V_5+\overline{V}_5\, .
\ee
The most general (local) kinetic term for an $\rm SU(2)$-based GFT field of rank-4 is\footnote{We indicate $\overline{\varphi^{j^{1}_\nu \iota_1}_{m^1_\nu}}$ by $\overline{\varphi}^{j^{1}_\nu \iota_1}_{m^1_\nu}$, for typographic reasons.}
\be\label{eq:kinetic}
K=\int\de\phi\;\sum_{j^a_{\nu},m^a_{\nu},\iota_a}\overline{\varphi}^{j^{1}_\nu\, \iota_1}_{m^1_\nu}\varphi^{j^2_\nu\, \iota_2}_{m^2_\nu}\mathcal{K}^{j^1_\nu \,\iota_1}_{m^1_\nu}\delta^{j^1_\nu j^2_\nu}\delta_{m^1_\nu m^2_\nu} \delta^{\iota_1 \iota_2},
\ee
and one has the interaction term corresponding to simplicial combinatorial structures given by\footnote{The authors would like to thank Marco Finocchiaro for pointing out an incorrect expression of the interaction potential that appeared in previous literature (private communication). The potential given here correctly implements the closure constraint of the 4-simplex obtained by `gluing' five tetrahedra.}
\be\label{eq:potential}
\begin{split}
V=\frac{1}{5}\int\de\phi\;\sum_{j_i,m_i,\iota_a}&\varphi^{j_1j_2j_3j_4\iota_1}_{m_1m_2m_3m_4}\varphi^{j_4j_5j_6j_7\iota_2}_{-m_4m_5m_6m_7}\varphi^{j_7j_3j_8j_9\iota_3}_{-m_7-m_3m_8m_9}\varphi^{j_9j_6j_2j_{10}\iota_4}_{-m_9-m_6-m_2m_{10}}\\&\times\varphi^{j_{10}j_8j_5j_1\iota_5}_{-m_{10}-m_8-m_5-m_1}\prod_{i=1}^{10}(-1)^{j_i-m_i}~\mathcal{V}_5(j_1,\dots,j_{10};\iota_1, \dots\iota_5) \; .
\end{split}
\ee
The details of the EPRL model would be encoded in the choice of kernels $\mathcal{K}$ and $\mathcal{V}_5$, and it is the interaction kernel that encodes the Lorentzian embedding of the theory and its full covariance, and what goes usually under the name of `spin foam vertex amplitude', here with boundary $\mbox{SU(2)}$-states. The explicit expression for such interaction kernel can be found in Ref.~\cite{Oriti:2016qtz,Oriti:2016ueo} and, in more details in Ref.~\cite{Speziale:2016axj}. We do not need to be explicit about the functional form of the interaction kernel for our purposes, while we will say more about the kinetic term in the following.
Some discrete symmetries of the interaction kernel will however be relevant for what follows. In fact, the coefficients $\mathcal{V}_5$ are invariant under permutations of the spins and of the intertwiners,  which preserve the combinatorial structure of the potential (\ref{eq:potential}).

Beside the general form ~(\ref{eq:kinetic}), in the following we will also use a specific case for the GFT kinetic term, \emph{i.e.} 
      \be\label{eq:freeaction}
   K=\int\de\phi\;\int_{\rm SU(2)^4}\de\mu_{\textrm{Haar}}\;\overline{\varphi}(g_{\nu},\phi)\mathcal{K}_{g_{\nu}}\varphi(g_{\nu},\phi)\; , 
   \ee
in the group representation, with
\be\label{eq:LocalKineticKernel}
\mathcal{K}_{g_{\nu}}=-\Big(\tau\pa_{\phi}^2+\sum_{i=1}^4\triangle_{g_i}\Big)+m^2 \; \;\; \tau, m^2 \, \in \mathbb{R},
\ee
which has been previously studied in the context of GFT cosmology in Refs.~\cite{Pithis:2016wzf,Pithis:2016cxg}, motivated by the renormalisation group analysis of GFT models (see Ref.~\cite{Carrozza:2013wda} and references therein). The same term can be given in the spin representation (using also the orthogonality of the intertwiners) as
\be\label{eq:freeactionSpin}
K=\int\de\phi\;\sum_{j_{\nu},m_{\nu},n_{\nu},\iota_1,\iota_2}\overline{\mathcal{I}}^{j_\nu \iota_1}_{n_{\nu}}\mathcal{I}^{j_\nu \iota_2}_{n_{\nu}}\overline{\varphi}^{j_{\nu}\iota_1}_{m_{\nu}}\hat{T}_{j_\nu}\varphi^{j_{\nu}\iota_2}_{m_{\nu}}=\int\de\phi\;\sum_{j_{\nu},m_{\nu},\iota}\overline{\varphi}^{j_{\nu}\iota}_{m_{\nu}}\hat{T}_{j_\nu}\varphi^{j_{\nu}\iota}_{m_{\nu}},
\ee
with
\be\label{eq:Toperator}
\hat{T}_{j_\nu}=-\tau\pa_{\phi}^2+\eta\sum_{i=1}^4j_i(j_i+1)+m^2.
\ee
Let us stress once more that the exact functional dependence on the discrete geometric data can be left more general for the EPRL model(s), since it is not uniquely fixed in the construction of the model, and it is only weakly constrained (mainly at large volumes) by the effective cosmological dynamics (which of course allows for the specific example above); the dependence on the scalar field variable $\phi$ is more important for obtaining the correct cosmological dynamics, at least in the isotropic case.

\subsection{Emergent Friedmann dynamics}\label{Sec:GFT:Background}
In this Section we derive the equations of motion of an isotropic cosmological background from the dynamics of the mean field for the GFT model. We reproduce in more detail the analysis of Ref.~\cite{Oriti:2016qtz,Oriti:2016ueo} and clearly spell out all the assumptions made in the derivation, including the necessary restrictions on the GFT field, such as isotropy (\emph{i.e.} considering equilateral tetrahedra) and left-invariance, which we now discuss. 
\subsubsection{Conditions on the mean field}\label{sec:LeftInv}
As we have recalled, the simplest effective cosmological dynamics is obtained as the mean field approximation of the full GFT quantum theory, for any specific model. The resulting hydrodynamic equations, which we will discuss in the next section, are basically the classical equations of motion of the given GFT model, subject to a few additional restrictions. One way to obtain such equations from the microscopic quantum dynamics is to consider operator equations of motion evaluated in mean value on simple field coherent states, \emph{i.e.} simple condensate states, and the resulting equations are going to be non-linear equations for the condensate wavefunctions, playing the role of classical GFT field. Such condensates have a geometric interpretation as homogeneous continuum spatial geometries and the condensate wavefunction is a probability distribution (a fluid density) on the space of such homogeneous geometries (\emph{i.e.} minisuperspace, or the corresponding conjugate space).

For this interpretation to be valid, however, one additional condition has to be imposed on the condensate wave function: {\it left-invariance} under the diagonal group action. If the wavefunction satisfies this additional condition, on top of the right invariance, and the domain is chosen to be ${\rm SU(2)}^4$, coming from the imposition of simplicity constraints on ${\rm SL}(2,\mathbb{C})$ data, like in the EPRL model, then the domain becomes isomorphic to minisuperspace of homogeneous geometries \cite{Gielen:2014ila}. Thus, this is not a symmetry of the GFT field, like the right-invariance, nor it is normally imposed on GFT quantum states. It is a property imposed on this specific class of states, in order to reduce the number of dynamical degrees of freedom and to guarantee the above interpretation.
Thus, we assume that the field components can be expressed as:
\be\label{eq:LeftInvarianceAnsatz}
\varphi^{j_{\nu}\iota}_{m_{\nu}}=\sum_{\iota^{\prime}}\varphi^{j_{\nu}\iota\iota^{\prime}}\mathcal{I}^{j_{\nu}\iota^{\prime}}_{m_{\nu}},
\ee
where $\iota^{\prime}$ is another intertwiner label, independent from $\iota$.  
Then, Eq.~(\ref{eq:kinetic}) becomes, using the assumption Eq.~(\ref{eq:LeftInvarianceAnsatz})
\be
K=\int\de\phi\;\sum_{j_{\nu},m_{\nu},\iota}\overline{\varphi}^{j_{\nu}\iota}_{m_{\nu}}\mathcal{K}^{j_\nu \iota}_{m_\nu}\varphi^{j_{\nu}\iota}_{m_{\nu}}=\int\de\phi\;\sum_{j_{\nu},\iota,\iota^{\prime},\iota^{\prime\prime}}\overline{\varphi}^{j_{\nu}\iota\iota^{\prime}}\tilde{\mathcal{K}}^{j_\nu \iota\iota^{\prime}\iota^{\prime\prime}}\varphi^{j_{\nu}\iota\iota^{\prime\prime}},
\ee
with
\be\label{eq:SimplificationGeneralLeftInv}
\tilde{\mathcal{K}}^{j_\nu \iota\iota^{\prime}\iota^{\prime\prime}}=\sum_{m_\nu}\overline{\mathcal{I}}^{j_{\nu}\iota^{\prime}}_{m_{\nu}}\mathcal{I}^{j_{\nu}\iota^{\prime\prime}}_{m_{\nu}}\mathcal{K}^{j_\nu \iota}_{m_\nu}\;\; , 
\ee
while the special case Eq.~(\ref{eq:freeactionSpin}) further simplifies to
\be\label{eq:freeactionSpinSimplified}
K=\int\de\phi\;\sum_{j_{\nu},m_{\nu},\iota}\overline{\varphi}^{j_{\nu}\iota}_{m_{\nu}}\hat{T}_{j_\nu}\varphi^{j_{\nu}\iota}_{m_{\nu}}=\int\de\phi\;\sum_{j_{\nu},\iota}\overline{\varphi}^{j_{\nu}\iota}\hat{T}_{j_\nu}\varphi^{j_{\nu}\iota}.
\ee
When the kernel $\mathcal{K}^{j_\nu \iota}_{m_\nu}$ does not depend on $m_\nu$, Eq.~(\ref{eq:SimplificationGeneralLeftInv}) simplifies considerably
\be
\tilde{\mathcal{K}}^{j_\nu \iota\iota^{\prime}\iota^{\prime\prime}}=\left(\sum_{m_\nu}\overline{\mathcal{I}}^{j_{\nu}\iota^{\prime}}_{m_{\nu}}\mathcal{I}^{j_{\nu}\iota^{\prime\prime}}_{m_{\nu}}\right)\mathcal{K}^{j_\nu \iota}=\delta^{\iota^{\prime}\iota^{\prime\prime}}\mathcal{K}^{j_\nu \iota}~,
\ee
leading us to the following expression for the kinetic term
\be
K=\int\de\phi\;\sum_{j_{\nu},\iota,\iota^{\prime}}\overline{\varphi}^{j_{\nu}\iota\iota^{\prime}}\mathcal{K}^{j_\nu \iota}\varphi^{j_{\nu}\iota\iota^{\prime}}~.
\ee

\subsubsection{Isotropic reduction and monochromatic tetrahedra}\label{sec:MaxVol}
Different definitions of `isotropy' are possible for GFT condensates, and have been used in the literature, depending on the chosen reconstruction procedure for the continuum geometry out of the discrete data associated to such GFT states (see Refs.~\cite{Gielen:2014ila,Gielen:2016dss}). For all of them, though, the result is qualitatively similar, as it should be: the condensate wavefunction has to depend on a single degree of freedom, \emph{e.g.} one single spin variable, corresponding to the volume information or the scale factor of the emergent universe. 
Here, as in  Refs.~\cite{Oriti:2016qtz,Oriti:2016ueo}, we adopt the simplest and most symmetric definition: we choose a condensate wavefunction such that the corresponding GFT quanta can be interpreted as equilateral tetrahedra. This implies that all of the spins labelling the quanta are equal $j_i=j, \;\forall i$, corresponding to tetrahedra with all triangle areas being equal. In this case, spin network vertices are said to be monochromatic. 
We further assume that the only non-vanishing coefficients for each $j$ are those which correspond to the largest eigenvalue of the volume and a fixed orientation of the vertex (which lifts the degeneracy of the volume eigenvalues). In this way, the label $\iota$ is uniquely determined in each intertwiner space following from right-invariance (see Appendix~\ref{sec:Volume}). We call this particular value $\iota^{\star}$. This means that we have fixed all the quantum numbers of a quantum tetrahedron. 
We are still left with the intertwiner label following from left-invariance. To fix this, we identify the two vectors in $\accentset{\circ}{\mathcal{H}}_{j_\nu}$ determined by the decomposition of $\varphi$ by assuming that the only non-vanishing components $\varphi^{j_{\nu}\iota\iota^{\prime}}$ are such that $\iota=\iota^{\prime}$, \emph{i.e.}
\be\label{eq:LeftRightIdentify}
\varphi^{j_{\nu}\iota\iota^{\prime}}=\varphi^{j_{\nu}\iota\iota} \delta^{\iota\iota^{\prime}}~\hspace{1em} \mbox{(no sums)}.
\ee
Thus, also this extra label is fixed by the maximal volume requirement. The geometric interpretation of this further step is unclear, at present, but it is at least compatible with what we know about the (quantum) geometry of GFT states.
Using Eqs.~(\ref{eq:LeftInvarianceAnsatz}),~(\ref{eq:LeftRightIdentify}), the expansion Eq.~(\ref{eq:GFTfield}) simplifies to
\be\label{eq:LeftRightInvField}
\varphi(g_\nu;\phi)=\sum_{j_\nu,m_\nu,n_\nu,\iota}\varphi^{j_{\nu}\iota}(\phi)\;\mathcal{I}^{j_{\nu}\,\iota}_{m_{\nu}}\mathcal{I}^{j_\nu\,\iota}_{n_\nu}\prod_{i=1}^4 \sqrt{d_{j_{i}}}D^{j_{i}}_{m_{i},n_{i}}(g_{i}).
\ee

Bringing all these conditions together (and dropping unnecessary repeated intertwiner labels), we get for the kinetic term
\be\label{eq:KineticSimplifiedABC}
K=\int\de\phi\;\sum_{j}\overline{\varphi}^{j\iota^{\star}}\tilde{\mathcal{K}}^{j\,\iota^{\star}}\varphi^{j\iota^{\star}}
\; , 
\ee
while the interaction term is given by
\be\label{eq:VwithVdoubleprime}
V=\frac{1}{5}\int\de\phi\;\sum_{j}\left(\varphi^{j\iota^\star}\right)^5\mathcal{V}^{\,\prime\prime}_5(j;\iota^\star).
\ee
In Eq.~(\ref{eq:VwithVdoubleprime}) we introduced the notation
\be\label{eq:IsotropicPotential}
\mathcal{V}^{\,\prime\prime}_5(j;\iota^\star)=\mathcal{V}^{\,\prime}_5(\underbrace{j,j\dots,j}_{10};\underbrace{\iota^\star,\iota^\star \dots\iota^\star}_{5})=
\mathcal{V}_5(j\dots j;\iota^\star \dots\iota^\star)\omega(j,\iota^\star)~,
\ee
with
\be\label{eq:OmegaDefinition}
\begin{split}
\omega(j,\iota^\star)=\sum_{m_i}\prod_{i=1}^{10}(-1)^{j_i-m_i}~&\mathcal{I}^{j\iota^\star}_{m_1m_2m_3m_4}\mathcal{I}^{j\iota^\star}_{-m_4m_5m_6m_7}\mathcal{I}^{j\iota^\star}_{-m_7-m_3m_8m_9}\\
&\times\mathcal{I}^{j\iota^\star}_{-m_9-m_6-m_2m_{10}}\mathcal{I}^{j\iota^\star}_{-m_{10}-m_8-m_5-m_1}~.
\end{split}
\ee
Thus, in the isotropic case, the effect of the interactions is contained in the diagonal of the potential and in the coefficient $\omega(j,\iota^\star)$ constructed out of the intertwiners. We also observe that distinct monochromatic components have independent dynamics.

Equation~(\ref{eq:OmegaDefinition}) can also be written in a different form, by expressing the intertwiner $\mathcal{I}^{j\iota^\star}_{m_\nu}$ in terms of the intertwiners $\alpha^{jJ}_{m_\nu}$ defined in Appendix~\ref{sec:Intertwiner} (see Eq.~(\ref{eq:DefAlphaIntertwiner}))
\be
\omega(j,\iota^\star)=\sum_{J_k}\left(\prod_{k=1}^5c^{J_k\iota^{\star}}\right)\{15j\}_{J_k}~,
\ee
where we identified the contraction of five intertwiners $\alpha^{jJ}_{m_\nu}$ with a 15j symbol of the first type
\be
\begin{split}
\{15j\}_{J_k}=\sum_{m_i}\prod_{i=1}^{10}(-1)^{j_i-m_i}~&\alpha^{jJ_1}_{m_1m_2m_3m_4}\alpha^{jJ_2}_{-m_4m_5m_6m_7}\alpha^{jJ_3}_{-m_7-m_3m_8m_9}\\
&\times\alpha^{jJ_4}_{-m_9-m_6-m_2m_{10}}\alpha^{jJ_5}_{-m_{10}-m_8-m_5-m_1}~.
\end{split}
\ee

\subsubsection{Background equation}\label{sec:Background}
The equations of motion for the background can be found by varying the action Eq.~(\ref{eq:Action}). Using Eqs.~(\ref{eq:KineticSimplifiedABC}),~(\ref{eq:IsotropicPotential}) we find (compare with ~\cite{Oriti:2016qtz,Oriti:2016ueo})
\be\label{eq:Background}
\tilde{\mathcal{K}}^{j\,\iota^{\star}}\varphi^{j\iota^{\star}}+\mathcal{V}^{\,\prime\prime}_5(j;\iota^\star) \left(\varphi^{j\iota^\star}\right)^4=0.
\ee
In the particular case given by Eq.~(\ref{eq:LocalKineticKernel}) we can write
\be\label{eq:GFTbackground:KineticFirstExample}
K=\int\de\phi\;\sum_{j}\overline{\varphi}^{j\iota^{\star}}\hat{T}_{j}\varphi^{j\iota^{\star}} \qquad 
\hat{T}_{j}=-\tau\pa_{\phi}^2+4\eta j(j+1)+m^2.
\ee
For the purpose of studying a concrete example, from now on we consider the special case in which $j=\frac{1}{2}$ of $\mbox{SU(2)}$. Then, we have, using the definition Eq.~(\ref{eq:OmegaDefinition})
\be\label{eq:OmegaVolumeEigenstatePlus}
\begin{split}
\omega\left(\frac{1}{2},\pm\right)=\sum_{m_i}\mathcal{I}^{\frac{1}{2}\,\iota_\pm}_{m_1m_2m_3m_4}\mathcal{I}^{\frac{1}{2}\,\iota_\pm}_{m_4m_5m_6m_7}\mathcal{I}^{\frac{1}{2}\,\iota_\pm}_{m_7m_3m_8m_9}\mathcal{I}^{\frac{1}{2}\,\iota_\pm}_{m_9m_6m_2m_{10}}\mathcal{I}^{\frac{1}{2}\,\iota_\pm}_{m_{10}m_8m_5m_1}\\
=\frac{3\mp i \sqrt{3}}{18 \sqrt{2}}~.
\end{split}
\ee
$\iota^\star=\iota_\pm$ means that we are considering as an intertwiner the volume eigenvector corresponding to a positive (resp. negative) orientation, see Appendix~\ref{sec:Volume}. %
Thus, the equation of motion for the background in this special case reads
\be\label{eq:BackgroundDynamics}
\left(-\tau\pa_{\phi}^2+3\eta+m^2\right)\varphi^{\frac{1}{2}\iota^\star}+\overline{\mathcal{V}^{\,\prime\prime}}_5\left(\frac{1}{2};\iota^\star\right)\left(\overline{\varphi}^{\frac{1}{2}\iota^\star}\right)^4=0,
\ee
with the coefficient of the interaction term given by Eqs.~(\ref{eq:IsotropicPotential}),~(\ref{eq:OmegaVolumeEigenstatePlus}). %
Note that, under the assumption that $j=\frac{1}{2}$, used in addition to the isotropic reduction, even the more general form of the EPRL GFT model coupled to a massless free scalar field (\ref{eq:kinetic}), as used in  Refs.~\cite{Oriti:2016qtz,Oriti:2016ueo} will collapse to the special case (\ref{eq:BackgroundDynamics}). Also, the dominance of a single (small) spin component in the cosmological dynamics of isotropic backgrounds can be shown to take place at late times \cite{Gielen:2016uft}, and it can be expected to be a decent approximation at earlier ones. Thus, even the special case we are considering is not too restrictive.

\section{Emergent Background: the non-interacting case}\label{Sec:NoInteractions}
In this section we study the properties of solutions of the modified
Friedmann equation, obtained in Ref.~\cite{Oriti:2016ueo}.
 The equation was derived in the previous section in the mean field approximation of the GFT dynamics.
The mean field, with the assumptions listed in Section~\ref{Sec:GFT:Background}, has non-vanishing components $\varphi^{j\iota^{\star}}$ with $j$ a spin representation index and $\iota^{\star}$ a fixed intertwiner label. We introduce the notation $\sigma_j=\varphi^{j\iota^{\star}}$ to be consistent with the one used in Refs.~\cite{deCesare:2016axk,deCesare:2016rsf}.
 The $j$-th component of the mean field can be written in the polar representation as $\sigma_j=\rho_j e^{i \theta_j}$. As shown above, different spin components of the mean field decouple. Therefore, we will study their evolution independently. 
 For convenience of the reader, we rewrite below the equation of motion for the background (\ref{eq:Background}), with kinetic operator given by Eq.~(\ref{eq:GFTbackground:KineticFirstExample}) and vanishing potential
\be
-\tau\pa_{\phi}^2\sigma_j+R_j\sigma_j=0~,
\ee
with $R_j=4\eta j(j+1)+m^2$. We also define the quantity $z_j$ as
\be\label{Eq:GFT:DefinitionSj}
z_j=\frac{R_j}{\tau}=\frac{4\eta j(j+1)+m^2}{\tau}~.
\ee
As in Refs.~\cite{Oriti:2016qtz,Oriti:2016ueo,deCesare:2016axk} this will be assumed to be positive. For $\tau>0$ this condition can be satistied for all spins, provided that $m^2>0$. However, this is not possible for $\tau<0$, since it is always possible to find a large enough value of $j$ such that the numerator in Eq.~(\ref{Eq:GFT:DefinitionSj}) is positive\footnote{In Ref.~\cite{Gielen:2016uft} it was shown that in this case the lowest spin component, \emph{i.e.} $j_0=1/2$ if one excludes the trivial representation $j_0=0$, gives the dominant contribution at late times.}.
 
Since evolution in gravity theories is purely relational, all dynamical quantities can be regarded as functions of a massless
scalar field $\phi$, used as an internal clock. With respect to this relational notion of evolution, there is a conserved charge associated to
$\theta_j$:
\be \rho_j^2\pa_\phi \theta_j =Q_{j}.  \ee 
Note that such a conserved charge only exists if the effective dynamics of the GFT mean field is invariant under global $\mbox{U(1)}$ transformations. In particular, there is no such quantity when the GFT field is real or for the EPRL model considered in the previous section (except when interactions are negligible, \emph{i.e.} near the bounce).

The modulus of the isotropic mean field satisfies the equation of motion\footnote{This has the form of Ermakov equation \cite{ermakov1880transformation}.}
\be\label{eq:EOM}
\pa_\phi^2 \rho_j-\frac{Q_j^2}{\rho_j^3}- z_j \rho_j =0, \ee 
leading to another conserved current, which will be referred to as `GFT energy'\footnote{Such quantity is conserved as a consequence of $\phi$-translation invariance of the GFT action with a massless scalar. It is not clear whether it would still be conserved in the case of a scalar with a potential or for different matter species, since none of these cases have been considered in the literature so far. However, it must be pointed out that the physical interpretation of the `GFT energy' $E$ is not clear. In fact, such term is motivated by the fact that it is by definition the mechanical energy of the one-dimensional mechanical system described by Eq.~(\ref{eq:EOM}). Nevertheless, the reader must be aware that such terminology is potentially misleading, due to the fact that in GFT there is no notion of energy of the system. The parameter $E$ plays an important role in the cosmology of GFT models, particularly the free model considered in this section, since it introduces departures from the LQC effective equations for a single-spin mean field, see Refs.~\cite{Oriti:2016qtz,Oriti:2016ueo}.} following Ref.~\cite{Oriti:2016ueo}
\be\label{eq:GFT energy}
E_j=(\pa_\phi\rho_j)^2+\frac{Q_j^2}{\rho_j^2}-z_j\rho_j^2~.
\ee
Equation (\ref{eq:EOM}) admits the following solution
\be\label{eq:rho} \rho_j(\phi)=\frac{e^{(-b-\phi ) \sqrt{z_j}}
  \Delta(\phi)}{2 \sqrt{z_j}}, \ee 
where
\be
 \Delta(\phi)=\sqrt{k^2-2 k e^{2 (b+\phi ) \sqrt{z_j}}+e^{4 (b+\phi )
      \sqrt{z_j}}+4 z_j^2 Q_j^2}
\ee
and $k$, $b$ are integration constants.  From Eq.~(\ref{eq:GFT
  energy}) follows 
\be E_j=k, \ee 
whereas the total $\mbox{U(1)}$ charge receives contributions from all the independent monochromatic componens of the condensate, and is identified with the canonical momentum of the scalar field (see Ref.~\cite{Oriti:2016ueo})  
\be\label{eq:Qmomentum} \sum_j Q_j=\pi_{\phi}.  \ee 
The dynamics of macroscopic observables is defined through that of the
expectation values of the corresponding quantum operators. As discussed in Section~\ref{Sec:GFT:KinematicalOperators}, in GFT (as
in LQG) the fundamental observables are geometric
operators, such as areas and volumes. The volume of space at a given
value of relational time $\phi$ can be computed using the definition (\ref{Eq:GFT:Volume2ndQuantized}) and replacing the field operators by the classical mean field\footnote{The result obtained in this way is equivalent to the one obtained by evaluating the expectation value of the GFT volume operator on a coherent state $|\sigma\rangle$, as done in Ref.~\cite{Oriti:2016qtz,Oriti:2016ueo}.}, subject to the restrictions discussed above  
\be\label{eq:VolumeExpectationValueDefinition}
V(\phi)=\sum_{j} V_{j}\rho_j^2(\phi)~,
\ee
 where $V_j\propto j^{3/2} \ell_{\rm \scriptscriptstyle Pl}$ is the eigenvalue of the
volume operator computed on a four-valent monochromatic vertex, with edges carrying spin $j$ representations of SU(2). Using
this as a definition and differentiating w.r.t. relational time $\phi$
one obtains, as in Ref.~\cite{Oriti:2016ueo} the following equations,
which play the role of effective Friedmann (and acceleration)
equations describing the dynamics of the Cosmos as it arises from that
of a condensate of spacetime quanta 
\begin{align}
 \frac{\pa_\phi V}{V}&=\frac{2
  \sum_{j} V_{j}\rho_j\pa_\phi\rho_j}{\sum_{j} V_{j}\rho_j^2}~,\\
  \frac{\pa_\phi^2 V}{V}&=\frac{2 \sum_{j} V_{j}\left(E_j+2z_j^2\rho_j^2\right)}{\sum_{j} V_{j}\rho_j^2}~.
\end{align} 
In the context of GFT, spacetime
is thus seen to emerge in the hydrodynamic limit of the theory; 
the evolution of a homogeneous
and isotropic universe is completely determined by that of its volume.
Note that the above equations are written in terms of functions of
$\phi$. In fact, as implied by the background independence of GFT, and
more in general of any theory of quantum geometry, \textit{a priori}
there is no spacetime at the level of the microscopic theory. In particular, there exists no coordinate
time. 
Nevertheless,
we will show how it is possible to introduce
a preferred choice of time, namely proper time, in order to study the 
dynamics of the model in a way similar to the one followed for standard
homogeneous and isotropic models.
This will be particularly useful for the study of the accelerated expansion
of the universe.
In the following we will restrict
our attention to the case in which the condensate belongs to one
particular representation of the symmetry group. This special case can
be obtained from the equations written above by considering a
condensate wave function $\sigma_j$ with support only on
$j=j_0$. Representation indices will hereafter be omitted. Hence, we
have 
\begin{align}
\frac{\pa_\phi V}{V}&=2\frac{\pa_\phi\rho}{\rho}\equiv2\zeta(\phi)~,\label{eq:effective dynamics}\\
\frac{\pa_\phi^2 V}{V}&=\left[\frac{\partial^2_{\phi}\rho}{\rho}+\left(\frac{\partial_{\phi}\rho}{\rho}\right)^2\right]=2
\left(\frac{E}{\rho^2}+2z\right)~.\label{eq:effective dynamics accel}
  \end{align}
These are effective Friedmann equations, giving the
relational evolution of the volume with respect to the scalar field
$\phi$. 
Eqs.~(\ref{eq:effective dynamics}),~(\ref{eq:effective dynamics accel})
give the evolution of the emergent spacetime, with $\rho$ playing the
role of an auxiliary field.

The large volume limit corresponds in this model to the limit of large (relational) time. As $\phi\to\pm\infty$, one has $\zeta(\phi)\to \sqrt{z}$. Thus, the standard
Friedmann and acceleration equations with a constant
gravitational coupling and a fluid with a stiff equation of state are recovered. 
 We will introduce proper time by means of
the relation between velocity and momentum of the scalar field
\be\label{eq:momentum}
\pi_{\phi}=\dot{\phi}V.
\ee
Furthermore, we can \textit{define} the scale factor as the cubic root
of the volume
\be\label{eq:scaleFactor}
a\propto V^{1/3}.
\ee
We can therefore write the evolution equation of the universe 
 obtained from GFT in the form of an \textit{effective Friedmann
  equation} ($H=\frac{\dot{V}}{3V}$ is the Hubble expansion rate and
$\varepsilon=\frac{\dot{\phi}^2}{2}$ the energy density\footnote{The energy densitity is denoted here by $\varepsilon$, unlike other chapters where it is denoted by $\rho$, to avoid confusion with the modulus of the condensate `wave function'.}) 
\be\label{eq:effective Friedmann}
H^2=\left(\frac{\pa_\phi V}{3V}\right)^2\dot{\phi}^2=\frac{8}{9}\zeta^2
\varepsilon.  \ee 
Using Eqs.~(\ref{eq:GFT energy}),~(\ref{eq:Qmomentum}),~(\ref{eq:momentum})
we can recast  Eq.~(\ref{eq:effective Friedmann}) in the following form
\be\label{eq:FriedmannEffFluids}
H^2=\frac{8}{9}Q^2\left(\frac{\gamma_R}{V^2}+\frac{\gamma_E}{V^3}+\frac{\gamma_Q}{V^4}\right),
\ee
where we introduced the quantities
\be\label{eq:fluids}
\gamma_R=\frac{z}{2},\hspace{1em}\gamma_E=\frac{V_jE}{2},\hspace{1em}\gamma_Q=-\frac{V_j^2Q^2}{2}.
\ee
The first term in Eq.~(\ref{eq:FriedmannEffFluids}) is, up to a constant factor, the energy density of a massless scalar field on a conventional FLRW background,
whereas the others represent the contribution of effective fluids with distinct equations of state and express
departures from the ordinary Friedmann dynamics. Respectively, the equations of state of the terms in Eq.~(\ref{eq:fluids})
are given by $w=1,2,3$, consistently with the (quantum corrected) Raychaudhuri equation Eq.~(\ref{eq:accelerationFluids}).
Effective fluids have been already considered in the context of LQC as a a way to encode quantum corrections, see \textit{e.g.} \cite{Singh:2005km}.

This equation  reduces to the conventional
Friedmann equation in the large $\phi$ limit, where the contributions of
the extra fluid components are negligible
\be\label{eq:Friedmann}
H^2=\frac{8\pi G_{\rm N}}{3}\varepsilon.  \ee 
Thus, consistency with the Friedmann equation of Standard Cosmology (Chapter~\ref{Chapter:StandardCosmology}) in this limit demands $z= 3\pi G$, which
allows to identify some particular combination of the parameters of the microscopic model with low energy quantities characteristic of macroscopic physics (see Ref.~\cite{Oriti:2016ueo} and Footnote~\ref{Footnote:SteffenGNewton} in this chapter). Specifically, we can write
\be
G_{\rm N}=\frac{R_{j_0}}{3\pi\tau}=\frac{4\eta j_0(j_0+1)+m^2}{3\pi\tau}~.
\ee
We observe in particular that higher values of the spin in this model correspond to a stronger gravitational attraction. However, if we allow for a general superposition of all monochromatic components and provided that the conditions spelled out in Ref.~\cite{Gielen:2016uft} are satisfied (in particular one must have $\tau<0$), in the limit $\phi\to\infty$ a low spin component $j=j_0$ will give the dominant contribution. If we assume that this corresponds to the fundamental representation of SU(2) $j_0=1/2$ we have
\be
G_{\rm N}=\frac{3\eta+m^2}{3\pi\tau}~,
\ee
provided $m^2<-3\eta$.

The interpretation of our model is made clear by Eq.~(\ref{eq:FriedmannEffFluids}).
In fact the dynamics has the usual Friedmann form with a classical background represented
by the scale factor $a$ and quantum geometrical corrections
given by two effective fluids, corresponding to the two conserved quantities $Q$ and $E$.
In the following we will consider for convenience
Eq.~(\ref{eq:effective Friedmann}) in order to study the properties of solutions.

\begin{figure}
  \begin{minipage}{.5\textwidth}
  \centering
 \includegraphics[width=\linewidth]{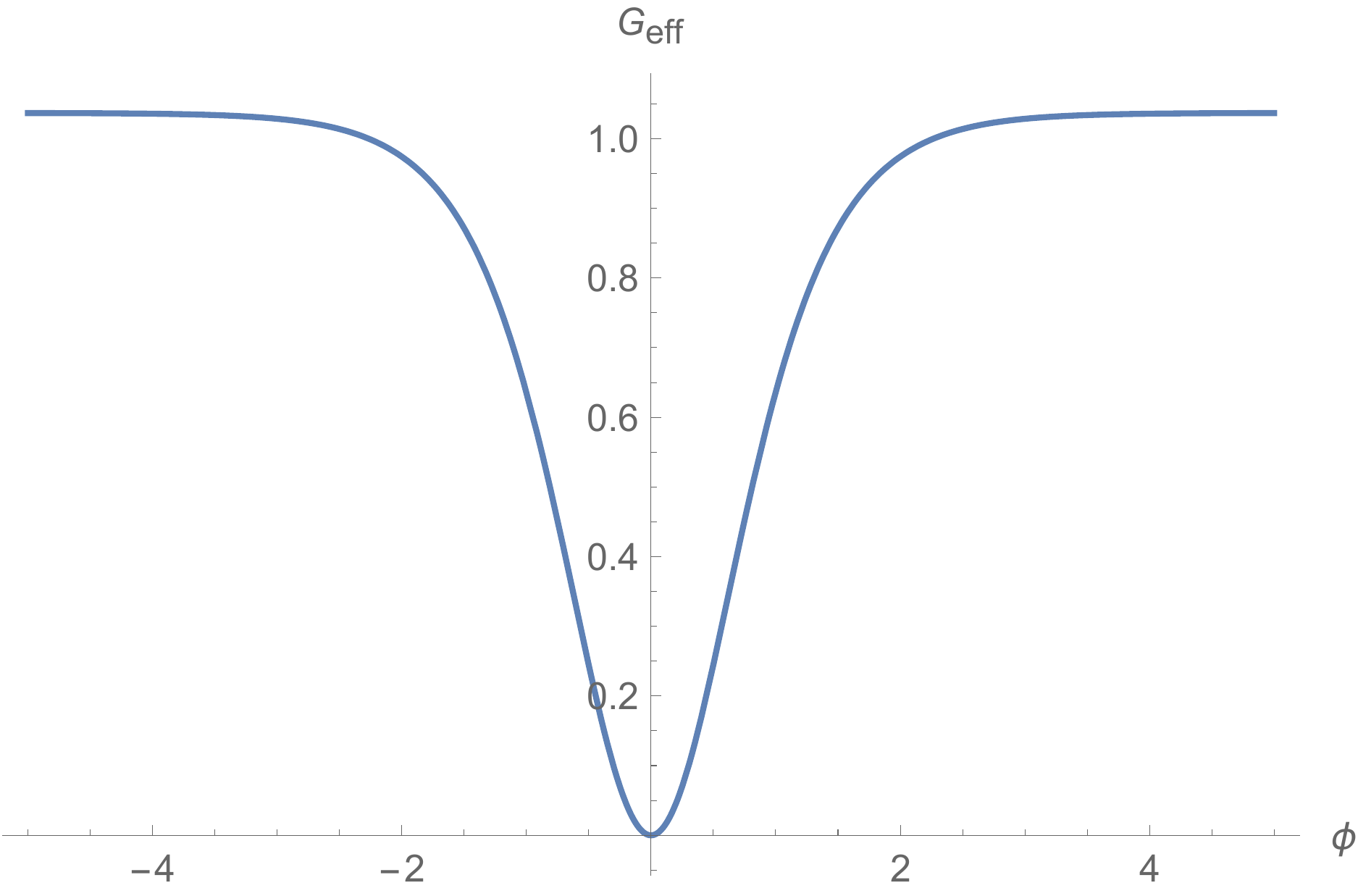}
 \subcaption{$E<0$}\label{fig:Geff E<0}
 \end{minipage}%
\begin{minipage}{.5\textwidth}
  \centering
 \includegraphics[width=\linewidth]{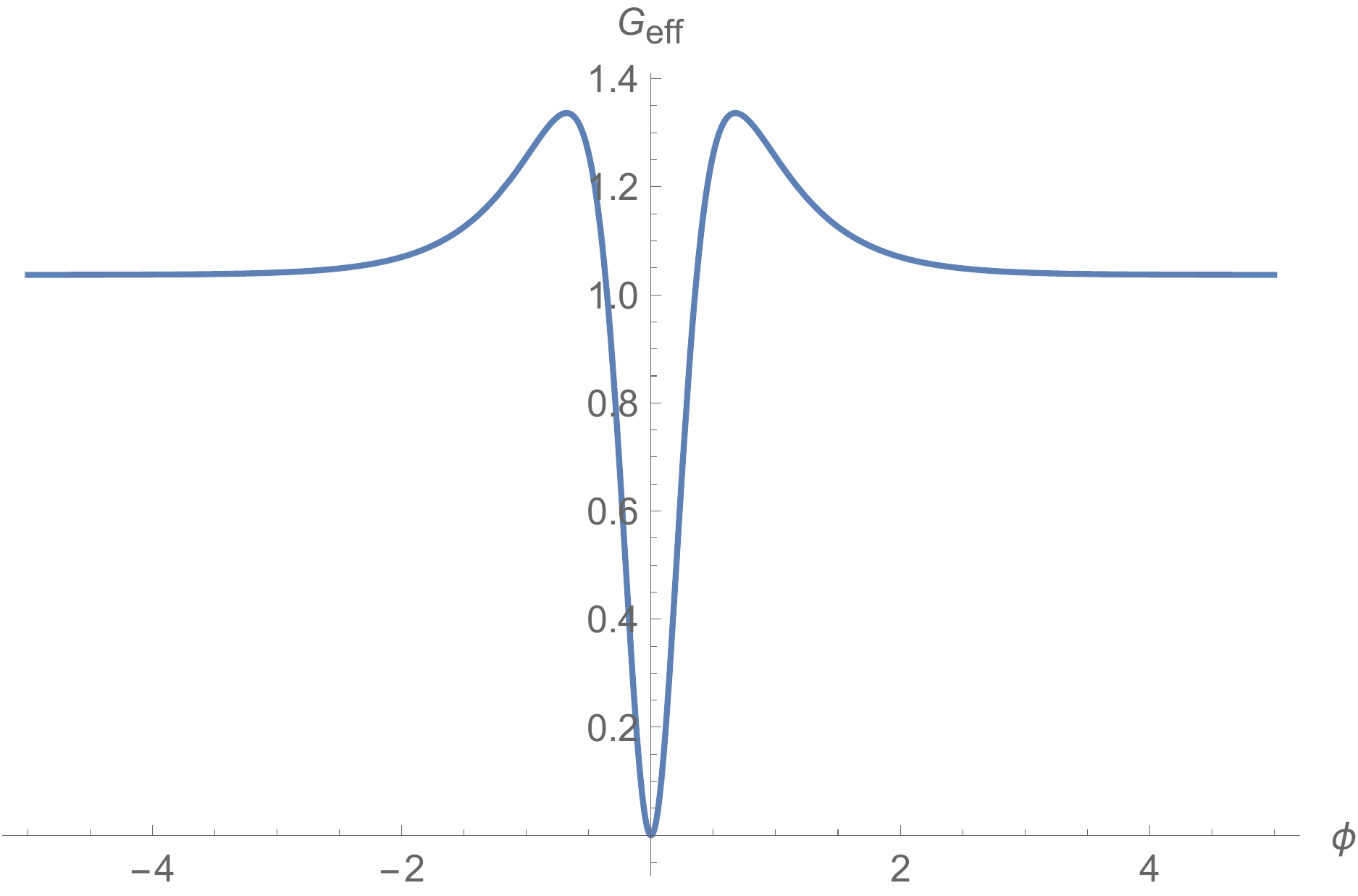}
 \subcaption{$E>0$}\label{fig:Geff E>0}
 \end{minipage}
 \caption[$G_{\rm eff}$ in the non-interacting GFT model]{The effective gravitational constant as a function of
   relational time $\phi$ for $E<0$ (Fig.~\ref{fig:Geff E<0}) and $E>0$ (Fig.~\ref{fig:Geff E>0}), in arbitrary units. There is a bounce
   replacing the classical singularity in both cases.The origin in the
   plots correponds to the bounce, occuring at $\phi=\Phi$. The
   asymptotic value for large $\phi$ is the same in both cases and
   coincides with Newton's constant. In the case $E<0$ this limit is
   also a supremum, whereas in the $E>0$ case $G_{\rm eff}$ has two
   maxima, equally distant from the bounce, and approaches Newton's $G_{\rm N}$
   constant from above.}
\end{figure}

Let us discuss in more detail the properties of the model at finite
(relational) times. Eq.~(\ref{eq:effective dynamics}) predicts a bounce
when $\zeta(\phi)$ vanishes. We denote by $\Phi$ the `instant' when the
bounce takes place. One can thus eliminate the integration
constant $b$ in favour of $\Phi$ 
\be b=\frac{\log \left(\sqrt{E^2+4
    z Q^2}\right)}{2 \sqrt{z}}-\Phi.  \ee 
We  define the \textit{effective} gravitational constant as
\be
G_{\rm eff}=\frac{1}{3\pi}\zeta^2, \ee
which can be expressed, using
Eqs.~(\ref{eq:rho}), (\ref{eq:effective dynamics}) as \be
G_{\rm eff}=\frac{G_{\rm N} \left(E^2+12 \pi G_{\rm N} Q^2\right)\sinh ^2\left(2 \sqrt{3
    \pi G_{\rm N}} (\phi -\Phi )\right)}{\left(E -\sqrt{E^2+12 \pi G_{\rm N} Q^2}
  \cosh \left(2 \sqrt{3 \pi G_{\rm N}} (\phi -\Phi )\right)\right)^2}.  \ee
Its profile is given in Fig.~\ref{fig:Geff E<0},\ref{fig:Geff E>0}, in
the cases $E<0$, $E>0$ respectively. Notice that it is symmetric about
the line $\phi=\Phi$, corresponding to the bounce.

The energy density has a maximum at the bounce, where the volume
reaches its minimum value 
\be
\varepsilon_{\rm max}=\frac{1}{2}\frac{Q^2}{V_{\rm\scriptscriptstyle bounce}^2}, \ee 
where 
\be
V_{\rm\scriptscriptstyle bounce}=\frac{V_{j_0} \left(\sqrt{E^2+12 \pi G_{\rm N} Q^2}-E\right)}{6 \pi
  G_{\rm N}}.  \ee 
  Clearly, the singularity is always avoided for $E<0$ and,
provided $Q\neq 0$, it is also avoided in the case
$E>0$. Moreover, if the GFT energy is negative, the energy density has
a vanishing limit at the bounce for vanishing $Q$:
 \be \lim_{Q\to
  0}\varepsilon_{\rm max}=0, \hspace{1em} E<0.  \ee 
Therefore, in this limiting case the energy density is zero at all
times. Nevertheless, the universe will still expand following the
evolution equations~(\ref{eq:effective dynamics}) and 
\be \lim_{Q\to0}V(\phi)=\frac{|E|
  V_{j_0} \cosh ^2\left(\sqrt{3 \pi G_{\rm N} } (\phi -\Phi )\right)}{3 \pi
  G_{\rm N}}, \hspace{1em} E<0.  \ee 
This is to be contrasted with the Friedmann equation in classical
cosmology~(\ref{eq:Friedmann}), which implies that the rate of expansion is zero
when the energy density vanishes.

\begin{figure}
  \begin{minipage}{.5\textwidth}
  \centering
 \includegraphics[width=\linewidth]{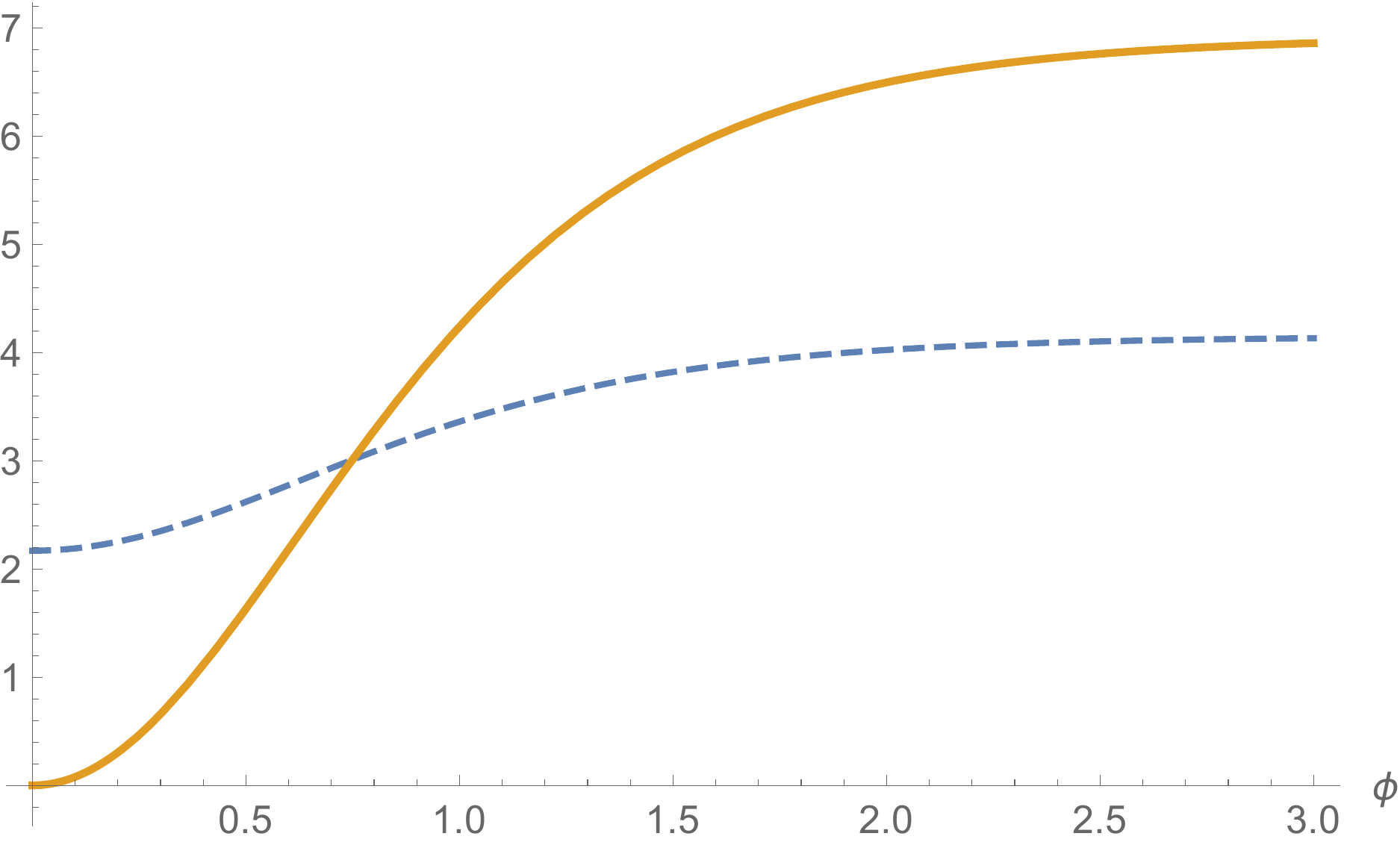}
 \subcaption{$E<0$}\label{fig:E<0}
 \end{minipage}%
\begin{minipage}{.5\textwidth}
  \centering
 \includegraphics[width=\linewidth]{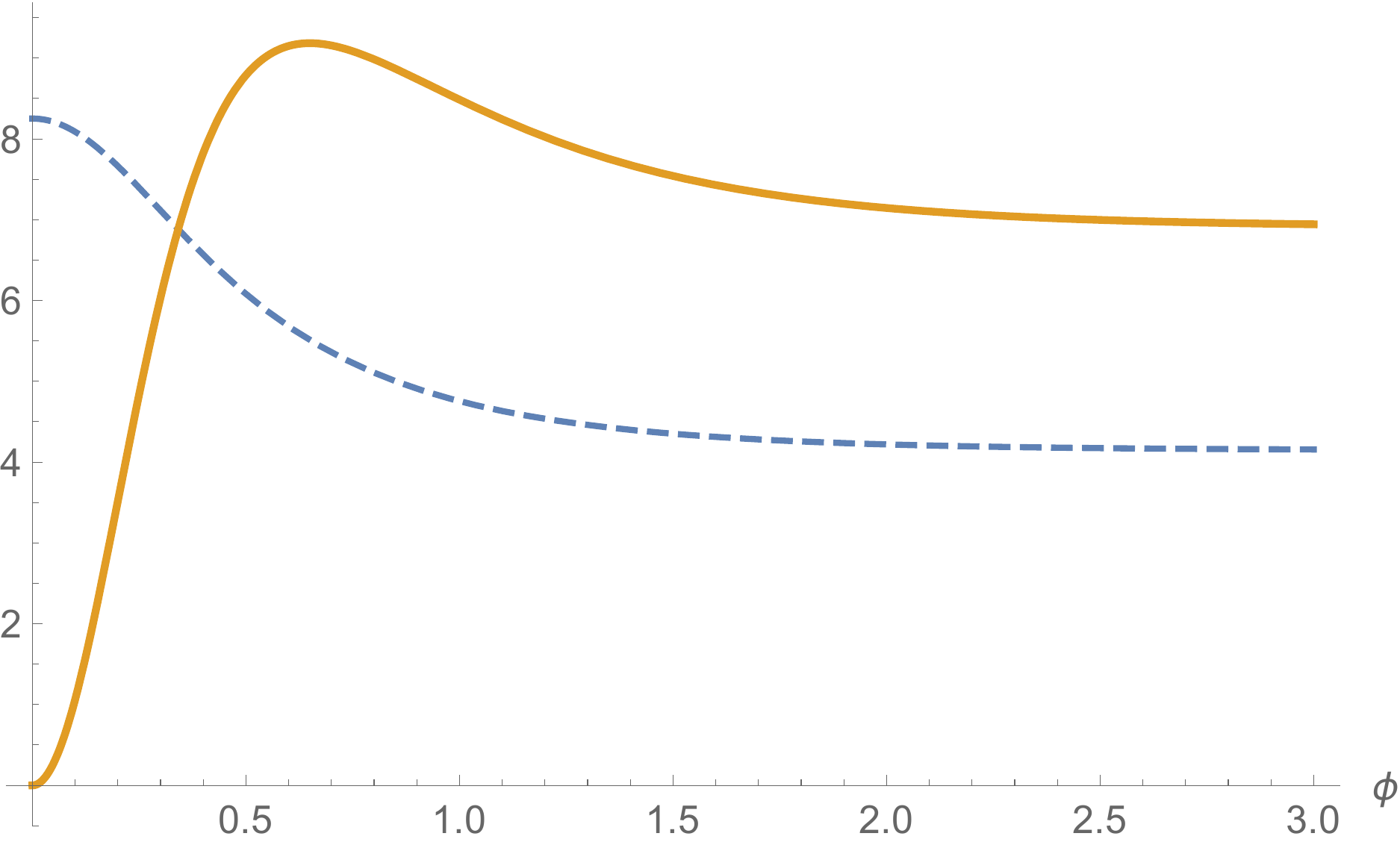}
 \subcaption{$E>0$}\label{fig:E>0}
   \end{minipage}
   \caption[Acceleration in the non-interacting GFT model]{The l.h.s.~and the r.h.s.~of inequality
   (\ref{ineq:acceleration}) as functions of the relational time
   $\phi$ correspond to the the dashed (blue) and thick (orange) curve
   respectively, in arbitrary units. When the dashed curve is above
   the thick one the universe is undergoing an epoch of accelerated
   expansion following the bounce. Fig.~\ref{fig:E<0} corresponds to
   the case $E<0$, whereas Fig.~\ref{fig:E>0}  is relative to the
   opposite case $E>0$. Notice that for the latter there is a stage of
   maximal deceleration after exiting the `inflationary' era. After
   that the acceleration takes less negative values until it relaxes
   to its asymptotic value. For $E<0$ instead the asymptote is
   approached from below.}
 \end{figure}

It is possible to express the condition that the universe has a
positive acceleration in purely relational terms. In fact this very
notion relies on the
choice of a particular time parameter, namely proper time, for its definition.
Introducing the scale factor
and proper time as in Eqs.~(\ref{eq:momentum}),~(\ref{eq:scaleFactor})
one finds
\be\label{eq:acceleration}
\frac{\ddot{a}}{a}=\frac{2}{3}\varepsilon\left[
  \frac{\pa_\phi^2 V}{V}-\frac{5}{3}\left(\frac{\pa_\phi V}{V}\right)^2\right].
\ee 
We observe that the last equation can also be rewritten as
\be\label{eq:accelerationFluids}
\frac{\ddot{a}}{a}=-\frac{4}{9}Q^2\left(4\frac{\gamma_R}{V^2}+7\frac{\gamma_E}{V^3}+10\frac{\gamma_Q}{V^4}\right).
\ee
We can trade the condition $\ddot{a}>0$
for having an accelerated expansion with the following one, which only
makes reference to relational evolution of observables (see Appendix~\ref{Appendix:EffectiveFriedmann})
\be
\frac{\pa_\phi^2 V}{V}>\frac{5}{3}\left(\frac{\pa_\phi V}{V}\right)^2~.
\ee 
The two conditions are obviously equivalent. 
However,
the second one has a wider range of applicability, since it is
physically meaningful also when the scalar field has
vanishing momentum. Making use of
  Eq.~(\ref{eq:effective dynamics}) the condition above can be
rewritten as 
\be\label{ineq:acceleration}
4z+\frac{2E}{\rho^2}>\frac{20}{3} \zeta^2.  \ee 
This is satisfied
trivially in a neighbourhood of the bounce since $g$ vanishes there
and the l.h.s. of the inequality is strictly positive, see
Figs.~\ref{fig:E<0},~\ref{fig:E>0}. It is instead violated at infinity,
consistently with a decelerating universe in the classical regime.
 
 Unfortunately, this epoch of accelerated expansion does not last long enough. In fact, the number of e-folds obtained in this model is much smaller than the one that is usually required in inflationary models, \emph{i.e.} $\mathcal{N}\ll50-60$ (see Section~\ref{Sec:InflationSolvesPuzzles}). We will show in Section~\ref{sec:GeometricInflation} how this problem can be addressed by the inclusion of suitable interactions in the GFT model, with no modifications to the matter sector.

 \subsection{Discussion of the non-interacting model}\label{sec:discussion}
 The dynamics of the universe predicted by the GFT model studied above is purely relational\footnote{These considerations are valid in general and are not limited to the non-interacting model considered in this section. In fact, the need for relational observables follows as a direct consequence from the background independence of the GFT approach. In particular, the dynamics of the emergent universe will be formulated in relational terms also in the case of the interacting model considered in the next section.},
 \textit{i.e.}, using the language of Ref.~\cite{Rovelli:2001bz}, it is expressed by the functional relation
 between \textit{partial observables}, here given by the volume $V$ and the
 scalar field $\phi$. According to this interpretation, physically meaningful statements
 about the predicted value of $V$ can only be made in conjunction with
 statements about the predicted outcome of a measurement of $\phi$.
 In fact, this interpretation is inspired by one of the main insights of GR,
 namely by the observation that coordinate time is purely gauge-dependent, and is therefore deprived
 of any physical meaning. Thus, it cannot be expected to play any role
 in the quantum theory either.
 In other words, the dynamics can be described entirely using the so called \textit{complete observables},
 which in this model  are exhausted by the functional relation $V(\phi)$. In a theory with gauge invariance, such quantities are the only ones having physical meaning;
 they can be seen as functions on the space of solutions modulo all gauges~\cite{Rovelli:2001bz}.
 Gauge invariance of $V(\phi)$ is trivially verified in classical cosmology;
 it is also valid at the quantum level, since gauge invariance of the volume operator
 follows from its general definition in GFT~\cite{Gielen:2016dss}. A discussion on the implementation of
  diffeomorphism invariance in GFT can be found in Ref.~\cite{Baratin:2011tg}.

We have shown that the bounce is accompanied by an early stage of accelerated expansion,
occurring for any values of the conserved quantities $E$, $Q$ (provided that the latter is non-vanishing)
 and despite the fact that the scalar field has no potential. This is a promising feature of the model, which  indicates that the
framework adopted allows for a mechanism leading to an accelerated expansion through quantum geometry effects.
From the point of view of Eq.~(\ref{eq:effective Friedmann}) one can say that the accelerated expansion 
is a consequence of $G_{\rm eff}$ not being constant. By looking at Eqs.~(\ref{eq:FriedmannEffFluids}),~(\ref{eq:fluids}),~(\ref{eq:accelerationFluids})
one sees that this phenomenon can be traced back to an effective fluid component with a negative energy density,
which arises from quantum geometry effects. In the next section we will see that the bounce is not spoiled by the inclusion of interactions between GFT quanta, whereas they have important effects for the duration of the acceleration era and for the  subsequent stages of expansion of the emergent universe.

\section{Emergent background: the effect of interactions}\label{Sec:GFT:Interactions}
In this Section, we study the cosmological implications of interactions between
GFT quanta from a phenomenological perspective. In
particular, we show how GFT interactions lead to a recollapse of the
universe while preserving the bounce replacing the initial
singularity, which has already been shown to occur in the free
case studied in the previous Section. It is remarkable that cyclic non-singular cosmologies are thus obtained in
this framework without any \emph{a priori} assumption on the geometry
of spatial sections of the emergent spacetime, unlike Standard Cosmology (Chapter~\ref{Chapter:StandardCosmology}). Furthermore, we show
how interactions make it possible to have an early epoch of
accelerated expansion, which can be made to last for an arbitrarily
large number of e-folds, without the need to introduce an \emph{ad
  hoc} potential for the scalar field.

\subsection{Non linear dynamics of a GFT condensate}\label{sec:Basics}

The dynamics of an isotropic GFT condensate can be described by means
of the effective action \cite{Oriti:2016qtz,Oriti:2016ueo}
\be\label{eq:EffectiveAction}
S=\int\de\phi\;\left(A~|\partial_{\phi}\sigma|^2+\mathcal{V}(\sigma)\right).
\ee
Extremising the effective action (\ref{eq:EffectiveAction}) one recovers the equations of motion of the
isotropic mean field, derived in Section~\ref{sec:2} for a simplicial GFT model. As in the previous section, we have defined $\sigma=\varphi^{j \iota^{\star}}$, where $j$ and $\iota^{\star}$ are fixed so that they can be dropped to make the notation lighter.
The mean field $\sigma$  is a complex scalar field that depends on the
relational time $\phi$.  As we showed in the previous section, the dynamics of the emergent universe
is given by the effective Friedmann equations, which can be written in relational form as
\begin{align}
\frac{\partial_{\phi}V}{V}&=2\frac{\partial_{\phi}\rho}{\rho},\label{eq:EmergingFriedmannI}\\
\frac{\partial^2_{\phi}V}{V}&=2\left[\frac{\partial^2_{\phi}\rho}{\rho}+\left(\frac{\partial_{\phi}\rho}{\rho}\right)^2\right]\label{eq:EmergingFriedmannII}~,
\end{align}
where $\rho=|\sigma|$ and $V$ is the volume.

The form of the effective potential
$\mathcal{V}(\sigma)$ can be motivated by means of the microscopic GFT
model and we require it to be bounded from below. For the applications considered in this section, its form will be determined by purely phenomenological considerations. There is an
ambiguity in the choice of the sign of $A$, which is not fixed by the
microscopic theory and will turn out to be particularly relevant for
the cosmological applications of the model\footnote{This ambiguity has also been discussed earlier in
  Ref.~\cite{Calcagni:2014tga} when exploring the possibility of embedding LQC
  in GFT.}. In particular, it can be
used to restrict the class of microscopic models by selecting only
those that are phenomenologically viable.  In fact, as we will show,
only models entailing $A<0$ are sensible from a phenomenological point
of view since otherwise one would have faster than exponential
expansion.

In this section we consider a model based on an effective potential of the following form
\be\label{eq:Potential}
\mathcal{V}(\sigma)=B|\sigma(\phi)|^2+\frac{2}{n}w|\sigma|^n+\frac{2}{n^{\prime}}w^{\prime}|\sigma|^{n^{\prime}},
\ee where we can assume $n^{\prime}>n$ without loss of generality. The
terms in the effective potential can be similarly motivated as in
Ref.~\cite{Oriti:2016qtz,Oriti:2016ueo}. The interaction terms appearing in
GFT actions are usually defined in such a way that the perturbative
expansion of the GFT partition function reproduces that of spin foam
models. Specifically, spin foam models for quantum gravity in $d=4$ are
mostly based on quintic interactions, called simplicial. An example is the EPRL vertex considered in Section~\ref{sec:2}.
In
the case that the GFT field is endowed with a particular tensorial
transformation property, other classes of models can be obtained whose
interaction terms, called tensorial, are based on even powers of the
modulus of the field. In this light, the particular type of
interactions considered here can be understood as mimicking such types
of interactions, which is the reason why we will refer to them as
pseudosimplicial and pseudotensorial, respectively. In the following
we will study their phenomenological consequences, and show how
interesting physical effects are determined as a result of the
interplay between two interactions of this type. The integer-valued
powers $n$, $n^{\prime}$ in the interactions are assumed non-negative and will be kept unspecified
throughout this section, thus making our analysis retain its full
generality. The particular values motivated by the above discussion
can be retrieved as particular cases. In the following we will show
how different ranges for such powers lead to phenomenologically
interesting features of the model, most notably concerning an early
era of accelerated expansion in Section~\ref{sec:MultiCriticalCase}.

Since $\mathcal{V}(\sigma)$ has to be bounded from below, we require
$w^{\prime}>0$. The equation of motion of the field $\sigma$ obtained
from Eqs. (\ref{eq:EffectiveAction}),~(\ref{eq:Potential}) is \be
-A\partial_{\phi}^2\sigma+B\sigma+w|\sigma|^{n-2}\sigma+w^{\prime}|\sigma|^{n^{\prime}-2}\sigma=0.
\ee Writing the complex field $\sigma$ in polar form 
$\sigma=\rho\;\e^{i\theta}$ one finds
(Ref.~\cite{Oriti:2016qtz,Oriti:2016ueo}) that the equation of motion for the
angular component leads to the conservation law
\be\label{eq:ConservationOfQ}
\partial_{\phi}Q=0,\;\mbox{with}\;Q\equiv\rho^2
\partial_{\phi}{\theta}, \ee while the radial component satisfies a
second order ODE\footnote{We note that Eq.~(\ref{eq:RadialEquation}) is Ermakov equation complemented with the contribution of two extra terms representing the interactions.}
 \be\label{eq:RadialEquation}
\partial^2_{\phi}\rho-\frac{Q^2}{\rho^3}-\frac{B}{A}\rho-\frac{w}{A}\rho^{n-1}-\frac{w^{\prime}}{A}\rho^{n^{\prime}-1}=0.
\ee 
The conserved charge $Q$ is identified with the momentum of the
scalar field $\pi_{\phi}= Q$ \cite{Oriti:2016qtz,Oriti:2016ueo}. One
immediately observes, that for large values of $\rho$ the term
$\rho^{n^{\prime}-1}$ becomes dominant.  In order to ensure that
Eq.~(\ref{eq:RadialEquation}) does not lead to drastic departures from
Standard Cosmology at late times (cf. Eq.~(\ref{eq:EmergingFriedmannI})),
the coefficient of such term has to be positive
\be\label{eq:DefinitionMu} \mu\equiv-\frac{w^{\prime}}{A}>0, \ee which
implies, since $w^{\prime}>0$, that one must have $A<0$. In fact, the
opposite case $\mu<0$ would lead to an open cosmology expanding at a
faster than exponential rate, which relates to a Big Rip.
 Thus, considering $A<0$, compatibility with the free case (see
 Refs.~\cite{Oriti:2016qtz,Oriti:2016ueo,deCesare:2016axk}) demands
 \be\label{eq:DefinitionMassSquared} z\equiv\frac{B}{A}>0, \ee which
 in turn implies $B<0$. The sign of $w$ is a priori not constrained,
 which leaves a considerable freedom in the model. Given the signs of
 the parameters $B$ and $w^{\prime}$, the potential in
 Eq.~(\ref{eq:Potential}) can be related to models with spontaneous
 symmetry breaking in Statistical Mechanics and Quantum Field
 Theory. The sub-leading term in the potential plays an important
 r{\^o}le in determining an inflationary-like era, as shown below in
 Section~\ref{sec:MultiCriticalCase}.

The connection to the theory of critical phenomena is to be expected
from the conjecture that GFT condensates arise through a phase
transition from a non-geometric to a geometric
phase~\cite{Oriti:2007qd,Oriti:2013jga}, which could be a possible realisation
of the geometrogenesis scenario~\cite{Konopka:2006hu}. Despite the
lack of a detailed theory near criticality and the fact that the
occurrence of the aforementioned phase transition is still a
conjecture, there are nonetheless encouraging results coming from the
analysis (both perturbative and non-perturbative) of the
Renormalisation Group flow, which shows the existence of non-trivial IR fixed
points in certain models, see Refs.~\cite{Benedetti:2014qsa,Geloun:2015qfa,Geloun:2016qyb,Benedetti:2015yaa,Carrozza:2016vsq}. On
this ground we will adopt, as a working hypothesis, the formation of a
condensate as a result of the phase transition, as in
Refs.~\cite{Oriti:2016qtz,Oriti:2016ueo,Calcagni:2014tga,Gielen:2016uft,Pithis:2016wzf,Gielen:2013kla,Gielen:2013naa,Gielen:2014ila,Gielen:2014uga,Gielen:2014usa,Gielen:2015kua,Sindoni:2014wya}.

\begin{figure}
\centering
\includegraphics[width=0.5\columnwidth]{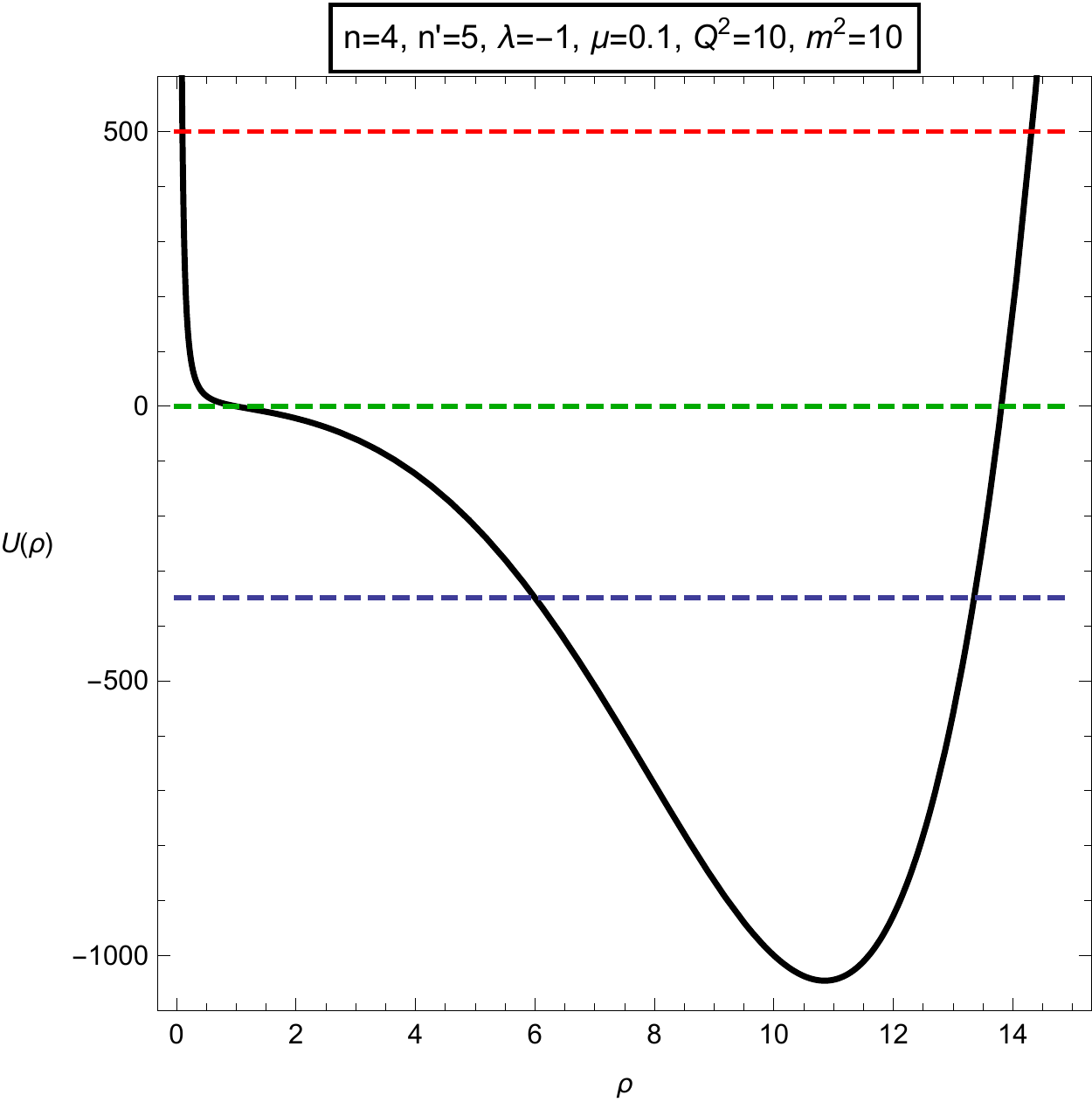}
\caption[Potential of the interacting GFT model]{Plot of the potential $U(\rho)$
  (Eq.~(\ref{eq:PointParticlePotentialU}) ) for the dynamical system
  described by Eq.~(\ref{eq:NewtonEquation}) and a particular choice
  of parameters. The three horizontal curves correspond to different
  values of the `GFT energy' $E$, in turn corresponding to different
  choices of initial conditions for $\rho$, $\rho^{\prime}$. The
  corresponding orbits in phase space are shown in
  Fig.~\ref{fig:PhasePortrait}. Recollapse is generic feature of the
  model and occurs for any values of the parameters, provided $\mu>0$
  and $Q\neq0$.}\label{fig:PotentialU}
\end{figure}

From Eqs.~(\ref{eq:DefinitionMu}),~(\ref{eq:DefinitionMassSquared})
and defining 
\be\label{eq:DefinitionLambda} \lambda\equiv-\frac{w}{A},
\ee 
we can rewrite Eq.~(\ref{eq:RadialEquation}) in the form
\be\label{eq:NewtonEquation} \partial^2_{\phi}\rho-z \rho
-\frac{Q^2}{\rho^3}+\lambda\rho^{n-1}+\mu\rho^{n^{\prime}-1}=0~, \ee
that will be used throughout the rest of this section.  The above
equation has the form of the equation of motion of a classical point
particle with potential (see Fig.~\ref{fig:PotentialU})
\be\label{eq:PointParticlePotentialU}
U(\rho)=-\frac{1}{2}z\rho^2+\frac{Q^2}{2\rho^2}+\frac{\lambda}{n}\rho^n+\frac{\mu}{n^{\prime}}\rho^{n^{\prime}}.
\ee 
Equation~(\ref{eq:NewtonEquation}) leads to another conserved
quantity, $E$, defined as 
\be\label{eq:Energy}
E=\frac{1}{2}(\partial_{\phi}\rho)^2+U(\rho), \ee 
which is referred to
as `GFT energy'~\cite{Oriti:2016qtz,Oriti:2016ueo,deCesare:2016axk}. Its
physical meaning, from a fundamental point of view, is yet to be
clarified.

\begin{figure}
\centering
\includegraphics[width=0.5\columnwidth]{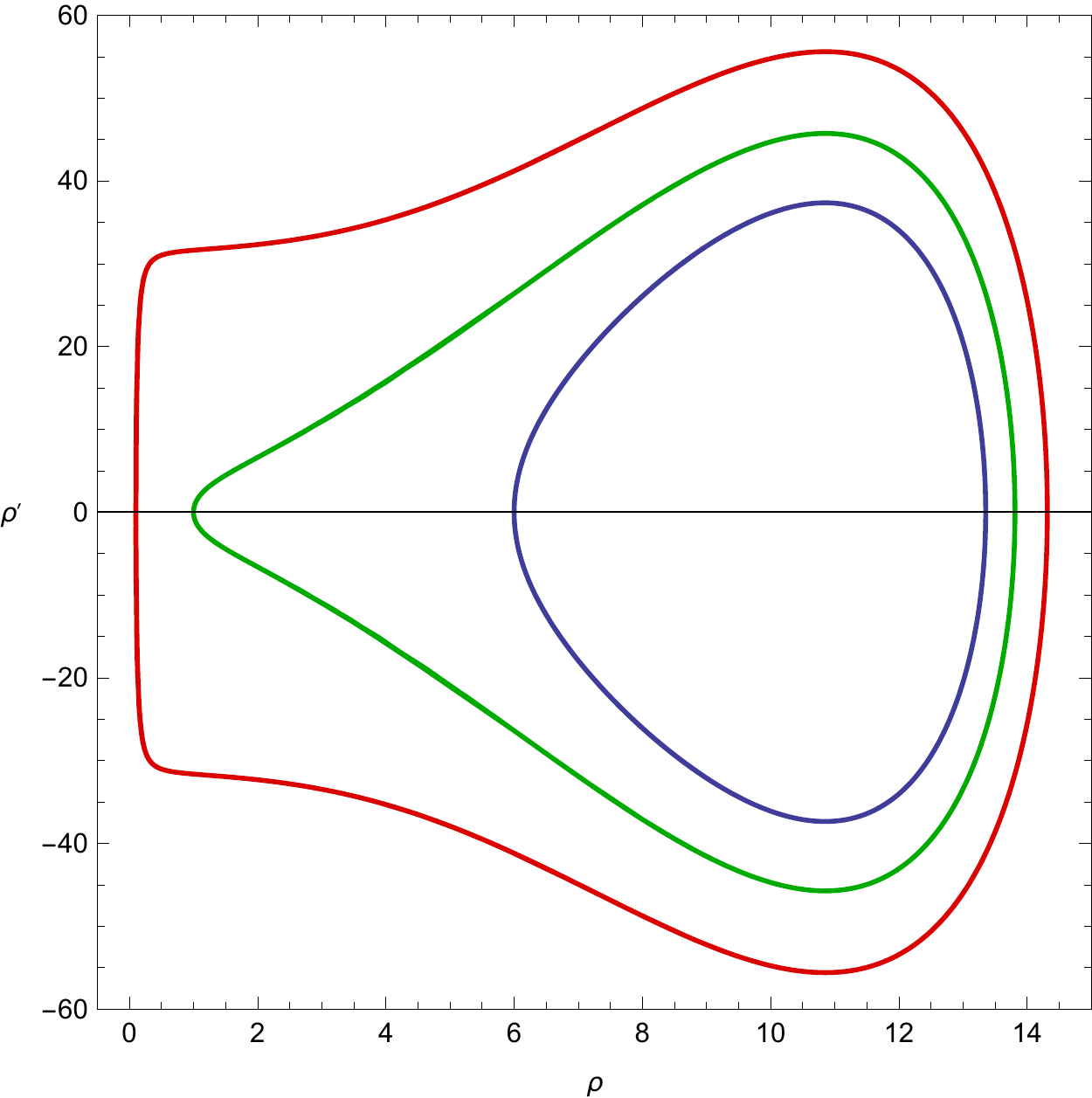}
\caption[Phase portrait interacting GFT model]{Phase portrait of the dynamical system given by
  Eq.~(\ref{eq:NewtonEquation}). Orbits have energy given by the
  corresponding color lines as Fig.~\ref{fig:PotentialU}. Orbits are
  periodic and describe oscillations around the stable equilibrium
  point (\emph{center fixed point}) given by the absolute minimum of
  the potential $U(\rho)$. This is a general feature of the model
  which does not depend on the particular choice of parameters,
  provided Eq.~(\ref{eq:DefinitionMu}) is
  satisfied.}\label{fig:PhasePortrait}
\end{figure}

\begin{figure}
\centering
\includegraphics[width=0.5\columnwidth]{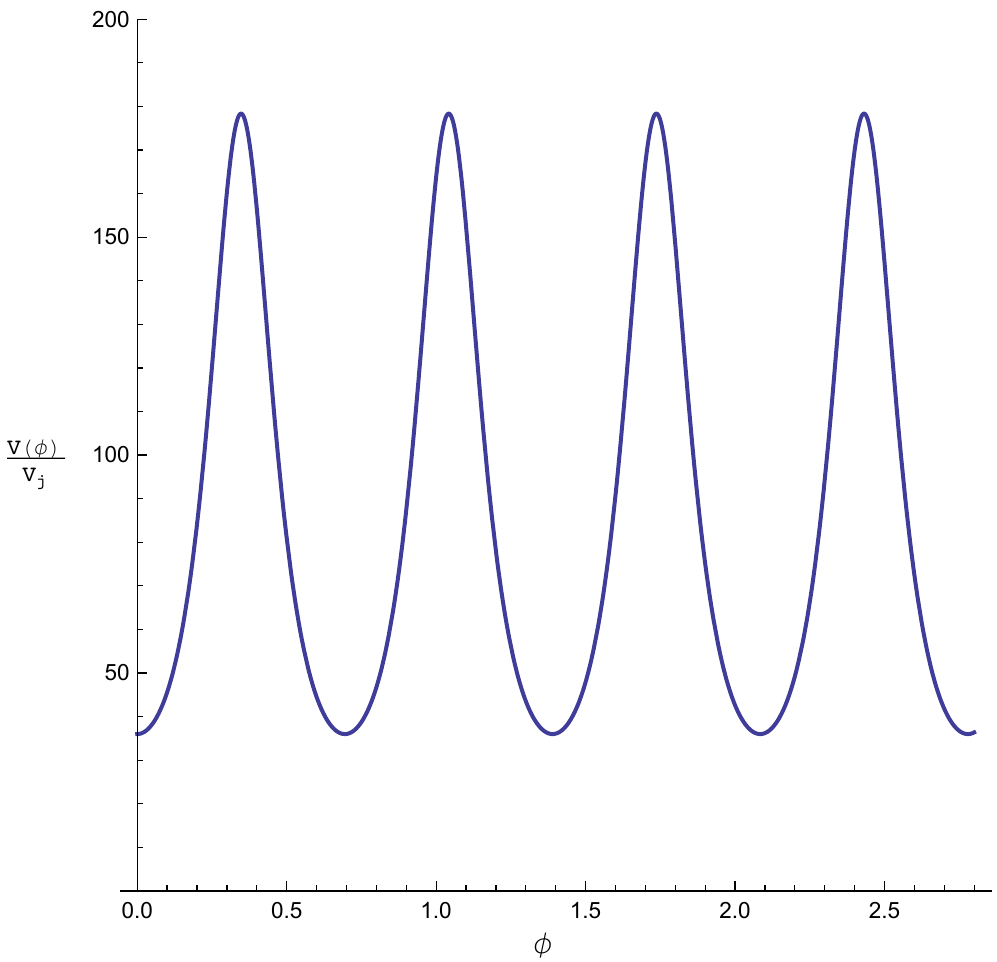}
\caption[Cyclic universe]{Plot of the volume of the universe as a function of
  relational time $\phi$ (in arbitrary units), corresponding to the
  blue orbit in Fig.~\ref{fig:PhasePortrait}. As a generic feature of
  the interacting model the universe undergoes a cyclic (and non-singular) evolution and
  its volume has a positive minimum, corresponding to a
  bounce.}\label{fig:Volume}
\end{figure}

\subsection{Recollapsing universe}\label{sec:Recollapse}

Some properties of the solutions of the model and its consequences for
cosmology can already be drawn by means of a qualitative analysis of
the solutions of the second order ODE in
Eq.~(\ref{eq:PointParticlePotentialU}). In fact, solutions are
confined to the positive half-line $\rho>0$, given the infinite
potential barrier at $\rho=0$ for $Q\neq 0$. Moreover, since $\mu>0$
the potential in Eq.~(\ref{eq:PointParticlePotentialU}) approaches
infinity as $\rho$ takes arbitrarily large values. Therefore, provided
that we fix the `GFT energy' $E$ at a value which is larger than
both the absolute minimum and that of (possible) local maxima of the
potential $U(\rho)$, the solutions of Eq.~(\ref{eq:NewtonEquation})
turn out to be cyclic motions (see Figs.~\ref{fig:PotentialU},
\ref{fig:PhasePortrait}) describing oscillations around a stable
equilibrium point. These, in turn, correspond via
Eq.~(\ref{eq:VolumeExpectationValueDefinition}) to cyclic solutions for the dynamics
of the universe Eqs.~(\ref{eq:EmergingFriedmannI}),
(\ref{eq:EmergingFriedmannII}) (see Fig.~\ref{fig:Volume}).

It is interesting to compare this result with what is known in the
case where interactions are
disregarded~\cite{Oriti:2016qtz,Oriti:2016ueo,deCesare:2016axk}. In that case one
has that the universe expands indefinitely and in the limit
$\phi\to\infty$ its dynamics follows the ordinary Friedmann equation
for a flat universe, filled with a massless and minimally coupled
scalar field. Therefore, we see that the given interactions in the GFT
model induce a recollapse of the universe, corresponding to the
turning point of the motion of $\rho$, as seen in
Fig.~\ref{fig:PhasePortrait}.

It is well-known in the classical theory that such a recollapse
follows as a simple consequence of the closed topology of 3-space. In
the GFT framework instead, the topology of space(time) is not fixed at
the outset, but should rather be reconstructed from the behavior of
the system in the macroscopic limit. In other words, the simple
condensate ansatz used here does not provide any information about the
topology of spatial sections of the emergent spacetime which, as it is
well known, play an important r{\^o}le in the dynamics of classical
cosmological models. Any topological information must therefore come
from additional input. A possible strategy one could follow is to work
with generalised condensates encoding such
information~\cite{Oriti:2015qva}. Here, instead, we propose that the
closedness of the reconstructed space need not be encoded in the
condensate ansatz as an input, but is rather determined by the
dynamics as a consequence of the GFT interactions. Hence, allowing
only interactions that are compatible with reproducing a given spatial
topology, one may recover the classical correspondence between closed
spatial topology and having a finitely expanding universe.

\subsection{Geometric inflation}\label{sec:GeometricInflation}

Cosmology obtained from GFT displays a number of interesting features
concerning the initial stage of the evolution of the universe, which
mark a drastic departure from the standard FLRW cosmologies. In
particular, the initial big-bang singularity is replaced by a regular
bounce (see Ref.~\cite{Oriti:2016qtz,Oriti:2016ueo,deCesare:2016axk}), followed
by an era of accelerated expansion \cite{deCesare:2016axk}. Similar
results were also obtained in the early LQC
literature, see \emph{e.g.}
Refs.~\cite{Bojowald:2001xe,Bojowald:2002nz}.  However, it is not
obvious \emph{a priori} that they must hold for GFT as
well. Nonetheless, it is remarkable that these two different
approaches yield qualitatively similar results for the dynamics of the
universe near the classical singularity, even though this fact by
itself does not necessarily point at a deeper connection between the
two.

In the model considered in this section, our results have a purely
quantum geometric origin and do not rely on the assumption of a
specific potential for the minimally coupled scalar
field\footnote{This is also the case in LQC, see
    Ref.~\cite{Bojowald:2002nz}. However, the number of e-folds computed in
    that framework turns out to be too small in order to supplant
    inflation~\cite{Bojowald:2003mc}.}, which is taken to be
massless and introduced for the sole purpose of having a relational
clock. This is quite unlike inflation, which instead heavily relies on
the choice of the potential and initial conditions for the inflaton in
order to predict an era of accelerated expansion with the desired
properties.

In this section we investigate under which conditions on the
interaction potential of the GFT model it is possible to obtain an
epoch of accelerated expansion that could last long enough, so as to
account for the minimum number of e-folds required by standard
arguments. The number of e-folds is given by
\be\label{eq:DefinitionNumberE-FoldsVolume}
\mathcal{N}=\frac{1}{3}\log\left(\frac{V_\mathrm{end}}{V_\mathrm{bounce}}\right),
\ee where $V_\mathrm{bounce}$ is the volume of the universe at the
bounce and $V_\mathrm{end}$ is its value at the end of the era of
accelerated expansion. A necessary condition for it to be called an
`inflationary' era is that the number of e-folds must be large enough,
namely $\mathcal{N}\gtrsim 60$.

Using Eq.~(\ref{eq:VolumeExpectationValueDefinition}), which in our case simplifies to
\be\label{eq:VolumeExpectationValueDefinition_SingleSpin}
V(\phi)= V_{j_0}\rho^2(\phi)~,
\ee
we can rewrite Eq.~(\ref{eq:DefinitionNumberE-FoldsVolume}) as
\be\label{eq:DefinitionNumberE-FoldsVolumeRho}
\mathcal{N}=\frac{2}{3}\log\left(\frac{\rho_\mathrm{end}}{\rho_\mathrm{bounce}}\right),
\ee with an obvious understanding of the notation. This formula is
particularly useful since it allows us to derive the number of e-folds
only by looking at the dynamics of $\rho$. We recall that $\rho$ denotes the norm of the `wave-function' $\sigma$; it must not be confused with the energy density.

Since there is no notion of proper time, a sensible definition of
acceleration can only be given in relational terms. In particular, we
seek a definition that agrees with the standard one given in ordinary
cosmology. As in Ref.~\cite{deCesare:2016axk} we can thus define the
acceleration as (see also the discussion following Eq.~(\ref{eq:acceleration}) and Appendix~\ref{Appendix:EffectiveFriedmann})
\be\label{eq:DefinitionAcceleration}
\mathfrak{a}(\rho)\equiv\frac{\partial^2_{\phi}V}{V}-\frac{5}{3}\left(\frac{\partial_{\phi}V}{V}\right)^2.
\ee Hence, from Eqs.~(\ref{eq:EmergingFriedmannI}),
(\ref{eq:EmergingFriedmannII}) one gets the following expression for
the acceleration $\mathfrak{a}$ as a function of $\rho$ for a generic
potential \be
\mathfrak{a}(\rho)=-\frac{2}{\rho^2}\left\{\partial_{\phi}U(\rho )
\rho +\frac{14}{3}\left[E-U(\rho )\right]\right\}.  \ee Using
Eq.~(\ref{eq:PointParticlePotentialU}) one finally has for our model
\be\label{eq:AccelerationVsRho}
\mathfrak{a}(\rho)=-\frac{2}{\rho^2}\left[\frac{14}{3}E+\left(1  -\frac{14
   }{3 n^{\prime}}\right) \mu\rho^{n^{\prime}}+\frac{4 z \rho ^2}{3}\right.
+\left.\left(1 -\frac{14  }{3n}\right)\lambda \rho
   ^n-\frac{10 Q^2}{3 \rho ^2}\right].
\ee 
Therefore, the sign of the acceleration is opposite to that of the
polynomial 
\be s(\rho)=P(\rho)+\left(3 -\frac{14 }{n}\right)\lambda
\rho ^{n+2}+\left(3 -\frac{14 }{ n^{\prime}}\right)
\mu\rho^{n^{\prime}+2}, \ee 
where we defined 
\be P(\rho)=4 z \rho
^4+14 E\rho^2-10 Q^2.  \ee In the following we will study in detail
the properties of the era of accelerated expansion. We will start by examining the free case, 
whereas the r{\^o}le of
interactions in allowing for an inflationary-like era will be studied later in this section.

\subsubsection{The non-interacting case}\label{sec:FreeCase}

In this case the acceleration is given by 
\be
\mathfrak{a}(\rho)=-\frac{2}{3\rho^4}P(\rho).  \ee
The bounce occurs
when $\rho$ reaches its minimum value, \emph{i.e.} when $U(\rho)=E$,
leading to 
\be\label{eq:RhoMinimumFreeCase}
\rho_\mathrm{bounce}^2=\frac{1}{S}\left(\sqrt{E^2+z Q^2}-E\right).
\ee 
A straightforward calculation shows that
$\mathfrak{a}(\rho_\mathrm{bounce})>0$ as expected. The era of
accelerated expansion ends when $P(\rho)$ vanishes, which happens at a
point $\rho_{\star}>\rho_\mathrm{bounce}$, which is given by
\be\label{eq:RhoEndFreeCase}
\rho_{\star}=\frac{1}{4S}\left(\sqrt{49E^2+40z Q^2}-7E\right).
\ee We can then use Eqs.~(\ref{eq:DefinitionNumberE-FoldsVolumeRho}),
(\ref{eq:RhoMinimumFreeCase}), (\ref{eq:RhoEndFreeCase}) to determine the
energy $E$ as a function of the number of e-folds $\mathcal{N}$. Reality of $E$
thus leads to the following bounds on $\mathcal{N}$ \be
\frac{1}{3}\log\left(\frac{10}{7}\right)\leq \mathcal{N}\leq
\frac{1}{3}\log\left(\frac{7}{4}\right), \ee that is \be 0.119\lesssim
\mathcal{N}\lesssim 0.186.  \ee Such tight bounds, holding for all values of the
parameters $z$ and $Q^2$, rule out the free case as a candidate to
replace the standard inflationary scenario in cosmology.

\subsubsection{The interacting case: the multicritical model}\label{sec:MultiCriticalCase}

In this subsection we investigate the consequences of interactions for
the evolution of the universe. In particular, we show how the
interplay between the two interaction terms in the effective potential
(Eq.~\ref{eq:Potential}) makes it possible to have an early epoch of
accelerated expansion, which lasts as long as in inflationary models.
Before studying their effect, we want to discuss how the occurrence of
such interaction terms could be motivated from the GFT perspective. In
principle, one could have infinitely many interaction terms given by
some power of the GFT field. However, only a finite number of them
will be of relevance at a specific scale, as dictated by the behaviour
of the fundamental theory under the Renormalisation Group (RG) flow.

In a continuum and large scale limit new terms in the action could be
generated, whereas others might become irrelevant. In this sense, one
might speculate that, \emph{e.g.}, in addition to the five-valent simplicial
interaction term the effective potential includes another term which
becomes relevant on a larger scale. Ultimately, rigorous RG arguments
will of course have the decisive word regarding the possibility to
obtain such terms from the fundamental theory. Nevertheless, by
studying the phenomenological features of such potentials and
extracting physical consequences from the corresponding cosmological
solutions, we aim at clarifying the map between the fundamental
microscopic and effective macroscopic dynamics of the theory. At the
same time, our results might help to shed some light onto the subtle
issue of the physical meaning of such interaction terms.

Hereafter we assume the hierarchy $\mu\ll |\lambda|$, since otherwise
an inflationary era cannot be easily accommodated. This means that the
higher order term in the interaction potential $\mathcal{V}(\sigma)$
becomes relevant only for very large values of the condensate field
$\sigma$, hence of the number of quanta representing the basic
building blocks of quantum spacetime. Consequently, the dynamics in
the immediate vicinity of the bounce is governed by the parameters of
the free theory and the sub-leading interaction term.

To begin with, let us start by fixing the value of the GFT energy. We
require the universe to have a Planckian volume at the bounce. Since
the volume is given by Eq.~(\ref{eq:VolumeExpectationValueDefinition_SingleSpin}), this is done
by imposing $\rho_\mathrm{bounce}=1$. Such a condition also fixes the
value of the GFT energy to \be E=U\left(\rho_\mathrm{bounce}=1\right).
\ee In fact, we demand that $\rho_\mathrm{bounce}$ is the minimal
value of $\rho$ which is compatible with the GFT energy $E$ available
to the system. Hence, we also have the condition
$\partial_{\rho}U\left(\rho_\mathrm{bounce}=1\right)\leq0$. Notice
that this is trivially satisfied in the free case. In the interacting
case (holding the hierarchy $\mu\ll |\lambda|$) one can therefore use
it to obtain a bound on $\lambda$ \be\label{first constraint on
  lambda} \lambda\leq z+Q^2.  \ee It is convenient for our purposes
and in order to carry over our analysis in full generality, to
introduce the definitions
\begin{align}
\alpha&\equiv\left(3-\frac{14}{n}\right)\lambda\label{eq:DefinitionAlpha},\\
\beta&\equiv\left(3-\frac{14}{n^{\prime}}\right)\mu\label{eq:DefinitionBeta}.
\end{align}
The acceleration Eq.~(\ref{eq:AccelerationVsRho}) can thus be written
as 
\be\label{eq:AccelerationAlphaBeta}
\mathfrak{a}(\rho)=-\frac{2}{\rho^4}\left[P(\rho)+\alpha
  \rho^{n+2}+\beta\rho^{n^{\prime}+2}\right].  \ee 
As pointed out
before, $\mathfrak{a}>0$ has to hold at the bounce. The first thing to
be observed is that $\alpha<0$ is a necessary condition in order to
have enough e-folds. In fact, if this were not the case, the bracket
in Eq.~(\ref{eq:AccelerationAlphaBeta}) would have a zero at a point
$\rho_\mathrm{end}<\rho_{\star}$ (cf. Eq.~(\ref{eq:RhoEndFreeCase})),
thus leading to a number of e-folds which is even smaller than the
corresponding one in the free case. Furthermore, it is possible to
constrain the value of $\mu$ in a way that leads both to the
aforementioned hierarchy and to the right value for $\mathcal{N}$, which we
consider as fixed at the outset. In order to do so, we solve
Eq.~(\ref{eq:DefinitionNumberE-FoldsVolumeRho})
w.r.t. $\rho_\mathrm{end}$, having fixed the bounce at
$\rho_\mathrm{bounce}=1$ \be\label{eq:EndBounceN}
\rho_\mathrm{end}=\rho_\mathrm{bounce}~\mbox{e}^{\frac{3}{2}\mathcal{N}}.  \ee
The end of geometric inflation occurs when the polynomial in the bracket in
Eq.~(\ref{eq:AccelerationAlphaBeta}) has a zero. Since
$\rho_\mathrm{end}\gg1$, it is legitimate to determine this zero by
taking into account only the two highest powers in the polynomial,
with respect to which all of the other terms are negligible. We
therefore have \be
\alpha\rho_\mathrm{end}^{n+2}+\beta\rho_\mathrm{end}^{n^{\prime}+2}\approx0,
\ee which, using Eq.~(\ref{eq:EndBounceN}), leads to
\be\label{eq:AlphaBetaN} \beta=-\alpha \; \e^{-\frac{3}{2} \mathcal{N}
  (n^{\prime}-n)}.  \ee The last equation is consistent with the
hierarchy $\mu\ll |\lambda|$ and actually fixes the value of $\mu$
once $\lambda$, $n$, $n^{\prime}$ and $\mathcal{N}$ are assigned. Furthermore
one has $\beta>0$ which, together with
Eqs.~(\ref{eq:DefinitionMu}),~(\ref{eq:DefinitionBeta}), implies
$n^{\prime}>\frac{14}{3}$. Importantly, this means that $n^{\prime}=
5$ is the lowest possible integer compatible with an inflationary-like
era. This particular value is also interesting in another respect,
since in GFT typically only specific combinatorially non-local
interactions characterized by such a power allow for an interpretation in
terms of simplicial quantum gravity (cf. Refs.~\cite{Oriti:2006se,Oriti:2014yla}).

Our considerations so far leave open two possibilities, namely:
 \begin{itemize}
 \item $\lambda<0$ and $n\geq5$ ($n^{\prime}>n$), which in the case of $n=5$ could correspond to the just mentioned simplicial interaction term and the higher order $n'$-term could possibly be generated in the continuum and large scale limit of the theory and becomes dominant for very large $\rho$. For even $n'$ it mimics so-called tensorial interactions.
\item $\lambda>0$ and $2<n<5$ ($n^{\prime}\geq5$), which for $n'=5$ could allow a connection to simplicial quantum gravity and would remain dominant for large $\rho$ over the $n$-term, which in the case $n=4$ is reminiscent of an interaction of tensorial type.
  \end{itemize}

\begin{figure}
\begin{minipage}{.3\textwidth}
\includegraphics[width=\linewidth]{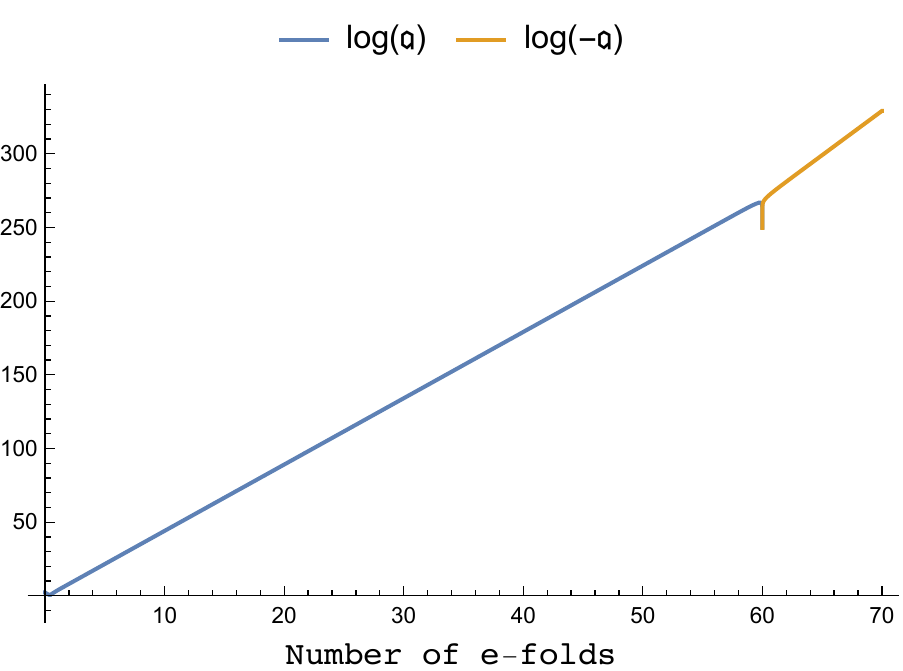}
\subcaption{}\label{fig:eFolds}
\end{minipage}
\begin{minipage}{.3\textwidth}
\vskip 16pt
\includegraphics[width=\linewidth]{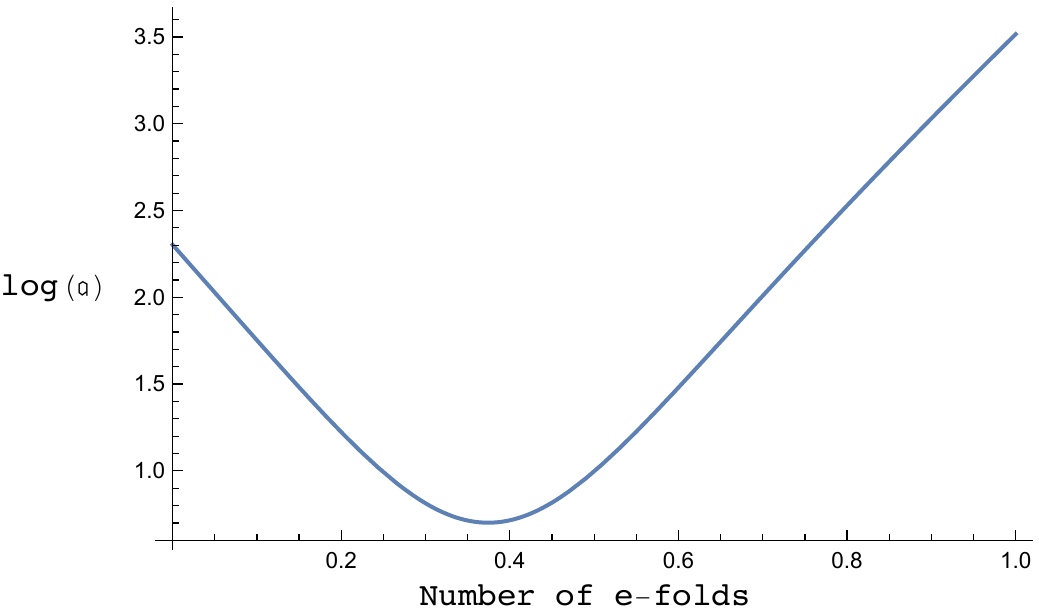}
\subcaption{}\label{fig:eFoldsBounce}
\end{minipage}
\begin{minipage}{.3\textwidth}
\includegraphics[width=\linewidth]{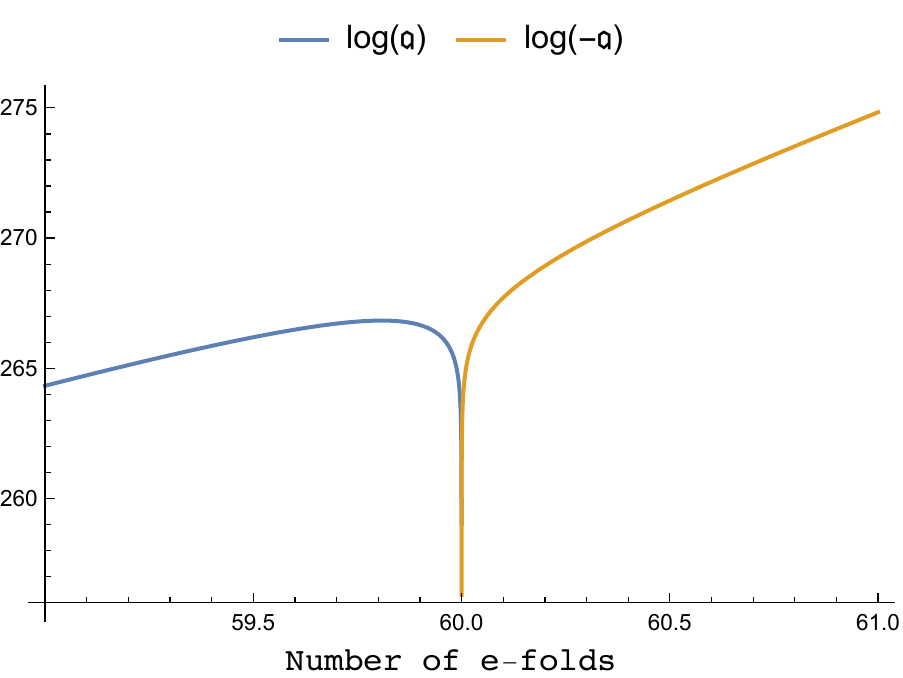}
\subcaption{}\label{fig:eFoldsEnd}
\end{minipage}

\caption[Number of e-folds in the interacting GFT model]{Inflationary era supported by GFT interactions in the
  multi-critical model. The blue (orange) curve in Fig.~(\ref{fig:eFolds}) represents the graph
  of the logarithm of the acceleration (minus the acceleration) as a
  function of the number of e-folds in the case $\lambda<0$. The plot
  refers to the particular choice of parameters $n=5$, $n^{\prime}=6$,
  $m=1$, $Q=1$, $\lambda=-3$. The value of $\mu$ is determined from
  Eq.~(\ref{eq:AlphaBetaN}) by requiring the number of e-folds to be
  $\mathcal{N}=60$. There is a logarithmic singularity at $\mathcal{N}\simeq 60$, marking
  the end of the accelerated
  expansion. 
  Figs.~\ref{fig:eFoldsBounce},~\ref{fig:eFoldsEnd} show
  the behavior of the acceleration close to the bounce and at the end
  of `inflation' respectively.}
\end{figure}

However, this is not yet enough in order to guarantee an
inflation-like era. In fact we have to make sure that there is no
intermediate stage of deceleration occurring between the bounce at
$\rho_b=1$ and $\rho_\mathrm{end}$, \emph{i.e.}, that
$\mathfrak{a}(\rho)$ stays positive in the interval between these two
points. In other words we want to make sure that $\rho_\mathrm{end}$
is \emph{the only zero} of the acceleration lying to the right of
$\rho_b$. In fact $\mathfrak{a}(\rho)$ starts positive at the bounce
and has a minimum when $P(\rho)$ becomes of the same order of
magnitude of the term containing the power $\rho^{n+2}$ (see
Eq.~(\ref{eq:AccelerationAlphaBeta})).  Thus we see that we have to
require that the local minimum of $\mathfrak{a}(\rho)$ (\emph{i.e.}~the maximum of the poynomial in brackets in
Eq.~(\ref{eq:AccelerationAlphaBeta})) is positive (resp.~negative). As
$\rho$ increases further, the acceleration increases again until it
reaches a maximum when the contribution coming from the term
containing $\rho^{n^{\prime}+2}$ becomes of the same order of
magnitude of the other terms. Thereafter the acceleration turns into a
decreasing function all the way until $\rho\to+\infty$ and therefore
has a unique zero. Positivity of the local minimum of
$\mathfrak{a}(\rho)$ translates into a further constraint on
parameters space. By direct inspection, it is possible to see that the
latter case listed above does not satisfy such condition for any value
of the parameters of the model. Therefore we conclude that
\emph{$\lambda$ must be negative} if the acceleration is to keep the
same sign throughout the inflationary era. The evolution of the
acceleration as a function of relational time $\phi$ is shown in
Figs.~\ref{fig:eFolds},~\ref{fig:eFoldsBounce},~\ref{fig:eFoldsEnd}
for some specific choice of the parameters. It is worthwhile stressing
that the behaviour of the model in the case $\lambda<0$ is
nevertheless generic and therefore does not rely on the specific
choice of parameters. Furthermore, by adjusting the value of $\mathcal{N}$ and
the other parameters in Eq.~(\ref{eq:AlphaBetaN}), it is possible to
achieve any desirable value of e-folds during geometric inflation.

All we said in this section applies to the multi-critical model with
the effective potential Eq.~(\ref{eq:Potential}) but \emph{does not
  hold} in a model with only one interaction term. In fact in that
case it is not possible to prevent the occurrence of an intermediate
era of deceleration between $\rho_{b}$ and $\rho_\mathrm{end}$, the
latter giving the scale at which the higher order interaction term
becomes relevant.

One last remark is in order: geometric inflation was shown to be a feature of
multicritical GFT models but only at the price of a fine-tuning in the
value of the parameter $\mu$ (see Eq.~(\ref{eq:AlphaBetaN})).

\subsection{Interactions and the final fate of the universe}\label{sec:EffectiveFriedmannEffFluids}

It is possible to recast the dynamical equations for the volume of the
universe in a form that bears a closer resemblance to the standard
Friedmann equation, as shown in Ref.~\cite{deCesare:2016axk}. In fact, the
Hubble expansion rate can be expressed as (see Appendix~\ref{Appendix:EffectiveFriedmann} for more
details) 
\be\label{eq:HubbleRate}
H=\frac{1}{3}\frac{\partial_{\phi}V}{V^2}\pi_{\phi}.  \ee 
From
Eq.~(\ref{eq:EmergingFriedmannI}) and the proportionality between the
momentum of the scalar field and $Q$ we have
\be\label{eq:FriedmannHubbleRate} H^2=\frac{4}{9}\frac{ Q^2}{
  V^2}\left(\frac{\partial_{\phi}\rho}{\rho}\right)^2.  \ee 
The term
in bracket can thus be interpreted as a dynamical effective
gravitational constant as in Ref.~\cite{deCesare:2016axk}.  Alternatively,
using
Eqs.~(\ref{eq:VolumeExpectationValueDefinition_SingleSpin}),~(\ref{eq:Energy}),~(\ref{eq:PointParticlePotentialU}),
the last equation Eq.~(\ref{eq:FriedmannHubbleRate}) becomes
\be\label{eq:FriedmannEffectiveFluids} H^2=\frac{8
   Q^2}{9}
\left[\frac{\varepsilon_z}{V^2}+\frac{\varepsilon_E}{V^3}+\frac{\varepsilon_Q}{V^4}+\frac{\varepsilon_\lambda}{V^{3-n/2}}+\frac{\varepsilon_\mu}{V^{3-n^{\prime}/2}}\right],
\ee 
where we defined
\begin{align}
\varepsilon_E&=V_j E,\\
\varepsilon_z&=\frac{z}{2},\\
\varepsilon_Q&=-\frac{Q^2}{2}V_j^2,\\
\varepsilon_\lambda&=-\frac{\lambda}{n}V_j^{1-n/2},\\
\varepsilon_{\mu}&=-\frac{\mu}{n^{\prime}}V_j^{1-n^{\prime}/2}.
\end{align}
The exponents of the denominators in
Eq.~(\ref{eq:FriedmannEffectiveFluids}) can be related to the $w$
coefficients in the equation of state $p=w \varepsilon$ of some
effective fluids, with energy density $\varepsilon$ and pressure
$p$. Each term scales with the volume as $\propto V^{-(w+1)}$.

It is worth pointing out that Eq.~(\ref{eq:FriedmannEffectiveFluids})
makes clearer the correspondence with the framework of ekpyrotic
models\footnote{We are thankful to Martin Bojowald for this
  observation (private communication).}, where one has the gravitational field coupled to
matter fields with $w>1$. Such models have been advocated as a
possible alternative to inflation, see Ref.~\cite{Lehners:2008vx}. We
observe that at early times (\emph{i.e.} small volumes) the occurrence
of the bounce is determined by the negative sign of $\varepsilon_{Q}$,
which is also the term corresponding to the highest $w$. However,
while this is sufficient to prevent the classical singularity, it is
not enough to guarantee that the minimum number of e-folds is reached
at the end of the accelerated expansion, as shown in
Section~\ref{sec:FreeCase}. In fact, the role of interactions is
crucial in that respect, as our analysis in
Section~\ref{sec:MultiCriticalCase} has shown.

At this stage we would like to focus instead on the consequences of
having interactions in the GFT model for the evolution of the universe
at late times. As we have already seen in Section~\ref{sec:Recollapse}, a positive $\mu$ entails a recollapsing
universe. This should also be clear from
Eq.~(\ref{eq:FriedmannEffectiveFluids}). In particular, we notice that
the corresponding term in the equation is an increasing function of
the volume for $n^{\prime}>6$. This is quite an unusual feature for a
cosmological model, where all energy components (with the exception of
the cosmological constant) are diluted by the expansion of the
universe. For $n^{\prime}=6$ one finds instead a cosmological constant
term. It is also possible to have the interactions reproduce the
classical curvature term $\propto \frac{K}{V^{2/3}}$ by choosing
$n^{\prime}=\frac{14}{3}$, which is however not allowed if one
restricts to integer powers in the interactions.

Our analysis shows that only $\lambda<0$ leaves room for an era
accelerated expansion analogous to that of inflationary models. In
order for this to be possible, one must also have $n^{\prime}>n\geq
5$. Moreover, if one rules out phantom energy (\emph{i.e.} $w<-1$), there
is only one case which is allowed, namely $n=5$, $n^{\prime}=6$. Then
during geometric inflation the universe can be described as dominated by a fluid
with equation of state $w=-\frac{1}{2}$. After the end of geometric inflation, the
energy density also receives contribution from a negative
cosmological constant, which eventually leads to a recollapse. It is
remarkable that this particular case selects an interaction term which
is in principle compatible with the simplicial interactions which have
been extensively considered in the GFT approach. However, it must be
pointed out that the realisation of the geometric inflation picture
imposes strong restrictions also on the type of interactions one can
consider, as well as on their relative strength.


%
\section{Stepping beyond isotropy}\label{sec:Perturbations}
The cosmologies that have been considered so far, with the exception of the partial analysis in Ref.~\cite{Pithis:2016cxg}, are `isotropic', thus governed by a single global degree of freedom, corresponding to the total volume (or scale factor) of the universe. Our aim in this section is to take a first step beyond the isotropic case. We will study GFT perturbations involving anisotropic degrees of freedom around isotropic background configurations, focusing on the cosmological evolution around the bouncing region, and remaining within the mean field approximation of the full quantum dynamics. 

A precise notion of isotropy of the emergent universe must rely on the definition of suitable observables that can quantify anisotropies. Nevertheless, such anisotropies are going to be encoded in the states describing a condensate of tetrahedra that deviate from perfect monochromaticity, by construction. In fact, the minisuperspace of homogeneous continuum spatial geometries can be identified with the (gauge-invariant) configuration space of a quantum tetrahedron (for details, see Ref.~\cite{Gielen:2014ila} and the review \cite{Gielen:2016dss}).
A detailed study in terms of physical observables will be needed to explore the precise relation between microscopic anisotropies (non-monochromaticity), and macroscopic ones, although the fact itself is not in question. In particular, if one is only interested in the specific issue of whether anisotropies become dominant during the effective cosmological  evolution or remain subdominant compared to the isotropic background, it is enough to study the dynamics of non-monochromatic GFT perturbations and their relative amplitude compared to the background condensate wavefunction. This is the issue we focus on.

Another difficulty we have to face (which affects the whole GFT cosmology programme, like all attempts to extract physics from GFT or Spin Foam models) is purely technical, and lies in the complication of the analytic expression of the interaction kernel for physically interesting models, here the EPRL model. The Lorentzian EPRL  vertex amplitude, in fact, has an analytic expression which can be expressed in integral form (where the integrals are over the Lorentz group and encode the covariant properties of the model) \cite{Speziale:2016axj}, but has not been put down explicitly as a function of its boundary data, which are usually given in terms of $\mbox{SU(2)}$ representations (to match the LQG form of quantum states). This is a serious limitation for the computation of transition amplitudes in the full theory, as well as for the solution of the classical dynamics of the corresponding GFT model, which would be our concern here. This is true both for the background dynamics and for that of the perturbations, so it is not sidestepped by our focus on the latter.

In the absence of an explicit closed expression for the Lorentzian EPRL vertex, there are two possible options. The first one is to adopt a more phenomenological approach and model the GFT interactions by simple functions that capture some of their essential features \cite{deCesare:2016axk,deCesare:2016rsf,Pithis:2016wzf,Pithis:2016cxg}. This is the path we followed in Section~\ref{Sec:GFT:Interactions}. The second option, which will be the focus of this section, is to treat the problem perturbatively in the regime where GFT interactions become subdominant compared to the kinetic term. This is what happens close to the cosmological bounce, or for small values of the GFT coupling constants.
The power of the perturbative scheme lies in the fact that only some general properties (\emph{e.g.} symmetries) of the interaction kernel are used, which makes our analysis applicable to a fairly broad class of GFT models.
Still, further results on the explicit evaluation of the fundamental dynamics will obviously be very important and will be needed to make the analysis more quantitative.

In Section~\ref{sec:Non-MonochromaticPerturbations} we derive the dynamics of non-monochromatic perturbations to first order, which is given by a system of four coupled linear differential equations. The terms arising from the linearisation of the interaction have non-constant coefficients depending on the background. Within this approximation, only one of the four faces of the tetrahedra can be perturbed. The others must match the spin of the background, due to a constraint given by the EPRL vertex. This is done in full generality (for EPRL-like models with a broad class of kinetic kernels). The dynamics can then be recast in a simpler and more compact form in the particular case of a local kinetic kernel and when considering a background with only spin $j=\frac{1}{2}$ being excited. Considering this specific case, in Section~\ref{sec:DynamicsAtBounce} we study the dynamics of non-monochromatic components around the bounce, where interactions are negligible. Hence, both the background and the perturbations satisfy linear equations of motion. In this regime, there is no need to impose the condition that perturbations are much smaller than the background. Thus, the perturbed geometry of the emergent spacetime can in principle be quite different from the one given by the background.

An important result we derive is the determination of a region of parameter space such that perturbations are bounded at all times while the background field grows unbounded. It is thus justified to neglect non-monochromaticity after the bounce, when the non-linear regime is entered. However, around the bounce the magnitude of the perturbations can be of the same order as the background, leading to interesting consequences. To illustrate this point, we compute geometric quantities such as the \emph{surface-area-to-volume ratio} and the \emph{effective volume per quantum}, which characterise the non-trivial corrections to the `mean geometry' of the elementary monochromatic constituents around the cosmological bounce. Our results do not depend on where we set the initial conditions for the non-monochromatic perturbations, \emph{i.e.} whether their amplitude is maximal at the bounce or in the contracting/expanding phase.

\subsection{Non-monochromatic perturbations}\label{sec:Non-MonochromaticPerturbations}
We can now derive the equations of motion for perturbations around an isotropic background GFT field configuration $\varphi_0$ satisfying Eq.~(\ref{eq:BackgroundDynamics}). 
\be
\varphi=\varphi_0+\delta\varphi.
\ee
Let us start by writing down the more general equation of motion for the background, by relaxing the isotropy assumption while retaining the other hypotheses of Sections~\ref{sec:LeftInv}.
\be
0=\frac{\delta S[\varphi]}{\delta\overline{\varphi}^{abcd\,\tilde{\iota}}}=\frac{\delta K[\varphi]}{\delta\overline{\varphi}^{abcd\,\tilde{\iota}}}+\frac{\delta \overline{V}_5[\varphi]}{\delta\overline{\varphi}^{abcd\,\tilde{\iota}}}
\ee
Above, we wrote explicitly all of the spin labels $j_\nu=(a,b,c,d)$. The first term is equal to
\be
\frac{\delta K[\varphi]}{\delta\overline{\varphi}^{abcd\,\tilde{\iota}}}=\mathcal{K}^{abcd\,\tilde{\iota}}\varphi^{abcd\,\tilde{\iota}} \; \; ,
\ee
while for the second term we have
\be\label{eq:InteractionTermGeneralEOM}
\begin{split}
\frac{\delta V_5[\varphi]}{\delta\varphi^{abcd\,\tilde{\iota}}}&=\frac{1}{5}\sum \left[ \varphi^{d567\,\iota_2}\varphi^{7c89\,\iota_3}\varphi^{96b10\,\iota_4}\varphi^{1085a\,\iota_5}\mathcal{V}^{\,\prime}_5\left(\scriptstyle{a,b,c,d,5,6,7,8,9,10;\tilde{\iota},\iota_2,\iota_3,\iota_4,\iota_5}\right)+\right.\\
&\phantom{=\frac{1}{5}\sum [ } \varphi^{123a\,\iota_1}\varphi^{d389\,\iota_3}\varphi^{9c210\,\iota_4}\varphi^{108b1\,\iota_5}\mathcal{V}^{\,\prime}_5\left(\scriptstyle{1,2,3,a,b,c,d,8,9,10;\iota_1,\tilde{\iota},\iota_3,\iota_4,\iota_5}\right)+\\
&\phantom{=\frac{1}{5}\sum [ }\varphi^{12b4\,\iota_1}\varphi^{456a\,\iota_2}\varphi^{d6210\,\iota_4}\varphi^{10c51\,\iota_5}\mathcal{V}^{\,\prime}_5\left(\scriptstyle{1,2,b,4,5,6,a,c,d,10;\iota_1,\iota_2,\tilde{\iota},\iota_4,\iota_5}\right)+\\
&\phantom{=\frac{1}{5}\sum [ }\varphi^{1c34\,\iota_1}\varphi^{45b7\,\iota_2}\varphi^{738a\,\iota_3}\varphi^{d851\,\iota_5}\mathcal{V}^{\,\prime}_5\left(\scriptstyle{1,c,3,4,5,b,7,8,a,d;\iota_1,\iota_2,\iota_3,\tilde{\iota},\iota_5}\right)+\\
&\phantom{=\frac{1}{5}\sum [ }\left. \varphi^{d234\,\iota_1}\varphi^{4c67\,\iota_2}\varphi^{73b9\,\iota_3}\varphi^{962a\,\iota_4}\mathcal{V}^{\,\prime}_5\left(\scriptstyle{d,2,3,4,c,6,7,b,9,a;\iota_1,\iota_2,\iota_3,\iota_4,\tilde{\iota}}\right)\right].
\end{split}
\ee
By just relabelling the indices, we obtain
\be
\begin{split}
\frac{\delta V_5[\varphi]}{\delta\varphi^{abcd\,\tilde{\iota}}}=&
\frac{1}{5}\sum\varphi^{d567\,\iota_2}\varphi^{7c89\,\iota_3}\varphi^{96b10\,\iota_4}\varphi^{1085a\,\iota_5}[\mathcal{V}^{\,\prime}_5\left(\scriptstyle{a,b,c,d,5,6,7,8,9,10;\tilde{\iota},\iota_2,\iota_3,\iota_4,\iota_5}\right)+\\&\mathcal{V}^{\,\prime}_5\left(\scriptstyle{10,8,5,a,b,c,d,6,7,9;\iota_5,\tilde{\iota},\iota_2,\iota_3,\iota_4}\right)+\mathcal{V}^{\,\prime}_5\left(\scriptstyle{9,6,b,10,8,5,a,c,d,7;\iota_4,\iota_5,\tilde{\iota},\iota_2,\iota_3}\right)+\\&\mathcal{V}^{\,\prime}_5\left(\scriptstyle{7,c,8,9,6,b,10,5,a,d;\iota_3,\iota_4,\iota_5,\tilde{\iota},\iota_2}\right)+\mathcal{V}^{\,\prime}_5\left(\scriptstyle{d,5,6,7,c,8,9,b,10,a;\iota_2,\iota_3,\iota_4,\iota_5,\tilde{\iota}}\right)] \;\; ,
\end{split}
\ee
which becomes, taking into account the discrete symmetries of the interaction kernel, the simpler expression 
\be\label{eq:InteractionTermGeneral}
\frac{\delta V_5[\varphi]}{\delta\varphi^{abcd\,\tilde{\iota}}}=\sum\varphi^{d567\,\iota_2}\varphi^{7c89\,\iota_3}\varphi^{96b10\,\iota_4}\varphi^{1085a\,\iota_5}\mathcal{V}^{\,\prime}_5\left(\scriptstyle{a,b,c,d,5,6,7,8,9,10;\tilde{\iota},\iota_2,\iota_3,\iota_4,\iota_5}\right).
\ee

Moreover, given the structure of the interaction term in Eq.~(\ref{eq:InteractionTermGeneral}), the only non-vanishing contributions to the first order dynamics of the perturbations around a monochromatic background come from terms having at least three identical spins among $(a,b,c,d)$. Therefore, depending on which of the four indices, labeled $j^\prime$, is singled out to be different from the other three, labeled $j$, we obtain four independent equations
\be
\mathcal{K}^{jjjj^{\prime}\,\tilde{\iota}}\delta\varphi^{jjjj^{\prime}\,\tilde{\iota}}+\sum_{\iota}\delta\overline{\varphi}^{j^{\prime}jjj\,\iota}\left(\overline{\varphi_0}^{j\,\iota^\star}\right)^3\mathcal{V}^{\,\prime}_5\left(\scriptstyle{j,j,j,j^{\prime},j,j,j,j,j,j;\tilde{\iota},\iota,\iota^\star,\iota^\star,\iota^\star}\right)=0.
\ee
\be
\mathcal{K}^{jj j^{\prime}j\,\tilde{\iota}}\delta\varphi^{jj j^{\prime}j\,\tilde{\iota}}+\sum_{\iota}\delta\overline{\varphi}^{jj^{\prime}jj\,\iota}\left(\overline{\varphi_0}^{j\,\iota^\star}\right)^3\mathcal{V}^{\,\prime}_5\left(\scriptstyle{j,j,j^{\prime},j,j,j,j,j,j,j;\tilde{\iota},\iota^\star,\iota,\iota^\star,\iota^\star}\right)=0.
\ee
\be
\mathcal{K}^{j j^{\prime}jj\,\tilde{\iota}}\delta\varphi^{j j^{\prime}jj\,\tilde{\iota}}+\sum_{\iota}\delta\overline{\varphi}^{jjj^{\prime}j\,\iota}\left(\overline{\varphi_0}^{j\,\iota^\star}\right)^3\mathcal{V}^{\,\prime}_5\left(\scriptstyle{j, j^{\prime},j,j,j,j,j,j,j,j;\tilde{\iota},\iota^\star,\iota^\star,\iota,\iota^\star}\right)=0.
\ee
\be
\mathcal{K}^{j^{\prime}jjj\,\tilde{\iota}}\delta\varphi^{j^{\prime}jjj\,\tilde{\iota}}+\sum_{\iota}\delta\overline{\varphi}^{jjjj^{\prime}\,\iota}\left(\overline{\varphi_0}^{j\,\iota^\star}\right)^3\mathcal{V}^{\,\prime}_5\left(\scriptstyle{j^{\prime},j,j,j,j,j,j,j,j,j;\tilde{\iota},\iota^\star,\iota^\star,\iota^\star,\iota}\right)=0.
\ee
We define a new function
\be
\mathcal{U}(j,j^{\prime},\iota,\iota^{\prime};n)\equiv\left(\overline{\varphi_0}^{j\,\iota^\star}\right)^3\mathcal{V}^{\,\prime}_5\left(\scriptstyle{\underbrace{\scriptstyle{j,\dots,j^{\prime}}}_\text{n},\dots,j,j,j,j,j,j,j;\underbrace{\scriptstyle{\iota,\dots,\iota^{\prime}}}_{5-n},\dots,\iota^\star}\right),
\ee
with $j^{\prime}$ in the $n$-th position ($n=1,2,3,4$) and $\iota^{\prime}$ appearing in position $5-n$ after $\iota$, which keeps the first place. For instance, one has for $n=1$
\be\label{eq:DefinitionEffectivePotential}
\mathcal{U}(j,j^{\prime},\iota^\star,\iota,\iota^{\prime};n)=\left(\overline{\varphi_0}^{j\,\iota^\star}\right)^3\mathcal{V}^{\,\prime}_5\left(\scriptstyle{j^{\prime},j,j,j,j,j,j,j,j,j;\iota,\iota^\star,\iota^\star,\iota^\star,\iota^{\prime}}\right).
\ee
Thus, the equations of motion for the perturbations can be rewritten more compactly as
\begin{align}\label{eq:PerturbationsFirstIndex}
\mathcal{K}^{j^{\prime}jjj\,\iota}\delta\varphi^{j^{\prime}jjj\,\iota}+\sum_{\iota^{\prime}}\delta\overline{\varphi}^{jjjj^{\prime}\,\iota^{\prime}}\mathcal{U}(j,j^{\prime},\iota^\star,\iota,\iota^{\prime};1)&=0\nonumber\\
\mathcal{K}^{jj^{\prime}jj\,\iota}\delta\varphi^{jj^{\prime}jj\,\iota}+\sum_{\iota^{\prime}}\delta\overline{\varphi}^{jjj^{\prime}j\,\iota^{\prime}}\mathcal{U}(j,j^{\prime},\iota^\star,\iota,\iota^{\prime};2)&=0\nonumber\\
\mathcal{K}^{jjj^{\prime}j\,\iota}\delta\varphi^{jjj^{\prime}j\,\iota}+\sum_{\iota^{\prime}}\delta\overline{\varphi}^{jj^{\prime}jj\,\iota^{\prime}}\mathcal{U}(j,j^{\prime},\iota^\star,\iota,\iota^{\prime};3)&=0\nonumber\\
\mathcal{K}^{jjjj^{\prime}\,\iota}\delta\varphi^{jjjj^{\prime}\,\iota}+\sum_{\iota^{\prime}}\delta\overline{\varphi}^{j^{\prime}jjj\,\iota^{\prime}}\mathcal{U}(j,j^{\prime},\iota^\star,\iota,\iota^{\prime};4)&=0 \;\; .
\end{align}

With the particular kinetic kernel~(\ref{eq:LocalKineticKernel}), one has that the kinetic operator acting on the perturbation does not depend on the position of the perturbed index $j^{\prime}$, neither it depends on the intertwiner label $\iota$. Hence, in that case we can define
\be
\mathcal{K}^{\prime}=\mathcal{K}^{j^{\prime}jjj\,\iota}=\mathcal{K}^{jj^{\prime}jj\,\iota}=\mathcal{K}^{jjj^{\prime}j\,\iota}=\mathcal{K}^{jjjj^{\prime}\,\iota}=-\tau\pa_{\phi}^2+\eta\left(3j(j+1)+j^{\prime}(j^{\prime}+1)\right)+m^2.
\ee

The above equations are generic. However, recoupling theory imposes several restrictions on our perturbations, due to the conditions imposed on the fields: 
a)  $j^{\prime}$ is an integer (half-integer) if the background spin $j$ is an integer (half-integer); b) $j^{\prime}$ cannot be arbitrarily large, since for $j^{\prime}>3j$ the closure (right-invariance) condition would be violated; c) of course,  the case $j^{\prime}=j$ is uninteresting since such perturbations can be reabsorbed into the monochromatic background. 

In the simplest example $j=\frac{1}{2}$ there is only one permitted value for the perturbed spin, namely $j^{\prime}=\frac{3}{2}$, and the perturbation is identified with the state such that the total spin of a pair is $J=1$. Any such state is trivially also a volume eigenstate since the volume operator is identically vanishing in such intertwiner space, as it is one-dimensional (see Appendix~\ref{sec:Volume}, in particular the comment after Eq.~(\ref{eq:VolumeSpectrum})~).  For this reason, we will omit the indices $\iota$, $\iota^{\prime}$ in the following.

Let us introduce some further notation for these specific perturbations. We define
\be\label{eq:PerturbedIndices}
\psi_1=\delta\varphi^{\frac{3}{2}\frac{1}{2}\frac{1}{2}\frac{1}{2}}, \hspace{1em} \psi_2=\delta\varphi^{\frac{1}{2}\frac{3}{2}\frac{1}{2}\frac{1}{2}}, \hspace{1em} \psi_3=\delta\varphi^{\frac{1}{2}\frac{1}{2}\frac{3}{2}\frac{1}{2}}, \hspace{1em} \psi_4=\delta\varphi^{\frac{1}{2}\frac{1}{2}\frac{1}{2}\frac{3}{2}}
\ee
and similarly
\be
\mathcal{K}_1=\mathcal{K}^{\frac{3}{2}\frac{1}{2}\frac{1}{2}\frac{1}{2}}, \hspace{1em} \mathcal{K}_2=\mathcal{K}^{\frac{1}{2}\frac{3}{2}\frac{1}{2}\frac{1}{2}}, \hspace{1em}\mathcal{K}_3=\mathcal{K}^{\frac{1}{2}\frac{1}{2}\frac{3}{2}\frac{1}{2}}, \hspace{1em} \mathcal{K}_4=\mathcal{K}^{\frac{1}{2}\frac{1}{2}\frac{1}{2}\frac{3}{2}}.
\ee
Hence, it follows from Eq.~(\ref{eq:PerturbationsFirstIndex}) that the dynamics of the perturbations is governed (to first order) by the following equations (omitting the perturbation variables $j^{\prime}$, $\iota$, $\iota^{\prime}$ and the background spin $j=\frac{1}{2}$ in the argument of $\mathcal{U}$, Eq.~(\ref{eq:DefinitionEffectivePotential})):
\begin{align}
\mathcal{K}_1 \psi_1+\mathcal{U}(\iota^{\star};1)\overline{\psi_4}&=0 \nonumber \\
\mathcal{K}_4 \psi_4+\mathcal{U}(\iota^{\star};4)\overline{\psi_1}&=0\nonumber \\
\mathcal{K}_2 \psi_2+\mathcal{U}(\iota^{\star};2)\overline{\psi_3}&=0 \nonumber \\
\mathcal{K}_3 \psi_3+\mathcal{U}(\iota^{\star};3)\overline{\psi_2}&=0 \;\; \label{eq:System1}.
\end{align}
 %
 %
The resulting equations for the perturbations are reasonably simple, thanks mainly to the isotropy assumption on the background, which simplifies considerably the contribution from the GFT interaction term $\mathcal{U}$. However, even the simplified functional form in which the Lorentzian EPRL vertex amplitude appears in these equations remains unknown in exact analytic terms. The above equations would have then to be studied numerically or in more phenomenological approach, in which the exact function $\mathcal{U}$ is replaced by some simpler trial function, or several ones in different ranges of the variable $j$, approximating it. 
Luckily, for our present concerns, which relate to the behaviour of perturbations close to the cosmological bounce, these difficulties can be sidestepped since the interaction term is generically subdominant in that regime of the theory, mainly due to the smallness of the background condensate wavefunction (in turn related to the smallness of the universe 3-volume). This will allow us to perform a study of this dynamics in the following section.

Before we turn to such dynamics, let us notice that the equations (\ref{eq:System1}) make manifest an asymmetry of the interaction terms of the GFT model we are considering, more specifically of the EPRL vertex amplitude, that is not apparent at first sight. The equations in fact couple perturbations in the first field argument with perturbations in the fourth, and perturbations in the second with perturbations in the third, with no other combination being present. This happens despite the isotropy assumption on the background  and the other symmetries of the model. One can trace this asymmetry back to the combinatorial structure of the vertex amplitude itself: it corresponds to a 4-simplex as projected down to the plane but it is not symmetric with respect to the face pairings, if such faces are ordered in their planar projection: it only couples first and fourth faces across common tetrahedra sharing them, or second and third ones, \emph{i.e.} exactly the type of asymmetry that is revealed in our perturbations equations. It is tempting to relate this asymmetry to an issue with orientability of the triangulations resulting from the Feynman expansion of the model, since the same type of issue has been identified in the Boulatov model for 3d gravity in Ref.~\cite{Freidel:2009hd}. It is unclear at this stage whether this is a problem or just a feature of the model; it is also unclear, in case one decides to remove such asymmetry, what is the best way to do so. The strategy followed in the 3d case  Ref.~\cite{Freidel:2009hd}, \emph{i.e.} to maintain the ordering of the GFT field arguments but modify the combinatorics of the interaction vertex to ensure orientability, does not seem available in this 4d case. An easy solution would be to impose that the GFT fields themselves are invariant under (even) permutations of their arguments, which also ensure orientability of the resulting triangulations. We leave this point, not directly relevant for the analysis of the next section, for further study.

 \subsection{Dynamics of the perturbations at the bounce}\label{sec:DynamicsAtBounce}
We now study the dynamics of the perturbations around a background homogeneous and isotropic solution of the fundamental QG dynamics, in the mean field approximation.
We focus on the bounce regime, since this is where typical bouncing models of the early universe have difficulties in controlling the dynamics of anisotropies. 
Luckily, as anticipated, this is also the regime where, in the GFT condensate cosmology framework we can have the best analytic control over the (quantum) dynamics of the theory, at least in the mean field approximation. 
In fact, the bouncing regime takes place, in the hydrodynamic approximation we are working in, for low densities, thus, intuitively, for low values of the modulus of the GFT mean field. 

Considering the kernel of Eq.~(\ref{eq:LocalKineticKernel}), assuming $j=\frac{1}{2}$ for the background, and neglecting the interaction term, the background equation reads as
\be\label{eq:BackgroundBounceFundRepr}
\left(-\tau\pa_{\phi}^2+3\eta+m^2\right)\varphi^{\frac{1}{2}\iota^\star}\simeq0.
\ee
On the other hand, to first order, perturbations satisfy the equation 
\be\label{eq:EquationPerturbationNoInteractions}
\mathcal{K}^{\prime}\psi\simeq0,
\ee
where
\be
\mathcal{K}^{\prime}=-\tau\pa_{\phi}^2+\eta\left(3j(j+1)+j^{\prime}(j^{\prime}+1)\right)+m^2=-\tau\pa_{\phi}^2+6\eta+m^2 \; ,
\ee
and we have indicated a generic perturbation by $\psi$, since there is no difference among them, in this approximation.

When the interaction term is no longer subdominant (\emph{i.e.}~after the universe exits the bouncing phase and after it has expanded enough), the dynamics of the perturbations is given by the systems of equations (\ref{eq:System1}), which remain valid until $\psi\simeq\varphi^{\frac{1}{2}\iota^\star}$. At that point, higher order corrections are needed. On the other hand, it is important to stress that since the equations of motion become linear  at the bounce, at that point we are no longer subject to the constraint that non-monochromatic components should be small. In other words, $\psi\simeq\varphi^{\frac{1}{2}\iota^\star}$ is allowed in that regime and perturbations can be large. This observation will be important in the following.

Using the analytic expression of the background solution, given in Ref.~\cite{deCesare:2016axk}, we have
\be\label{eq:AnalyticBackground}
|\varphi^{\frac{1}{2}\iota^\star}|=\frac{e^{\sqrt{\frac{3 \eta +m^2}{\tau} } (\Phi-\phi )}
   \sqrt{-2 E_0 e^{2 \sqrt{\frac{3 \eta +m^2}{\tau} } (\phi
   -\Phi)} \sqrt{\Omega_0}+e^{4 \sqrt{\frac{3 \eta +m^2}{\tau} } (\phi
   -\Phi)} \Omega_0+\Omega_0}}{2 \sqrt{\frac{3 \eta +m^2}{\tau} } \sqrt[4]{\Omega_0}},
\ee
where
\be\label{eq:OmegaParameterBackground}
\Omega_0=E_0^2+4 Q_0^2 \left(\frac{3\eta +m^2}{\tau} \right)
\ee
and $E_0$, $Q_0$ are conserved quantities\footnote{The conservation of $Q_0$ is not exact, as it follows from an approximate $\mbox{U(1)}$-symmetry, which holds as long as interactions are negligible.}. $E_0$ is referred to as the `GFT energy' \cite{Oriti:2016qtz,Oriti:2016ueo} of the monochromatic background\footnote{The \lq GFT energy\rq~is the total mechanical energy of an associated one-dimensional mechanical problem which governs the evolution of the modulus of the GFT mean field. Its relation to any macroscopic conserved quantity is not known at this stage.}, defined in Eq.~(\ref{eq:GFT energy}).
Reality of the expression in Eqs.~(\ref{eq:AnalyticBackground}) implies
\be
\Omega_0\geq 0.
\ee
In order to have the same dynamics for the background as in Refs.~\cite{Oriti:2016qtz,Oriti:2016ueo,deCesare:2016axk}, we demand that
\be\label{eq:BackgroundInequality}
\frac{3 \eta +m^2}{\tau}>0.
\ee
In this case, the modulus of the backround $|\varphi^{\frac{1}{2}\iota^\star}|$ has a unique global minimum at $\phi=\Phi$, corresponding to the quantum bounce.
We will consider two possible cases in which condition (\ref{eq:BackgroundInequality}) is still satisfied, but different conditions are imposed on the parameters governing the dynamics of the perturbations. This gives qualitatively the same evolution of the background but two radically different pictures for the evolution of the perturbations.

\begin{enumerate}[label=$\diamond$]
\item {\bf Case \textit{i}) }~The first possibility is that $\tau,~ m^2 \geq0$. In this case, also the perturbations satisfy an analogous condition
\be\label{eq:PerturbationInequality}
\frac{6 \eta +m^2}{\tau}>0.
\ee
The analytic solution of the equation for the perturbations has the same form as Eq.~(\ref{eq:AnalyticBackground})
\be\label{eq:AnalyticPerturbations}
|\psi|=\frac{e^{\sqrt{\frac{6 \eta +m^2}{\tau} } (\Phi_1-\phi )}
   \sqrt{-2 E_1 e^{2 \sqrt{\frac{6 \eta +m^2}{\tau} } (\phi
   -\Phi_1)} \sqrt{\Omega_1}+e^{4 \sqrt{\frac{6 \eta +m^2}{\tau} } (\phi
   -\Phi_1)} \Omega_1+\Omega_1}}{2 \sqrt{\frac{6 \eta +m^2}{\tau} } \sqrt[4]{\Omega_1}}.
\ee
We introduced the quantity $\Omega_1$, in analogy with Eq.~(\ref{eq:OmegaParameterBackground})
\be\label{eq:DefOmega1}
\Omega_1=E_1^2+4 Q_1^2 \left(\frac{6\eta +m^2}{\tau} \right).
\ee
$E_1$ and $Q_1$ are two conserved quantities. We will refer to $E_1$ as to the `GFT energy' of the perturbations. Reality of Eq.~(\ref{eq:AnalyticPerturbations}) requires that $\Omega_1\geq0$. $|\psi|$ has a minimum at $\Phi_1$.
From Eqs.~(\ref{eq:AnalyticBackground}),~(\ref{eq:AnalyticPerturbations}), we find in the limit of large $\phi$
\be
\frac{|\psi|}{|\varphi^{\frac{1}{2}\iota^\star}|}\sim e^{ \Big(\sqrt{\frac{6 \eta +m^2}{\tau}} - \sqrt{\frac{3 \eta +m^2}{\tau}}\Big)\phi},
\ee
which means that perturbations cannot be neglected in this limit, \emph{i.e.} away from the bounce occurring at $\phi=\Phi$ (see Eq.~(\ref{eq:AnalyticBackground}) and discussion below Eq.~(\ref{eq:BackgroundInequality})). Therefore, when we are in this region of parameter space, they should be properly taken into account. Depending on the values of the parameters, they can become dominant already close to the bounce. At the same time, the value of the `time' $\phi$ at which this approximation is usable cannot be too large, because then we expect the GFT interactions to grow in importance, breaking the approximation on the background and perturbation dynamics we have assumed to be valid so far.

\item {\bf Case \textit{ii})}~A second possibility is represented by the case in which condition (\ref{eq:BackgroundInequality}) is still satisfied while inequality (\ref{eq:PerturbationInequality}) is not. This can be accomplished with $\tau<0$ and $-6\eta<m^2<-3\eta$. In this case, the modulus of the perturbations oscillates around the minimum of a one-dimensional mechanical potential. 

Writing $|\psi|=\rho$, its dynamics is given by (see Refs.~\cite{Oriti:2016qtz,Oriti:2016ueo,deCesare:2016axk,deCesare:2016rsf})
\be\label{eq:RadialPartPerturbations}
\partial_{\phi}^2 \rho=-\pa_\rho U,
\ee
where
\be\label{eq:PotentialMechanics}
U(\rho)=\frac{Q^2_1}{2\rho^2}-\left(\frac{6\eta+m^2}{\tau}\right)\frac{\rho^2}{2}.
\ee
What happens in this case is that, away from the bounce, the perturbations are always dominated by the background.

In order to see this in a more quantitative way, we can make the simplifying assumption that the minimum of $U$ and the amplitude of the oscillations of $\rho$ are such that the interactions between quanta are always negligible for the perturbations. This can be realised by making an appropriate choice for the values of the parameters of the model. 

In this case, Eq.~(\ref{eq:RadialPartPerturbations}) describes the evolution of the perturbations at all times. Their qualitative behaviour around the bounce is illustrated in Fig.~\ref{Fig:AtoV}. Non-monochromatic perturbations are relevant at the bounce but drop off quickly away from it.\footnote{This is reminiscent of the results obtained in the context of a different model in Ref.~\cite{Pithis:2016cxg}, also suggesting that such non-monochromatic modes are only relevant in the regime of small volumes.} 

The behaviour of the perturbations is oscillatory, since $|\psi|$ is trapped in the potential well $U$ of Eq.~(\ref{eq:PotentialMechanics}). As a consequence of this, the number of non-monochromatic quanta $N_1=|\psi|^2$ has an upper bound. Conversely, the number of quanta in the background grows unbounded. 

We conclude that, in this window of parameter space, perturbations can be relevant at the bounce but are negligible for large numbers of quanta in the background. 
For a suitable strength of the interactions, non-monochromatic perturbations can become completely irrelevant for the dynamics before interactions kick in.

\end{enumerate}

 \subsection{Measures of deviations from monochromatic GFT condensates}
It is interesting to explore further the deviations from perfect monochromaticity, by computing some quantities which can characterise the dynamics of the perturbed condensate and distinguish it from the purely monochromatic case. We do so in the following. The quantities we compute do not have a clear cosmological meaning, and do not correspond to specific gauge-invariant observables characterising anisotropies in relativistic cosmology. They are however well-defined formal observables for GFT condensates. 

The first one we consider is a \emph{surface-area-to-volume ratio}. A first quantised area operator in GFT can be defined for a tetrahedron as in Ref.~\cite{Gielen:2015kua}: $\hat{A}=\kappa\sum_{i=1}^4\sqrt{-\Delta_i}$ (with $\kappa=8\pi G_{\rm N}\beta$), where the sum runs over all the faces of the tetrahedron, in analogy with the LQG area operator (cf. Eq.~(\ref{Eq:GFT:Area})). 

We have for its expectation value on a single monochromatic (equilateral) quantum:
\be
A_0=2\kappa\sqrt{3}
\ee
and for a perturbed non-monochromatic quantum:
\be
A_1=\frac{\kappa}{2}\sqrt{3}\left(3+\sqrt{5}\right) \qquad .
\ee

This operator can then be turned into a second quantised counterpart of the same (see \emph{e.g.} Ref.~\cite{Thiemann:2007zz,Rovelli:2004tv}), to be applied to ensembles of tetrahedra, as in Eq.~(\ref{Eq:GFT:Area}).  One can then easily compute the expectation value of this operator, as well as the expectation value of the total volume operator, in both an unperturbed and in a perturbed condensate state (of the simplest type considered in this work). The resulting quantity is heuristically the sum of the areas of the four faces of each tetrahedron times the number of tetrahedra with the same areas.

The area-to-volume ratio for the example considered can then be expressed as
\be\label{eq:AreaToVolume}
\frac{A}{V}=\frac{A_0 N_0+A_1 N_1}{V_0 N_0}=\frac{A_0}{V_0}\left(1+\frac{A_1}{A_0}~\frac{N_1}{N_0}\right).
\ee
$A_0$ is the surface area of an unperturbed quantum and $A_1$ that of a perturbed one. $N_0$ and $N_1$ are the corresponding number of quanta, which can be computed using Eq.~(\ref{eq:NumberOfQuanta})
\begin{align}
N_0&=|\varphi^{\frac{1}{2}\iota^\star}|^2,\\
N_1&=|\psi|^2.
\end{align}
$V_0$ is the volume of a quantum of space in the background. We recall that the perturbed quanta considered in this example have vanishing volume (see Appendix~\ref{sec:Volume}). This has significant consequences which we illustrate in the following.
Hence, Eq.~(\ref{eq:AreaToVolume}) leads to
\be
\frac{A}{V}=\frac{A_0}{V_0}\left(1+\frac{3+\sqrt{5}}{4}~\frac{N_1}{N_0}\right).
\ee
Since $\frac{N_1}{N_0}\geq0$, we have
\be
\frac{A}{V}\geq\frac{A_0}{V_0}.
\ee
This inequality means that, for a given volume, quanta have \emph{on average} more surface than they would in a purely monochromatic (isotropic) background. 

\begin{figure}
\begin{center}
\includegraphics[width=0.6\columnwidth]{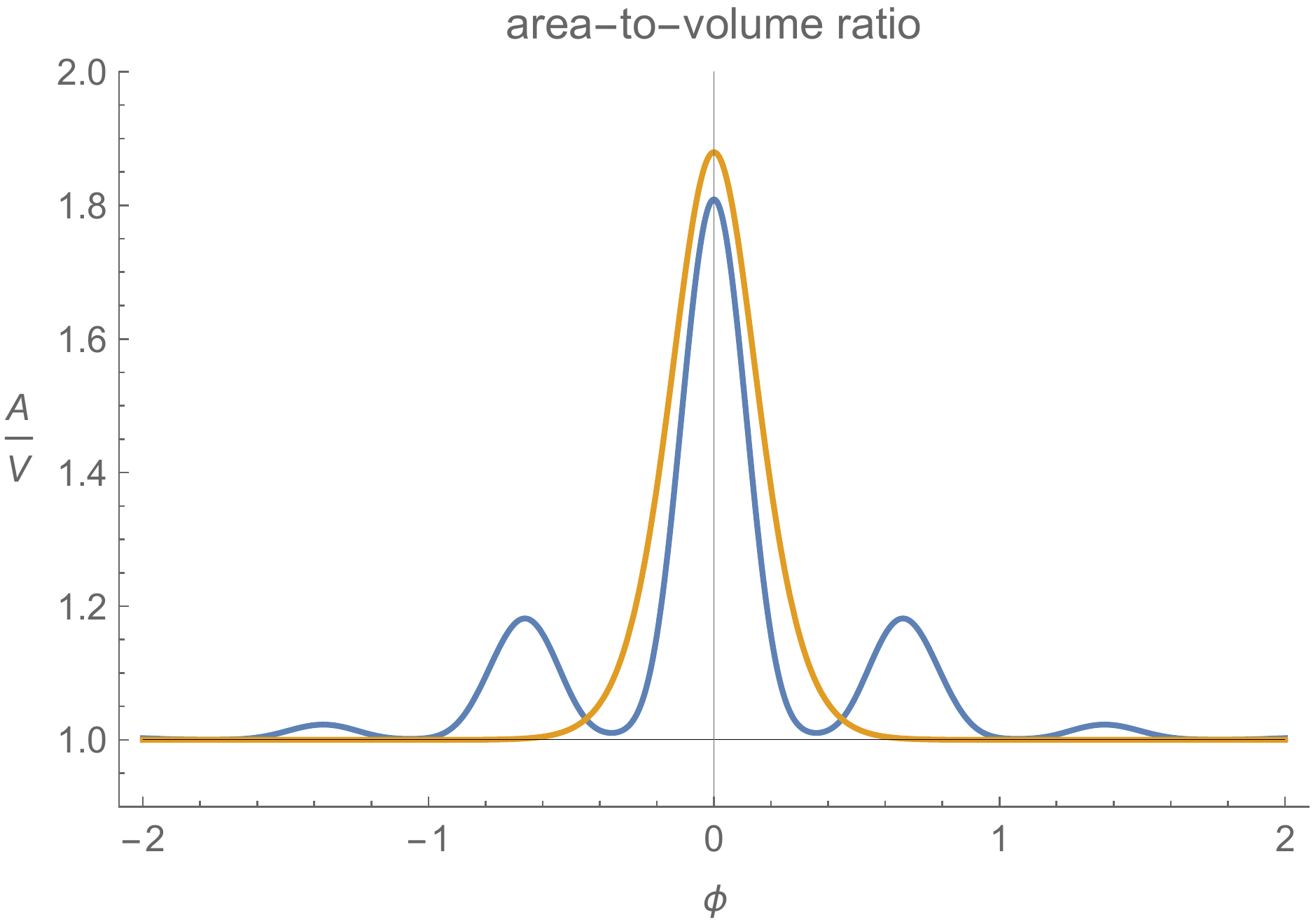}
\caption[Surface-area-to-volume ratio]{Plot of the \emph{surface-area-to-volume ratio} as a function of relational time $\phi$ in the case $\tau<0$, $-6\eta<m^2<-3\eta$. The vertical axis is in units of $\frac{A_0}{V_0}$. The orange curve $(\tau=-1,m^2=-42, Q_0=1, Q_1=1, E_0=-70, E_1=3)$  is obtained by considering perturbations sitting at the minimum of the potential $U$. Although the initial conditions can be chosen so that the surface-area-to-volume ratio $\frac{A}{V}$ is significantly different from its value for a single tetrahedron at the bounce, it decays exponentially moving away from it. The blue curve $(\mu=-24, Q_0=3, Q_1=1.5, E_0=2, E_1=14)$ represents instead the case in which the energy of the perturbations is above the minimum of the potential, but the amplitude of the oscillations is small enough so as to justify the harmonic approximation. Initial conditions are chosen such that perturbations start oscillating with maximum amplitude at the bounce. The ratio $\frac{A}{V}$ undergoes damped oscillations as we move away from the bounce. The value $\frac{A_0}{V_0}$ corresponds to that of a single tetrahedron and is always a lower bound, which is asymptotically attained at infinite relational time $\phi$. Although the values of the parameters for the two curves were chosen by hand, the qualitative behaviour represented by the blue curve is generic. When the kinetic energy of the perturbations is negligible compared to the potential energy, one obtains the behaviour represented by the orange curve.}\label{Fig:AtoV}.
\end{center}
\end{figure}

The evolution of $\frac{A}{V}$ in case \textit{ii}) is shown in Fig.~\ref{Fig:AtoV}. If the perturbations have minimum `GFT energy' (introduced in Eq.~(\ref{eq:DefOmega1}) as $E_1$), \emph{i.e.} they sit at the minimum of $U$, $\frac{A}{V}$ drops off monotonically as we move away from the bounce, due to the growth of the background. One could say that anisotropies, to the extent in which they are captured by the non-monochromatic perturbations, are diluted away by the expansion of the isotropic background. The background value $\frac{A_0}{V_0}$ is a lower bound, which is asymptotically attained in the infinite volume limit (obviously, before too large volumes can be attained, one expects GFT interactions to kick in, breaking the approximation we have employed here). On the other hand, if the `GFT energy' of the perturbations is above the minimum of the potential $U$, the perturbations will start to oscillate around such minimum. Therefore, $\frac{A}{V}$ will oscillate as it drops off. The asymptotic properties are unchanged.

\begin{figure}
\begin{center}
\includegraphics[width=0.6\columnwidth]{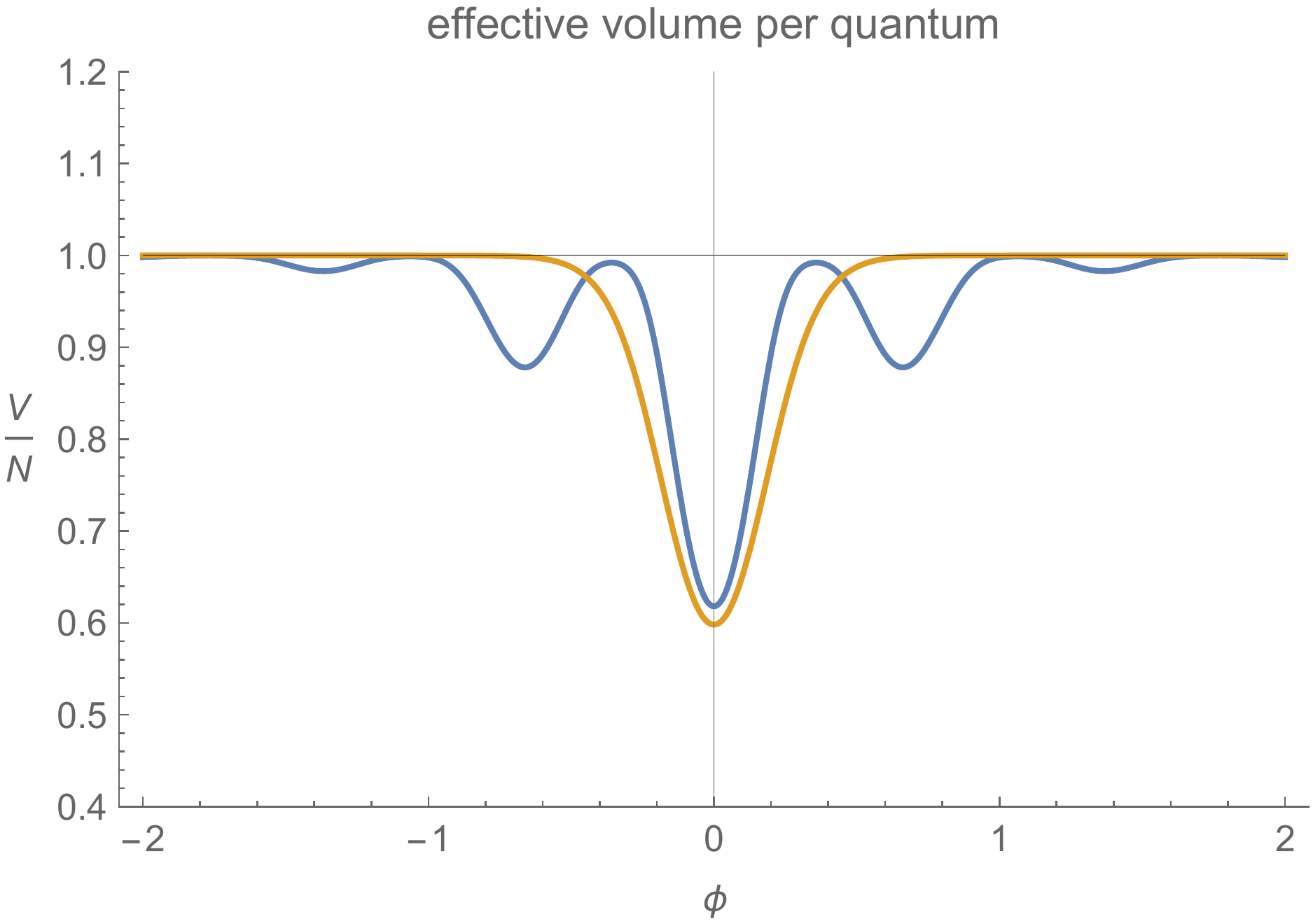}
\caption[Effective volume per quantum]{Evolution of the effective volume $\frac{V}{N}$ of a quantum over relational time for $\tau<0$, $-6\eta<m^2<-3\eta$. The vertical axis is in units of $V_0$. The parameters chosen for the two curves correspond to those of Fig.~\ref{Fig:AtoV}. $\frac{V}{N}$ relaxes to the volume $V_0$ of a quantum in the backgound away from the bounce. However, at the bounce it can be significantly different from such value.}\label{Fig:EffectiveVolume}
\end{center}
\end{figure}

\begin{figure}
\begin{center}
\includegraphics[width=0.6\columnwidth]{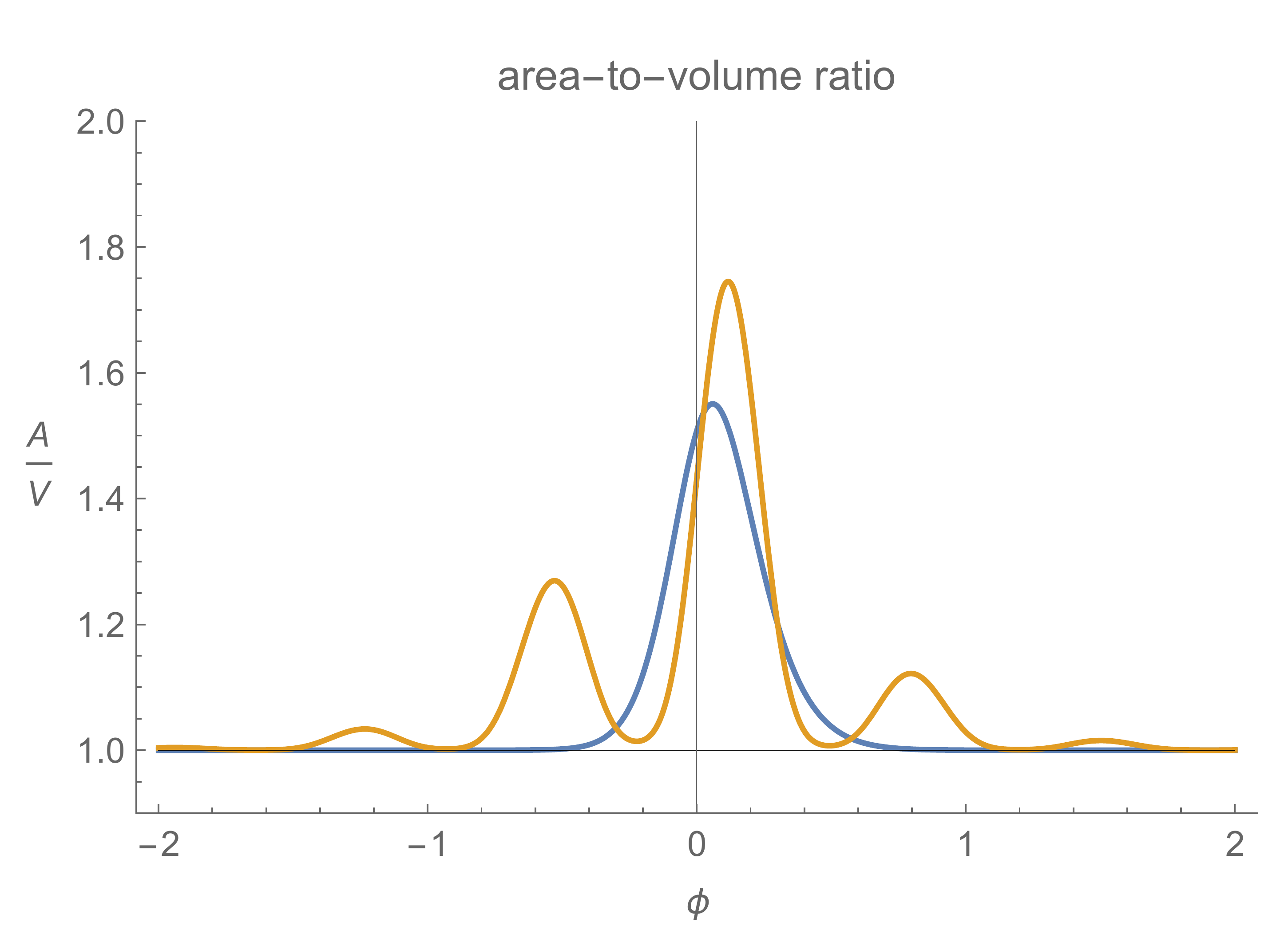}
\caption[Surface-area-to-volume ratio: asymmetric initial conditions]{Plots of $\frac{A}{V}$ for the case of a non-maximal amplitude of the non-monochromatic perturbations at the bounce. The curves correspond to the same values considered in Fig.~\ref{Fig:AtoV}. An initial `velocity' $\pa_{\phi}|\psi|$ is given to the perturbations at the bounce $\phi=\Phi$, with the value 0.6 for the blue curve and 0.4 for the orange curve. The non-symmetric initial conditions result in a deformation of the profile of $\frac{A}{V}$. 
}\label{Fig:PreBounceIC}
\end{center}
\end{figure}

Another interesting quantity to compute is the \emph{effective volume per quantum}, defined as
\be
\frac{V}{N}=\frac{ N_0~V_0}{N_0+N_1}=\frac{V_0 }{1+\frac{N_1}{N_0}}\qquad ,
\ee
where again all quantities entering the above formula are expectation values of 2nd quantised GFT observables in the (perturbed) GFT condensate state. 
It satisfies the bounds
\be
0\leq \frac{V}{N}\leq V_0.
\ee
Its profile for the example given above is shown in Fig.~\ref{Fig:EffectiveVolume}. The ratio $\frac{V}{N}$ represents the average volume of a quantum of space. Its value is generally lower than $V_0$, \emph{i.e.}~the volume of an equilateral quantum tetrahedron with minimal areas. In fact, zero volume quanta\footnote{See Appendix~\ref{sec:Volume}.} can change the total number of quanta $N$, leaving $V$ unchanged. Explicit calculations show that, in the limit of large $N$, the ratio $\frac{V}{N}$ approaches the value $V_0$ (see Fig.~\ref{Fig:EffectiveVolume}). In Fig.~\ref{Fig:PreBounceIC} we show the plot relative to the case where perturbations do not reach their maximum amplitude at the bounce, resulting in a deformation of the profile of $\frac{A}{V}$. This corresponds to setting initial conditions for the microscopic anisotropies (non-monochromaticity) before the bounce.

To summarise, in the region of parameter space corresponding to case ii) above, our results confirm that, from a bouncing phase, where the quantum geometry can be rather degenerate and  anisotropies (encoded in non-monochromatic perturbations of the simplest GFT condensate state) quite large, a cosmological background emerges whose dynamics can be cast into the form of an effective Friedmann equation for a homogeneous, isotropic universe.

 \subsection{Conclusion}
 In this chapter we considered the Group Field Theory (GFT) approach to Quantum Gravity and its emergent Cosmology, obtained from the dynamics of a condensate of isotropic quanta. Such quanta are interpreted as the fundamental building blocks of geometry. After reviewing the fundamentals of GFT and how the condensate dynamics is obtained in the mean field approximation, we considered some cosmological applications of GFT models, focussing in particular on the results obtained by the author in Refs.~\cite{deCesare:2016axk,deCesare:2016rsf,deCesare:2017ynn}
 
The first model we considered (Section~\ref{Sec:NoInteractions}) is a non-interacting one, first obtained in Ref.~\cite{Oriti:2016ueo} in the mean field approximation of the GFT dynamics. We discussed properties of the solution of the model and showed that there are significant departures from the dynamics of a classical
 FLRW spacetime. The emergent classical background satisfies an evolution equation
 of the Friedmann type with quantum corrections appearing in the r.h.s as
 effective fluids with distinct equations of state. Such correction terms
 vanish in the limit of infinite volume, where the standard Friedmann dynamics
 is recovered.
 
 The main results obtained in this model are two. First, confirming the result of Ref.~\cite{Oriti:2016ueo},
 we have shown that there
 is a \textit{bounce} which resolves the classical singularity, taking place regardless of the particular values
 of the conserved charges $Q$ and $E$. It should be pointed out that
 the origin of this bounce is quite different from the one given by
 Loop Quantum Cosmology (see Appendix~\ref{Appendix:LQC}).
The second result is the occurence of an era of \textit{accelerated
  expansion} without the need for introducing \textit{ad hoc} potentials
and initial conditions for a scalar field. However, in this model the number of e-folds achieved during the era of accelerated
expansion is not large enough compared to its typical value in inflationary models.

 We have seen that the interesting features of the model arise
 from quantum geometry corrections which are captured by
 a description in terms of effective fluids defined on the emergent classical
 background. A similar phenomenon was already observed in LQC (see \textit{e.g.} Ref.~\cite{Singh:2005km}).
   We also showed that there is another
 way of formulating the dynamics, which
 makes no reference to such effective fluids,
 but instead differs from the standard Friedmann
 equation in that the gravitational constant is
 replaced by a dynamical quantity.
  In fact, another interesting result is that, even though Newton's constant $G_{\rm N}$
 is related to, and actually constrains, the parameters of the
 microscopic GFT theory, the dynamics of the
 expansion of the universe is actually determined by the
 \textit{effective gravitational constant} $G_{\rm eff}$. 
 We should stress that such
  quantity was introduced in first place for the only purpose of
  studying the properties of the solutions of the model.

The second model we considered (Section~\ref{Sec:GFT:Interactions}) is based on an effective action for the GFT condensate which includes two interaction terms, and is motivated by phenomenological considerations. 
A
general prediction of the model is the occurrence of a recollapse when
the higher order interaction term becomes codominant. Results that
have been previously obtained in the free theory
survive in the
interacting case, in particular the occurrence of a
bounce and an early epoch of accelerated expansion. The former result,
together with the recollapse induced by interactions, leads to cyclic
cosmologies with no big-bang or big-crunch singularities. These are obtained under fairly general assumptions on
the effective potential, namely its boundedness from
below.

A detailed analysis showed that in the interacting model one can
attain an arbitrary number of e-folds as the universe accelerates
after the bounce. Furthermore, having an inflationary-like expansion
imposes a restriction on the class of viable models. In fact, this is
possible only for some ranges of values of the parameters, specifically $\lambda<0$ and $5\leq n<n^{\prime}$ and when one
has the hierarchy $\mu\ll |\lambda|$. Reasonable phenomenological
arguments lead to select only the case $n=5$, $n^{\prime}=6$ as
physical. The two powers can be related, respectively, to simplicial
interactions, commonly considered in the GFT framework, and to a
negative cosmological constant.

While such results are encouraging as a
first step towards a quantum geometric description of the inflationary
era, a few remarks are in order. In fact, it must be pointed out that
it comes at the price of a fine tuning in the coupling constant of the
higher order interaction term. From this point of view, it shares one
of the major difficulties of ordinary inflationary
models. Furthermore, an inflationary-like era that lasts for a sufficient number of e-folds does not seem to be a
generic property of GFT models, but in fact requires interactions of a
suitable form.

In Section~\ref{sec:Perturbations}, considering the GFT formulation of the Lorentzian EPRL model, we derived the equations of motion for non-monochromatic perturbations of the mean field. The dynamics is given by a system of coupled differential equations, relating the perturbations on the four faces of a tetrahedron. As an application, we considered the particularly interesting case of an isotropic background, with all its faces labelled by the fundamental representation of SU(2). We introduced quantities such as \emph{surface-area-to-volume ratio} and the effective volume of a quantum in order to study the evolution of such deviations from perfect microscopic isotropy in a quantitative way. Studying the dynamics of perturbations around the bounce, we determined a region in the parameters space of the model such that perturbations exhibit an oscillatory behaviour in a neighbourhood of the bounce, but are rapidly dominated by the background away from the bounce. 

Future work must be devoted to identify suitable observables that can relate non-monochromaticity of GFT quanta to anisotropies of the emergent spacetime. Progress is also needed to incorporate other degrees of freedom in the effective description of spacetime dynamics obtained from GFT.

\part{Effective Theories of Gravity}

\newpage
\chapter{Varying Fundamental Constants and $G$ as a Stochastic Process}\label{Chapter:VariableG}

In this chapter we discuss the physical meaning and the implications of having variable fundamental `constants', with particular attention to the gravitational constant. 
In a work by the author \cite{deCesare:2016dnp}, the gravitational constant is modelled as a stochastic process; this is suggested to provide a phenomenological description of quantum gravity effects taking place at the Planck scale. Our main interest is in the consequences of such an assumption for cosmological evolution. We will cast the discussion of our model within the broader framework of theories with varying constants, recalling the meaning and classification of physical constants, the historical roots of such models and modern theoretical developments.

In section~\ref{Sec:VariableG:Dirac} we briefly review the first proposal based on varying fundamental constants of Physics, namely Dirac's \emph{large number hypothesis}; we discuss those ideas underlying it which are also common to modern theories with varying constants. The definition and the role of the fundamental physical constants is reviewed in Section~\ref{Sec:VariableG:FundamentalConstants}. The link between varying constants and violation of the equivalence principle is discussed in Section~\ref{Sec:VariableG:EP}. In Section~\ref{Sec:VariableG:Units} we examine the relation between fundamental constants and systems of physical units. We give a brief overview of theories with varying $c$, $\hbar$ and $G$ in Section~\ref{Sec:VariableG:VSLetc}. In Section~\ref{Sec:PhenoModifiedGR} we assume a phenomenological perspective and modify the Einstein equations by allowing for a dynamical gravitational constant. We show that a correction term to the stress-energy tensor of matter must be included in order to satisfy the Bianchi identities. Finally, in Section~\ref{Sec:VariableG:StochasticG} we analyze a particular phenomenological model with varying $G$, originally proposed by the author in Ref.~\cite{deCesare:2016dnp}. In this model, the late time accelerated expansion of the universe is linked to stochastic fluctuations of the gravitational `constant'.

\section{Dirac's large number hypothesis}\label{Sec:VariableG:Dirac}
In a 1937 paper \cite{Dirac:1938mt}, Dirac observed a numerical coincidence between some dimensionless quantities that can be constructed using the age of the Universe and the magnitude of basic dimensionful quantities in Cosmology and atomic physics\footnote{It must be stressed that the coincidence observed is only a rough one, since numerical factors involving numbers such as, \emph{e.g.}, the fine structure constant $\alpha=e^2/ \hbar c\sim1/137$ or the ratio of the the proton and the electron mass $m_p/m_e\sim 1.8\times 10^3$ are considered of order unity \cite{DICKE:1961aa}  (also cf. \cite{Dirac:1938mt}).
} (see also Refs.~\cite{Dirac:1974aa,Dirac:1937ti}). The age of the Universe $T\simeq 13.7\times 10^9~\mbox{yrs}$ can be expressed in atomic units by comparing it with the Compton frequency of the electron
\be\label{Eq:VariableG:UniverseAge}
T\left(\frac{\hbar}{m_e c^2}\right)^{-1}\simeq3.4\times 10^{38}~.
\ee
Comparing the ratio of the electric to the gravitational force between an electron and a proton one finds the small number
\be\label{Eq:VariableG:RatioGravToElectric}
\frac{F_{\rm g}}{F_{\rm e}}=G_N~\frac{m_p m_e}{e^2}\sim4.4\times10^{-40}~.
\ee
In fact, up to small integer powers of $m_p/m_e$ and $\alpha$, this number is equal to the ratio of the electron mass to the Planck mass, squared
\be
\left(\frac{m_e}{M_{\rm\scriptscriptstyle Pl}}\right)^2=m_e^2 \left(\frac{\hbar c}{G}\right)^{-1}\simeq 1.8\times 10^{-45}~.
\ee
Dirac argued that the extreme closeness of the number in Eq.~(\ref{Eq:VariableG:UniverseAge}) with the inverse of the ratio (\ref{Eq:VariableG:RatioGravToElectric}) must be more than a mere coincidence and could be due to a deeper connection between cosmology and the laws of microscopic physics. Dirac suggested that all such very large or very small numbers must be connected by simple mathematical relations, in which the coefficients are of order unity\footnote{With the same caveat as in footnote 1.}. We will refer to this statement as to the \emph{Dirac hypothesis}. Thus, since the age of the Universe $T$ is changing in an evolutionary Universe, the large number $\sim 10^{40}$ in Eq.~(\ref{Eq:VariableG:UniverseAge}) must also change with time. Hence, if we assume the Dirac hypothesis, all numbers of the order $(10^{40})^n$, where $n$ is an integer, are changing with time. From this, we draw the conclusion that the `gravitational constant', which appears in Eq.~(\ref{Eq:VariableG:RatioGravToElectric}), must also change with time. It would scale with the inverse of the age of the Universe $G\propto T^{-1}$, which would explain the smallness of the gravitational force compared to the other fundamental interactions. Therefore, the Dirac hypothesis naturally leads to physical laws that are changing over cosmological scales. 
  
Dirac's assumption seems to find support in simple statistical arguments \cite{Dirac:1937ti}. In fact, if we regard the age of the Universe $T$ as a random variable which a priori can take  any positive value, an accidental correspondence between (\ref{Eq:VariableG:UniverseAge}) and (\ref{Eq:VariableG:RatioGravToElectric}) would be highly improbable. However, not all values of $T$ are allowed, as assumed in this simplistic way of reasoning. In fact, only certain values of $T$ are compatible with the existence of observers, as remarked by Dicke in Ref.~\cite{DICKE:1961aa}. Neglecting this crucial fact introduces an observation selection effect, sometimes also called anthropic bias, which invalidates the argument\footnote{See Ref.~\cite{bostrom2013anthropic}. Different interpretations of the anthropic principle and its applications to Cosmology are examined in Ref.~\cite{Barrow:1988yia}. Note that not all of them are necessarily connected to observation selection effects \cite{bostrom2013anthropic}.}.

It is fairly straightforward to make an order of magnitude estimate of the range of values of $T$ 
that is compatible with the existence of observers \cite{DICKE:1961aa}. The starting point of Dicke's argument is that the existence of observers (as living beings, not just abstract entities) requires the existence of a planet orbiting a luminous star. An upper bound for $T$ is thus obtained as the maximum age of a star producing energy through nuclear processes. It is found to satisfy the following relation\footnote{Again, dropping small integer powers of $m_p/m_e$ and $\alpha$.}
\be
T_{\rm max}\left(\frac{\hbar}{m_e c^2}\right)^{-1}\sim  \frac{\hbar c}{m_e^2G}~.
\ee
A lower bound $T_{\rm min}$ is determined instead from requirements of stellar stability and is found to be of the same order of magnitude as $T_{\rm max}$. Thus, according to Dicke, the \emph{Dirac hypothesis} is not justified. Rather than a hint of a new fundamental physical principle, the numerical coincidences observed by Dirac should be interpreted as conditions for the existence of living observers. In particular, they do not hold for any value of $T$ in an evolving Universe, but only during the epoch where the laws of physics are compatible with life. As a consequence, it is not possible to conclude from this argument that $G$ changes over time.
 
Although the Dirac hypothesis has been severely criticised by Dicke, pointing at the incorrectness of the underlying assumptions, it represented nonetheless an important turning point in theoretical physics and opened a window of new possibilities. In fact, regardless of the original statement of the Dirac hypothesis and the status of the `large numbers', there are two essential speculative elements that we can extract from Dirac's proposal. Firstly, fundamental constants could not be actual constants, but their actual value can be different in different spacetime regions. This includes, but is not limited to, the `gravitational constant' and can be extended beyond the cosmological case, considered by Dirac, to allow for more general configurations of the gravitational field. Secondly, the values of the fundamental `constants'  would be determined by some connection between micro-physics and macro-physics. These two elements are very general and can be realised in different ways, as we will discuss further in the next sections.

\section{What is a fundamental constant?}\label{Sec:VariableG:FundamentalConstants}
According to Weinberg, a fundamental constant is defined as a physical quantity which appears in a given theory, and whose value cannot be determined within that theory \cite{Weinberg:1983ph}. This definition is quite broad and clearly makes the number of fundamental constants theory-dependent. As such, the definition of what physical constants should be regarded as fundamental depends on the current status of our knowledge. As our understanding of physics advances, some constants that were previously regarded as fundamental would be expressed in terms of other constants. This must be possible at least in principle for a constant that is not fundamental, although the actual computation can be so complicated to be beyond the present ability to perform such calculations. Examples in this sense would be given by the viscosity of a fluid in hydrodynamics or the mass of the pion in low energy strong interaction physics \cite{Weinberg:1983ph}. The value of a fundamental constant instead can only be measured by experiments.

One can be more specific than above and introduce different classes of constants in order of increasing generality. Following L\'evy-Leblond \cite{LevyLeblond:1977qm} (see also Ref.~\cite{Uzan:2010pm}), we will distinguish among three classes of fundamental constants: \emph{class A} is the class of the constants characteristic of a specific system, \emph{class B} is the class of constants characteristic of certain sets of physical phenomena, \emph{class C} is the class of universal constants. The status of a fundamental constant can change over time, with the development of physical theories. This is indeed the case of the speed of light in vacuo $c$, which was initially considered as a constant specific of light propagation (class A). With the unification of electricity and magnetism, it was then recognised as a constant characteristic of all electromagnetic phenomena (class B). It was only later, with the formulation of the theory of special relativity that its status as a universal constant of Physics (class C) was recognised (see Ref.~\cite{Uzan:2010pm}). The independent parameters in the Standard Model of particle phsyics are class B constants. The only three constants that qualify as class C are $G$, $\hbar$ and $c$ \cite{LevyLeblond:1977qm}. In the following, we will use the \emph{fundamental constant} only to refer to such universal constants.

According to Levy-Le Blond \cite{LevyLeblond:1977qm}, fundamental constants are \emph{concept synthesisers}, \emph{i.e.} they are introduced to express the relation between quantities that would otherwise be thought of as incommensurable. Indeed, the speed of light\footnote{More specifically, its properties of universality and its finiteness, and the fact that it represents a \emph{limit speed} for all signals.} $c$ leads to the introduction of a unified description of space and time in terms of a single geometric entity. Similarly, the Planck constant $\hbar$ is introduced in order to establish a link between the wave and corpuscular properties of a particle, through the Einstein-DeBroglie relations\footnote{$E$ denotes the energy of the particle and ${\bf p}$ its three-momentum, whereas $\omega$ and ${\bf k}$ are the frequency and the wave-vector of the corresponding wave.}  $E=\hbar\omega$, ${\bf p}=\hbar {\bf k}$. Finally, the gravitational constant $G$ relates the mechanical properties of matter to the geometric properties of spacetime, via the Einstein equations.

\begin{figure}
\begin{center}
\includegraphics[width=0.5\columnwidth]{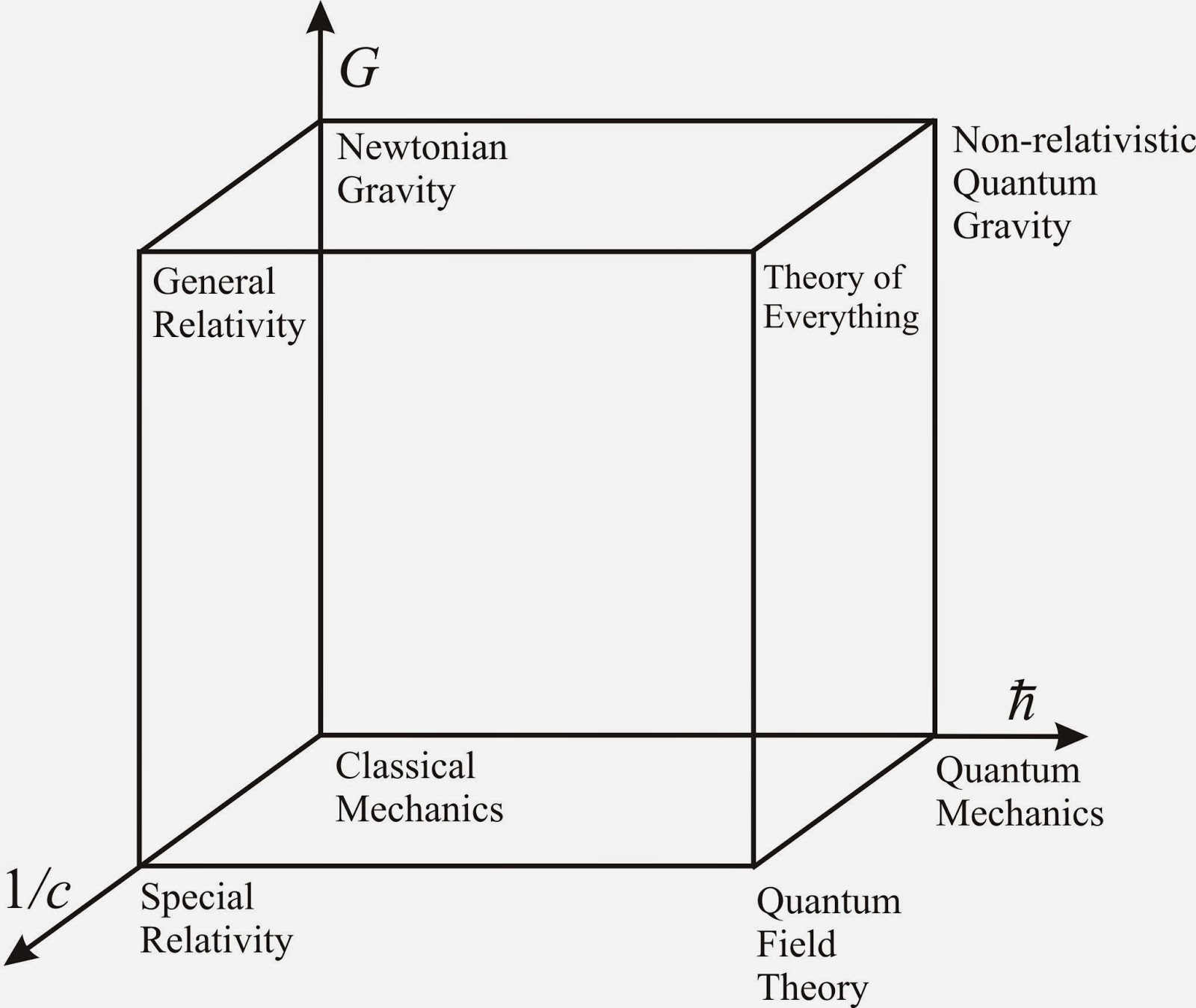}
\caption[The Bronstein cube]{The Bronstein cube of physical theories \cite{Okun:1991wm}.
Fundamental constants $G$, $\hbar$, $1/c$ are used to label the axis of a three-dimensional Cartesian frame. The edges of the cube correspond to particular limiting procedures, in which at least one of the fundamental constants is formally sent to zero. However, it must be noted that the meaning of the cube must not be taken too literally, since it only serves classification (and perhaps heuristic) purposes. For instance, the $G\to0$ limit of General Relativity does not yield Special Relativity, since it is well-known that non-trivial vacuum solutions exist. Moreover, the correspondence of the top front right corner with a theory of everything (TOE) is still matter of debate. The top back right corner would represent the non-relativistic limit of such theory, assuming it exists. (Image taken from \href{http://backreaction.blogspot.co.uk/2011/05/cube-of-physical-theories.html}{\texttt{backreaction.blogspot.co.uk}})}\label{Fig:VariableG:Cube}
\end{center}
\end{figure}

The three fundamental constants also play an important epistemological role (see Okun's contribution in Ref.~\cite{Duff:2001ba}). In fact, they are related to the regimes where peculiar physical phenomena emerge in a physical theory. For instance, relativistic effects become relevant when velocities approach $c$. Similarly, the quantum nature of a physical system becomes manifest when the action is not much larger than $\hbar$. Moreover, the fundamental constants show the relation between different limits of physical theories and reveal their regime of validity. The last point can be illustrated by means of the so-called Bronstein\emph{cube of physical theories}, introduced by Okun in Ref.~ \cite{Okun:1991wm}. The \emph{cube} is depicted in Fig.~\ref{Fig:VariableG:Cube} and can be used for classification purposes.

As a final remark, we would like to mention that the three fundamental constants seem to play a slightly different role. In fact, it can be argued\footnote{See Deser's remark reported at the end of Duff's contribution to Ref.~\cite{Duff:2001ba}.} that the role of $\hbar$ and $c$ is `kinematical', since they express the relation between kinematical variables such as position and momentum, or space and time, respectively. Note that $G$ plays a `dynamical' role in GR, since it only appears in the Einstein equations. However, it is not clear how fundamental would be such a distinction between `kinematical' and `dynamical' fundamental constants. In fact, for a given theory of gravity, such distinction may be an artefact which depends on the particular formulation adopted. To be specific, we observe that in the canonical formulation of GR with Ashtekar variables, $G$ explicitly enters kinematics (either through the Poisson brackets \cite{Thiemann:2007zz} or in the definition of the connection one-form \cite{Kiefer:2007ria}) also in the case of pure gravity. 

\section{Fundamental constants and physical units system}\label{Sec:VariableG:Units}
Given their universality, fundamental constants can be used to define a distinguished set of dimensionful constant quantities, such that all dimensionful physical quantities can be expressed in terms of them. In other words, they provide a basic system of units \cite{Duff:2001ba}.
In fact, the number of fundamental constants coincides with the number of basic \emph{physical dimensions}\footnote{It must be stressed that the coincidence with the number of \emph{spatial dimensions} is only accidental. We would have the same number of fundamental constants in any dimensions, regardless of the number and nature of fundamental interactions, as stressed by Okun in Ref.~\cite{Duff:2001ba}.}: space, time and matter (\emph{i.e.}~mass). This is the standard point of view, expressed \emph{e.g.} by Okun in Ref.~\cite{Duff:2001ba}. Such physical dimensions, which we quantify by the introduction of suitable units, are \emph{a priori} irreducible given our experience with low energy physical phenomena. It is nevertheless possible that more fundamental physical theories, yet to be fully understood or discovered, may allow for a reduction of the number of physical dimensions\footnote{Some authors argued for the complete elimination of the concept of dimensionful quantities from physical theories, see \emph{e.g.} Duff's contribution in Ref.~\cite{Duff:2001ba}. We do not share this point of view.}. This was argued by Veneziano in the context of superstring theory \cite{Veneziano:1986zf,Duff:2001ba}. However, since any such UV complete theory should, by definition, be able to recover low energy physics in some limit, we will assume the standard point of view on physical dimensions without loss of generality.

Fundamental units can be used to build a system of physical units. These are the so-called natural (or Planckian) units. Such unit system is the most fundamental one since, rather than relying on the physical laws of a specific system\footnote{An example of such a unit system is offered by Stoney's units, based on the physical constants characterising the electron \cite{Duff:2001ba}. Units of length, time and mass are defined, respectively as follows $\ell_S=\frac{e\sqrt{G}}{c^2}$, $t_S=\frac{e\sqrt{G}}{c^3}$, $m_S=\frac{e}{\sqrt{G}}$.}, it is defined using the most general laws of Physics. Indeed its universality follows from the universality of quantum mechanics, special relativity and of the gravitational interaction. Planck units of length, time and mass are defined in terms of $\hbar$, $c$, $G$ as follows:
\be\label{Eq:VariableG:PlanckUnits}
\ell_{\rm\scriptscriptstyle Pl}=\frac{\hbar}{m_{\rm\scriptscriptstyle Pl}c}~,\hspace{1em}t_{\rm\scriptscriptstyle Pl}=\frac{\hbar}{m_{\rm\scriptscriptstyle Pl}c^2},\hspace{1em}m_{\rm\scriptscriptstyle Pl}=\sqrt{\frac{\hbar c}{G}}~.
\ee

Planck units can be conveniently used to define the different physical regimes of a given theory. In fact, when the value of a given dimensionful physical quantity becomes $\mathcal{O}(1)$ in Planck units, characteristic phenomena of that regime are expected to take place. This is best illustrated with some examples, that should be well-known to the reader. When the speed of a particle $v$ becomes of order unity in Planck units, \emph{i.e.} $\frac{v}{c}\sim1$, relativistic effects become dominant. Similarly, when the action of a physical system (or its angular momentum) is of order $\hbar$, the correct physical description of the system is given by quantum mechanics. For these two examples peculiar physical effects appear in such regimes which, on the contrary, become irrelevant in the non-relativistic limit or in the classical limit, respectively. 
We would like to remark that in a theory of quantum geometry and for pure gravity only $c$ and $\ell_{Pl}$ would feature as fundamental constants (\emph{i.e.} $G$ and $\hbar$ appear only in a particular combination in all of the equations of the theory), as observed by Zel'dovich in Ref.~\cite{zeldovich1975structure}.
A similar observation was made by Veneziano in the context of superstring theory. In Refs.~\cite{Veneziano:1986zf,Duff:2001ba} he argued that the number of fundamental constants needed is reduced from three to only two: the speed of light $c$ and the string length $\lambda_{\rm s}$. They can be used to define fundamental units of time and length in a unified theory of all interactions. In fact, considering the Nambu-Goto action $S/\hbar=\tau/\hbar \int\de\left(Area\right)\equiv \lambda_{\rm s}^{-2} \int\de\left(Area\right)$, $\hbar$ combines with the string tension $\tau$ to give a new fundamental constant with dimensions of an area $\lambda_{\rm s}^2$. According to Veneziano, the Planck constant $\hbar$ has not disappeared from the theory, but it has been promoted to a universal UV cut-off which resolves at the same time the divergences of QFT and the singularities of GR. From this point of view, both the number and the nature of the fundamental constants required in the fundamental theory could be different from those needed in the low energy effective descriptions. This would have important physical consequences. In fact, if this was the case, the unit of mass would not be arbitrary (as it must be for all basic units) but would rather be dependent on the choices made for time and length units. Hence, all mass scales (including the masses of elementary particles, those of bound states and the Planck mass $m_{\rm \scriptscriptstyle Pl}$) should be computable in superstring theory and have an expression in terms of the above string units. Although there are some indications that this could be achieved in Quantum String Theory (QST), there are no definite results at present and the problem is far from being settled. Incidentally, we observe that if $m_{\rm \scriptscriptstyle Pl}$ could be computed in terms of $\lambda_{\rm s}$ and $c$, this would also determine the value of the Planck constant.

In some sense, if the claim made by Veneziano turned out to be correct, the situation in superstring theory would be analogous to that of pure quantum gravity, discussed above. In fact, the framework of superstring theory offers a unified description of matter and geometry. Hence, mechanical quantities such as momenta and masses would have an expression in terms of purely geometric quantities. The reader is referred to Ref.~\cite{Duff:2001ba} for a similar discussion by Veneziano in the context of M-theory, leading to similar conclusions.

\section{Varying fundamental constants}\label{Sec:VariableG:VSLetc}
The possibility that fundamental constants may evolve over cosmic history was originally motivated from Dirac's large numbers hypothesis, which led him to the conclusion that the gravitational `constant' may be a function of time. However, as we discussed in Section~\ref{Sec:VariableG:Dirac}, this is a possibility that could be realised independently of Dirac's proposal. In fact, we observed that the conclusions of Dirac's idea are more general than the arguments he used to prove them. The basic idea of a gravitational `constant' evolving over cosmological timescales can be generalised to allow for the variation of other constants of physics, more in general over both time and space, and not necessarily on cosmological scales. Different cases have been considered in the literature, and attention has been paid both to the case of the variability of fundamental constants and that of physical constants of a lesser status (\emph{class A} or \emph{class B} according to L\'evy-Leblond's classification). Some examples with the appropriate references to the literature will be given later in this section. 

The variation of a fundamental constant sounds like an oximoron. In fact, the constancy of universal quantities such as $c$, $G$, $\hbar$ is a cornerstone of modern physics, which justifies their use as basic dimensional quantities defining a natural system of physical units, the Planck units. It is therefore crucial to put their constancy on a basis that is as solid as possible, by means of experimental tests. Such experiments are actually investigating the validity of the currently accepted theoretical frameworks of GR and QFT. The measurement of any such variations would have a huge impact in our understanding of the physical world. In fact, spacetime variation of the fundamental constants would imply the existence of new degrees of freedom, providing an insight in the laws of Nature lying beyond the present level of our knowledge. 

Before discussing in more detail how the variation of fundamental constants could be realised in a physical theory, we would like to discuss what is meant by the variation of a dimensionful physical quantity. The following considerations may be elementary, but we find it nonetheless appropriate to include them for the sake a clarity, also in light of an ongoing debate on this topic (see Ref.~\cite{Duff:2001ba}). Although it is certainly true that only dimensionless quantities can be actually measured in physical experiments, as it has been pointed out in particular by Duff in Ref.~\cite{Duff:2001ba}, we agree with the point of view expressed by Okun in Ref.~\cite{Duff:2001ba} that physical dimensions are nonetheless useful (if not necessary) to fully characterise physical quantities. In fact, the measurement of any physical quantity is nothing but a comparison between two physical systems with the same physical dimensions \cite{Uzan:2010pm}. Therefore, every measurement is by definition a relative measurement, which will give as a result a pure number. 
A given dimensionful quantity $q$ can be expressed in terms of some homogeneous physical quantity (\emph{i.e.} a quantity of the same kind) as (see Ref.~\cite{Duff:2001ba})
\be\label{Eq:VariableG:Units}
q=(q/u_q) u_q~.
\ee
Here $u_q$ represents the physical units of the quantity $q$ and the ratio $q/u_q$ (which is dimensionless) is the value of $q$ in those units. Clearly, units are arbitrary and can be changed at will
\be
q=(q/u_{q_1})u_{q_1}=\frac{q}{u_{q_1}}\frac{u_{q_1}}{u_{q_2}}u_{q_2}~.
\ee
The ratio $u_{q_1}/u_{q_2}$ is a conversion factor between the two units.

It is clear that it does not make sense to talk about the variation of some physical quantity if we do not specify what is kept fixed. For example, any measurement of a variation in the ratio $q/u_q$ Eq.~(\ref{Eq:VariableG:Units}) could either be interpreted as a change in $q$ or as a change in $u_q$. The two situations would be physically indistinguishable. If a convention is adopted such that, \emph{e.g.}, $u_q$ is chosen as a reference, the measurement of a variation of $q/u_q$ can be interpreted as a variation of $q$. Hence, when a statement about the variation of a dimensionful constant is made, it is implicitly assumed that a consistent choice of such references is made, \emph{i.e.} a system of units is fixed. The possibility of talking about the spacetime variation of dimensionful quantities is thus equivalent to the possibility of choosing (as a matter of convention) a system of units to be used as a reference. Without such references Physics would be impossible. A good unit system must be such that the laws of Physics are as simple as possible (cf. Ref.~\cite{Magueijo:2003gj}).

As discussed above, $c$, $G$, $\hbar$ can be used to define a system of natural units. This is based on our current theoretical understanding, which attributes to these quantities the status of fundamental constants. Obviously, in Planck units, these quantities are unitary by definition; hence, their variation, if any, could not be measured. In fact, the variation of the different fundamental constants is measurable and well posed from an operational point of view, provided that suitable physical references can be determined. This has to be discussed on a case by case basis for the three fundamental constants, since their status is different both from a theoretical and an experimental point of view.
The current bounds on the variation of the fundamental constants are given in Ref.~\cite{Uzan:2010pm}.
 We discuss in the following the variation of the three fundamental constants, with reference to the experimental tests and to the theoretical models that could allow for such variations. Particular attention will be given to the case of a variable gravitational constant, in view of the applications considered later in this chapter.

\subsection{Varying constants and the equivalence principle}\label{Sec:VariableG:EP}
Before moving on to the discussion of the special role played by fundamental constants (\emph{i.e.}  class C constants according to L\'evy-Leblond, see above) and of the theories entailing their variation, we would like to discuss some general consequences of the variation of \emph{any} constants entering the physical laws.
In fact, it was first pointed out by Dicke \cite{Dicke:1964pna} that any variation of the constants of Physics would imply a violation of the universality of free fall, \emph{i.e.} of the weak equivalence principle (WEP). This is not limited to what we called \emph{fundamental constants}; it is, in fact, more general and applies to class A and class B constants as well. For this reason, their constancy represents a crucial test of GR. The link to WEP is due to the fact that the mass of any composite body can (at least in principle) be expressed as a function of the masses of the elementary particles constituting it, and of the binding energies of the fundamental interactions holding them together \cite{Uzan:2010pm}. We will denote it by $m(\alpha_i)$, where $\alpha_i$ are the `constants' it depends on (such as the fine structure constant, the Yukawa couplings to the Higgs, the Planck constant, the gravitational constant, etc.). The precise functional form will be different for different bodies, and will depend in particular on the chemical structure of the body under consideration.

The action of a massive test particle reads as\footnote{In case also the speed of light is a spacetime function, the $c$ in Eq.~(\ref{Eq:VariableG:ActionTestParticle}) must be understood as a \emph{fiducial value}, introduced merely as a conversion factor. This must be clearly distinguished from the \emph{physical} speed of light. See Section~\ref{Sec:VariableG:VSL} for a discussion of the different aspects of the speed of light concept.}
\be\label{Eq:VariableG:ActionTestParticle}
S_{\rm m}=-c\int\de t\; m(\alpha_i)\sqrt{g_{\mu\nu}(x)u^{\mu}u^{\nu}}~.
\ee
Variation of the action gives  \cite{Uzan:2010pm}
\be\label{Eq:VariableG:VioationEP}
u^\nu\nabla_\nu u^\mu=-\sum_i\frac{\pa \log m}{\pa \alpha_i}\frac{\pa \alpha_i}{\pa x^\nu}\left(g^{\nu\mu}+u^\nu u^\mu\right)~.
\ee
Hence, if the constants of Physics are varying, \emph{i.e.} the $\alpha_i$ are spacetime dependent, freely falling bodies will not move along geodesics as in GR. They will instead experience an anomalous acceleration, given by the r.h.s.~of Eq.~(\ref{Eq:VariableG:VioationEP}). This would in general lead to a violation of the WEP. Notice that, in case only the gravitational constant is varying, one has
\be\label{Eq:VariableG:AnomalousAccelerationVarG}
u^\nu\nabla_\nu u^\mu=-\left(G\frac{\pa \log m}{\pa G}\right)G^{-1}\frac{\pa G}{\pa x^\nu}\left(g^{\nu\mu}+u^\nu u^\mu\right)~.
\ee
The mass of the body includes the contribution of gravitational self-energy. The quantity $G\frac{\pa \log m}{\pa G}$ is the compactness factor of the massive body\footnote{More generally, the compactness factor is written as $\frac{G}{m}\frac{\delta m}{\delta G}$. In fact, $m$ would also depend on the other parameters of the gravitational theory, such as post-Newtonian parameters $\gamma$, $\beta$, etc. \cite{Nordtvedt:1990zz}. In principle, all such parameters can evolve in time.}; it depends on the gravitational theory one considers, on the equation of state of the body's matter, and on the quantity of matter \cite{Nordtvedt:1990zz}. The effects of the anomalous cosmological acceleration of massive bodies due to the time variation of $G$ were studied in Ref.~\cite{Nordtvedt:1990zz}, where it was also shown that they alter the relation between the rate of change of the orbital period of a binary system (\emph{e.g.}~a binary pulsar) and that of $G$.

\subsection{Varying speed of light}\label{Sec:VariableG:VSL}
The constancy of the speed of light plays a distinguished role in modern Physics. The formulation of Special Relativity and then of General Relativity elevated this property of light propagation to the status of a fundamental physical principle. Its experimental foundations are on a solid basis, and so much so that it is usually regarded just as a conversion factor between units of time and units of space\footnote{Nevertheless, its role as a fundamental constant is not diminished by that since, as stressed in the previous section, when velocities become of order $c$ new phenomena take place.}. In fact, the constancy of the speed of light in vacuo $c$ is assumed in order to define the units of length in the international system of units SI.

When talking about the speed of light one should clearly distinguish between the different roles that it plays in Physics. A systematic discussion of the several facets of the speed of light $c$ is given in Ref.~\cite{Ellis:2003pw}, where the different speeds of light are examined based on the context in which they appear along with the relations between them. The speed of light features in fact as the propagation speed of electromagnetic waves in Maxwell's theory $c_{\rm\scriptscriptstyle EM}$, in the spacetime metric $c_{\rm\scriptscriptstyle ST}$, as the speed of propagation of gravitational waves $c_{\rm\scriptscriptstyle GW}$, combined with the gravitational constant to give the coupling between geometry and matter in the Einstein equations $c_{\rm\scriptscriptstyle E}$. In GR one has $c_{\rm\scriptscriptstyle GW}=c_{\rm\scriptscriptstyle ST}=c_{\rm\scriptscriptstyle E}$. If the electromagnetic field is minimally coupled to gravity one also finds $c_{\rm\scriptscriptstyle EM}=c_{\rm\scriptscriptstyle ST}$. However, the different `speeds of light' can be numerically different \emph{a priori}, since they refer to different physical quantities. In fact, the relations between them are modified in theories of gravity other than GR and for non-minimally coupled electromagnetic field.

A broad class of theories has been considered in the literature under the name of \emph{varying speed of light} theories (VSL). The consistent implementation of VSL in a physical theory is discussed in Ref.~\cite{Magueijo:2003gj}. The discussion of VSL requires the specification of a particular proposal for the dynamics, also determining which speed of light $c$ is being varied (see above) \cite{Ellis:2003pw,Magueijo:2007gf}. The analysis must be carried out on a case by case basis. However, it is true in general that the $c$ that varies in VSL proposals is \emph{never} a coordinate speed of light \cite{Magueijo:2007gf}, which we denoted by $c_{\rm ST}$ above. Hence, its variation cannot be undone by a coordinate transformation and its consequences are physical. Those have been studied in different frameworks. In particular, we mention the realisations of VSL in the following theoretical frameworks: broken Lorentz symmetry (see \emph{e.g.} for `soft breaking' Refs.~\cite{Moffat:1992ud,Magueijo:2000zt}, and for `hard breaking' Refs.~\cite{Albrecht:1998ir,Barrow:1999aa}), bimetric theories (where there is a metric for matter and one for gravity and the speed of gravity waves is different from that of other massless fields) \cite{Clayton:1998hv,Clayton:2001rt,Clayton:2000xt,Drummond:2001rj}, deformed special relativity\footnote{Also known as doubly special relativity (DSR), characterised by a non-linear realisation of the Lorentz group and leading to a frequency-dependent speed of photon propagation \cite{AmelinoCamelia:2002wr,KowalskiGlikman:2004qa}.}.

VSL has several important applications to Cosmology. In fact, it has been proposed as an alternative solution to the classic cosmological puzzles (reviewed in Section~\ref{Sec:CosmologicalPuzzles}) \cite{Albrecht:1998ir,Barrow:1999aa,Moffat:1992ud}. The generation of a spectrum of nearly scale-invariant spectrum of primordial fluctuations was studied in Refs.~\cite{Moffat:1992ud,Magueijo:2003gj,Albrecht:1998ir,Barrow:1999aa,Magueijo:2008sx}. Other realisations of VSL as an alternative to the inflationary paradigm have been considered in \cite{Franzmann:2017nsc,Costa:2017abc} under the name of c-flation.

\subsection{Varying the Planck constant}
A variable Planck constant has not been considered in the literature as often as theories with varying $G$ or $c$. The (reduced) Planck constant $\hbar$ represents the natural unit of action and angular momentum and it is also related to the fundamental area element in phase space.
In Ref.~\cite{Barrow:1998df} it was noted that variations in the fine-structure constant $\alpha$ could also be interpreted as variations in $\hbar$, as well as in $c$ or the electron charge $e$.

The commutator of canonically conjugated variables is proportional to $\hbar$, which leads to the Heisenberg uncertainty relations. Deformations of quantum phase space have been proposed, leading to a \emph{generalised uncertainty principle} (GUP) compatible with the existence of a minimal length in Quantum Gravity (see \emph{e.g.} Refs.~\cite{Kempf:1994su,Maggiore:1993kv}). Considering the algebra of position and momentum operators, one has\footnote{The structure of the other commutators, which are needed in order to close the algebra, depends on additional physical assumptions. We refer the reader to the literature, \emph{e.g.} Refs.~\cite{Kempf:1994su,Maggiore:1993kv}). Locally, the other commutators (\emph{i.e.} $[x^i,x^j]$ and $[p_i,p_j]$) can always be made trivial by means of a Darboux map, see Ref.~\cite{Mangano:2015pha}.}
\be\label{Eq:VariableG:Commutator}
[x^i,p_j]=i\hbar \left(\delta^i_j+f^i_j(x,p)\right)~.
\ee
The idea is in the same spirit of non-commutative geometry \cite{connes1985non}. The original motivation for GUP from a quantum gravity proposal was given in the context of String Theory \cite{Amati:1988tn,Konishi:1989wk}.
General considerations based on a thought experiment for the measurement of the area of a black hole horizon in Quantum Gravity are in agreement with GUP \cite{Maggiore:1993rv}.

We notice that the commutation relations (\ref{Eq:VariableG:Commutator}) can also be rewritten introducing a phase-space dependent $\hbar$. Trivially, if $f^i_j$ is assumed isotropic, \emph{i.e.} $f^i_j\propto \delta^i_j$, one has
\be
[x^i,p_j]=i\hbar(x,p)\delta^i_j~.
\ee
It is worth appreciating the analogy with the varying $G$ theory discussed in Chapter~\ref{Chapter:Weyl} (see Ref.~\cite{deCesare:2016dnp}). In fact, in both cases, modifications of the geometric (or algebraic) structures (phase-space versus Riemannian geometry) manifest themselves through the variation of some fundamental constants.
In Ref.~\cite{Mangano:2015pha} it was suggested that the corrections to $\hbar$ are dominated by high frequency modes and can be described as classical noise. The authors then considered a simple model in which $\hbar$ depends only on time and is characterised by white noise stochastic fluctuations around its observed value. A similar phenomenological model was put forward in Ref.~\cite{deCesare:2016dnp} in the case of the gravitational constant.

\subsection{Varying the gravitational constant}

The gravitational constant made its first historical appearance in Newton's law, which describes the gravitational interaction of two bodies of masses $m_1$ and $m_2$ at a distance $r$ from each other
\be
{\bf F}=G \frac{m_1 m_2}{r^2}\hat{\bf r}~.
\ee
The value of $G$ does not depend on the internal constitution or the state of motion of the two bodies, implying the universality of free fall which lies at the basis of the formulation of the weak equivalence principle. Hence, already in classical mechanics it has the status of a universal constant of Nature.
In GR the gravitational constant appears in the Einstein equations as the coupling between matter and the geometry of spacetime. Also in this case, its constancy and its universality enforce the weak equivalence principle. Newtonian gravity is obtained as the low-curvature, low-speed limit of GR.

Although sufficient to implement the weak equivalence principle in a generally covariant theory of the gravitational field, the constancy of $G$ is not necessary. This was realised by Brans and Dicke \cite{Brans:1961sx}, who introduced a new scalar degree of freedom $\phi$ (also called dilaton) playing the role of a dynamical gravitational constant. In their theory, test particles in free fall move along geodesics of the Levi-Civita connection, exactly as in GR. However, unlike GR the strength of the gravitational interaction is determined from the motion of all matter fields in the Universe. This is seen as a realisation of Mach's principle. The action of Brans-Dicke theory\footnote{This is the Brans-Dicke action in the so-called \emph{Jordan frame}, \emph{i.e.} with matter fields minimally coupled to the metric. By means of a conformal transformation it is possible to recast the action in the \emph{Einstein frame}, in which the gravitational sector is the same as in GR but the coupling of matter fields to gravity is non-minimal. Despite the mathematical equivalence, which allows one to choose the frame that is most convenient for the applications one has in mind, it is still debated whether the two frames are also physically equivalent and, if they are not, which is the physical one. For different perspectives on this issue the reader is referred to \emph{e.g.} Refs.~\cite{Artymowski:2013qua,Faraoni:1999hp,Postma:2014vaa,Capozziello:2010sc,Capozziello:1996xg}}
 is
\be
S_{\rm\scriptscriptstyle BD}=\frac{1}{16\pi G}\int\de^4x\;\left(\phi R-\omega \frac{\pa^\mu\phi\pa_\mu\phi}{\phi}-V(\phi)\right)+S_{\rm m}[g_{\mu\nu},\psi]~.
\ee
The matter sector depends on the matter fields (denoted here by $\psi$) and the metric $g_{\mu\nu}$, although it \emph{does not} depend on the Brans-Dicke scalar $\phi$. This is in order to ensure that the equivalence principle is respected\footnote{The equivalence principle clearly \emph{does not} hold in the Einstein frame.}.

Brans-Dicke theory is generally considered as experimentally not viable, since a large value of the Brans-Dicke parameter $\omega\gg 1$ is required in order to pass solar system tests, which is viewed as unnatural (see \emph{e.g.} Refs.~\cite{Bertotti:2003rm,Sotiriou:2007yd}). 
Brans-Dicke theory is also related to some modified gravity theories $f(R)$ for specific values of the Brans-Dicke parameter $\omega$. The relation (\emph{i.e.} the precise value of $\omega$) depends on whether the metric or the Palatini formulation of such theories is considered \cite{Sotiriou:2007yd,Sotiriou:2006hs}. The dilaton $\phi$ also arises from Kaluza-Klein theory and String Theory, where it is seen as originating from compactification of the extra dimensions. Several generalisations of Brans-Dicke theory have been proposed, commonly known as scalar-tensor theories. Particularly relevant among those are Horndeski theories \cite{horndeski1974second,Gleyzes:2014qga}, which describe the most general class of theories whose equations of motion are second order\footnote{Horndeski theories can indeed be shown to be equivalent to the Generalised Galileons theory \cite{Deffayet:2011gz}.}, thus avoiding the Ostrogradski instability which is typical of higher derivatives theories. The literature on scalar-tensor gravity theories is prolific and reviewing it would be beyond our purposes.

From an experimental point of view, $G$ is among the fundamental constants the one with the less stringent bounds on its variability \cite{Uzan:2010pm}. Its value according to CODATA\footnote{CODATA is the Commitee on Data for Science and Technology, see \href{http://www.codata.org}{www.codata.org}.} is $G= 6.67428(67)\times10^{-11}~{\rm m}^3~{\rm kg}^{-1}~{\rm s}^{-2}$ and the relative standard uncertainty is 0.01\% (ibid.). The best constraints come from solar system observations, and specifically from lunar laser ranging (LLR) experiments. Such experiments involve the comparison of a gravitational timescale (\emph{e.g.} the orbital period) and an atomic timescale; it is thus assumed that the atomic constants do not vary over the time of the experiment \cite{Uzan:2010pm}. It must be pointed out that the determination of such constraints is typically model-dependent. The constraint on the present time variation of the gravitational constant using LLR and assuming Brans-Dicke theory yields \cite{Williams:2004qba}
\be
\left.\frac{\dot{G}}{G}\right|_0 = (4\pm9) \times 10^{-13}~{\rm yr}^{-1}~.
\ee

Model independent bounds have been obtained from binary pulsars timing \cite{Kaspi:1994hp,Damour:1988zz,Damour:1990wz} (also compare with the model-dependent analysis of Ref.~\cite{Nordtvedt:1990zz}), although they are (at best) one order of magnitude less accurate than those coming from LLR (see Ref.~\cite{Uzan:2010pm}). Theory-dependent bounds from pulsar timing observations and based on scalar-tensor gravity were obtained in Ref.~\cite{Freire:2012mg}. For other astrophysical constraints on the variability of $G$ we refer the reader to Rev.~\cite{Uzan:2010pm} an references therein. Although less accurate than LLR, the other bounds are important as independent tests of the constancy of $G$.
It is more difficult to obtain bounds on $\dot{G}/G$ from cosmological data. This is due to the fact that cosmological observations depend on the whole history of $G$ as a function of time \cite{Uzan:2010pm}. From the point of view of cosmology, observable signatures of a variable $G$ can be found in Big Bang Nucleosynthesis (BBN) data or in the CMB. In particular, since $G$ enters the Friedmann equation, its variation would modify the age of the Universe as well as the relation between time and redshift. The implications of these effects for CMB observations are discussed in Ref.~\cite{Uzan:2010pm}.

At this stage we would like to make some general comments on the formulation of theories accounting for a variable $G$, which may lead to the effects discussed above. 
It is well-known that in a classical theory the action principle represents one of the most convenient ways to formulate a physical theory in which all fields are dynamical, as pointed out in Ref.~\cite{Ellis:2003pw}. In particular, for the case at hand, scalar-tensor theories can be formulated by writing down a suitable action functional which generalises the Einstein-Hilbert action. This is a much more effective way of model building rather than modifying the Einstein equations in order to accommodate for a variable $G$. In fact, it guarantees that the scalar field satisfies dynamical laws which are compatible with the gravitational field equations, which would be hard (albeit not impossible) to achieve otherwise. This is the path followed in the construction of scalar-tensor theories such as \emph{e.g.} Brans-Dicke and Horndeski theories (see above).
However, there are also some drawbacks. In fact, this procedure requires a knowledge of the exact nature of the extra degrees of freedom in the modified gravity theory to be constructed, plus additional assumptions which depend on the model. 
For this reason, starting from an action principle may not represent the most economical choice. Moreover, some model-independent features that are typical of theories with a variable gravitational constant\footnote{The same considerations also apply to VSL, see Ref.~\cite{Magueijo:2003gj}.} may not be captured in their full generality when a particular action is specified. To this end, it may be more convenient to work with the Einstein equations and only impose some basic physical and mathematical requirements, such as \emph{e.g.} consistency with Riemannian geometry, in order to study the phenomenological consequences of the variability of fundamental constants in gravitational physics. The problem of determining the correct action functional giving rise to such dynamics, which is clearly crucial for the consistency of the effective description, is thus only postponed to a later stage. In the next section we will adopt a phenomenological point of view and formulate a simple model entailing a varying gravitational constant. 

\section{Phenomenological approach to a dynamical $G$}\label{Sec:PhenoModifiedGR}
We start from the observation that naively turning $G$ into a dynamical quantity in Einstein's equations would be inconsistent with the Bianchi identities \cite{deCesare:2016dnp}. In fact, assuming that the stress-energy tensor of matter is covariantly conserved $\nabla^{\mu}T_{\mu\nu}=0$, the contracted Bianchi identities $\nabla^{\mu}G_{\mu\nu}=0$ are satisfied if and only if $G={\rm const}$~. Thus, if we want to consider a dynamical $G$, we can do so by introducing a correction term to the r.h.s. of Einstein equations, which we can interpret as an effective stress-energy tensor $\tau_{\mu\nu}$. In Ref.~\cite{deCesare:2016dnp} we proposed the following extension of Einstein's equations
\be\label{Eq:VariableG:Einstein}
G_{\mu\nu}=\frac{8\pi G}{c^4}\left(T_{\mu\nu}+\tau_{\mu\nu}\right)~.
\ee
Hence, we have from the Bianchi identities
\be\label{Eq:VariableG:EnergyNonConservation}
\left(\nabla^\mu \frac{G}{c^4}\right)\left(T_{\mu\nu}+\tau_{\mu\nu}\right)+\frac{G}{c^4}\nabla^\mu\tau_{\mu\nu}=0~.
\ee
We do not specify the detailed form of $\tau_{\mu\nu}$ at this stage, only demanding that it satisfies Eq.~(\ref{Eq:VariableG:EnergyNonConservation}). Notice that, by only looking at Eq.~(\ref{Eq:VariableG:EnergyNonConservation}), there is no way to differentiate between the case of a variable $c$ and that of a variable $G$, or whether both are varying.
 In the following, we will set $c=1$ and only consider $G$ as varying. Although it must be stressed that unless further physical input is given, it is the ratio $G/c^4$ that is actually varying in spacetime. Modified continuity equations entailing source terms depending on derivatives of fundamental constants, like in Eq.~(\ref{Eq:VariableG:EnergyNonConservation}), have appeared previously in the literature, see Refs.~\cite{Albrecht:1998ir,Fritzsch:2015lua}. In particular, in Ref.~\cite{Albrecht:1998ir} the Bianchi identities applied to a VSL model were shown to lead to a modified continuity equation (in a cosmological context) with source terms proportional to $\dot{c}/c$ and $\dot{G}/G$. In that paper it was pointed out that, despite its generality, the model is not equivalent to Brans-Dicke theory when $\dot{G}\neq0$. In fact, Brans-Dicke theory respects the equivalence principle; hence, it does not lead to any modifications in the continuity equation. Similar arguments apply for the extra source of stress-energy $\tau_{\mu\nu}$ in the model considered here, see Eq.~(\ref{Eq:VariableG:EnergyNonConservation}). In fact, physical bodies are still freely falling since we assumed $\nabla^{\mu}T_{\mu\nu}=0$. In Ref.~\cite{Fritzsch:2015lua} the variation of the gravitational constant has been related to the cosmological constant and to the non-conservation of matter. The modified continuity equation (\ref{Eq:VariableG:EnergyNonConservation}) was obtained also in Ref.~\cite{Franzmann:2017nsc} in the case in which $\tau_{\mu\nu}$ represents a dark energy term. In the case $\nabla^{\mu}T_{\mu\nu}=0$ is also dropped, one would have a more general continuity equation, implying violation of the local conservation laws of energy and momentum (see Ref.~\cite{Franzmann:2017nsc}), similarly to the situation encountered in Ref.~\cite{Albrecht:1998ir}.
 
Given our considerations above, we rewrite Eq.~(\ref{Eq:VariableG:EnergyNonConservation}) as
\be
\nabla^\mu\tau_{\mu\nu}=-\nabla^\mu \log G \left(T_{\mu\nu}+\tau_{\mu\nu}\right)~.
\ee
This equation expresses the non-conservation of energy and momentum in $\tau_{\mu\nu}$. Notice that the stress-energy non-conserving term depends on the spacetime variation of $G$ and on the stress-energy of physical matter fields $T_{\mu\nu}$. In a cosmological context, both $T_{\mu\nu}$ and $\tau_{\mu\nu}$ can be considered as perfect fluids. In particular, we would have
\be
\tau_{\mu\nu}=(\tilde{\rho}+\tilde{p}) u_\mu u_\nu+\tilde{p}g_{\mu\nu}~.
\ee
Notice that we have three unknowns $\tilde{\rho}$, $\tilde{p}$, $G$ and only one equation to relate them. An additional equation is given in the form of an equation of state relating $\tilde{\rho}$ and $\tilde{p}$. In the following we will make the assumption
\be\label{Eq:VariableG:TauAnsatzCC}
\tilde{\rho}=-\tilde{p}=\lambda~,
\ee
which corresponds to a dark energy term. 
The reader must be aware that Eq.~(\ref{Eq:VariableG:TauAnsatzCC})
is not the most general ansatz. However, it is the most economical one which is also phenomenologically viable. In fact, we know from observations
that dark energy must have an equation of state compatible with $w=-1$ at late times\footnote{See Section~\ref{Sec:StandardCosmology:LCDM}.}. 
We would like to mention that, from a purely theoretical standpoint, modifications of GR
in which the gravitational and the cosmological constants are regarded as
canonically conjugate variables have been recently considered in Ref.~\cite{Smolin:2015fqa}.

Assuming Eq.~(\ref{Eq:VariableG:TauAnsatzCC}) the number of unknowns is reduced to two and we have an equation relating the dynamics of dark energy to that of the dynamical gravitational `constant'.
\be
\nabla_\mu\lambda=(\nabla^\nu \log G) T_{\mu\nu}-(\nabla_\mu \log G)\lambda~.
\ee
Assuming homogeneity and isotropy we have
\be\label{Eq:VariableG:EOMdarkEnergy}
\dot{\lambda}=-\frac{\de \log G}{\de t}(\lambda+\rho)~.
\ee
The physical fluid satisfies instead the continuity equation
\be\label{Eq:VariableG:ContinuityEquation}
\dot{\rho}+3H(\rho+p)=0~.
\ee
The solution to Eq.~(\ref{Eq:VariableG:EOMdarkEnergy}) reads as
\be\label{Eq:VariableG:SolutionLambda}
\lambda(t)=\frac{G(t_i)}{G(t)}\left(\lambda(t_i)-\int_{t_i}^{t}\mbox{d}t^{\prime}\;\frac{\dot{G}(t^{\prime})}{G(t_i)}\rho(t^{\prime})\right)~,
\ee
where $t_i$ is an initial time where initial conditions are fixed. The precise functional form of $\lambda(t)$ can be obtained once the initial condition  $\lambda(t_i)$ is assigned and $G(t)$ is specified. The time dependence of the energy density $\rho(t)$ is obtained by solving the continuity equation (\ref{Eq:VariableG:ContinuityEquation}) for a fluid with a given equation of motion $\rho=w p$. We would like to stress at this point that, while the conclusions reached so far are fairly general, the functional form of $G(t)$ is model dependent and has to be assigned at the outset. In section~\ref{Sec:VariableG:StochasticG} we will consider a particular phenomenological model, in which $G(t)$ is modelled as a stochastic process.

\section{The gravitational constant as a stochastic process}\label{Sec:VariableG:StochasticG}
We consider a phenomenological model in which the dynamical gravitational `constant' is modelled as a stochastic process\footnote{See Appendix \ref{Appendix:Stochastic} for a review of stochastic processes and stochastic ordinary differential equations (ODEs).}. More specifically, we regard it as a dynamical variable whose mean value is given by Newton's constant, here denoted by $\overline{G}=G_{\rm N}$, and subject to white noise fluctuations about its mean value
\be\label{Eq:VariableG:StochasticG}
G(t)=\overline{G}(1+\sigma\xi(t))~.
\ee
$\sigma$ is a free parameter of the model and represents the noise strength. With our hypothesis, we expect to be able to account (at least qualitatively) for the effects of quantum fluctuations of the gravitational field and of new degrees of freedom that may emerge from a theory of Quantum Gravity\footnote{This is the case for example in String Theory, where the gravitational multiplet described by the low-energy spacetime effective action contains two new fields beside the graviton. These are the dilaton and the Kalb-Ramond antisymmetric tensor field. The Kalb-Ramond field may have important consequences for cosmology; in particular, it could play a role in leptogenesis, see \emph{e.g.} Ref.~\cite{deCesare:2014dga}.}\textsuperscript{,}\footnote{It was argued in Ref.~\cite{deCesare:2016axk}, based on the Group Field Theory approach, that the gravitational constant could be a a dynamical quantity whose behaviour is determined by the underlying microscopic dynamics of the fundamental degrees of freedom of Quantum Gravity.}
, in a way which is sufficiently general and model independent\footnote{This heuristic assumption could be justified on the basis of general properties of quantum systems related to their stochastic limit, see Ref.~\cite{accardi2013quantum}. In the limit, fast degrees of freedom can be approximated by stochastic noise.}
. The fluctuations and the typical timescale of the stochastic process must be suitably small, \textit{e.g.} ${\cal {O}}(t_{\rm Pl})$, to ensure that there is no contradiction with observations.
We will show in this section that in this simplified model the observed accelerated expansion of the universe is related to the assumed stochastic behaviour of the gravitational `constant'.

In the following we will assume that the function $\xi(t)$ in Eq.~(\ref{Eq:VariableG:StochasticG}) has a white noise distribution. Hence, it satisfies the properties
\begin{align}
\langle\xi(t)\rangle&=0~,\\
\langle\xi(t)\xi(t^{\prime})\rangle&=\delta(t-t^{\prime})~,
\end{align}
where $\langle~\rangle$ denotes the average over a statistical ensemble. 
 Loosely speaking white noise is
the derivative of a Wiener process $W_{t}$, which provides the mathematically rigorous description of a random walk. Thus, we have\footnote{For reasons of mathematical rigour it is preferable to use a differential notation when studying stochastic ODEs, as in Eq.~(\ref{Eq:VariableG:NoiseDifferential}). In fact, one must be aware that stochastic processes such as the Wiener process (Brownian motion) are  (a.s.) continuous but nowhere differentiable functions. The differential notation must be understood as an `informal' way of writing SDEs, which are more rigorously formulated in the form of integral equations, entailing a choice of differential calculus (\emph{e.g.} \^Ito or Stratonovich). For details see Appendix \ref{Appendix:Stochastic}. }
\be\label{Eq:VariableG:NoiseDifferential}
\de W_{t}=\xi(t)\de t~.
\ee
We introduce a time discretisation in order to define the Wiener process $W_t$. For the sake of simplicity we consider a partition of the time interval\footnote{This is only for convenience of our presentation of the model, and to introduce the notation that will be used later in this section. A more general definition of the Wiener process is given in Appendix \ref{Appendix:Stochastic}.} $t_i\equiv t_0 <t_1<\dots <t_k<\dots<t_N\equiv t$ with uniform time step $h$.
A Wiener process $W_t$ must satisfy the following properties:
\begin{enumerate}[label=$\roman*.$]
\item $W_{t_i}=0$ with probability $1$;
\item The increments $W_{t_{k+1}}-W_{t_{k}}$ are statistically independent Gaussian variables
with mean $0$ and variance $t_{k+1}-t_{k}=h$.
\end{enumerate}
In Appendix \ref{Appendix:Stochastic} we give more details on stochastic differential calculus and stochastic integration, which generalise the rules of ordinary calculus to stochastic processes.

We can now study the cosmological consequences of the hypothesis (\ref{Eq:VariableG:StochasticG}). Using the results of the previous section, we can write down the Friedmann equation as
\be\label{Eq:VariableG:FriedmannStochastic}
\begin{split}
H^2(t)=\frac{8\pi}{3}G(t)\left(\rho(t)+\lambda(t)\right)=&\frac{8\pi}{3}\left(G(t)\rho(t)+G(t_i)\lambda(t_i)-\int\de t^{\prime}\;\dot{G}(t^{\prime})\rho(t^{\prime})\right)\\
=&\frac{8\pi}{3}\left(G(t_i)\left(\rho(t_i)+\lambda(t_i)\right)+\int_{t_i}^{t}\de t^{\prime}\; G(t^{\prime})\dot{\rho}(t^{\prime})\right)
\end{split}
\ee
Considering $t_i$ in the matter- or radiation-dominated era,
$\lambda(t_i)\ll\rho(t_i)$ and we can safely neglect the second term
in the r.h.s.\ of the above equation. Notice that in this case,
Eq.~(\ref{Eq:VariableG:FriedmannStochastic}) does not make any explicit reference to $\lambda$, but all
dark energy contributions are instead included in the time
evolution of $G$ via an integral operator. 

By differentiating Eq.~(\ref{Eq:VariableG:FriedmannStochastic}) with respect to time and using the continuity equation (\ref{Eq:VariableG:ContinuityEquation}), we find Eq.~(\ref{Eq:DerivativeHubbleRate}) with Newton's constant replaced by the dynamical $G(t)$
\be\label{Eq:VariableG:DerivativeHubble}
\dot{H}(t)=-4\pi G(t) (\rho(t) +p(t))=-4\pi G(t)(1+w)\rho(t)~.
\ee
The set given by the two equations (\ref{Eq:VariableG:DerivativeHubble}) and (\ref{Eq:VariableG:ContinuityEquation}) gives the coupled dynamics of $H(t)$ and $\rho(t)$. Their initial conditions are not independent, but are related by the Friedmann equation (\ref{Eq:VariableG:FriedmannStochastic}), which implies
\be\label{Eq:VariableG:InitialConditionHubble}
H^2(t_i)=\frac{8\pi}{3}G(t_i)\rho(t_i)~.
\ee

Equations (\ref{Eq:VariableG:DerivativeHubble}) and (\ref{Eq:VariableG:ContinuityEquation}) can be given a precise mathematical sense also in the case of a stochastic $G(t)$. To this end, we rewrite them in differential notation and use Eqs.~(\ref{Eq:VariableG:StochasticG}),~(\ref{Eq:VariableG:NoiseDifferential}). We have
\begin{align}\label{Eq:VariableG:StochasticSystemODE}
\mbox{d}H&=-4\pi \overline{G} (1+w) \rho (\mbox{d}t+\sigma\mbox{d}W_t)~,\\
\de{\rho}&=-3H(1+w)\rho\de t~.
\end{align}
Thus, $H(t)$ and $\rho(t)$ are stochastic processes, which we interpret as {\^I}to processes (see Ref.~\cite{oksendal2003stochastic}). Their dynamics is given by the above system of coupled differential equations, with $\mbox{d}W_t$ playing the role of a stochastic driving term.

\begin{figure}
\begin{center}
\includegraphics[width=0.5\columnwidth]{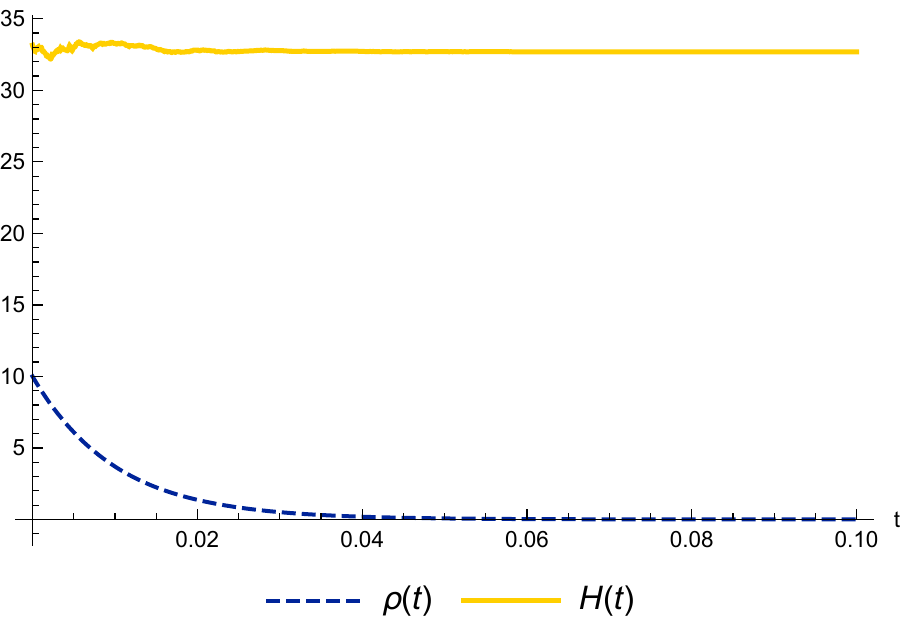}
\caption[Numerical solution of the SDEs system]{Plot of the numerical solution of the model, formulated as a system of two SDEs (\ref{Eq:VariableG:StochasticSystemODE}), with initial conditions satisfying the constraint (\ref{Eq:VariableG:InitialConditionHubble}). The solution represented here corresponds to a particular choice of parameters $\sigma=0.1$, $\rho_0=10$ and time step $h=10^{-4}$, in units
  such that $\overline{G}_{\rm N}=1$. The horizontal asymptote of $H(t)$ for $t\to\infty$ corresponds to a late era of accelerated expansion, when the universe is dominated by a `cosmological constant'. Stochastic noise has an effect which is relevant at early times, whereas it is negligible at late times since it is suppressed by the small value of the energy density, as one can also see from the second equation in (\ref{Eq:VariableG:StochasticSystemODE}). The qualitative behaviour of the solutions is a general feature, which does not depend on the particular choice of parameters.}\label{Fig:VariableG:EffectiveLambda}
\end{center}
\end{figure}

In Ref.~\cite{deCesare:2016dnp}, we studied numerically the solution of the system of stochastic ODEs (\ref{Eq:VariableG:StochasticSystemODE}) in the case of pressureless dust ($w=0$). This was done by using the stochastic Euler integration scheme (also called Euler-Maruyama scheme) \cite{higham2001algorithmic,talay1994numerical}, which is sufficient for our purposes. We explored the range of parameters $0\leq\sigma\leq1$, $10^{-6}\leq h\leq 10^{-2}$, in units such that $\overline{G}=1$, for different choices of the initial condition $\rho(t_i)$. The initial condition for the Hubble rate $H(t_i)$ is computed using Eq.~(\ref{Eq:VariableG:InitialConditionHubble}). We note the dependence of $H(t_i)$ on $\xi(t_i)$. The qualitative behaviour of the solutions  is the same for different values of the parameters $\sigma$, $h$ and is represented in Fig.~\ref{Fig:VariableG:EffectiveLambda} for a given path $W_t$, where the plot of the energy density $\rho(t)$ and the Hubble rate $H(t)$ is shown.

As $t\to\infty$, $\rho$ decays as in standard cosmology, while $H$
attains a nonvanishing positive limit. Thus, the universe enters a de Sitter phase at late times. The value of the asymptotic limit does not vary much with the random sequence, and is not particularly sensitive to the parameters. Since we are dealing with a simplified model, this semi-qualitative analysis is sufficient. Our results show the occurrence of an exponential
expansion at late times without the need to introduce a cosmological
constant by hand or to originate it from the matter sector,
\textit{e.g.} as vacuum energy. Note that this is a \textit{general feature
of the model} that does not depend
on the particular values chosen for $\rho_0$ or the noise strength
$\sigma$. Furthermore, when $\sigma=0$ one recovers the standard
cosmology in the matter-dominated era, namely $\rho
\propto 1/t^2$ and $H \propto 1/t$. It is important to note that similar results can be obtained
for fluids characterised by a different equation of state parameter $w$. For instance,
in the case $w=1/3$ an initial radiation dominated era gives way to a de Sitter era at late times,
as in the case $w=0$ studied above.

The effective cosmological constant in the r.h.s.\ of
Eq.~(\ref{Eq:VariableG:FriedmannStochastic}), neglecting $\lambda(t_i)$, is given by the formula 
\be\label{eq:effective Lambda}
\Lambda_{\rm eff}(t)=\sigma\left(\xi(t_i)\rho(t_i)+\int_{t_i}^{t}\dot{\rho}\;
\mbox{d}W_t\right)~, \ee 
where the second term is a stochastic
integral\footnote{See Appendix \ref{Appendix:Stochastic} for its mathematical definition.}. Considering a universe with more than one matter component, the
effective cosmological constant receives contribution from each of
them, with the dominant contribution still coming from the initial
value of the white noise $\xi(t_i)$ multiplied by the \textit{total energy density}
as in (\ref{eq:effective Lambda}). In other words, the effective cosmological
constant depends on the initial value of the total energy density,
but not on the species populating the universe.

The first term in Eq.~(\ref{eq:effective Lambda}) dominates over the second one with probability close to $1$,
as shown in the following. Hence, when the first step of the
random walk is positive, the universe is exponentially expanding at
late times, whereas for a negative initial step $H$ takes large
negative values\footnote{A negative cosmological constant corresponds to an anti-de Sitter (AdS) universe. This is incompatible with observations, that show a positive value of the cosmological constant. Although unphysical in cosmology, it is nevertheless worth mentioning that AdS spacetime has interesting applications in holography (AdS/CFT correspondence).}. We are therefore led to assume the initial condition
\be\label{eq:initial noise}
\xi(t_i)>0~, \ee 
which is necessary to guarantee the stability of the universe and is compatible with
the observed accelerated expansion.

Even assuming the initial condition (\ref{eq:initial noise}), there is a possibility that the universe may enter an instability at late times, corresponding to $\Lambda_{\rm eff}(t)<0$. Therefore, we compute the probability to observe a negative value of the
effective cosmological constant given the initial condition on the noise (\ref{eq:initial noise}).
This actually forces the first step of the random walk to follow a half-normal
distribution, whereas all the other steps of the random walk
are normally distributed, as it is customary.
Using the statistical properties of the increments of the Wiener process
$W_t$ one finds
\ba
\left\langle \frac{\Lambda_{\rm eff}(t)}{\sigma}\right\rangle&=&\left\langle \frac{W_{t_i}}{h}\right\rangle\rho(t_i)=\sqrt{\frac{2}{\pi h}}\rho(t_i)\\
\left\langle \frac{\Lambda_{\rm eff}^2(t)}{\sigma^2}\right\rangle&=&\left\langle \left(\frac{W_{t_i}}{h}\right)^2\right\rangle\rho^2(t_i)+\sum_k (\rho^{\prime}(t_k))^2 h+2\rho(t_i)\rho^{\prime}(t_i)\nonumber\\
&\approx&\left\langle \left(\frac{W_{t_i}}{h}\right)^2\right\rangle\rho^2(t_i)=\frac{\rho^2(t_i)}{h}
\ea
Therefore the value $\Lambda_{\rm eff}=0$ is $x$ standard deviations away from its mean value,
with
\be
x=\sqrt\frac{2\pi}{\pi-2}.
\ee
Hence, the probability of observing a non-positive value of $\Lambda_{\rm eff}$
(leading to a collapsing universe) is $\lesssim0.0095$, regardless
of the initial condition on $\rho$, the noise strength $\sigma$ and the time step.
In a more complete treatment which may take also spatial variations of $G$ into account, these unstable cases will not be an issue, since the instabilities will be just covered by expanding patches in the universe.

The simplifications introduced in the model and the sensitivity
of the effective cosmological constant $\Lambda_{\rm eff}$ to the initial conditions (\emph{i.e.} the value of the total energy density
at the Planck time $t_{\rm Pl}$)
prevent us from using the model to
extract quantitative predictions for the observed value of the cosmological
constant and its probability.
It is our hope that this preliminary work could open up
new avenues to tackle the cosmological constant problem.

If the picture obtained from our model could be consistently obtained
from a fundamental theory of quantum gravity, it would show that
quantum gravitational phenomena are already being observed in our
Universe today, \emph{i.e.} when looking at the acceleration of its expansion rate.
Moreover, it would
provide indirect evidence for the variation over time of the
gravitational constant. More work has to be done in order to gain a better
understanding of the role played by the initial conditions
on the energy content of the universe,
since the cosmological constant obtained from our toy
model seems to have a strong dependence on them.

The way in which time-reparametrization invariance can be implemented in this scenario will be studied in a future work.
The generalization of our toy-model to a fully relativistic theory of gravity entailing a stochastic gravitational constant will also be the subject of further research.

\newpage
\chapter{Weyl Geometry and Scalar-Tensor-Vector Gravity}\label{Chapter:Weyl}
In this chapter we discuss an extension of the Standard Model and General Relativity proposed by the author in Ref.~\cite{deCesare:2016mml}, which is built upon the principle of local conformal invariance. The model represents a generalisation of a previous work by Bars, Steinhardt and Turok \cite{Bars:2013yba}.
This is naturally realised by adopting as a geometric framework a particular class of non-Riemannian geometries, first studied by Weyl. The gravitational sector is enriched by a scalar and a vector field. The latter has a geometric origin and represents the novel feature of our approach. We argue that physical scales could emerge from a theory with no dimensionful parameters, as a result of the spontaneous breakdown of conformal and electroweak symmetries. We study the dynamics of matter fields in this modified gravity theory and show that test particles follow geodesics of the Levi-Civita connection, thus resolving an old criticism raised by Einstein against Weyl's original proposal.

This chapter is organised as follows. In Section~\ref{sec:Weyl} we recall the fundamentals of Weyl geometry and introduce the notation. In Section~\ref{Theory} we formulate our effective field theory and discuss how the Higgs and the scalar fields couple to gravity. In Section~\ref{EW SSB} we discuss the EW symmetry breaking and show how the dimensionful couplings which govern low energy physics are determined from the parameters of the model and from the scale of `broken' conformal symmetry. In Section~\ref{Sec:CouplingSM} we address the important issue of coupling the other SM fields in a conformal invariant way. In Section~\ref{Sec:WeylFluids} we consider the approximate description of matter as a fluid, following from the underlying field theory of Section~\ref{Sec:CouplingSM}, and use it to derive the equations of motion of
test bodies.
In Section~\ref{Conclusions} we examine the relation between our proposal and earlier ones in the literature. Finally, in the Conclusions, we review our results and point at directions for future work.

\section{Conformal Symmetry in Physics}
The classical action of the Standard Model (SM) of particle physics is close to being conformally invariant. The only dimensionful coupling constants it features are given by the Higgs mass and its vacuum expectation value (vev), the latter setting the scale of electroweak (EW) symmetry breaking at $v\sim$ 246~GeV. Such a value is remarkably small compared to the Planck mass $M_{\rm P}\sim10^{19}$~GeV, which is set by the strength of the gravitational coupling. The huge gap between the two scales defines the hierarchy problem. A fourth dimensionful parameter, the cosmological constant, is responsible for the observed late acceleration of the Universe. The cosmological constant scale is $10^{-123}$ smaller than the Planck scale, leading to a second hierarchy problem in the SM coupled to gravity.

In this work, we embed the SM and General Relativity (GR) in a larger theory which exhibits local scale invariance classically. All couplings are therefore dimensionless. A mass scale arises through gauge fixing the conformal symmetry, from which all dimensionful couplings can be derived. Thus, all couplings which characterise fundamental physics at low energy scales are shown to have a common origin, in the same spirit as in Ref.~\cite{Bars:2013yba}. The role of EW symmetry breaking is crucial in this respect and is realised by means of a potential having the same form as the Higgs-dilaton potential, which was considered in Refs.~\cite{Bars:2013yba,Shaposhnikov:2008xb}.

A natural framework in which scale invariance can be realised as a local symmetry is given by a generalisation of Riemannian geometry, known as Weyl geometry. A Weyl manifold is defined as an equivalence class of conformally equivalent Riemannian manifolds, equipped with a notion of parallel transport which preserves the metric only up to local rescalings~\cite{calderbank1997einstein}. Such non-Riemannian structures were first introduced by Weyl in pursuit of a unification of gravity and electromagnetism \cite{weyl:1918}. They were later reconsidered in an early paper by Smolin~\cite{Smolin:1979uz} in an attempt to reformulate gravity as a renormalizable quantum field theory. In this chapter, as in Ref.~\cite{Smolin:1979uz}, Weyl geometry and conformal invariance are used to motivate the occurrence of new degrees of freedom in the gravitational sector and as guiding principles to build the action functional. Weyl geometry was later rediscovered independently by Cheng \cite{Cheng:1988zx}, who used it to formulate a model with no Higgs particle.

Conformal invariance imposes strong constraints on the terms that can appear in the action and enriches the gravitational sector with a scalar and a vector field. The theory thus obtained is a generalisation of Brans-Dicke theory and of conformally invariant gravity theories, such as the one considered in Ref.~\cite{Bars:2013yba}. When the Weyl vector is pure gauge, the theory is equivalent to Brans-Dicke, of which it provides a geometric interpretation. This particular case has appeared in the literature under the name of Weyl Integrable Space-Time (WIST)~\cite{Romero:2012hs,Almeida:2013dba,Salim:1996ei}. However, in those works an additional assumption motivated by Ref.~\cite{Ehlers2012} is made about the free fall of test bodies, which marks a difference with Brans-Dicke. For applications of WIST to cosmology and to the study of spacetime singularities, see \emph{e.g.} Refs.~\cite{Lobo:2015zaa,Gannouji:2015vva}. Generalised scale invariant gravity theories were also obtained in Ref.~\cite{Padilla:2013jza}, by gauging the global conformal symmetry of (a subset of) the Horndeski action with the introduction of the Weyl vector.

Our framework is distinct from conformal gravity~\cite{Mannheim:1988dj,Maldacena:2011mk,Mannheim:2011ds}, where the affine connection is the Levi-Civita one also in the gravity sector. In that case, conformal symmetry is implemented by taking the square of the Weyl tensor as the Lagrangian. The Weyl tensor squared also appears in the bosonic spectral action in the context of noncommutative geometry~\cite{Kurkov:2014twa,Sakellariadou:2016dfl} and in the computation of the (formal) functional integral for quantum gravity~\cite{Hooft:2010ac}. 

In this chapter we construct an effective field theory with local conformal invariance and show how the SM of particle physics and GR are recovered from it, by means of a two-stage spontaneous symmetry breaking. Our proposal is based on a generalisation of Riemannian geometry, namely Weyl geometry, which leads to the introduction of new degrees of freedom, namely a scalar field $\phi$ and the Weyl vector $B_{\mu}$. There has been a recent surge of interest in the role of conformal symmetry in gravitational physics, see \emph{e.g.} Refs.~\cite{Bars:2013yba,Hooft:2010ac,Gielen:2015uaa,Gielen:2016fdb}, suggesting that it may play a role in Quantum Gravity. It is therefore possible that the gravitational theory emerging in the classical limit would also display such a symmetry. In this sense, our work is motivated by similar considerations to the ones usually put forward for the introduction of modified gravity theories, see \emph{e.g.} Refs.~\cite{Sotiriou:2007yd,Sotiriou:2008rp,Capozziello:2011et}. In addition, we adopt local conformal invariance as a guiding principle in selecting the action functional \emph{and} the geometric structure of spacetime. The enriched gravitational sector is to be interpreted as purely classical. SM fields are quantised as usual on the classical curved background defined by $g_{\mu\nu}$ \emph{and} $\phi$, $B_\mu$. This can be considered as a generalisation of what is usually done in conventional quantum field theory on curved spacetimes.

We would like to mention that the same geometric setting and symmetry breaking process were considered in an unpublished work by Nishino and Rajpoot \cite{Nishino:2004kb}, although their motivations were different. In Ref.~\cite{Nishino:2004kb} the authors point out issues with renormalisability and unitarity in their model. Other aspects of the quantum theory are discussed in Refs.~\cite{Nishino:2009in,Nishino:2011zz}. Furthermore, the authors of Ref.~\cite{Nishino:2004kb} claim that local conformal invariance ``inevitably leads to the introduction of General Relativity". We disagree with their statement. Local conformal invariance of the SM sector only leads to the introduction of the Weyl vector, which is also not enough to determine the affine connection of a Weyl spacetime. Moreover, in our approach there are no issues with renormalisability and unitarity since our model is a \emph{classical} effective field theory.

Lastly, we would like to remark that modified gravity theories are sometimes advocated as an alternative solution to resolve tension between observational data and the predictions of General Relativity, which is customarily done by introducing dark matter. In the case of MOND \cite{Milgrom:1983ca}
and in MOG \cite{Moffat:2005si}, this is achieved by modifying the law of gravitational attraction. In other words, test particles no longer follow geodesics. The model proposed in this chapter sits midway between dark matter models and modified gravity. In fact, in our approach the law of gravitational attraction is not modified, but the gravitational sector is extended and includes a dark matter candidate. The main difference with mainstream dark matter models is that in our model dark matter is not postulated \emph{ad hoc} but has a geometric origin.

\section{Weyl geometry}\label{sec:Weyl} 
In this section we introduce the mathematical framework that will be used to formulate the model. We stress that the concepts and the geometrical objects of Weyl geometry do not have an immediate physical interpretation; they will only serve as tools for the construction of a modified theory of gravity compatible with local conformal invariance. This is an important point, since attempting a direct physical interpretation of the mathematical objects introduced in this section can be misleading and leads to incorrect conclusions. The physical interpretation of the model will come only later, when studying the dynamics.

We follow Ref.~\cite{Smolin:1979uz} to introduce the basic concepts and notation\footnote{We remark that our conventions for the Riemann tensor are different from those of Smolin in Ref.~\cite{Smolin:1979uz}.}. A Weyl manifold is a conformal manifold\footnote{A conformal manifold is an equivalence class of Riemannian manifolds, related by conformal transformations.} equipped with a torsionless connection, the Weyl connection, that preserves the conformal structure\footnote{The conformal structure of a spacetime is a smoothly varying family of light-cones in the tangent space at each point \cite{Ward:1990vs}. As such, it determines the causal structure of spacetime. All representatives in the equivalence class of a Weyl manifold have the same conformal structure.}\textsuperscript{,}\footnote{For a mathematical presentation of Weyl manifolds see Ref.~\cite{folland1970weyl}.  We refer the reader to Ref.~\cite{ehlers2012republication} for an elegant discussion of  Weyl structures following a constructive axiomatic approach.}.
We thus consider a torsion-free affine connection which satisfies the condition
\be\label{eq:DefWeylConnection}
\nabla^{\rm \scriptscriptstyle W}_{\lambda}g_{\mu\nu}=B_{\lambda}\;g_{\mu\nu}\; .
\ee
Equation~(\ref{eq:DefWeylConnection}) defines the Weyl connection $\nabla^{\rm \scriptscriptstyle W}_{\lambda}$, which is a particular case of a connection with non-metricity (see \emph{e.g.} Ref.~\cite{Sotiriou:2006qn}). The Levi-Civita connection will instead be denoted by $\nabla^{\rm \scriptscriptstyle LC}_{\lambda}$.
The connection coefficients are given by
\be\label{eq:ConnectionCoefficients}
\Gamma^{\sigma}_{\mu\nu}=\left\{ {\sigma \atop \mu\;\nu} \right\}-\frac{1}{2}\left(\delta^{\sigma}_{\mu}\,B_{\nu}+\delta^{\sigma}_{\nu}\,B_{\mu}-g_{\mu\nu}\,B^{\sigma}\right)\;.
\ee
Notice that the l.h.s. of Eq.~(\ref{eq:ConnectionCoefficients}) is conformally invariant, even though both terms in the r.h.s.~are not, when considered separately. 
The derivation of Eq.~(\ref{eq:ConnectionCoefficients}) is entirely analogous to that of the Christoffel symbols from the metricity conditions satisfied by the Levi-Civita connection. In this case, one should consider cyclic permutations of non-metricity condition (\ref{eq:DefWeylConnection}) and expand the connection coefficients as Christoffel symbols plus a rank-three tensor field. The result follows from simple algebraic manipulations.

Under a local conformal transformation, also called Weyl rescaling, the metric obeys the following transformation rule
\be\label{eq:LocalConformalMetric}
g_{\mu\nu}\rightarrow\tilde{g}_{\mu\nu}=\Omega^2g_{\mu\nu}~.
\ee
Using Eq.~(\ref{eq:LocalConformalMetric}) and the conformal invariance of the Weyl connection, one finds that the Weyl one-form $B_\mu$ transforms as an Abelian gauge field
\be\label{eq:TransformationLawWeylVector}
B_\mu\rightarrow\tilde{B}_\mu=B_\mu+2\Omega^{-1}\nabla^{\rm \scriptscriptstyle W}_{\mu}\Omega~,
\ee
so that the condition given by Eq.~(\ref{eq:DefWeylConnection}) is preserved. The connection coefficients in Eq.~(\ref{eq:ConnectionCoefficients}) are by definition conformally invariant.

In a local chart, the Riemann curvature tensor of the Weyl connection has components
\be\label{eq:defineRiemann}
R_{\mu\nu\rho}^{\ph\ph\ph\sigma}=-\pa_\mu\Gamma^{\sigma}_{\nu\rho}+\pa_\nu\Gamma^{\sigma}_{\mu\rho}-\Gamma^{\sigma}_{\mu\kappa}\Gamma^{\kappa}_{\nu\rho}+\Gamma^{\sigma}_{\nu\kappa}\Gamma^{\kappa}_{\mu\rho}\;.
\ee
The Riemann tensor satisfies the following properties, as in the standard (metric) case:
\begin{enumerate}[label=\alph*)]
\item ~$R_{\mu\nu\rho}^{\ph\ph\ph\sigma}=-R_{\nu\mu\rho}^{\ph\ph\ph\sigma}$~;
\item ~$R_{[\mu\nu\rho]}^{\ph\ph\ph\ph\sigma}=0$, which follows from the symmetry of the connection coefficients, \emph{i.e.} the vanishing of the torsion~;
\item ~$\nabla_{[\lambda}R_{\mu\nu]\rho}^{\ph\ph\ph\ph\sigma}=0$~.
\end{enumerate}
Antisymmetry over the last two indices, which holds in the metric case, is replaced by 
\be\label{eq:FourhtPropertyRiemann}
R_{\mu\nu\rho\sigma}=-R_{\mu\nu\sigma\rho}+H_{\mu\nu}\;g_{\rho\sigma}\;,
\ee
where $H_{\mu\nu}$ is the field strength of $B_{\mu}$, defined as in electromagnetism
\be\label{eq:FieldStrength}
H_{\mu\nu}=\nabla^{\rm \scriptscriptstyle W}_{\mu}B_{\nu}-\nabla^{\rm \scriptscriptstyle W}_{\nu}B_{\mu}=
\pa_{\mu}B_{\nu}-\pa_{\nu}B_{\mu}\; .
\ee
The Riemann curvature of the Weyl connection, defined by Eq.~(\ref{eq:DefWeylConnection}), has the following expression\footnote{Square brackets denote antisymmetrisation, as in $T_{[\mu\nu]}=\frac{1}{2}\left(T_{\mu\nu}-T_{\nu\mu}\right)$.}
\be\label{eq:RiemannDefinition}
\begin{split}
R_{\mu\nu\rho}^{\ph\ph\ph\sigma}=&R_{\mu\nu\rho}^{0\ph\ph\sigma}+\delta^\sigma_{[\nu} \nabla^{\rm \scriptscriptstyle LC}_{\mu]} B_\rho+\delta^\sigma_\rho \nabla^{\rm \scriptscriptstyle LC}_{[\mu}B_{\nu]}-g_{\rho[\nu}\nabla^{\rm \scriptscriptstyle LC}_{\mu]}B^{\sigma}\\ &-\frac{1}{2}\left(B_{[\mu}\,g_{\nu]\rho}B^{\sigma}+\delta^{\sigma}_{[\mu}\,B_{\nu]}B_{\rho}+g_{\rho[\mu}\,\delta^{\sigma}_{\nu]}B_\lambda B^\lambda \right)~.
\end{split}
\ee
In the last equation, $R_{\mu\nu\rho}^{0\ph\ph\sigma}$ is the Riemann tensor of the Levi-Civita connection. It can be computed from Eq.~(\ref{eq:defineRiemann}), using the Christoffel symbols as the connection coefficients
\be\label{eq:defineOrdinaryRiemann}
R_{\mu\nu\rho}^{0\ph\ph\sigma}=-\pa_\mu\left\{ {\sigma \atop \nu\;\rho} \right\}+\pa_\nu\left\{ {\sigma \atop \mu\;\rho} \right\}-\left\{ {\sigma \atop \mu\;\kappa} \right\}\left\{ {\kappa \atop \nu\;\rho} \right\}+\left\{ {\sigma \atop \nu\;\kappa} \right\}\left\{ {\kappa \atop \mu\;\rho} \right\}\;.
\ee
Defining the Ricci tensor by contracting the second and the fourth indices of the Riemann curvature in Eq.~(\ref{eq:RiemannDefinition})
\be
R_{\mu\nu}=R_{\mu\sigma\nu}^{\ph\ph\ph\sigma}\; ,
\ee
one has
\be\label{eq:RicciTensorExpand}
R_{\mu\nu}=R^0_{\mu\nu}+\nabla^{\rm \scriptscriptstyle LC}_\mu B_\nu +\frac{1}{2}H_{\mu\nu}+\frac{1}{2}g_{\mu\nu}\nabla^{\rm \scriptscriptstyle LC}_{\sigma}B^{\sigma}
+\frac{1}{2}\left(B_\mu B_\nu-g_{\mu\nu}B_\sigma B^\sigma\right)~.
\ee
Note that, as a consequence of Eq.~(\ref{eq:FourhtPropertyRiemann}), the Ricci tensor is not symmetric. In fact, one has
\be
R_{[\mu\nu]}=H_{\mu\nu}\; .
\ee
The Riemann and the Ricci tensors are by definition conformally invariant.
The Ricci scalar is then defined as
\be\label{eq:RicciScalarDefine}
R=g^{\mu\nu}R_{\mu\nu}~.
\ee
Under a conformal transformation the Ricci scalar reads
\be\label{eq:RescalingScalarCurvature}
R\rightarrow\tilde{R}=\Omega^{-2}R~.
\ee
Substituting  Eq.~(\ref{eq:RicciTensorExpand}) into Eq.~(\ref{eq:RicciScalarDefine}), the Ricci scalar is
\be\label{eq:RicciScalarExpand}
R=R^0+3\nabla^{\rm \scriptscriptstyle LC}_\mu B^\mu-\frac{3}{2}B_\mu B^\mu~,
\ee
where $R^0$ is the Ricci scalar computed from the ordinary Riemann curvature, Eq.~(\ref{eq:defineOrdinaryRiemann}).

We would like to give a geometric interpretation of the Weyl connection introduced above, following Ref.~\cite{Romero:2012hs}. To this end, let us consider a vector field $V^{\mu}$ and two vector fields $X^{\mu}$, $Y^{\mu}$ that are parallel transported (with the Weyl connection) along the integral curves of $V^{\mu}$
\be\label{Eq:ParallelTransportWeyl}
V^{\mu}\nabla^{\rm \scriptscriptstyle W}_{\mu} X^{\nu}=V^{\mu}\nabla^{\rm \scriptscriptstyle W}_{\mu} Y^{\nu}=0~.
\ee
Using Eqs.~(\ref{Eq:ParallelTransportWeyl}) and (\ref{eq:DefWeylConnection}) one finds that the scalar product $X_{\mu}Y^{\mu}$ satisfies the following equation
\be\label{Eq:ScalarProductWeylODE}
V^{\mu}\nabla^{\rm \scriptscriptstyle W}_{\mu}\left(X_{\nu}Y^{\nu}\right)=V^{\mu}B_{\mu}\left(X_{\nu}Y^{\nu}\right)~.
\ee
Considering an integral curve $\gamma$ of $V_{\mu}$ and two points $x_0,~x_1\in\gamma$, one has from the solution of Eq.~(\ref{Eq:ScalarProductWeylODE})
\be\label{Eq:ScalarProductWeylODESolution}
\left(X_{\mu}Y^{\mu}\right)(x_1)=\left(X_{\mu}Y^{\mu}\right)(x_0)\,\exp\left(\int_{0}^{1}\de\lambda\; \left(V^{\mu}B_{\mu}\right)(\gamma(\lambda))\right)~.
\ee
This equation states the path dependence of the scalar product of any two vectors that are parallel transported along $\gamma$.
The parameter $\lambda$ is such that $\frac{\de x^{\mu}}{\de\lambda}=V^\mu$ and $\gamma(0)=x_0$, $\gamma(1)=x_1$. We can define $f(x)=\left(X_{\mu}Y^{\mu}\right)(x)$ and introduce differential forms notation to rewrite Eq.~(\ref{Eq:ScalarProductWeylODESolution}) more compactly as
\be
f(x_1)=f(x_0)\,\exp\int_\gamma B~.
\ee
Considering a closed curve $\gamma$, \emph{i.e.} such that $\gamma(0)=\gamma(1)=x_0$, we have
\be\label{Eq:ScalarProductWeylStokes}
f(\lambda=1)=f(\lambda=0)\,\exp\oint_\gamma B=f(\lambda=0)\,\exp\int_\Sigma H~.
\ee
In the last step of the chain of equalities (\ref{Eq:ScalarProductWeylStokes}) we used Stokes' theorem and $H=\de B$ to transform the integral over the closed curve $\gamma$ into a surface integral over the two-dimensional submanifold $\Sigma$, such that $\pa\Sigma=\gamma$. Eq.~(\ref{Eq:ScalarProductWeylStokes}) gives the analogue of the Aharonov-Bohm effect in Weyl geometry, where the Weyl one-form $B$ plays an analogous role to the gauge connection in Yang-Mills theories. We conclude that, in general, scalar products of parallel-transported vectors are path-dependent in Weyl geometry. Only in the case $H=0$, \emph{i.e.} when the Weyl one-form is pure gauge $B=\de \omega$, is path-independence recovered. This is the case considered in Ref.~\cite{Romero:2012hs}. Note that when this happens the Weyl spacetime is related to a Riemannian spacetime (\emph{i.e.} one with a metric-compatible connection) by a conformal transformation.

Similar observations were made by Einstein in his criticism of Weyl's original proposal. In fact, if we interpret the Weyl connection as a physical connection, \emph{i.e.} the one that prescribes how physical objects are parallel transported from one spacetime event to another, Eq.~(\ref{Eq:ScalarProductWeylStokes}) implies the so-called `second clock effect'. This amounts to the fact that clock rates are influenced by their history (unless $H=0$, see above). Einstein also pointed out that such an effect would be against the principle of identity of particles of the same type. In fact, since a particle with rest mass $m$ has a natural clock rate $mc^2\hbar^{-1}$, the mass itself would be path-dependent in Weyl's proposal \cite{Penrose:RoadReality}.
However, we remark that the Weyl connection is not necessarily the physical one, as it will be clarified in the following sections of this chapter. In fact, the physical connection is the one determined from the action principle and can a priori be different from the Weyl connection. The latter will only be used as a tool to write down an action functional that has local conformal invariance built in.
\section{A geometric scalar-vector-tensor theory}\label{Theory}
\subsection{The simplest model}\label{sec:SimpleModel}
Our aim is to build an action functional for gravity which is conformally invariant. We will follow Smolin for its derivation~\cite{Smolin:1979uz}. From Eq.~(\ref{eq:RescalingScalarCurvature}) we see that the simplest action displaying such property is
\be\label{eq:Action}
S_{\rm g}=\int \sqrt{-g}\; \xi_\phi\phi^2 R~,
\ee
where $\xi_\phi$ is a coupling constant and $\phi$ is a real scalar field transforming under local rescalings, Eq.~(\ref{eq:LocalConformalMetric}), according to its canonical dimensions
\be
\phi\rightarrow\tilde{\phi}=\Omega^{-1}\phi~.
\ee
We impose the further requirements that the equations of motion shall contain no derivatives higher than second order and no inverse powers of the scalar field $\phi$ shall appear in the action.  Equation~(\ref{eq:Action}) is therefore singled out as the unique action satisfying the above conditions, in the case of a single non-minimally coupled real scalar field. The scalar field contributes another term to the action
\be\label{eq:PhiSector}
S_{\rm s}=\int \sqrt{-g}\; \left[-\frac{\omega}{2}\; g^{\mu\nu}\left(\pa_\mu\phi+\frac{1}{2}B_{\mu}\phi\right) \left(\pa_\nu\phi+\frac{1}{2}B_{\nu}\phi\right)\right]~,
\ee
where a minimal coupling to the Weyl one-form $B_{\mu}$ has been considered in order to make the action consistent with the principle of local conformal invariance, and $\omega$ is the Brans-Dicke parameter. Lastly, $B_{\mu}$ is made dynamical by adding a kinetic term to the action
\be\label{eq:YMaction}
S_{\rm v}=\int \sqrt{-g}\; \left[-\frac{1}{4f^2}\; H_{\mu\nu}H^{\mu\nu}\right]~,
\ee
in complete analogy with electrodynamics. The field strength $H_{\mu\nu}$  of $B_\mu$ is defined as in Eq.~(\ref{eq:FieldStrength}). The action (\ref{eq:YMaction}) is the Yang-Mills action for an Abelian gauge field. It represents the most natural choice which is compatible with local scale invariance, since the Yang-Mills action is conformally invariant in four dimensions. The parameter $f$ is a universal coupling costant. The action $S_{\rm g}+S_{\rm s}+S_{\rm v}$ defines the extended gravitational sector of the theory.

The scalar field $\phi$ introduced above can be interpreted as a dilaton. In fact, its expectation value gives the strength of the gravitational coupling. However, since we are considering \emph{local} conformal symmetry, the dilaton $\phi$ can be eliminated by an appropriate gauge fixing, as we will show in the next section. Gauge fixing also yields a massive vector $B_\mu$ in the spectrum, thus preserving the total number of degrees of freedom. We should point out that there are other gauge choices in which $\phi$ is instead dynamical, such as those considered in Ref.~\cite{Bars:2015trh}.
\subsection{Coupling the Higgs field to gravity}
The theory given in the previous section can be immediately extended to include the Standard Model Higgs field. In fact, we will show that it is possible to embed the SM in a theory with local conformal invariance. As a result, all dimensionful parameters such as the gravitational constant, the Higgs vev, as well as the Higgs mass, and the cosmological constant, will all have a common origin. The tensor sector is given by
\be\label{eq:PhiHiggsTensorSector}
S_{\rm g}=\int \sqrt{-g}\; \left(\xi_{\phi}\;\phi^2+2\xi_H\; H^{\dagger}H\right) R~,
\ee
where $\xi_{\phi}$, $\xi_H$ are dimensionless couplings. The Higgs kinetic term, including a minimal coupling to the Weyl one-form, is given by
\be\label{eq:HiggsSector}
S_{\rm H}=\int \sqrt{-g}\; \left[-g^{\mu\nu}\left(\pa_\mu H^{\dagger}+ \frac{1}{2}B_{\mu} H^{\dagger}\right)\left(\pa_\nu H+ \frac{1}{2}B_{\nu} H\right)\right]~.
\ee
When introducing Yang-Mills connections corresponding to the SM gauge group, partial and covariant derivatives are replaced by gauge covariant derivatives.
 
We can then introduce a Higgs-dilaton potential as in Ref.~\cite{Shaposhnikov:2008xb},
\be\label{eq:HiggsDilatonPotential}
V(\phi,H)=\frac{\lambda}{4}\left(H^{\dagger}H-\kappa^2\phi^2\right)^2+\lambda^
{\prime}\phi^4~,
\ee
where $\lambda$, $\lambda^{\prime}$, $\kappa$ are dimensionless parameters. 

Fixing the gauge in such a way that $\phi$ takes a constant value $\phi_0$ everywhere in spacetime, the Higgs-dilaton potential takes the form of the usual Mexican hat potential, including a cosmological constant term, namely
\be
V(\phi,H)=\frac{\lambda}{4}\left(H^{\dagger}H-\kappa^2\phi_0^2\right)^2+\lambda^{\prime}\phi_0^4~.
\ee
We can write the Higgs doublet in the unitary gauge
\be
H=\frac{1}{\sqrt{2}}\begin{pmatrix} 0\\ h\end{pmatrix}~.
\ee
It is then readily seen that EW symmetry breaking fixes the values of the gravitational coupling $G$, the Higgs vev $v$, as well as the Higgs mass $\mu$, and the cosmological constant $\Lambda$, in terms of the scale of conformal symmetry breaking $\phi_0$, as (cf.~Ref.~\cite{Bars:2013yba})
\be
\frac{\Lambda}{8\pi G}=\lambda^{\prime}\phi_0^4~
, \hspace{1em} \frac{v^2}{2}=\kappa^2\phi_0^2~,
 \hspace{1em} \frac{1}{16\pi G}=\xi_{\phi}\;\phi_0^2+\xi_H\; v^2~
 , \hspace{1em} \mu^2=-\lambda\kappa^2\phi_0^2~.
\ee

The conformally invariant theory of gravity given here can be seen as a generalisation of other theories with local conformal invariance proposed in the literature. Considering Eq.~(\ref{eq:RicciScalarExpand}), we can rewrite the total action given by the sum of the $S_{\rm g}$, $S_s$, $S_{\rm H}$ and $S_{\rm v}$ contributions from Eqs.~(\ref{eq:PhiHiggsTensorSector}), (\ref{eq:PhiSector}), (\ref{eq:HiggsSector}) and (\ref{eq:YMaction}), respectively, and including the potential Eq.~(\ref{eq:HiggsDilatonPotential}) as
\be
\begin{split}
S=&\int \sqrt{-g}\; \bigg[\left(\xi_{\phi}\;\phi^2+2\xi_H\; H^{\dagger}H\right) R^0-\frac{\omega}{2}\pa^\mu\phi \pa_\mu\phi-\frac{1}{2}(\omega+12\xi_{\phi})\phi B^\mu\pa_\mu\phi \\&-\frac{1}{8}(\omega+12\xi_\phi)\phi^2B_\mu B^\mu -\pa^\mu H^{\dagger}\pa_\mu H-\frac{1}{2}(1+12\xi_H)B^\mu(H^{\dagger}\pa_\mu H+\pa_\mu H^{\dagger} H)\\&-\frac{1}{4}(1+12\xi_H)H^{\dagger}H\; B_\mu B^\mu-\frac{1}{4f^2}\; H_{\mu\nu}H^{\mu\nu}
- \frac{\lambda}{4}\left(H^{\dagger}H-\kappa^2\phi^2\right)^2-\lambda^
{\prime}\phi^4\bigg]~,
\end{split}
\ee
up to a surface term.
\section{EW symmetry breaking and the scalar-tensor-vector gravity}\label{EW SSB}
As a consequence of the spontaneous breakdown of conformal and EW symmetries, the vector $B^{\mu}$ acquires a mass. This can be seen by looking at Eqs.~(\ref{eq:PhiSector}),~(\ref{eq:PhiHiggsTensorSector}),~(\ref{eq:HiggsSector}) and taking into account Eq.~(\ref{eq:RicciScalarExpand}). In fact, excluding interactions with other matter fields and with the Higgs boson, the action of $B^{\mu}$ reads
\be\label{eq:MassiveVectorAction}
S_{\rm v}=\int \sqrt{-g}\; \left[-\frac{1}{4f^2}\; H_{\mu\nu}H^{\mu\nu}-\frac{1}{2}m_B^2\; B_{\mu}B^{\mu}\right]~,
\ee
with
\be
m_B^2= 3\left(\xi_{\phi}\;\phi_0^2+\xi_H\; v^2\right)+\frac{\omega}{4}\phi_0^2+\frac{v^2}{4}=\frac{3}{16\pi G}+\frac{v^2}{4}\left(\frac{\omega}{2\kappa^2}+1\right)~.
\ee
It is possible to rewrite the action of the vector field in canonical form, by expanding the first term in Eq.~(\ref{eq:MassiveVectorAction}) and rescaling the field as $B^{\mu}\rightarrow f\; B^{\mu}$. We have
\be\label{eq:ProcaActionBField}
S_{\rm v}= \int \sqrt{-g}\; \bigg[ -\frac{1}{2}\Big( (D^{\mu}B^{\nu})(D_{\mu}B_{\nu})-(D^{\nu}B^{\mu})(D_{\mu}B_{\nu})\Big) -\frac{1}{2}f^2 m_B^2\; B_{\mu}B^{\mu}\bigg]~.
\ee
Hence, the physical mass squared of the vector is given by
\be
m_{\rm v}^2=f^2 m_B^2~.
\ee
Equation~(\ref{eq:ProcaActionBField}) is the Proca action in a curved spacetime. Sources $j^{\mu}$ for the field $B_{\mu}$ come from the other sectors of the theory; they are covariantly conserved, $D_{\mu}j^{\mu}=0$, as a consequence of the minimal coupling prescription. From the equations of motion one gets the subsidiary condition $D_{\mu}B^{\mu}=0$ (since $m_v^2\neq0$), which restricts the number of degrees of freedom of the vector field to three, namely two transverse modes and a longitudinal mode. Hence, counting degrees of freedom before and after the breaking of conformal invariance gives the same result. In analogy with the Higgs mechanism, we can say that the vector field $B^{\mu}$ acquires a mass and a longitudinal polarisation mode as a result of conformal symmetry breaking. The dilaton $\phi$ can be completely decoupled from the theory by choosing a suitable gauge, as it happens for the Goldstone boson in the unitary gauge (see however the remark at the end of Section~\ref{sec:SimpleModel}). In fact, a stronger result holds: the kinetic term of $\phi$ is identically vanishing, which makes the field non-dynamical. Only its constant value $\phi_0$ appears in all equations written in this gauge. 

Before closing this section, we want to specify the connection between our model and the ones in the literature about conformal invariance in gravity and cosmology. Choosing the particular values of the parameters $\xi_H=\frac{\xi_\phi}{\omega}=-\frac{1}{12}$, the Higgs and the dilaton fields are completely decoupled from the vector field, which yields the action
\be\label{eq:BarsTurokAction}
S=\int \sqrt{-g}\; \bigg[-\left(\frac{\omega}{12}\phi^2+\frac{1}{6}\; H^{\dagger}H\right) R^0-\frac{\omega}{2}\pa^\mu\phi \pa_\mu\phi -\pa^\mu H^{\dagger}\pa_\mu H
-V(\phi,H)\bigg]~,
\ee
with $V(\phi,H)$ the Higgs-dilaton potential given from Eq.~(\ref{eq:HiggsDilatonPotential}).
Equation~(\ref{eq:BarsTurokAction}) is the action of two scalar fields with conformal coupling to curvature; it is the model considered in Ref.~\cite{Bars:2013yba}, for $\omega=-1$. Writing the Higgs field in the unitary gauge, the action Eq.~(\ref{eq:BarsTurokAction})  can also be seen as equivalent to the conformally invariant two-field model of Ref.~\cite{Kallosh:2013hoa} with $\mbox{SO(1,1)}$ symmetry.
\section{Coupling to SM fields}\label{Sec:CouplingSM}
So far, we have focused our attention on the gravitational sector of the theory, given by the fields $g_{\mu\nu}$, $B_\mu$ and $\phi$, and considered their couplings to the Higgs doublet. In this section we will focus on their couplings to SM fields and study whether the framework of Weyl geometry introduces any modifications to such sectors. We will discuss separately the cases of gauge bosons and spin-$1/2$ fermions (leptons and quarks).

Let us consider a gauge field $A_\mu^a$, where $a$ is an internal index labelling components in the Lie algebra of the gauge group. Its kinetic term is given by the square of its field strength\footnote{$g$ is the gauge coupling constant, $f^{abc}$ are the structure constants of the gauge group. In the Abelian case the second term in Eq.~(\ref{eq:GaugeFieldStrength}) vanishes.},
defined using the affine connection $\nabla^{\rm \scriptscriptstyle W}_\mu$
\be\label{eq:GaugeFieldStrength}
F_{\mu\nu}^a=\nabla^{\rm \scriptscriptstyle W}_\mu A^a_\nu - \nabla^{\rm \scriptscriptstyle W}_\nu A^a_\mu + g f^{abc} A_\mu^bA_\nu^c.
\ee
It is well known that for all symmetric (\emph{i.e.} torsion-free) connections $\nabla^{\rm \scriptscriptstyle W}_\mu$ the above can be rewritten as
\be
F_{\mu\nu}^a=\nabla^{\rm \scriptscriptstyle LC}_\mu A^a_\nu - \nabla^{\rm \scriptscriptstyle LC}_\nu A^a_\mu + g f^{abc} A_\mu^bA_\nu^c
=\pa_\mu A^a_\nu - \pa_\nu A^a_\mu + g f^{abc} A_\mu^bA_\nu^c~.
\ee
In particular, this is true in the case when $\nabla^{\rm \scriptscriptstyle W}_\mu$ is the Weyl connection.  Hence, there is no direct coupling between the Weyl vector and gauge bosons. The kinetic term of the gauge boson $A_\mu^a$ is given by the standard Yang-Mills action
\be
S_{\rm YM} =-\frac{1}{4}\int \sqrt{-g}\; F^a_{\mu\nu}F^{a\,\mu\nu}~,
\ee
which is conformally invariant in four dimensions. 
 The scalar field $\phi$ is real in our model, therefore it does not couple to ordinary gauge fields through the minimal coupling prescription. Although it is certainly possible to generalise the model to allow for non minimal couplings, they can potentially spoil conformal invariance or renormalizability of the SM (or both).

The description of the dynamics of fermions on curved spacetime requires the introduction of a tetrad and of a spin connection. The action of a massless Dirac spinor is given by (see \emph{e.g.} \cite{Parker:2009uva})
\be\label{eq:ActionDirac}
S_{\rm Dirac}=\int  \sqrt{-g}\; i \overline{\psi}\gamma^c e^{\mu}_c\left(\pa_\mu+\frac{1}{8}[\gamma^a,\gamma^b]\,e_a^{\;\nu}(\nabla^{\rm \scriptscriptstyle LC}_\mu e_{b\,\nu})\right)\psi~.
\ee
Observe that Eq.~(\ref{eq:ActionDirac}) uses the Levi-Civita connection $\nabla^{\rm \scriptscriptstyle LC}_\mu$. The reason for this choice will be clear from the following.
Latin indices are used for the Lorentzian frame defined pointwise by the tetrad $e^a_\mu$
\be
e^a_{\,\mu} e_{a\,\nu}=g_{\mu\nu},\hspace{1em} e^a_{\,\mu} e^{b\,\mu}=\eta^{ab}~.
\ee
$\eta_{ab}$ is the Minkowski metric ${\rm diag}$(-1,1,1,1). The gamma matrices in Eq.~(\ref{eq:ActionDirac}) are the flat ones $\{\gamma^a,\gamma^b\}=2\eta^{ab}$.
Under a conformal transformation, each field in Eq.~(\ref{eq:ActionDirac}) transforms according to its conformal weight
\be
\psi\rightarrow \tilde{\psi}=\Omega^{-3/2}\psi,\quad \overline{\psi}\rightarrow \tilde{\overline{\psi}}=\Omega^{-3/2}\overline{\psi},\quad e^a_{\,\mu}\rightarrow\tilde{e}^a_{\,\mu}=\Omega\, e^a_{\,\mu}~.
\ee
It is possible to check by explicit computation that, under such a transformation, all terms involving derivatives of the function $\Omega$ cancel in Eq.~(\ref{eq:ActionDirac}). The reader is referred to Appendix \ref{Appendix:ConformalDirac} for details. Hence, the action of a Dirac fermion defined using the Levi-Civita connection is conformally invariant. The same conclusion  can also be reached by looking at the square of the Dirac operator defined by Eq.~(\ref{eq:ActionDirac}). In this way, one finds a generalisation of the Klein-Gordon equation with a non-minimal coupling to curvature, which turns out to be conformally invariant \cite{Parker:2009uva,konno:1988}.
In Ref.~\cite{Cheng:1988zx} the action of a Dirac particle was defined by considering a generalisation of Eq.~(\ref{eq:ActionDirac}) which makes both terms in the bracket separately conformally invariant, when acting on $\psi$. Namely, the Weyl connection is considered instead of the Levi-Civita connection and the coupling to the Weyl vector is included, with the appropriate coupling constant given by the conformal weight of the spinor
\be\label{Eq:ActionDirac2}
\int  \sqrt{-g}\; i \overline{\psi}\gamma^c e^{\mu}_c\left(\pa_\mu+\frac{3}{4}B_\mu+\frac{1}{8}[\gamma^a,\gamma^b]\,e_a^{\;\nu}(\nabla^{\rm \scriptscriptstyle W}_\mu e_{b\,\nu})\right)\psi~.
\ee
However, it turns out that this action is equal to the one in Eq.~(\ref{eq:ActionDirac}), since the terms involving the Weyl vector cancel exactly. 

We conclude this section by stressing that the requirement of local conformal invariance does not introduce new direct couplings of the elementary matter fields (with the only exception of the Higgs) with the new fields $\phi$ and $B_\mu$. Their interactions with leptons, quarks and gauge bosons can only be mediated by the gravitational field $g_{\mu\nu}$ or the Higgs field. This has important implications for the dynamics of matter in a gravitational field. 

\section{Motion of fluids and test particles}\label{Sec:WeylFluids}
In the previous section we showed that the dynamics of free vector and spinor fields is determined solely by the Levi-Civita connection. The only field in the gravitational sector with whom they can interact directly is the metric tensor $g_{\mu\nu}$. A description of matter which is particularly convenient for applications to macroscopic physics (\emph{e.g.} astrophysics, cosmology) in certain regimes, is in terms of perfect fluids. Following Ref.~\cite{Brown:1992kc}, the matter action for a perfect and isentropic fluid is given by (see Section~\ref{Sec:Fluid})
\be\label{eq:ActionMatter}
S_{\rm matter}=-\int\sqrt{-g} \left[\rho\left(\frac{|J|}{\sqrt{-g}}\right)+J^{\mu}(\pa_\mu\chi+\beta_A\pa_\mu\alpha^A)\right]~.
\ee
The Lagrange multiplier $\chi$ enforces particle number conservation. Additional constraints can be imposed. In fact, interpreting $\alpha^A$ ($A=1,2,3$) as Lagrangian coordinates for the fluid, the Lagrange multipliers $\beta_A$ impose the condition that the fluid flows along lines of constant $\alpha^A$. The stress-energy tensor obtained from the action Eq.~(\ref{eq:ActionMatter}) has the standard form
\be\label{eq:StressEnergy}
T_{\mu\nu}=-\frac{2}{\sqrt{-g}}\frac{\delta S_{\rm matter}}{\delta g^{\mu\nu}}= (\rho+p)\, U_\mu U_\nu + p\, g_{\mu\nu}~,
\ee

The dynamics of the fluid is obtained by looking at the stationary points of the action (\ref{eq:ActionMatter}). In particular, diffeomorphism invariance implies that the stress-energy tensor is covariantly conserved
\be\label{eq:ConservationStressEnergy}
\nabla^{\rm \scriptscriptstyle LC\,\mu} T_{\mu\nu}=0. 
\ee
Notice that the local conservation law Eq.~(\ref{eq:ConservationStressEnergy}) is formulated in terms of the Levi-Civita connection. We remark that, as it is well-known, Eq.~(\ref{eq:ConservationStressEnergy}) holds for all types of matter (including elementary matter fields, which we considered in the previous section) as long as interactions with other matter species are negligible. The Higgs field represents an exception, since it has direct couplings to the Weyl vector.

Different regimes have to be considered for the dynamics of matter, depending on the energy scale. Above the scale of EW symmetry breaking (and regardless of the fact that conformal symmetry is broken or unbroken), all particles are massless and can be described as a perfect radiation fluid $\rho_{\rm rad}(n)\propto n^{4/3}$. At lower scales and after the spontaneous breakdown of EW symmetry, photons and neutrinos remain massless, while baryonic matter\footnote{As in the ordinary usage of the word by cosmologists, \emph{i.e.} including leptons and actual baryons.} is characterised by $\rho_{\rm bar}(n)\propto n$. As far as the dynamics of matter fields alone\footnote{Again, with the exception of the Higgs field.} is concerned, there is no difference with the corresponding equations obtained in GR. Interactions with $B_\mu$ and $\phi$ can only be mediated by the gravitational field $g_{\mu\nu}$ or the Higgs field $H$. As it is well known, the dynamics of a small test body can be obtained from the conservation law Eq.~(\ref{eq:ConservationStressEnergy}) \cite{Geroch:1975uq}. This is readily seen for dust ($p=0$), in which case the worldline of each dust particle is a geodesic of the Levi-Civita connection, \emph{i.e.} the four-velocity satisfies the equation
\be
U^\mu \nabla^{\rm \scriptscriptstyle LC}_\mu U^\nu=0~.
\ee

Geodesic motion of test bodies is a consequence of the coupled dynamics of the gravitational field and matter \cite{Geroch:1975uq}, not an independent physical principle. Hence, the connection that is used to define the parallel transport of \emph{physical} objects is \emph{not an independent prescription} fixed at the outset, but it is instead a consequence of the dynamics. Although this is a well-known result in General Relativity (see Ref.~\cite{Geroch:1975uq}), to the best of the authors' knowledge it has not been stressed previously in a non-Riemannian framework. In our case, the dynamics follows from an action principle which we built using local conformal invariance as an additional guiding principle. The Weyl connection is used as a tool to implement this principle in a natural way in the gravitational sector. It turns out that local conformal invariance in the sector of gauge bosons and spin-$1/2$ fermions does not require using a non-metric connection. The standard minimal coupling to the gravitational field is enough to ensure that conformal invariance holds as a local symmetry.

We would like to stress at this point that, although our approach is based on Weyl geometry as a framework for a dynamical theory of gravity, it differs from Weyl's original formulation in certain important respects. The main objection against Weyl geometry as a framework for gravitational physics is based on a criticism moved by Einstein against Weyl's original proposal. 
 Einstein's argument is the following. If a vector is parallel transported along a closed path, with parallel transport defined by the Weyl connection $\nabla^{\rm \scriptscriptstyle W}_\mu$ instead of the Levi-Civita connection $\nabla^{\rm \scriptscriptstyle LC}_\mu$, the norm of the vector changes as a result. This would have obvious physical consequences. In fact, considering any two paths in spacetime having the same starting and end points, rods lengths and clocks rates would depend on their histories\footnote{The same argument would also apply for parallel transport given by other non-metric connections $\nabla^{\rm \scriptscriptstyle W}_\lambda g_{\mu\nu}=Q_{\lambda\mu\nu}$ with non-vanishing Weyl vector, defined as the trace of the non-metricity $B_\mu=\frac{1}{4}Q^{\phantom{a}\phantom{a}\lambda}_{\mu\lambda}$.}. This is known as the \emph{second clock effect}. Any theory leading to such effects is clearly non-physical\footnote{The Aharonov-Bohm effect is an analogue of this effect which is instead physical. In that case though, the gauging is not done in physical space, as in Weyl's original proposal, but in the internal space given by the phase of the wave-function.}.

It is worth stressing that this is an argument against the use of the Weyl connection as the one defining parallel transport of \emph{physical} objects, such as rods and clocks. This is clearly not the case in our model. In fact, the dynamics of all elementary matter fields (with the important exception of the Higgs) only involves the Levi-Civita connection $\nabla^{\rm \scriptscriptstyle LC}_\mu$. Hence, it does not entail any direct coupling to the new fields in the gravitational sector. Classical test particles move along geodesics defined by $\nabla^{\rm \scriptscriptstyle LC}_\mu$, as in GR.

\section{Conclusion}\label{Conclusions}
We considered an extension of GR and SM with local conformal invariance. The purpose is to provide a new framework for the study of conformal symmetry and its relation to fundamental physics at high energy scales. This is achieved by considering a generalisation of Riemannian geometry, first introduced by Weyl and later proposed by Smolin~\cite{Smolin:1979uz}. The affine connection is no longer given by the Levi Civita connection, as only the conformal structure of the metric is preserved by parallel transport. This leads to the introduction of a gauge vector $B_{\mu}$ in the gravitational sector: the Weyl vector. A scalar field $\phi$ is also needed in order to build a conformally invariant action functional. The framework is that of a classical effective field theory of gravity. The interpretation of our model is similar to that of quantum field theory in curved spacetime. SM fields are quantised as usual, with $g_{\mu\nu}$, $B_{\mu}$ and $\phi$ representing \emph{classical} background fields\footnote{This is clearly the case for the metric $g_{\mu\nu}$ and the Weyl vector $B_{\mu}$ since they define the classical geometric structure of spacetime, see Eq.~(\ref{eq:DefWeylConnection}). In fact, either they are both classical or both quantum. The status of the field $\phi$ is a more subtle issue and both cases are possible \emph{a priori}. Only a careful analysis of the implications of the two possibilities can determine which one is correct.}.

Our model is a generalisation of previous works in the scientific literature on conformal symmetry in gravity theories \cite{Bars:2013yba,Kallosh:2013hoa}, which can be recovered as a particular case of our model. The main difference in our approach is due to the introduction of a new geometric degree of freedom, represented by the Weyl vector field entering the definition of the Weyl connection. Suitable choices of some parameters of the theory lead to the decoupling of the Weyl vector from the Higgs and the scalar fields. Although, in the general case its dynamics cannot be neglected. After gauge fixing the conformal symmetry (which can be interpreted as a spontaneous symmetry breaking) and EW symmetry breaking, the Weyl vector acquires a mass and the scalar is completely decoupled from the theory. The relevance of the scalar for low energy physics lies in the fact that, through gauge fixing, it leads to the introduction of a physical scale in a theory which is scale-free at the outset. All dimensionful parameters of the SM and gravity can be expressed in terms of it and of the dimensionless parameters of the theory.

\section{Discussion}
We would like to stress that the present model does not necessarily offer a resolution of the naturalness (or hierarchy) problem. In fact, such problem is now translated in the fine-tuning of its dimensionless parameters. Namely, the hierarchy of the Planck versus EW scale leads to $\frac{v^2}{M_{Pl}^2}=\frac{\xi_\phi}{2\kappa^2}+\xi_H\sim 10^{-34}$. Nevertheless, classical conformal invariance of the extended SM sector is important as a guideline for model building, since it restricts the class of allowed couplings to those having dimensionless coupling constants \cite{Heikinheimo:2013fta}. Furthermore, the possibility of addressing the hierarchy problem in conformally invariant extensions of SM has been considered in \emph{e.g.} Ref.~\cite{Meissner:2006zh} and in earlier works Refs.~\cite{Buchmuller:1988cj,Buchmuller:1990pz}. In the models considered in those works, the EW and the Planck scales are determined by non-trivial minima of the one-loop effective potential in the Higgs-dilaton sector\footnote{The mechanism is a generalisation of the one originally proposed by Coleman and Weinberg in Ref.~\cite{Coleman:1973jx}.}. It will be the subject of future work to study whether a similar mechanism could be implemented consistently within our framework. In fact, whereas it is clear that the Weyl vector cannot be quantised without also quantising the metric, one may speculate that the scalar field $\phi$ should be treated on the same footing of matter fields and be regarded as quantum. Hence, similar analysis as in the works cited above should be carried out to check the viability of such working hypothesis. In the affirmative case, it would be possible to address the important point concerning the exact value of the scale $\phi_0$, which we regarded as a free parameter in this work\footnote{Similarly, the scale of `conformal symmetry breaking' (gauge fixing) $\phi_0$ is a free parameter in all models with classical conformal invariance see \emph{e.g.} Refs.~\cite{Bars:2013yba,Kallosh:2013hoa}.}.

Einstein's criticism to Weyl's original proposal is addressed in our model, which is not affected by the \emph{second clock effect}. In fact, the affine connection that defines parallel transport of physical obejcts, such as \emph{e.g.} clocks and rods, is the Levi-Civita connection. Test particles move along Levi-Civita geodesics as in GR. We remark that this is not prescribed at the outset. It is instead a consequence of the dynamics, which has been formulated using conformal invariance as a guiding principle. The Weyl connection \emph{does} play a role in determining the gravitational sector of the theory, although it does not determine the motion of test particels\footnote{It is remarkable that essentially the same observation was made by Weyl in a reply to Einstein's comment to his original paper. We quote from the English translation contained in Ref.~\cite{ORaifeartaigh:1997dvq}: \emph{``It is to be observed that the mathematical ideal of vector-transfer {\rm (Author's Note: \emph{i.e.}, parallel transport)}, on which the construction of the geometry is based, has nothing to do with the real situation regarding the movement of a clock, which is determined by the equations of motion''}.}. Furthermore, the introduction of the $B_\mu$ field is necessary in order to build a conformally invariant action functional for scalar fields in four dimensions, but has no (direct) effects on radiation and baryonic matter.

SM fields do not couple to the new fields in the gravitational sector, with the exception of the Higgs. Their interactions with $\phi$ and $B_\mu$ can only be mediated by gravity or the Higgs field. In future work we will explore the physical consequences of our model for cosmology and astrophysics. In particular, it would be interesting to study whether $\phi$ and $B_\mu$ could represent valid dark matter candidates. This was first hinted in Ref.~\cite{Cheng:1988zx} for the Weyl vector. If this was the case, the interpretation would be substantially different  from standard dark matter. In fact, the Weyl vector should not be regarded, strictly speaking, as matter but as a property of the spacetime geometry. Important viability checks for the model require the determination of constraints on the couplings of $\phi$ and $B_\mu$ to the Higgs that may come from collider physics. The Weyl vector $B_\mu$ is a classical background field; hence, it can only contribute external lines to the diagrams describing known processes. This would be true also for the scalar $\phi$, if this is to be regarded as classical. If, on the other hand, $\phi$ is treated as a quantum field, there will be a new scalar entering loop diagrams. In this case, it is crucial to determine which values of the coupling constants (such as \emph{e.g.} $\xi_\phi$, $\xi_H$) in the bare action are such that renormalizability of SM is not spoiled (see \emph{e.g.} the analysis in Ref.~\cite{Coriano:2012nm}). Addressing this question may also help to shed some light on the `naturalness' of the particular choice of parameters\footnote{Commonly known as conformal couplings, since in the standard framework of Riemannian geometry these are the unique values that make the kinetic terms of $\phi$ and $H$ conformally invariant.} $\xi_H=\frac{\xi_\phi}{\omega}=-\frac{1}{12}$ within the broader framework of Weyl geometry. The phenomenology of the model should be studied in detail both in the case where $\phi$ is quantised and when it represents instead a classical background.

Detailed studies of the consequences for gravitational experiments are also in order and will be the subject of future work. In particular, we plan to explore in a future work the possible observable consequences of the enriched gravity sector and its implications for astrophysics and cosmology. It is also worth studying the possible relations between our model and modified gravity theories such as the one considered in Ref.~\cite{Moffat:2005si}.

\part{Conclusion}

\chapter{Concluding Remarks and Outlook}
In this thesis we studied the consequences of quantum gravitational effects in cosmology. The problem was tackled using two alternative plans of attack, which are complementary to each other. More specifically, in the first part of this work (Chapters 2 and 3) we considered non-perturbative and background independent approaches to Quantum Gravity (or Quantum Cosmology). Their implications for the dynamics of the cosmological sector in the quantum theory were then studied in detail. Depending on the approach, particular attention was paid to the effective dynamics of the emergent spacetime, or to semiclassical dynamics. In the second part (chapters 4 and 5), we considered classical effective theories of gravity. In this case, an attempt was made to relate the departures from general relativity introduced by these theories to the phenomenology of gravitational physics on large scales, generally ascribed to a dark sector.

In the first chapter we reviewed the formulation of the standard cosmological model, which is based on classical General Relativity and assumes FLRW spacetime as a background. We focused our attention on the dynamics of the background, which is most relevant for comparison with our results presented in the next chapters. Our exposition is intentionally non-standard, and the emphasis is on the physical concepts and mathematical tools needed for our applications.

In the second chapter we studied an extension of the standard minisuperspace quantum cosmology in the geometrodynamics formulation, also known as Wheeler-DeWitt approach. The Wheeler-DeWitt equation is generalised by the inclusion of extra terms (interactions) which preserve the linearity of the theory. We considered a closed universe with a massless scalar field, for which wave packet solutions with suitable semiclassical properties exist and are known in the literature. We developed fully general perturbative methods, that can be applied to study the perturbations of such wave packets caused by the new terms.

Using the massless scalar as internal time, we divide the interaction terms in two broad classes: those that depend on the internal time, and those that are independent from it. In both cases, we determined the Green function of the Wheeler-DeWitt operator, subject to suitable boundary conditions. In the `time'-independent case the Green function is obtained from a fundamental solution of the one-dimensional Helmoltz equation with an exponential potential. The solution can be computed exactly and depends on a real parameter, which is linked to the choice of boundary conditions at the singularity. In the `time'-dependent case, the Green function of the Wheeler-DeWitt operator was computed non-perturbatively by means of a conformal transformation in two-dimensional minisuperspace, and using an appropriate generalisation of the method of image charges in electrostatics. The boundary conditions satisfied by the propagator at the big-bang and big-crunch singularities can be seen as a generalisation of the Feynman causal boundary conditions in ordinary quantum field theory.

As an application, we considered the case of a white noise interaction, which we argued can be used to give an effective description of the influence of the microscopic degrees of freedom on the background. The perturbed wave packet retains the properties of semiclassicality of the unperturbed state. Moreover, the cosmological arrow of time is unaffected and can be identified (at the background level) with the direction of increasing scale factor, as previously suggested in the literature for the unperturbed solutions.

It will be interesting to study in detail  in a future work the connection between different boundary conditions imposed on the solutions of the Wheeler-DeWitt equation and integration contours in the path-integral formulation of Lorentzian quantum cosmology. We remark that the techniques we used are fully general and do not rely on specific choices of the extra interaction terms, which makes them suitable for applications in different contexts. It is worth observing that the Green functions obtained in this work can find an application also in areas of physics that are unrelated to quantum cosmology, and where the dynamics is governed by the Klein-Gordon equation with an exponential potential. Possible examples may come from atomic and condensed matter physics.

\

In the third chapter we considered the Group Field Theory approach to Quantum Gravity and its emergent cosmology. In this scenario the dynamics of the expansion of the universe is governed by an effective Friedmann equation including quantum gravitational corrections. Such effective cosmological dynamics is obtained by considering a particular class of states, namely coherent states, in the Fock space of the theory. Equivalently, the dynamics can be derived by means of a mean field approximation of the microscopic theory.

It was shown in the literature that in this scenario the classical initial singularity is avoided and the solutions of the effective Friedmann equation are characterised by a bounce taking place at a small volume. By considering a mean field with only one monochromatic component, we showed that such behaviour of the solutions is generic and does not depend on the initial conditions, nor on the inclusion of interaction terms. In particular, in the non-interacting case we found exact analytic solutions of the model and identified an era of accelerated expansion which accompanies the bounce.

The number of e-folds during the era of accelerated expansion turns out to be much smaller compared to inflationary models. This shortcome of the non-interacting model is overcome by the inclusion, in the effective dynamics of Group Field Theory, of suitable interactions, which in our work are phenomenologically motivated. Thus, we showed that by an appropriate choice of the interactions it is possible to achieve an arbitrarily large number of e-folds, without having to introduce an inflaton field. In particular, the strength of the interactions can be adjusted so as to match typical values obtained in inflationary models.

In Group Field Theory, interactions are especially important due to their geometric interpretation in the microscopic theory. Thus, our results seem to be hinting at a deeper connection between phenomenological aspects of the emergent cosmology and fundamental ones of the underlying Quantum Gravity model. We showed that interactions also lead to a recollapse of the universe after it reaches a maximum volume, which happens under fairly general conditions imposed on the potential. This result, together with the bounce taking place at small volumes, implies a cyclic evolution of the emergent universe.

Considering the Group Field Theory formulation of the EPRL model for Quantum Gravity, we studied the dynamics of non-monochromatic perturbations of the mean field. Such perturbations represent microscopic anisotropies of the fundamental building blocks of quantum geometry. We identified a region of parameter space such that non-monochromatic perturbations decay rapidly away from the bounce. Furthermore, we showed that for  suitable  values  of  the  initial  conditions  and  of  the interaction  strength, perturbations  can  become  negligible  before  the  interactions  kick  in.

Future work should aim at making the model more realistic, in order to extract quantitative predictions and establish a link with cosmological observations. More precisely, more general types of matter need to be introduced, going beyond the simplest case of a minimally coupled massless scalar field. More work is also needed to understand the precise role and the physical meaning of graph connectivity, \emph{i.e.} correlations between different nodes, from the point of view of macroscopic physics. In fact, in the mean field approximation that we considered such information is lost, and the geometry of the emergent spacetime is essentially determined by that of a single tetrahedron.

The anisotropies considered in our work are microscopic ones (non-monochromaticity). In principle, we expect them to be related to macroscopic anisotropies, although no clear correspondence between the two has been established so far. To this end, a suitable set of observables must be identified in the quantum theory, that can quantify the anisotropies of a given geometry. This would enable us to put the correspondence between microscopic and macroscopic anisotropies on more solid ground. At the same time, we will be able to understand in more precise terms the geometric properties of the emergent spacetime for generic group field theory condensate states. 

\

In the fourth chapter we modified the Einstein equations of classical General Relativity by allowing for a dynamical gravitational constant. Consistency with the Bianchi identities is achieved by introducing an extra source of stress-energy, which we interpreted as dark energy. We considered a model in which the gravitational constant is subject to stochastic fluctuations. The dynamics of a FLRW universe in this model was studied in detail. The energy density of the universe at late times is dominated by an effective cosmological constant, which is defined by a stochastic integral. Our results show that the probability of having a positive value for the cosmological constant in this model is very close to one.

In a future work, we plan to extend such a toy model and to go beyond the strictly homogenous and isotropic case considered here. In order to do so,
more input is needed to specify the stochastic dynamics of the effective gravitational `constant'. Importantly, a generalisation of our model will necessarily entail a generalisation of fluctuation-dissipation theorems to the gravitational case. In fact, dissipation was not considered in our model as a first approximation. However, it can potentially play an important role in determining the evolution of macroscopic physical variables and may have an impact on the behaviour of gravity on large scales.

\

In the fifth chapter we reconsidered Weyl geometry as a framework for an extension of General Relativity that is compatible with the principle of local conformal invariance. Conformal symmetry is spontaneously broken in a way that is compatible with electroweak symmetry breaking via the Higgs mechanism. All the dimensionful parameters in the Standard Model and gravitational physics are determined by the scale of conformal symmetry breaking and dimensionless couplings in the model.

The new geometric framework leads to the introduction of new fields in the gravitational sector of the theory, namely a scalar field (dilaton) and the Weyl vector. Standard Model fields have no direct couplings to the Weyl vector, whose interactions with matter can only be mediated by the Higgs and gravitational fields. Test particles in free-fall move along geodesics of the Levi-Civita connection as in General Relativity. Deviations from General Relativity in the gravitational field equations are introduced by the non-minimal coupling of the scalars (Higgs and dilaton) and by the Weyl vector. After the spontaneous breakdown of the conformal and electroweak symmetries, the Weyl vector becomes massive. We argued that it can be interpreted as dark matter.

The most intriguing aspect of the model is the fact that it provides (at least) one dark matter candidate, namely the Weyl vector. There is a similarity with so-called Higgs portal model which deserves further investigation. In future work we expect to put constraints on the parameters of the model which may arise from particle physics experiments and astrophysical observations. If such a scenario is viable, dark matter would have quite a different origin from what is currently believed. In fact, it would represent a geometric property of spacetime and an observable signature of the departure from the Riemannian case.

\

For all approaches considered in this thesis, our aim has been to work out the consequences of the underlying quantum nature of spacetime for macroscopic physics. In particular, we focused on the implications for cosmology. In fact, this is the arena in which quantum gravitational effects have a possibility of being observed for the first time, due to the rapid development of the field of precision cosmology. There is a possibility that a better understanding of Quantum Gravity may shed light on the main open questions in the standard $\Lambda$CDM cosmological model. It may, for instance, offer an explanation on the origin of dark matter and dark energy, and provide a justification of the inflationary paradigm from a more fundamental point of view.

The development of Quantum Gravity is primarily motivated by the need to address some long-standing problems in theoretical physics, such as the incompatibility of quantum mechanics and general relativity, and the resolution of spacetime singularities. A fundamental theory of Quantum Gravity has yet to be established, although there are many competing candidates. Progress in this field ultimately rests on the possibility of recovering General Relativity (and the Standard Model) in some limit, and predicting new phenomena beyond the regime of validity of such theories. Therefore, it is vital to make contact with macroscopic physics, particularly in those regimes where new phenomena are expected to take place, or when the classical description breaks down. The work presented in this thesis represents a first step in this direction, although more progress is needed to extract quantitative predictions from the models we considered.

Research in Quantum Gravity is multi-faceted and, as such, it does not proceed only in one direction. However, there is a hope that the combined efforts made in different directions may converge at some point, and will thus be beneficial for the field as a whole. Insights and even partial results obtained in one approach can motivate further progress in another one. Ultimately, progress in our understanding of the laws of physics at a fundamental level is driven by the need to explain experimental data. Thus, present day cosmology offers a unique opportunity to put Quantum Gravity theories to test and probe the properties of the fabric of spacetime at the smallest length scales.


\begin{spacing}{0.9}


\bibliographystyle{habbrv}
\bibliography{References/references} 

\begin{thebibliography}{100}

\bibitem{accardi2013quantum}
L.~Accardi, Y.~G. Lu, and I.~Volovich.
\newblock {\em Quantum theory and its stochastic limit}.
\newblock Springer, 2013.

\bibitem{Ade:2013zuv}
P.~A.~R. Ade et~al.
\newblock {Planck 2013 results. XVI. Cosmological parameters}.
\newblock {\em Astron. Astrophys.}, 571:A16, 2014, 1303.5076.

\bibitem{Ade:2015xua}
P.~A.~R. Ade et~al.
\newblock {Planck 2015 results. XIII. Cosmological parameters}.
\newblock {\em Astron. Astrophys.}, 594:A13, 2016, 1502.01589.

\bibitem{Albrecht:1998ir}
A.~Albrecht and J.~Magueijo.
\newblock {A Time varying speed of light as a solution to cosmological
  puzzles}.
\newblock {\em Phys. Rev.}, D59:043516, 1999, astro-ph/9811018.

\bibitem{Alesci:2012md}
E.~Alesci and F.~Cianfrani.
\newblock {A new perspective on cosmology in Loop Quantum Gravity}.
\newblock {\em Europhys. Lett.}, 104:10001, 2013, 1210.4504.

\bibitem{Alesci:2015nja}
E.~Alesci and F.~Cianfrani.
\newblock {Quantum reduced loop gravity: Universe on a lattice}.
\newblock {\em Phys. Rev.}, D92:084065, 2015, 1506.07835.

\bibitem{Alesci:2016rmn}
E.~Alesci and F.~Cianfrani.
\newblock {Improved regularization from Quantum Reduced Loop Gravity}.
\newblock 2016, 1604.02375.

\bibitem{Alesci:2016gub}
E.~Alesci and F.~Cianfrani.
\newblock {Quantum Reduced Loop Gravity and the foundation of Loop Quantum
  Cosmology}.
\newblock {\em Int. J. Mod. Phys.}, D25(08):1642005, 2016, 1602.05475.

\bibitem{Alesci:2013xya}
E.~Alesci, F.~Cianfrani, and C.~Rovelli.
\newblock {Quantum-Reduced Loop-Gravity: Relation with the Full Theory}.
\newblock {\em Phys. Rev.}, D88:104001, 2013, 1309.6304.

\bibitem{Alexandrino2015}
M.~M. Alexandrino and R.~G. Bettiol.
\newblock {\em Lie Groups with Bi-invariant Metrics}, pages 27--47.
\newblock Springer International Publishing, Cham, 2015.

\bibitem{Almeida:2013dba}
T.~S. Almeida, M.~L. Pucheu, C.~Romero, and J.~B. Formiga.
\newblock {From Brans-Dicke gravity to a geometrical scalar-tensor theory}.
\newblock {\em Phys. Rev.}, D89(6):064047, 2014, 1311.5459.

\bibitem{Amati:1988tn}
D.~Amati, M.~Ciafaloni, and G.~Veneziano.
\newblock {Can Space-Time Be Probed Below the String Size?}
\newblock {\em Phys. Lett.}, B216:41--47, 1989.

\bibitem{Ambjorn:1990ge}
J.~Ambjorn, B.~Durhuus, and T.~Jonsson.
\newblock {Three-dimensional simplicial quantum gravity and generalized matrix
  models}.
\newblock {\em Mod. Phys. Lett.}, A6:1133--1146, 1991.

\bibitem{Ambjorn:2011cg}
J.~Ambjorn, S.~Jordan, J.~Jurkiewicz, and R.~Loll.
\newblock {A Second-order phase transition in CDT}.
\newblock {\em Phys. Rev. Lett.}, 107:211303, 2011, 1108.3932.

\bibitem{Ambjorn:2004qm}
J.~Ambjorn, J.~Jurkiewicz, and R.~Loll.
\newblock {Emergence of a 4-D world from causal quantum gravity}.
\newblock {\em Phys. Rev. Lett.}, 93:131301, 2004, hep-th/0404156.

\bibitem{AmelinoCamelia:2002wr}
G.~Amelino-Camelia.
\newblock {Doubly special relativity}.
\newblock {\em Nature}, 418:34--35, 2002, gr-qc/0207049.

\bibitem{Anastopoulos:2004gk}
C.~Anastopoulos and N.~Savvidou.
\newblock {Minisuperspace models in histories theory}.
\newblock {\em Class. Quant. Grav.}, 22:1841--1866, 2005, gr-qc/0410131.

\bibitem{Arnowitt:1959lin}
R.~Arnowitt and S.~Deser.
\newblock Quantum theory of gravitation: General formulation and linearized
  theory.
\newblock {\em Phys. Rev.}, 113:745--750, Jan 1959.

\bibitem{Arnowitt:1959ah}
R.~L. Arnowitt, S.~Deser, and C.~W. Misner.
\newblock {Dynamical Structure and Definition of Energy in General Relativity}.
\newblock {\em Phys. Rev.}, 116:1322--1330, 1959.

\bibitem{Arnowitt:1962hi}
R.~L. Arnowitt, S.~Deser, and C.~W. Misner.
\newblock {The Dynamics of general relativity}.
\newblock {\em Gen. Rel. Grav.}, 40:1997--2027, 2008, gr-qc/0405109.

\bibitem{Artymowski:2013qua}
M.~Artymowski, Y.~Ma, and X.~Zhang.
\newblock {Comparison between Jordan and Einstein frames of Brans-Dicke gravity
  a la loop quantum cosmology}.
\newblock {\em Phys. Rev.}, D88(10):104010, 2013, 1309.3045.

\bibitem{Arvanitaki:2009fg}
A.~Arvanitaki, S.~Dimopoulos, S.~Dubovsky, N.~Kaloper, and J.~March-Russell.
\newblock {String Axiverse}.
\newblock {\em Phys. Rev.}, D81:123530, 2010, 0905.4720.

\bibitem{Ashtekar:1986yd}
A.~Ashtekar.
\newblock {New Variables for Classical and Quantum Gravity}.
\newblock {\em Phys. Rev. Lett.}, 57:2244--2247, 1986.

\bibitem{Ashtekar:2003hd}
A.~Ashtekar, M.~Bojowald, and J.~Lewandowski.
\newblock {Mathematical structure of loop quantum cosmology}.
\newblock {\em Adv. Theor. Math. Phys.}, 7(2):233--268, 2003, gr-qc/0304074.

\bibitem{Ashtekar:2007}
A.~Ashtekar, A.~Corichi, and P.~Singh.
\newblock {Robustness of key features of loop quantum cosmology}.
\newblock {\em Phys. Rev.}, D77:024046, 2008, 0710.3565.

\bibitem{Ashtekar:1996eg}
A.~Ashtekar and J.~Lewandowski.
\newblock {Quantum theory of geometry. 1: Area operators}.
\newblock {\em Class. Quant. Grav.}, 14:A55--A82, 1997, gr-qc/9602046.

\bibitem{Ashtekar:1997fb}
A.~Ashtekar and J.~Lewandowski.
\newblock {Quantum theory of geometry. 2. Volume operators}.
\newblock {\em Adv. Theor. Math. Phys.}, 1:388--429, 1998, gr-qc/9711031.

\bibitem{Ashtekar:2001xp}
A.~Ashtekar and J.~Lewandowski.
\newblock {Relation between polymer and Fock excitations}.
\newblock {\em Class. Quant. Grav.}, 18:L117--L128, 2001, gr-qc/0107043.

\bibitem{Ashtekar:2002vh}
A.~Ashtekar, J.~Lewandowski, and H.~Sahlmann.
\newblock {Polymer and Fock representations for a scalar field}.
\newblock {\em Class. Quant. Grav.}, 20:L11--1, 2003, gr-qc/0211012.

\bibitem{Ashtekar:2006rx}
A.~Ashtekar, T.~Pawlowski, and P.~Singh.
\newblock {Quantum nature of the big bang}.
\newblock {\em Phys. Rev. Lett.}, 96:141301, 2006, gr-qc/0602086.

\bibitem{Ashtekar:2006D73}
A.~Ashtekar, T.~Pawlowski, and P.~Singh.
\newblock {Quantum Nature of the Big Bang: An Analytical and Numerical
  Investigation. I.}
\newblock {\em Phys. Rev.}, D73:124038, 2006, gr-qc/0604013.

\bibitem{Ashtekar:2006improved}
A.~Ashtekar, T.~Pawlowski, and P.~Singh.
\newblock {Quantum Nature of the Big Bang: Improved dynamics}.
\newblock {\em Phys. Rev.}, D74:084003, 2006, gr-qc/0607039.

\bibitem{Ashtekar:2006}
A.~Ashtekar, T.~Pawlowski, P.~Singh, and K.~Vandersloot.
\newblock {Loop quantum cosmology of k=1 FRW models}.
\newblock {\em Phys.Rev.}, D75:024035, 2007, gr-qc/0612104.

\bibitem{Ashtekar:2011ni}
A.~Ashtekar and P.~Singh.
\newblock {Loop Quantum Cosmology: A Status Report}.
\newblock {\em Class. Quant. Grav.}, 28:213001, 2011, 1108.0893.

\bibitem{Ashtekar:2009vc}
A.~Ashtekar and E.~Wilson-Ewing.
\newblock {Loop quantum cosmology of Bianchi I models}.
\newblock {\em Phys. Rev.}, D79:083535, 2009, 0903.3397.

\bibitem{Ashtekar:2009um}
A.~Ashtekar and E.~Wilson-Ewing.
\newblock {Loop quantum cosmology of Bianchi type II models}.
\newblock {\em Phys. Rev.}, D80:123532, 2009, 0910.1278.

\bibitem{Banerjee:2011qu}
K.~Banerjee, G.~Calcagni, and M.~Martin-Benito.
\newblock {Introduction to loop quantum cosmology}.
\newblock {\em SIGMA}, 8:016, 2012, 1109.6801.

\bibitem{Baratin:2011tg}
A.~Baratin, F.~Girelli, and D.~Oriti.
\newblock {Diffeomorphisms in group field theories}.
\newblock {\em Phys. Rev.}, D83:104051, 2011, 1101.0590.

\bibitem{Baratin:2010wi}
A.~Baratin and D.~Oriti.
\newblock {Group field theory with non-commutative metric variables}.
\newblock {\em Phys. Rev. Lett.}, 105:221302, 2010, 1002.4723.

\bibitem{Baratin:2011tx}
A.~Baratin and D.~Oriti.
\newblock {Quantum simplicial geometry in the group field theory formalism:
  reconsidering the Barrett-Crane model}.
\newblock {\em New J. Phys.}, 13:125011, 2011, 1108.1178.

\bibitem{Baratin:2011hp}
A.~Baratin and D.~Oriti.
\newblock {Group field theory and simplicial gravity path integrals: A model
  for Holst-Plebanski gravity}.
\newblock {\em Phys. Rev.}, D85:044003, 2012, 1111.5842.

\bibitem{Baratin:2011aa}
A.~Baratin and D.~Oriti.
\newblock {Ten questions on Group Field Theory (and their tentative answers)}.
\newblock {\em J. Phys. Conf. Ser.}, 360:012002, 2012, 1112.3270.

\bibitem{Barbero:1994ap}
J.~F. Barbero~G.
\newblock {Real Ashtekar variables for Lorentzian signature space times}.
\newblock {\em Phys. Rev.}, D51:5507--5510, 1995, gr-qc/9410014.

\bibitem{FernandoBarbero:2010qy}
J.~F. Barbero~G. and E.~J.~S. Villasenor.
\newblock {Quantization of Midisuperspace Models}.
\newblock {\em Living Rev. Rel.}, 13:6, 2010, 1010.1637.

\bibitem{Barbieri:1997ks}
A.~Barbieri.
\newblock {Quantum tetrahedra and simplicial spin networks}.
\newblock {\em Nucl. Phys.}, B518:714--728, 1998, gr-qc/9707010.

\bibitem{Barrow:1999aa}
J.~D. Barrow.
\newblock Cosmologies with varying light speed.
\newblock {\em Physical Review D}, 59(4), 1999.

\bibitem{Barrow:1998df}
J.~D. Barrow and J.~Magueijo.
\newblock {Varying alpha theories and solutions to the cosmological problems}.
\newblock {\em Phys. Lett.}, B443:104--110, 1998, astro-ph/9811072.

\bibitem{Barrow:1988yia}
J.~D. Barrow and F.~J. Tipler.
\newblock {\em {The Anthropic Cosmological Principle}}.
\newblock Oxford University Press, 1988.

\bibitem{Bars:2015trh}
I.~Bars and A.~James.
\newblock {Physical Interpretation of Antigravity}.
\newblock {\em Phys. Rev.}, D93(4):044029, 2016, 1511.05128.

\bibitem{Bars:2013yba}
I.~Bars, P.~Steinhardt, and N.~Turok.
\newblock {Local Conformal Symmetry in Physics and Cosmology}.
\newblock {\em Phys. Rev.}, D89(4):043515, 2014, 1307.1848.

\bibitem{Barvinsky:1986bt}
A.~O. Barvinsky.
\newblock {Quantum Geometrodynamics: The Wheeler-de Witt Equations for the Wave
  Function of the Universe}.
\newblock {\em Phys. Lett.}, B175:401--404, 1986.

\bibitem{Barvinsky:1985jt}
A.~O. Barvinsky.
\newblock {Operator ordering in theories subject to constraints of the
  gravitational type}.
\newblock {\em Class. Quant. Grav.}, 10:1985--1999, 1993.

\bibitem{Barvinsky:1986qn}
A.~O. Barvinsky and V.~N. Ponomarev.
\newblock {Quantum Geometrodynamics: The Path Integral and the Initial Value
  Problem for the Wave Function of the Universe}.
\newblock {\em Phys. Lett.}, 167B:289--294, 1986.

\bibitem{Batalin:1977pb}
I.~A. Batalin and G.~A. Vilkovisky.
\newblock {Relativistic S Matrix of Dynamical Systems with Boson and Fermion
  Constraints}.
\newblock {\em Phys. Lett.}, 69B:309--312, 1977.

\bibitem{Battefeld:2014uga}
D.~Battefeld and P.~Peter.
\newblock {A Critical Review of Classical Bouncing Cosmologies}.
\newblock {\em Phys. Rept.}, 571:1--66, 2015, 1406.2790.

\bibitem{Baumann:2009ds}
D.~Baumann.
\newblock {Inflation}.
\newblock In {\em {Physics of the large and the small, TASI 09, proceedings of
  the Theoretical Advanced Study Institute in Elementary Particle Physics,
  Boulder, Colorado, USA, 1-26 June 2009}}, pages 523--686, 2011, 0907.5424.

\bibitem{Becchi:1975nq}
C.~Becchi, A.~Rouet, and R.~Stora.
\newblock {Renormalization of Gauge Theories}.
\newblock {\em Annals Phys.}, 98:287--321, 1976.

\bibitem{becker2006string}
K.~Becker, M.~Becker, and J.~H. Schwarz.
\newblock {\em String theory and M-theory: A modern introduction}.
\newblock Cambridge University Press, 2006.

\bibitem{Belinsky:1970ew}
V.~A. Belinsky, I.~M. Khalatnikov, and E.~M. Lifshitz.
\newblock {Oscillatory approach to a singular point in the relativistic
  cosmology}.
\newblock {\em Adv. Phys.}, 19:525--573, 1970.

\bibitem{Geloun:2015qfa}
J.~Ben~Geloun, R.~Martini, and D.~Oriti.
\newblock {Functional Renormalization Group analysis of a Tensorial Group Field
  Theory on $\mathbb{R}^3$}.
\newblock {\em Europhys. Lett.}, 112(3):31001, 2015, 1508.01855.

\bibitem{Geloun:2016qyb}
J.~Ben~Geloun, R.~Martini, and D.~Oriti.
\newblock {Functional Renormalisation Group analysis of Tensorial Group Field
  Theories on $\mathbb{R}^d$}.
\newblock {\em Phys. Rev.}, D94(2):024017, 2016, 1601.08211.

\bibitem{Benedetti:2014qsa}
D.~Benedetti, J.~Ben~Geloun, and D.~Oriti.
\newblock {Functional Renormalisation Group Approach for Tensorial Group Field
  Theory: a Rank-3 Model}.
\newblock {\em JHEP}, 03:084, 2015, 1411.3180.

\bibitem{Benedetti:2015yaa}
D.~Benedetti and V.~Lahoche.
\newblock {Functional Renormalization Group Approach for Tensorial Group Field
  Theory: A Rank-6 Model with Closure Constraint}.
\newblock {\em Class. Quant. Grav.}, 33(9):095003, 2016, 1508.06384.

\bibitem{Bergmann:1961zz}
P.~G. Bergmann.
\newblock {`Gauge-Invariant' Variables in General Relativity}.
\newblock {\em Phys. Rev.}, 124:274--278, 1961.

\bibitem{bergmann1972coordinate}
P.~G. Bergmann and A.~Komar.
\newblock The coordinate group symmetries of general relativity.
\newblock {\em International Journal of Theoretical Physics}, 5(1):15--28,
  1972.

\bibitem{Bergmann:1981fc}
P.~G. Bergmann and A.~Komar.
\newblock {The phase space formulation of General Relativity and approaches
  toward its canonical quantization}.
\newblock In A.~Held, editor, {\em General Relativity and Gravitation, vol. 1,
  One Hundred Years after the Birth of Albert Einstein}. Plenum, 1981.

\bibitem{Bertone:2004pz}
G.~Bertone, D.~Hooper, and J.~Silk.
\newblock {Particle dark matter: Evidence, candidates and constraints}.
\newblock {\em Phys. Rept.}, 405:279--390, 2005, hep-ph/0404175.

\bibitem{Bertotti:2003rm}
B.~Bertotti, L.~Iess, and P.~Tortora.
\newblock {A test of general relativity using radio links with the Cassini
  spacecraft}.
\newblock {\em Nature}, 425:374--376, 2003.

\bibitem{Bianchi:2010gc}
E.~Bianchi, P.~Dona, and S.~Speziale.
\newblock {Polyhedra in loop quantum gravity}.
\newblock {\em Phys. Rev.}, D83:044035, 2011, 1009.3402.

\bibitem{Bini:2013fea}
D.~Bini, G.~Esposito, C.~Kiefer, M.~Kraemer, and F.~Pessina.
\newblock {On the modification of the cosmic microwave background anisotropy
  spectrum from canonical quantum gravity}.
\newblock {\em Phys. Rev.}, D87(10):104008, 2013, 1303.0531.

\bibitem{Bojowald:2001xe}
M.~Bojowald.
\newblock {Absence of singularity in loop quantum cosmology}.
\newblock {\em Phys. Rev. Lett.}, 86:5227--5230, 2001, gr-qc/0102069.

\bibitem{Bojowald:2002nz}
M.~Bojowald.
\newblock {Inflation from quantum geometry}.
\newblock {\em Phys. Rev. Lett.}, 89:261301, 2002, gr-qc/0206054.

\bibitem{Bojowald2006}
M.~Bojowald.
\newblock Loop quantum cosmology and inhomogeneities.
\newblock {\em General Relativity and Gravitation}, 38(12):1771--1795, Dec
  2006.

\bibitem{Bojowald:2008zzb}
M.~Bojowald.
\newblock {Loop quantum cosmology}.
\newblock {\em Living Rev. Rel.}, 11:4, 2008.

\bibitem{Bojowald2011}
M.~Bojowald.
\newblock {\em General Aspects of Effective Descriptions}, pages 275--299.
\newblock Springer New York, New York, NY, 2011.

\bibitem{Bojowald:2012xy}
M.~Bojowald.
\newblock {Quantum Cosmology: Effective Theory}.
\newblock {\em Class. Quant. Grav.}, 29:213001, 2012, 1209.3403.

\bibitem{Bojowald:2012we}
M.~Bojowald.
\newblock {Mathematical structure of loop quantum cosmology: Homogeneous
  models}.
\newblock {\em SIGMA}, 9:082, 2013, 1206.6088.

\bibitem{Bojowald:2007ra}
M.~Bojowald, D.~Cartin, and G.~Khanna.
\newblock {Lattice refining loop quantum cosmology, anisotropic models and
  stability}.
\newblock {\em Phys. Rev.}, D76:064018, 2007, 0704.1137.

\bibitem{Bojowald:2002}
M.~Bojowald and F.~Hinterleitner.
\newblock {Isotropic loop quantum cosmology with matter}.
\newblock {\em Phys. Rev.}, D66:104003, 2002, gr-qc/0207038.

\bibitem{Bojowald:2010xp}
M.~Bojowald, P.~A. Hoehn, and A.~Tsobanjan.
\newblock {An Effective approach to the problem of time}.
\newblock {\em Class. Quant. Grav.}, 28:035006, 2011, 1009.5953.

\bibitem{Bojowald:2003}
M.~Bojowald and K.~Vandersloot.
\newblock {Loop quantum cosmology and boundary proposals}.
\newblock In {\em {On recent developments in theoretical and experimental
  general relativity, gravitation, and relativistic field theories.
  Proceedings, 10th Marcel Grossmann Meeting, MG10, Rio de Janeiro, Brazil,
  July 20-26, 2003. Pt. A-C}}, pages 1089--1103, 2003, gr-qc/0312103.

\bibitem{Bojowald:2003mc}
M.~Bojowald and K.~Vandersloot.
\newblock {Loop quantum cosmology, boundary proposals, and inflation}.
\newblock {\em Phys. Rev.}, D67:124023, 2003, gr-qc/0303072.

\bibitem{Bombelli:1987aa}
L.~Bombelli, J.~Lee, D.~Meyer, and R.~Sorkin.
\newblock {Space-Time as a Causal Set}.
\newblock {\em Phys. Rev. Lett.}, 59:521--524, 1987.

\bibitem{Bondi:1960}
H.~Bondi.
\newblock {\em Cosmology}.
\newblock Cambridge University Press, 1960.

\bibitem{bostrom2013anthropic}
N.~Bostrom et~al.
\newblock Anthropic bias: Observation selection effects in science and
  philosophy, 2013.

\bibitem{Boulatov:1992vp}
D.~V. Boulatov.
\newblock {A Model of three-dimensional lattice gravity}.
\newblock {\em Mod. Phys. Lett.}, A7:1629--1646, 1992, hep-th/9202074.

\bibitem{Bousso:2002ju}
R.~Bousso.
\newblock {The Holographic principle}.
\newblock {\em Rev. Mod. Phys.}, 74:825--874, 2002, hep-th/0203101.

\bibitem{Brandenberger:2016vhg}
R.~Brandenberger and P.~Peter.
\newblock {Bouncing Cosmologies: Progress and Problems}.
\newblock {\em Found. Phys.}, 47(6):797--850, 2017, 1603.05834.

\bibitem{Brandenberger2000}
R.~H. Brandenberger.
\newblock Inflationary cosmology: Progress and problems.
\newblock In R.~Mansouri and R.~Brandenberger, editors, {\em Large Scale
  Structure Formation}, pages 169--211, Dordrecht, 2000. Springer Netherlands.

\bibitem{Brandenberger:2003vk}
R.~H. Brandenberger.
\newblock {Lectures on the theory of cosmological perturbations}.
\newblock {\em Lect. Notes Phys.}, 646:127--167, 2004, hep-th/0306071.
\newblock [,127(2003)].

\bibitem{Brandenberger:2008nx}
R.~H. Brandenberger.
\newblock {String Gas Cosmology}.
\newblock In {\em {String Cosmology, J.Erdmenger (Editor). Wiley, 2009.
  p.193-230}}, pages 193--230, 2008, 0808.0746.

\bibitem{Brans:1961sx}
C.~Brans and R.~H. Dicke.
\newblock {Mach's principle and a relativistic theory of gravitation}.
\newblock {\em Phys. Rev.}, 124:925--935, 1961.

\bibitem{Brizuela:2015tzl}
D.~Brizuela, C.~Kiefer, and M.~Kr{\"a}mer.
\newblock {Quantum-gravitational effects on gauge-invariant scalar and tensor
  perturbations during inflation: The de Sitter case}.
\newblock {\em Phys. Rev.}, D93(10):104035, 2016, 1511.05545.

\bibitem{Brizuela:2016gnz}
D.~Brizuela, C.~Kiefer, and M.~Kr{\"a}mer.
\newblock {Quantum-gravitational effects on gauge-invariant scalar and tensor
  perturbations during inflation: The slow-roll approximation}.
\newblock {\em Phys. Rev.}, D94(12):123527, 2016, 1611.02932.

\bibitem{Brown:1992kc}
J.~D. Brown.
\newblock {Action functionals for relativistic perfect fluids}.
\newblock {\em Class. Quant. Grav.}, 10:1579--1606, 1993, gr-qc/9304026.

\bibitem{Brunnemann:2004xi}
J.~Brunnemann and T.~Thiemann.
\newblock {Simplification of the spectral analysis of the volume operator in
  loop quantum gravity}.
\newblock {\em Class. Quant. Grav.}, 23:1289--1346, 2006, gr-qc/0405060.

\bibitem{Buchmuller:1990pz}
W.~Buchmuller and C.~Busch.
\newblock {Symmetry breaking and mass bounds in the standard model with hidden
  scale invariance}.
\newblock {\em Nucl. Phys.}, B349:71--90, 1991.

\bibitem{Buchmuller:1988cj}
W.~Buchmuller and N.~Dragon.
\newblock {Dilatons in Flat and Curved Space-time}.
\newblock {\em Nucl. Phys.}, B321:207--231, 1989.

\bibitem{Butterfield:1998dd}
J.~Butterfield and C.~J. Isham.
\newblock {On the emergence of time in quantum gravity}.
\newblock In J.~Butterfield, editor, {\em The Arguments of Time}. Oxford
  University Press, 1999, gr-qc/9901024.

\bibitem{Calcagni:2014tga}
G.~Calcagni.
\newblock {Loop quantum cosmology from group field theory}.
\newblock {\em Phys. Rev.}, D90(6):064047, 2014, 1407.8166.

\bibitem{Calcagni:2012vb}
G.~Calcagni, S.~Gielen, and D.~Oriti.
\newblock {Group field cosmology: a cosmological field theory of quantum
  geometry}.
\newblock {\em Class. Quant. Grav.}, 29:105005, 2012, 1201.4151.

\bibitem{Calcagni:2012}
G.~Calcagni, S.~Gielen, and D.~Oriti.
\newblock {Group field cosmology: a cosmological field theory of quantum
  geometry}.
\newblock {\em Class.Quant.Grav.}, 29:105005, 2012, 1201.4151.

\bibitem{calderbank1997einstein}
D.~M. Calderbank and H.~Pederson.
\newblock {\em Einstein-Weyl geometry}, page 387.
\newblock Odense Universitet. Institut for Matematik og Datalogi, 1997.

\bibitem{Calzetta:1992gv}
E.~Calzetta and M.~Sakellariadou.
\newblock {Inflation in inhomogeneous cosmology}.
\newblock {\em Phys. Rev.}, D45:2802--2805, 1992.

\bibitem{Calzetta:1992bp}
E.~Calzetta and M.~Sakellariadou.
\newblock {Semiclassical effects and the onset of inflation}.
\newblock {\em Phys. Rev.}, D47:3184--3193, 1993, gr-qc/9209007.

\bibitem{Capozziello:2011et}
S.~Capozziello and M.~De~Laurentis.
\newblock {Extended Theories of Gravity}.
\newblock {\em Phys. Rept.}, 509:167--321, 2011, 1108.6266.

\bibitem{Capozziello:1996xg}
S.~Capozziello, R.~de~Ritis, and A.~A. Marino.
\newblock {Some aspects of the cosmological conformal equivalence between
  `Jordan frame' and `Einstein frame'}.
\newblock {\em Class. Quant. Grav.}, 14:3243--3258, 1997, gr-qc/9612053.

\bibitem{Capozziello:2010sc}
S.~Capozziello, P.~Martin-Moruno, and C.~Rubano.
\newblock {Physical non-equivalence of the Jordan and Einstein frames}.
\newblock {\em Phys. Lett.}, B689:117--121, 2010, 1003.5394.

\bibitem{Carrozza:2016vsq}
S.~Carrozza.
\newblock {Flowing in Group Field Theory Space: a Review}.
\newblock {\em SIGMA}, 12:070, 2016, 1603.01902.

\bibitem{Carrozza:2013wda}
S.~Carrozza, D.~Oriti, and V.~Rivasseau.
\newblock {Renormalization of a SU(2) Tensorial Group Field Theory in Three
  Dimensions}.
\newblock {\em Commun. Math. Phys.}, 330:581--637, 2014, 1303.6772.

\bibitem{Cartin:2005an}
D.~Cartin and G.~Khanna.
\newblock {Absence of pre-classical solutions in Bianchi I loop quantum
  cosmology}.
\newblock {\em Phys. Rev. Lett.}, 94:111302, 2005, gr-qc/0501016.

\bibitem{Lamb-shift-spontaneous-emission-SED}
A.~M. Cetto and L.~de~la Pe{\~n}a.
\newblock Environmental effects on spontaneous emission and lamb shift,
  according to stochastic electrodynamics.
\newblock {\em Physica Scripta}, 1988(T21):27, 1988.

\bibitem{Lamb-shift-SED}
A.~M. Cetto and L.~de~la Pe{\~n}a.
\newblock Environmental effects on the lamb shift according to stochastic
  electrodynamics.
\newblock {\em Phys. Rev. A}, 37:1952--1959, Mar 1988.

\bibitem{QM-SED}
A.~M. Cetto and L.~de~la Pe{\~n}a‐Auerbach.
\newblock Derivation of quantum mechanics from stochastic electrodynamics.
\newblock {\em Journal of Mathematical Physics}, 18(8):1612--1622, 1977.

\bibitem{Charmousis:2009tc}
C.~Charmousis, G.~Niz, A.~Padilla, and P.~M. Saffin.
\newblock {Strong coupling in Horava gravity}.
\newblock {\em JHEP}, 08:070, 2009, 0905.2579.

\bibitem{Cheng:1988zx}
H.~Cheng.
\newblock {The Possible Existence of Weyl's Vector Meson}.
\newblock {\em Phys. Rev. Lett.}, 61:2182, 1988.

\bibitem{Clayton:1998hv}
M.~A. Clayton and J.~W. Moffat.
\newblock {Dynamical mechanism for varying light velocity as a solution to
  cosmological problems}.
\newblock {\em Phys. Lett.}, B460:263--270, 1999, astro-ph/9812481.

\bibitem{Clayton:2001rt}
M.~A. Clayton and J.~W. Moffat.
\newblock {A Scalar - tensor cosmological model with dynamical light velocity}.
\newblock {\em Phys. Lett.}, B506:177--186, 2001, gr-qc/0101126.

\bibitem{Clayton:2000xt}
M.~A. Clayton and J.~W. Moffat.
\newblock {Vector field mediated models of dynamical light velocity}.
\newblock {\em Int. J. Mod. Phys.}, D11:187--206, 2002, gr-qc/0003070.

\bibitem{Coleman:1988cy}
S.~R. Coleman.
\newblock {Black Holes as Red Herrings: Topological Fluctuations and the Loss
  of Quantum Coherence}.
\newblock {\em Nucl. Phys.}, B307:867--882, 1988.

\bibitem{Coleman:1988tj}
S.~R. Coleman.
\newblock {Why There Is Nothing Rather Than Something: A Theory of the
  Cosmological Constant}.
\newblock {\em Nucl. Phys.}, B310:643--668, 1988.

\bibitem{Coleman:1973jx}
S.~R. Coleman and E.~J. Weinberg.
\newblock {Radiative Corrections as the Origin of Spontaneous Symmetry
  Breaking}.
\newblock {\em Phys. Rev.}, D7:1888--1910, 1973.

\bibitem{connes1985non}
A.~Connes.
\newblock Non-commutative differential geometry.
\newblock {\em Publications math{\'e}matiques de l'IH{\'E}S}, 62(1):41--144,
  1985.

\bibitem{Connes:2017oxm}
A.~Connes.
\newblock {Geometry and the Quantum}.
\newblock In J.~Kouneiher, editor, {\em Foundations of Mathematics and Physics
  one century after Hilbert}. Springer, 2017, 1703.02470.

\bibitem{Coriano:2012nm}
C.~Coriano, L.~Delle~Rose, A.~Quintavalle, and M.~Serino.
\newblock {Dilaton interactions and the anomalous breaking of scale invariance
  of the Standard Model}.
\newblock {\em JHEP}, 06:077, 2013, 1206.0590.

\bibitem{Corichi:2007tf}
A.~Corichi, T.~Vukasinac, and J.~A. Zapata.
\newblock {Polymer Quantum Mechanics and its Continuum Limit}.
\newblock {\em Phys. Rev.}, D76:044016, 2007, 0704.0007.

\bibitem{Corichi:2007}
A.~Corichi, T.~Vukasinac, and J.~A. Zapata.
\newblock {On a continuum limit for loop quantum cosmology}.
\newblock {\em AIP Conf. Proc.}, 977:64--71, 2008, 0711.0788.

\bibitem{Costa:2017abc}
R.~Costa, R.~R. Cuzinatto, E.~M.~G. Ferreira, and G.~Franzmann.
\newblock {Covariant c-flation}.
\newblock 2017, 1705.03461.

\bibitem{Craig:2010ai}
D.~Craig and P.~Singh.
\newblock {Consistent Histories in Quantum Cosmology}.
\newblock {\em Found. Phys.}, 41:371--379, 2011, 1001.4311.

\bibitem{Damour:1988zz}
T.~Damour, G.~W. Gibbons, and J.~H. Taylor.
\newblock {Limits on the Variability of G Using Binary-Pulsar Data}.
\newblock {\em Phys. Rev. Lett.}, 61:1151--1154, 1988.

\bibitem{Damour:1990wz}
T.~Damour and J.~H. Taylor.
\newblock {On the orbital period change of the binary pulsar PSR-1913+16}.
\newblock {\em Astrophys. J.}, 366:501--511, 1991.

\bibitem{Date:2005ik}
G.~Date.
\newblock {Pre-classical solutions of the vacuum Bianchi I loop quantum
  cosmology}.
\newblock {\em Phys. Rev.}, D72:067301, 2005, gr-qc/0505030.

\bibitem{deCarlos:1993wie}
B.~de~Carlos, J.~A. Casas, F.~Quevedo, and E.~Roulet.
\newblock {Model independent properties and cosmological implications of the
  dilaton and moduli sectors of 4-d strings}.
\newblock {\em Phys. Lett.}, B318:447--456, 1993, hep-ph/9308325.

\bibitem{deCesare:2015vca}
M.~de~Cesare, M.~V. Gargiulo, and M.~Sakellariadou.
\newblock {Semiclassical solutions of generalized Wheeler-DeWitt cosmology}.
\newblock {\em Phys. Rev.}, D93(2):024046, 2016, 1509.05728.

\bibitem{deCesare:2016dnp}
M.~de~Cesare, F.~Lizzi, and M.~Sakellariadou.
\newblock {Effective cosmological constant induced by stochastic fluctuations
  of Newton's constant}.
\newblock {\em Phys. Lett.}, B760:498--501, 2016, 1603.04170.

\bibitem{deCesare:2014dga}
M.~de~Cesare, N.~E. Mavromatos, and S.~Sarkar.
\newblock {On the possibility of tree-level leptogenesis from Kalb--Ramond
  torsion background}.
\newblock {\em Eur. Phys. J.}, C75(10):514, 2015, 1412.7077.

\bibitem{deCesare:2016mml}
M.~de~Cesare, J.~W. Moffat, and M.~Sakellariadou.
\newblock {Local conformal symmetry in non-Riemannian geometry and the origin
  of physical scales}.
\newblock {\em Eur. Phys. J.}, C77(9):605, 2017, 1612.08066.

\bibitem{deCesare:2017xhv}
M.~de~Cesare, R.~Oliveri, and J.~W. van Holten.
\newblock {Field theoretical approach to gravitational waves}.
\newblock {\em Fortsch. Phys.}, 65(5):1700012, 2017, 1701.07794.

\bibitem{deCesare:2017ynn}
M.~de~Cesare, D.~Oriti, A.~G.~A. Pithis, and M.~Sakellariadou.
\newblock {Dynamics of anisotropies close to a cosmological bounce in quantum
  gravity}.
\newblock {\em Class. Quant. Grav.}, 35(1):015014, 2018, 1709.00994.

\bibitem{deCesare:2016rsf}
M.~de~Cesare, A.~G.~A. Pithis, and M.~Sakellariadou.
\newblock {Cosmological implications of interacting Group Field Theory models:
  cyclic Universe and accelerated expansion}.
\newblock {\em Phys. Rev.}, D94(6):064051, 2016, 1606.00352.

\bibitem{deCesare:2016axk}
M.~de~Cesare and M.~Sakellariadou.
\newblock {Accelerated expansion of the Universe without an inflaton and
  resolution of the initial singularity from Group Field Theory condensates}.
\newblock {\em Phys. Lett.}, B764:49--53, 2017, 1603.01764.

\bibitem{de1983stochastic}
L.~de~la Pe{\~n}a.
\newblock Stochastic electrodynamics: Its development, present situation, and
  perspectives.
\newblock {\em Stochastic Processes Applied to Physics and other Related
  Fields}, pages 428--581, 1983.

\bibitem{DePietri:2000ke}
R.~De~Pietri.
\newblock {Matrix model formulation of four-dimensional gravity}.
\newblock {\em Nucl. Phys. Proc. Suppl.}, 94:697--700, 2001, hep-lat/0011033.
\newblock [,697(2000)].

\bibitem{DePietri:1999bx}
R.~De~Pietri, L.~Freidel, K.~Krasnov, and C.~Rovelli.
\newblock {Barrett-Crane model from a Boulatov-Ooguri field theory over a
  homogeneous space}.
\newblock {\em Nucl. Phys.}, B574:785--806, 2000, hep-th/9907154.

\bibitem{DePietri:2000ii}
R.~De~Pietri and C.~Petronio.
\newblock {Feynman diagrams of generalized matrix models and the associated
  manifolds in dimension 4}.
\newblock {\em J. Math. Phys.}, 41:6671--6688, 2000, gr-qc/0004045.

\bibitem{Deffayet:2011gz}
C.~Deffayet, X.~Gao, D.~A. Steer, and G.~Zahariade.
\newblock {From k-essence to generalised Galileons}.
\newblock {\em Phys. Rev.}, D84:064039, 2011, 1103.3260.

\bibitem{Deser:1970hs}
S.~Deser.
\newblock {Scale invariance and gravitational coupling}.
\newblock {\em Annals Phys.}, 59:248--253, 1970.

\bibitem{DeWitt:1962cg}
B.~S. DeWitt.
\newblock {The Quantization of geometry}.
\newblock In L.~Witten, editor, {\em Gravitation: An introduction to current
  research}, pages 266--381. Wiley, 1962.

\bibitem{DeWitt:1967yk}
B.~S. DeWitt.
\newblock {Quantum Theory of Gravity. 1. The Canonical Theory}.
\newblock {\em Phys. Rev.}, 160:1113--1148, 1967.

\bibitem{DiFrancesco:1993cyw}
P.~Di~Francesco, P.~H. Ginsparg, and J.~Zinn-Justin.
\newblock {2-D Gravity and random matrices}.
\newblock {\em Phys. Rept.}, 254:1--133, 1995, hep-th/9306153.

\bibitem{disessa}
A.~Di~Sessa.
\newblock {Quantization of a field with mass in two-dimensional Euclidean
  space}.
\newblock {\em Phys. Rev.}, D9:2926--2932, 1974.

\bibitem{DiazDorronsoro:2017hti}
J.~Diaz~Dorronsoro, J.~J. Halliwell, J.~B. Hartle, T.~Hertog, and O.~Janssen.
\newblock {The Real No-Boundary Wave Function in Lorentzian Quantum Cosmology}.
\newblock {\em Phys. Rev.}, D96(4):043505, 2017, 1705.05340.

\bibitem{Dicke:1979}
R.~Dicke and P.~Peebles.
\newblock The big bang cosmology - enigmas and nostrums.
\newblock In S.~Hawking and W.~Israel, editors, {\em General Relativity -- an
  Einstein Centenary Survey}. Cambridge University Press, 1979.

\bibitem{DICKE:1961aa}
R.~H. Dicke.
\newblock {Dirac's Cosmology and Mach's Principle}.
\newblock {\em Nature}, 192(4801):440--441, 11 1961.

\bibitem{Dicke:1964pna}
R.~H. Dicke.
\newblock {Experimental relativity}.
\newblock In {\em {Relativit{\'e}, Groupes et Topologie: Proceedings, Ecole
  d'{\'e}t{\'e} de Physique Th{\'e}orique, Session XIII, Les Houches, France,
  Jul 1 - Aug 24, 1963}}, pages 165--316, 1964.

\bibitem{Dirac:1974aa}
P.~Dirac.
\newblock Cosmological models and the large numbers hypothesis.
\newblock {\em Proceedings of the Royal Society of London A: Mathematical,
  Physical and Engineering Sciences}, 338(1615):439--446, 1974.

\bibitem{dirac1964lectures}
P.~A. Dirac.
\newblock {Lectures on quantum mechanics, Belfer graduate school of Science,
  Yeshiva University}.
\newblock {\em New York}, 1964.

\bibitem{Dirac:1937ti}
P.~A.~M. Dirac.
\newblock {The Cosmological constants}.
\newblock {\em Nature}, 139:323, 1937.

\bibitem{Dirac:1938mt}
P.~A.~M. Dirac.
\newblock {New basis for cosmology}.
\newblock {\em Proc. Roy. Soc. Lond.}, A165:199--208, 1938.

\bibitem{Dittrich:2013xwa}
B.~Dittrich and S.~Steinhaus.
\newblock {Time evolution as refining, coarse graining and entangling}.
\newblock {\em New J. Phys.}, 16:123041, 2014, 1311.7565.

\bibitem{Dona:2010hm}
P.~Dona and S.~Speziale.
\newblock {Introductory lectures to loop quantum gravity}.
\newblock In {\em {Gravitation Th{\'e}orie et Exp{\'e}rience.Proceedings,
  Troisi{\`e}me {\'e}cole de physique th{\'e}orique de Jijel: Jijel, Algeria,
  September 26--October 03, 2009}}, pages 89--140, 2013, 1007.0402.

\bibitem{Donnelly:2016auv}
W.~Donnelly and L.~Freidel.
\newblock {Local subsystems in gauge theory and gravity}.
\newblock {\em JHEP}, 09:102, 2016, 1601.04744.

\bibitem{Drummond:2001rj}
I.~T. Drummond.
\newblock {Bimetric gravity and ``dark matter''}.
\newblock {\em Phys. Rev.}, D63:043503, 2001, astro-ph/0008234.

\bibitem{Duff:2001ba}
M.~J. Duff, L.~B. Okun, and G.~Veneziano.
\newblock {Trialogue on the number of fundamental constants}.
\newblock {\em JHEP}, 03:023, 2002, physics/0110060.

\bibitem{Duruisseau:1986ga}
J.~P. Duruisseau and R.~Kerner.
\newblock {The Effective Gravitational Lagrangian and the Energy Momentum
  Tensor in the Inflationary Universe}.
\newblock {\em Class. Quant. Grav.}, 3:817--824, 1986.

\bibitem{ehlers2012republication}
J.~Ehlers, F.~A. Pirani, and A.~Schild.
\newblock Republication of: The geometry of free fall and light propagation.
\newblock {\em General Relativity and Gravitation}, 44(6):1587--1609, 2012.

\bibitem{Ehlers2012}
J.~Ehlers, F.~A.~E. Pirani, and A.~Schild.
\newblock Republication of: The geometry of free fall and light propagation.
\newblock {\em General Relativity and Gravitation}, 44(6):1587--1609, 2012.

\bibitem{Eichhorn:2017egq}
A.~Eichhorn.
\newblock {Status of the asymptotic safety paradigm for quantum gravity and
  matter}.
\newblock In {\em Workshop on Black Holes, Gravitational Waves and Spacetime
  Singularities, Specola Vaticana}, 2017, 1709.03696.

\bibitem{Ellis:2003pw}
G.~F.~R. Ellis and J.~P. Uzan.
\newblock {`c' is the speed of light, isn't it?}
\newblock {\em Am. J. Phys.}, 73:240--247, 2005, gr-qc/0305099.

\bibitem{Engle:2007wy}
J.~Engle, E.~Livine, R.~Pereira, and C.~Rovelli.
\newblock {LQG vertex with finite Immirzi parameter}.
\newblock {\em Nucl. Phys.}, B799:136--149, 2008, 0711.0146.

\bibitem{ermakov1880transformation}
V.~Ermakov.
\newblock Transformation of differential equations.
\newblock {\em Univ. Izv. Kiev}, 20(1), 1880.

\bibitem{Esposito:1992xz}
G.~Esposito.
\newblock {Quantum gravity, quantum cosmology and Lorentzian geometries}.
\newblock {\em Lect. Notes Phys. Monogr.}, 12:1--326, 1992.

\bibitem{Faraoni:2009aa}
V.~Faraoni.
\newblock The lagrangian description of perfect fluids and modified gravity
  with an extra force.
\newblock {\em Phys. Rev.}, D80:124040, 12 2009, 0912.1249.

\bibitem{Faraoni:1999hp}
V.~Faraoni and E.~Gunzig.
\newblock {Einstein frame or Jordan frame?}
\newblock {\em Int. J. Theor. Phys.}, 38:217--225, 1999, astro-ph/9910176.

\bibitem{faraut2008analysis}
J.~Faraut.
\newblock {\em Analysis on Lie groups: an introduction}, volume 110.
\newblock Cambridge University Press, 2008.

\bibitem{Feldbrugge:2017kzv}
J.~Feldbrugge, J.-L. Lehners, and N.~Turok.
\newblock {Lorentzian Quantum Cosmology}.
\newblock {\em Phys. Rev.}, D95(10):103508, 2017, 1703.02076.

\bibitem{Feldbrugge:2017fcc}
J.~Feldbrugge, J.-L. Lehners, and N.~Turok.
\newblock {No smooth beginning for spacetime}.
\newblock {\em Phys. Rev. Lett.}, 119(17):171301, 2017, 1705.00192.

\bibitem{Fernandez-Mendez:2014}
M.~Fern\'{a}ndez-M\'{e}ndez, G.~A. Mena~Marug\'{a}n, and J.~Olmedo.
\newblock {Effective dynamics of scalar perturbations in a flat
  Friedmann-Robertson-Walker spacetime in Loop Quantum Cosmology}.
\newblock {\em Phys. Rev.}, D89(4):044041, 2014, 1401.5256.

\bibitem{Feynman:1948ur}
R.~P. Feynman.
\newblock {Space-time approach to nonrelativistic quantum mechanics}.
\newblock {\em Rev. Mod. Phys.}, 20:367--387, 1948.

\bibitem{Finocchiaro}
M.~Finocchiaro and D.~Oriti.
\newblock (to appear).
\newblock 2017.

\bibitem{folland1970weyl}
G.~B. Folland et~al.
\newblock Weyl manifolds.
\newblock {\em Journal of differential geometry}, 4(2):145--153, 1970.

\bibitem{Fradkin:1975cq}
E.~S. Fradkin and G.~A. Vilkovisky.
\newblock {Quantization of relativistic systems with constraints}.
\newblock {\em Phys. Lett.}, 55B:224--226, 1975.

\bibitem{Franzmann:2017nsc}
G.~Franzmann.
\newblock {Varying fundamental constants: a full covariant approach and
  cosmological applications}.
\newblock 2017, 1704.07368.

\bibitem{Freidel:2005qe}
L.~Freidel.
\newblock {Group field theory: An Overview}.
\newblock {\em Int. J. Theor. Phys.}, 44:1769--1783, 2005, hep-th/0505016.

\bibitem{Freidel:2009hd}
L.~Freidel, R.~Gurau, and D.~Oriti.
\newblock {Group field theory renormalization - the 3d case: Power counting of
  divergences}.
\newblock {\em Phys. Rev.}, D80:044007, 2009, 0905.3772.

\bibitem{Freidel:2007py}
L.~Freidel and K.~Krasnov.
\newblock {A New Spin Foam Model for 4d Gravity}.
\newblock {\em Class. Quant. Grav.}, 25:125018, 2008, 0708.1595.

\bibitem{Freidel:2002tg}
L.~Freidel and D.~Louapre.
\newblock {Nonperturbative summation over 3-D discrete topologies}.
\newblock {\em Phys. Rev.}, D68:104004, 2003, hep-th/0211026.

\bibitem{Freire:2012mg}
P.~C.~C. Freire, N.~Wex, G.~Esposito-Farese, J.~P.~W. Verbiest, M.~Bailes,
  B.~A. Jacoby, M.~Kramer, I.~H. Stairs, J.~Antoniadis, and G.~H. Janssen.
\newblock {The relativistic pulsar-white dwarf binary PSR J1738+0333 II. The
  most stringent test of scalar-tensor gravity}.
\newblock {\em Mon. Not. Roy. Astron. Soc.}, 423:3328, 2012, 1205.1450.

\bibitem{Friedman:1922kd}
A.~Friedman.
\newblock {On the Curvature of space}.
\newblock {\em Z. Phys.}, 10:377--386, 1922.
\newblock [Gen. Rel. Grav.31,1991(1999)].

\bibitem{Friedmann:1924bb}
A.~Friedmann.
\newblock {On the Possibility of a world with constant negative curvature of
  space}.
\newblock {\em Z. Phys.}, 21:326--332, 1924.
\newblock [Gen. Rel. Grav.31,2001(1999)].

\bibitem{Fritzsch:2015lua}
H.~Fritzsch and J.~Sola.
\newblock {Fundamental constants and cosmic vacuum: the micro and macro
  connection}.
\newblock {\em Mod. Phys. Lett.}, A30(22):1540034, 2015, 1502.01411.

\bibitem{Gannouji:2015vva}
R.~Gannouji, H.~Nandan, and N.~Dadhich.
\newblock {FLRW Cosmology in Weyl-integrable Space-time}.
\newblock In {\em {Proceedings, 13th Marcel Grossmann Meeting on Recent
  Developments in Theoretical and Experimental General Relativity,
  Astrophysics, and Relativistic Field Theories (MG13): Stockholm, Sweden, July
  1-7, 2012}}, pages 1285--1287, 2015.

\bibitem{Garay:2010sk}
L.~J. Garay, M.~Martin-Benito, and G.~A. Mena~Marugan.
\newblock {Inhomogeneous Loop Quantum Cosmology: Hybrid Quantization of the
  Gowdy Model}.
\newblock {\em Phys. Rev.}, D82:044048, 2010, 1005.5654.

\bibitem{gerlach1969derivation}
U.~H. Gerlach.
\newblock Derivation of the ten einstein field equations from the semiclassical
  approximation to quantum geometrodynamics.
\newblock {\em Physical Review}, 177(5):1929, 1969.

\bibitem{Geroch:1975uq}
R.~P. Geroch and P.~S. Jang.
\newblock {Motion of a body in general relativity}.
\newblock {\em J. Math. Phys.}, 16:65--67, 1975.

\bibitem{Giddings:1988cx}
S.~B. Giddings and A.~Strominger.
\newblock {Loss of Incoherence and Determination of Coupling Constants in
  Quantum Gravity}.
\newblock {\em Nucl. Phys.}, B307:854--866, 1988.

\bibitem{Giddings:1988}
S.~B. Giddings and A.~Strominger.
\newblock {Baby Universes, Third Quantization and the Cosmological Constant}.
\newblock {\em Nucl. Phys.}, B321:481, 1989.

\bibitem{Gielen:2014ila}
S.~Gielen.
\newblock {Quantum cosmology of (loop) quantum gravity condensates: An
  example}.
\newblock {\em Class. Quant. Grav.}, 31:155009, 2014, 1404.2944.

\bibitem{Gielen:2015kua}
S.~Gielen.
\newblock {Identifying cosmological perturbations in group field theory
  condensates}.
\newblock {\em JHEP}, 08:010, 2015, 1505.07479.

\bibitem{Gielen:2014usa}
S.~Gielen.
\newblock {Perturbing a quantum gravity condensate}.
\newblock {\em Phys. Rev.}, D91(4):043526, 2015, 1411.1077.

\bibitem{Gielen:2016uft}
S.~Gielen.
\newblock {Emergence of a low spin phase in group field theory condensates}.
\newblock {\em Class. Quant. Grav.}, 33(22):224002, 2016, 1604.06023.

\bibitem{Gielen:2014uga}
S.~Gielen and D.~Oriti.
\newblock {Quantum cosmology from quantum gravity condensates: cosmological
  variables and lattice-refined dynamics}.
\newblock {\em New J. Phys.}, 16(12):123004, 2014, 1407.8167.

\bibitem{Gielen:2013kla}
S.~Gielen, D.~Oriti, and L.~Sindoni.
\newblock {Cosmology from Group Field Theory Formalism for Quantum Gravity}.
\newblock {\em Phys. Rev. Lett.}, 111(3):031301, 2013, 1303.3576.

\bibitem{Gielen:2013naa}
S.~Gielen, D.~Oriti, and L.~Sindoni.
\newblock {Homogeneous cosmologies as group field theory condensates}.
\newblock {\em JHEP}, 06:013, 2014, 1311.1238.

\bibitem{Gielen:2016dss}
S.~Gielen and L.~Sindoni.
\newblock {Quantum Cosmology from Group Field Theory Condensates: a Review}.
\newblock {\em SIGMA}, 12:082, 2016, 1602.08104.

\bibitem{Gielen:2015uaa}
S.~Gielen and N.~Turok.
\newblock {Perfect Quantum Cosmological Bounce}.
\newblock {\em Phys. Rev. Lett.}, 117(2):021301, 2016, 1510.00699.

\bibitem{Gielen:2016fdb}
S.~Gielen and N.~Turok.
\newblock {Quantum propagation across cosmological singularities}.
\newblock {\em Phys. Rev.}, D95(10):103510, 2017, 1612.02792.

\bibitem{Girelli:2008gc}
F.~Girelli, S.~Liberati, and L.~Sindoni.
\newblock {Gravitational dynamics in Bose Einstein condensates}.
\newblock {\em Phys. Rev.}, D78:084013, 2008, 0807.4910.

\bibitem{Giulini:2006yg}
D.~Giulini.
\newblock {Some remarks on the notions of general covariance and background
  independence}.
\newblock {\em Lect. Notes Phys.}, 721:105--120, 2007, gr-qc/0603087.

\bibitem{Gleyzes:2014qga}
J.~Gleyzes, D.~Langlois, F.~Piazza, and F.~Vernizzi.
\newblock {Exploring gravitational theories beyond Horndeski}.
\newblock {\em JCAP}, 1502:018, 2015, 1408.1952.

\bibitem{Godfrey:1990dt}
N.~Godfrey and M.~Gross.
\newblock {Simplicial quantum gravity in more than two-dimensions}.
\newblock {\em Phys. Rev.}, D43:1749--1753, 1991.

\bibitem{Govaerts:1988ch}
J.~Govaerts.
\newblock {The Nambu-goto String: Its Phase Space Path Integral}.
\newblock {\em Int. J. Mod. Phys.}, A4:173, 1989.

\bibitem{Green:1987sp}
M.~B. Green, J.~H. Schwarz, and E.~Witten.
\newblock {\em {Superstring Theory. Vol. 1: Introduction}}.
\newblock Cambridge Monographs on Mathematical Physics. Cambridge University
  Press, 1988.

\bibitem{Griffiths:1984rx}
R.~B. Griffiths.
\newblock {Consistent histories and the interpretation of quantum mechanics}.
\newblock {\em J. Statist. Phys.}, 36:219--272, 1984.

\bibitem{Gupta:1997nd}
R.~Gupta.
\newblock {Introduction to lattice QCD: Course}.
\newblock In {\em {Probing the standard model of particle interactions.
  Proceedings, Summer School in Theoretical Physics, NATO Advanced Study
  Institute, 68th session, Les Houches, France, July 28-September 5, 1997. Pt.
  1, 2}}, pages 83--219, 1997, hep-lat/9807028.

\bibitem{Guth:1981aa}
A.~H. Guth.
\newblock Inflationary universe: A possible solution to the horizon and
  flatness problems.
\newblock {\em Physical Review D}, 23(2):347--356, 1981.

\bibitem{Haggard:2011qvx}
H.~M. Haggard.
\newblock {\em {Asymptotic Analysis of Spin Networks with Applications to
  Quantum Gravity}}.
\newblock PhD thesis, UC, Berkeley, 2011.

\bibitem{Halliwell}
J.~Halliwell.
\newblock Time in quantum cosmology.
\newblock In {Ashtekar, A. and Stachel, J.}, editor, {\em {Conceptual problems
  of quantum gravity. Proceedings, Osgood Hill Conference}}, pages 204--210.
  Birkh{\"a}user, May 15-18 1988.

\bibitem{Halliwell:1988wc}
J.~J. Halliwell.
\newblock {Derivation of the Wheeler-De Witt Equation from a Path Integral for
  Minisuperspace Models}.
\newblock {\em Phys. Rev.}, D38:2468, 1988.

\bibitem{Halliwell:2002th}
J.~J. Halliwell and J.~Thorwart.
\newblock {Life in an energy eigenstate: Decoherent histories analysis of a
  model timeless universe}.
\newblock {\em Phys. Rev.}, D65:104009, 2002, gr-qc/0201070.

\bibitem{Han:2011rf}
M.-X. Han and M.~Zhang.
\newblock {Asymptotics of Spinfoam Amplitude on Simplicial Manifold: Euclidean
  Theory}.
\newblock {\em Class. Quant. Grav.}, 29:165004, 2012, 1109.0500.

\bibitem{hanson1976constrained}
A.~Hanson, T.~Regge, and C.~Teitelboim.
\newblock {\em Constrained hamiltonian systems}.
\newblock Accademia Nazionale dei Lincei, Roma, 1976.

\bibitem{Hartle:1992as}
J.~B. Hartle.
\newblock {Space-time quantum mechanics and the quantum mechanics of
  space-time}.
\newblock In {\em {Gravitation and quantizations. Proceedings, 57th Session of
  the Les Houches Summer School in Theoretical Physics, NATO Advanced Study
  Institute, Les Houches, France, July 5 - August 1, 1992}}, pages 0285--480,
  1992, gr-qc/9304006.

\bibitem{Hartle:1983}
J.~B. Hartle and S.~W. Hawking.
\newblock {Wave Function of the Universe}.
\newblock {\em Phys. Rev.}, D28:2960--2975, 1983.

\bibitem{Hartle:1997dc}
J.~B. Hartle and D.~Marolf.
\newblock {Comparing formulations of generalized quantum mechanics for
  reparametrization - invariant systems}.
\newblock {\em Phys. Rev.}, D56:6247--6257, 1997, gr-qc/9703021.

\bibitem{Hawking:1966sx}
S.~Hawking.
\newblock {The Occurrence of singularities in cosmology}.
\newblock {\em Proc. Roy. Soc. Lond.}, A294:511--521, 1966.

\bibitem{Hawking:1981}
S.~W. Hawking.
\newblock {The Boundary Conditions of the Universe}.
\newblock {\em Pontif. Acad. Sci. Scr. Varia}, 48:563--574, 1982.

\bibitem{Hawking:1983hj}
S.~W. Hawking.
\newblock {The Quantum State of the Universe}.
\newblock {\em Nucl. Phys.}, B239:257, 1984.

\bibitem{Hawking:1985}
S.~W. Hawking.
\newblock {The Arrow of Time in Cosmology}.
\newblock {\em Phys. Rev.}, D32:2489, 1985.

\bibitem{Hawking:1973uf}
S.~W. Hawking and G.~F.~R. Ellis.
\newblock {\em {The Large Scale Structure of Space-Time}}.
\newblock Cambridge Monographs on Mathematical Physics. Cambridge University
  Press, 2011.

\bibitem{Hawking:1993tu}
S.~W. Hawking, R.~Laflamme, and G.~W. Lyons.
\newblock {The Origin of time asymmetry}.
\newblock {\em Phys. Rev.}, D47:5342--5356, 1993, gr-qc/9301017.

\bibitem{Hawking:1969sw}
S.~W. Hawking and R.~Penrose.
\newblock {The Singularities of gravitational collapse and cosmology}.
\newblock {\em Proc. Roy. Soc. Lond.}, A314:529--548, 1970.

\bibitem{Hehl:1994ue}
F.~W. Hehl, J.~D. McCrea, E.~W. Mielke, and Y.~Ne'eman.
\newblock {Metric affine gauge theory of gravity: Field equations, Noether
  identities, world spinors, and breaking of dilation invariance}.
\newblock {\em Phys. Rept.}, 258:1--171, 1995, gr-qc/9402012.

\bibitem{Heikinheimo:2013fta}
M.~Heikinheimo, A.~Racioppi, M.~Raidal, C.~Spethmann, and K.~Tuominen.
\newblock {Physical Naturalness and Dynamical Breaking of Classical Scale
  Invariance}.
\newblock {\em Mod. Phys. Lett.}, A29:1450077, 2014, 1304.7006.

\bibitem{Henneaux:1992ig}
M.~Henneaux and C.~Teitelboim.
\newblock {\em {Quantization of gauge systems}}.
\newblock Princeton University Press, 1992.

\bibitem{higham2001algorithmic}
D.~J. Higham.
\newblock An algorithmic introduction to numerical simulation of stochastic
  differential equations.
\newblock {\em SIAM review}, 43(3):525--546, 2001.

\bibitem{Hinterbichler:2011tt}
K.~Hinterbichler.
\newblock {Theoretical Aspects of Massive Gravity}.
\newblock {\em Rev. Mod. Phys.}, 84:671--710, 2012, 1105.3735.

\bibitem{Hohm:2010jy}
O.~Hohm, C.~Hull, and B.~Zwiebach.
\newblock {Background independent action for double field theory}.
\newblock {\em JHEP}, 07:016, 2010, 1003.5027.

\bibitem{Hojman:1976vp}
S.~A. Hojman, K.~V. Kucha{\v r}, and C.~Teitelboim.
\newblock {Geometrodynamics Regained}.
\newblock {\em Annals Phys.}, 96:88--135, 1976.

\bibitem{horndeski1974second}
G.~W. Horndeski.
\newblock Second-order scalar-tensor field equations in a four-dimensional
  space.
\newblock {\em International Journal of Theoretical Physics}, 10(6):363--384,
  1974.

\bibitem{Hubble:1929ig}
E.~Hubble.
\newblock {A relation between distance and radial velocity among extra-galactic
  nebulae}.
\newblock {\em Proc. Nat. Acad. Sci.}, 15:168--173, 1929.

\bibitem{Hull:2009mi}
C.~Hull and B.~Zwiebach.
\newblock {Double Field Theory}.
\newblock {\em JHEP}, 09:099, 2009, 0904.4664.

\bibitem{HUSAIN1989205}
V.~Husain and L.~Smolin.
\newblock {Exactly solvable quantum cosmologies from two Killing field
  reductions of general relativity}.
\newblock {\em Nuclear Physics B}, 327(1):205 -- 238, 1989.

\bibitem{Isham:1992}
C.~J. Isham.
\newblock {Canonical quantum gravity and the problem of time}.
\newblock In {\em {19th International Colloquium on Group Theoretical Methods
  in Physics (GROUP 19) Salamanca, Spain, June 29-July 5, 1992}},
  gr-qc/9210011.

\bibitem{Isham:1995wr}
C.~J. Isham.
\newblock {Structural issues in quantum gravity}.
\newblock In {\em {General relativity and gravitation. Proceedings, 14th
  International Conference, Florence, Italy, August 6-12, 1995}}, pages
  167--209, gr-qc/9510063.

\bibitem{Jackson:2014nla}
S.~Jackson, L.~McGough, and H.~Verlinde.
\newblock {Conformal Bootstrap, Universality and Gravitational Scattering}.
\newblock {\em Nucl. Phys.}, B901:382--429, 2015, 1412.5205.

\bibitem{Kallosh:2013hoa}
R.~Kallosh and A.~Linde.
\newblock {Universality Class in Conformal Inflation}.
\newblock {\em JCAP}, 1307:002, 2013, 1306.5220.

\bibitem{Kaminski:2009fm}
W.~Kaminski, M.~Kisielowski, and J.~Lewandowski.
\newblock {Spin-Foams for All Loop Quantum Gravity}.
\newblock {\em Class. Quant. Grav.}, 27:095006, 2010, 0909.0939.
\newblock [Erratum: Class. Quant. Grav.29,049502(2012)].

\bibitem{Kaspi:1994hp}
V.~M. Kaspi, J.~H. Taylor, and M.~F. Ryba.
\newblock {High - precision timing of millisecond pulsars. 3: Long - term
  monitoring of PSRs B1855+09 and B1937+21}.
\newblock {\em Astrophys. J.}, 428:713, 1994.

\bibitem{Kegeles:2015oua}
A.~Kegeles and D.~Oriti.
\newblock {Generalized conservation laws in non-local field theories}.
\newblock {\em J. Phys.}, A49(13):135401, 2016, 1506.03320.

\bibitem{Kegeles:2016wfg}
A.~Kegeles and D.~Oriti.
\newblock {Continuous point symmetries in Group Field Theories}.
\newblock {\em J. Phys.}, A50(12):125402, 2017, 1608.00296.

\bibitem{Kegeles:2017ems}
A.~Kegeles, D.~Oriti, and C.~Tomlin.
\newblock {Inequivalent coherent state representations in Group Field Theory}.
\newblock 2017, 1709.00161.

\bibitem{Kempf:1994su}
A.~Kempf, G.~Mangano, and R.~B. Mann.
\newblock {Hilbert space representation of the minimal length uncertainty
  relation}.
\newblock {\em Phys. Rev.}, D52:1108--1118, 1995, hep-th/9412167.

\bibitem{Kiefer:1988}
C.~Kiefer.
\newblock {Wave Packets in Minisuperspace}.
\newblock {\em Phys.Rev.}, D38:1761--1772, 1988.

\bibitem{Kiefer:1990ms}
C.~Kiefer.
\newblock {On the Meaning of Path Integrals in Quantum Cosmology}.
\newblock {\em Annals Phys.}, 207:53--70, 1991.

\bibitem{Kiefer:2007ria}
C.~Kiefer.
\newblock {\em {Quantum Gravity}}.
\newblock Oxford University Press, 2007.

\bibitem{Kiefer:2008bs}
C.~Kiefer.
\newblock {Quantum geometrodynamics: whence, whither?}
\newblock {\em Gen. Rel. Grav.}, 41:877--901, 2009, 0812.0295.

\bibitem{Kiefer:2009xr}
C.~Kiefer.
\newblock {Can the Arrow of Time be understood from Quantum Cosmology?}
\newblock In L.~Mersini-Houghton and R.~Vaas, editors, {\em The Arrow of Time}.
  Springer, 2012, 0910.5836.

\bibitem{Kiefer:2012sy}
C.~Kiefer and M.~Kraemer.
\newblock {Can effects of quantum gravity be observed in the cosmic microwave
  background?}
\newblock {\em Int. J. Mod. Phys.}, D21:1241001, 2012, 1205.5161.

\bibitem{Kiefer:2011cc}
C.~Kiefer and M.~Kraemer.
\newblock {Quantum Gravitational Contributions to the CMB Anisotropy Spectrum}.
\newblock {\em Phys. Rev. Lett.}, 108:021301, 2012, 1103.4967.

\bibitem{Kiefer:2014qpa}
C.~Kiefer and M.~Kr{\"a}mer.
\newblock {On the Observability of Quantum-Gravitational Effects in the Cosmic
  Microwave Background}.
\newblock {\em Springer Proc. Phys.}, 157:531--538, 2014.

\bibitem{Kiefer:2006je}
C.~Kiefer, I.~Lohmar, D.~Polarski, and A.~A. Starobinsky.
\newblock {Pointer states for primordial fluctuations in inflationary
  cosmology}.
\newblock {\em Class. Quant. Grav.}, 24:1699--1718, 2007, astro-ph/0610700.

\bibitem{Kiefer:1998qe}
C.~Kiefer, D.~Polarski, and A.~A. Starobinsky.
\newblock {Quantum to classical transition for fluctuations in the early
  universe}.
\newblock {\em Int. J. Mod. Phys.}, D7:455--462, 1998, gr-qc/9802003.

\bibitem{Kiefer:1994gp}
C.~Kiefer and H.~D. Zeh.
\newblock {Arrow of time in a recollapsing quantum universe}.
\newblock {\em Phys. Rev.}, D51:4145--4153, 1995, gr-qc/9402036.

\bibitem{Kimpton:2010xi}
I.~Kimpton and A.~Padilla.
\newblock {Lessons from the decoupling limit of Horava gravity}.
\newblock {\em JHEP}, 07:014, 2010, 1003.5666.

\bibitem{talay1994numerical}
P.~E. Kloeden and E.~Platen.
\newblock Numerical solution of stochastic differential equations.
\newblock {\em Springer-Verlag}, 1992.

\bibitem{komar1983generalized}
A.~Komar.
\newblock Generalized constraint structure for gravitation theory.
\newblock {\em Physical Review D}, 27(10):2277, 1983.

\bibitem{Konishi:1989wk}
K.~Konishi, G.~Paffuti, and P.~Provero.
\newblock {Minimum Physical Length and the Generalized Uncertainty Principle in
  String Theory}.
\newblock {\em Phys. Lett.}, B234:276--284, 1990.

\bibitem{konno:1988}
K.~Konno.
\newblock {Gravitational Effects on Dirac Particles}.
\newblock M.Sc. thesis, Hirosaki University, 1988.

\bibitem{Konopka:2006hu}
T.~Konopka, F.~Markopoulou, and L.~Smolin.
\newblock {Quantum Graphity}.
\newblock 2006, hep-th/0611197.

\bibitem{KowalskiGlikman:2004qa}
J.~Kowalski-Glikman.
\newblock {Introduction to doubly special relativity}.
\newblock {\em Lect. Notes Phys.}, 669:131--159, 2005, hep-th/0405273.

\bibitem{Krajewski:2012aw}
T.~Krajewski.
\newblock {Group field theories}.
\newblock {\em PoS}, QGQGS2011:005, 2011, 1210.6257.

\bibitem{Krajewski:2010yq}
T.~Krajewski, J.~Magnen, V.~Rivasseau, A.~Tanasa, and P.~Vitale.
\newblock {Quantum Corrections in the Group Field Theory Formulation of the
  EPRL/FK Models}.
\newblock {\em Phys. Rev.}, D82:124069, 2010, 1007.3150.

\bibitem{kuchavr1972bubble}
K.~V. Kucha{\v{r}}.
\newblock A bubble-time canonical formalism for geometrodynamics.
\newblock {\em Journal of Mathematical Physics}, 13(5):768--781, 1972.

\bibitem{kuchar1991problem}
K.~V. Kucha{\v r}.
\newblock The problem of time in canonical quantization.
\newblock In A.~Ashtekar and J.~Stachel, editors, {\em Conceptual Problems of
  Quantum Gravity}. Boston, MA: Birkhauser, 1991.

\bibitem{Kuchar:1991qf}
K.~V. Kucha{\v r}.
\newblock {Time and interpretations of quantum gravity}.
\newblock {\em Int. J. Mod. Phys. Proc. Suppl.}, D20:3--86, 2011.

\bibitem{1981qugr.conf..329K}
K.~V. {Kucha{\v r}}.
\newblock {Canonical Methods of Quantization}.
\newblock In C.~J. {Isham}, R.~{Penrose}, and D.~W. {Sciama}, editors, {\em
  Quantum Gravity II}, page 329. Clarendon Press, 1981.

\bibitem{Kurkov:2014twa}
M.~A. Kurkov, F.~Lizzi, M.~Sakellariadou, and A.~Watcharangkool.
\newblock {Spectral action with zeta function regularization}.
\newblock {\em Phys. Rev.}, D91(6):065013, 2015, 1412.4669.

\bibitem{Lehners:2008vx}
J.-L. Lehners.
\newblock {Ekpyrotic and Cyclic Cosmology}.
\newblock {\em Phys. Rept.}, 465:223--263, 2008, 0806.1245.

\bibitem{Lemaitre:1927zz}
G.~Lemaitre.
\newblock {A Homogeneous Universe of Constant Mass and Growing Radius
  Accounting for the Radial Velocity of Extragalactic Nebulae}.
\newblock {\em Annales Soc. Sci. Brux. Ser. I Sci. Math. Astron. Phys.},
  A47:49--59, 1927.

\bibitem{LevyLeblond:1977qm}
J.~M. Levy-Leblond.
\newblock {The Importance of Being (a) Constant}.
\newblock In G.~Toraldo~di Francia, editor, {\em Problems In The Foundations Of
  Physics}, pages 237--263. North-Holland, 1977.

\bibitem{Li:2017uao}
Y.~Li, D.~Oriti, and M.~Zhang.
\newblock {Group field theory for quantum gravity minimally coupled to a scalar
  field}.
\newblock {\em Class. Quant. Grav.}, 34(19):195001, 2017, 1701.08719.

\bibitem{Livine:2017xww}
E.~R. Livine.
\newblock {From Coarse-Graining to Holography in Loop Quantum Gravity}.
\newblock 2017, 1704.04067.

\bibitem{Lizzi:1997yr}
F.~Lizzi and R.~J. Szabo.
\newblock {Noncommutative geometry and space-time gauge symmetries of string
  theory}.
\newblock {\em Chaos Solitons Fractals}, 10:445--458, 1999, hep-th/9712206.

\bibitem{Lobo:2015zaa}
I.~P. Lobo, A.~B. Barreto, and C.~Romero.
\newblock {Space-time singularities in Weyl manifolds}.
\newblock {\em Eur. Phys. J.}, C75(9):448, 2015, 1506.02180.

\bibitem{Maggiore:1993rv}
M.~Maggiore.
\newblock {A Generalized uncertainty principle in quantum gravity}.
\newblock {\em Phys. Lett.}, B304:65--69, 1993, hep-th/9301067.

\bibitem{Maggiore:1993kv}
M.~Maggiore.
\newblock {The Algebraic structure of the generalized uncertainty principle}.
\newblock {\em Phys. Lett.}, B319:83--86, 1993, hep-th/9309034.

\bibitem{Magueijo:2000zt}
J.~Magueijo.
\newblock {Covariant and locally Lorentz invariant varying speed of light
  theories}.
\newblock {\em Phys. Rev.}, D62:103521, 2000, gr-qc/0007036.

\bibitem{Magueijo:2003gj}
J.~Magueijo.
\newblock {New varying speed of light theories}.
\newblock {\em Rept. Prog. Phys.}, 66:2025, 2003, astro-ph/0305457.

\bibitem{Magueijo:2008sx}
J.~Magueijo.
\newblock {Bimetric varying speed of light theories and primordial
  fluctuations}.
\newblock {\em Phys. Rev.}, D79:043525, 2009, 0807.1689.

\bibitem{Magueijo:2007gf}
J.~Magueijo and J.~W. Moffat.
\newblock {Comments on `Note on varying speed of light theories'}.
\newblock {\em Gen. Rel. Grav.}, 40:1797--1806, 2008, 0705.4507.

\bibitem{Maldacena:2011mk}
J.~Maldacena.
\newblock {Einstein Gravity from Conformal Gravity}.
\newblock 2011, 1105.5632.

\bibitem{Maldacena:1997re}
J.~M. Maldacena.
\newblock {The Large N limit of superconformal field theories and
  supergravity}.
\newblock {\em Int. J. Theor. Phys.}, 38:1113--1133, 1999, hep-th/9711200.
\newblock [Adv. Theor. Math. Phys.2,231(1998)].

\bibitem{Mangano:2015pha}
G.~Mangano, F.~Lizzi, and A.~Porzio.
\newblock {Inconstant Planck's constant}.
\newblock {\em Int. J. Mod. Phys.}, A30(34):1550209, 2015, 1509.02107.

\bibitem{Mannheim:2011ds}
P.~D. Mannheim.
\newblock {Making the Case for Conformal Gravity}.
\newblock {\em Found. Phys.}, 42:388--420, 2012, 1101.2186.

\bibitem{Mannheim:1988dj}
P.~D. Mannheim and D.~Kazanas.
\newblock {Exact Vacuum Solution to Conformal Weyl Gravity and Galactic
  Rotation Curves}.
\newblock {\em Astrophys. J.}, 342:635--638, 1989.

\bibitem{Marnelius:1982bz}
R.~Marnelius.
\newblock {Canonical Quantization of Polyakov's String in Arbitrary
  Dimensions}.
\newblock {\em Nucl. Phys.}, B211:14--28, 1983.

\bibitem{Martin:2001aa}
J.~Martin.
\newblock Trans-planckian problem of inflationary cosmology.
\newblock {\em Physical Review D}, 63(12), 2001.

\bibitem{MartinBenito:2008ej}
M.~Martin-Benito, L.~J. Garay, and G.~A. Mena~Marugan.
\newblock {Hybrid Quantum Gowdy Cosmology: Combining Loop and Fock
  Quantizations}.
\newblock {\em Phys. Rev.}, D78:083516, 2008, 0804.1098.

\bibitem{MartinBenito:2008wx}
M.~Martin-Benito, G.~A. Mena~Marugan, and T.~Pawlowski.
\newblock {Loop Quantization of Vacuum Bianchi I Cosmology}.
\newblock {\em Phys. Rev.}, D78:064008, 2008, 0804.3157.

\bibitem{Meissner:2006zh}
K.~A. Meissner and H.~Nicolai.
\newblock {Conformal Symmetry and the Standard Model}.
\newblock {\em Phys. Lett.}, B648:312--317, 2007, hep-th/0612165.

\bibitem{MenaMarugan:1997us}
G.~A. Mena~Marugan.
\newblock {Canonical quantization of the Gowdy model}.
\newblock {\em Phys. Rev.}, D56:908--919, 1997, gr-qc/9704041.

\bibitem{Milgrom:1983ca}
M.~Milgrom.
\newblock {A Modification of the Newtonian dynamics as a possible alternative
  to the hidden mass hypothesis}.
\newblock {\em Astrophys. J.}, 270:365--370, 1983.

\bibitem{Misner:1974qy}
C.~W. Misner, K.~S. Thorne, and J.~A. Wheeler.
\newblock {\em {Gravitation}}.
\newblock W. H. Freeman, San Francisco, 1973.

\bibitem{Moffat:1992bf}
J.~W. Moffat.
\newblock {Quantum gravity, the origin of time and time's arrow}.
\newblock {\em Found. Phys.}, 23:411--437, 1993, gr-qc/9209001.

\bibitem{Moffat:1992ud}
J.~W. Moffat.
\newblock {Superluminary universe: A Possible solution to the initial value
  problem in cosmology}.
\newblock {\em Int. J. Mod. Phys.}, D2:351--366, 1993, gr-qc/9211020.

\bibitem{Moffat:2005si}
J.~W. Moffat.
\newblock {Scalar-tensor-vector gravity theory}.
\newblock {\em JCAP}, 0603:004, 2006, gr-qc/0506021.

\bibitem{Mukhanov:2005sc}
V.~Mukhanov.
\newblock {\em {Physical Foundations of Cosmology}}.
\newblock Cambridge University Press, 2005.

\bibitem{Mukhanov:1990me}
V.~F. Mukhanov, H.~A. Feldman, and R.~H. Brandenberger.
\newblock {Theory of cosmological perturbations. Part 1. Classical
  perturbations. Part 2. Quantum theory of perturbations. Part 3. Extensions}.
\newblock {\em Phys. Rept.}, 215:203--333, 1992.

\bibitem{Nelson:2008bx}
W.~Nelson and M.~Sakellariadou.
\newblock {Numerical techniques for solving the quantum constraint equation of
  generic lattice-refined models in loop quantum cosmology}.
\newblock {\em Phys. Rev.}, D78:024030, 2008, 0803.4483.

\bibitem{Nelson:2008vz}
W.~Nelson and M.~Sakellariadou.
\newblock {Unique factor ordering in the continuum limit of LQC}.
\newblock {\em Phys. Rev.}, D78:024006, 2008, 0806.0595.

\bibitem{Niemi:1986kc}
A.~J. Niemi.
\newblock {Closed Bosonic Strings in a Covariant Phase Space}.
\newblock {\em Phys. Lett.}, B179:228--234, 1986.

\bibitem{Nishino:2004kb}
H.~Nishino and S.~Rajpoot.
\newblock {Broken scale invariance in the standard model}.
\newblock 2004, hep-th/0403039.

\bibitem{Nishino:2009in}
H.~Nishino and S.~Rajpoot.
\newblock {Implication of Compensator Field and Local Scale Invariance in the
  Standard Model}.
\newblock {\em Phys. Rev.}, D79:125025, 2009, 0906.4778.

\bibitem{Nishino:2011zz}
H.~Nishino and S.~Rajpoot.
\newblock {Weyl's scale invariance for the standard model, renormalizability
  and the zero cosmological constant}.
\newblock {\em Class. Quant. Grav.}, 28:145014, 2011.

\bibitem{Nordtvedt:1990zz}
K.~Nordtvedt.
\newblock {$\dot{G} /G$ and a cosmological acceleration of gravitationally
  compact bodies}.
\newblock {\em Phys. Rev. Lett.}, 65:953--956, 1990.

\bibitem{oksendal2003stochastic}
B.~{\O}ksendal.
\newblock Stochastic differential equations.
\newblock In {\em Stochastic differential equations}, pages 65--84. Springer,
  2003.

\bibitem{Okun:1991wm}
L.~B. Okun.
\newblock {Fundamental constants of physics}.
\newblock In {\em {Sakharov Conf.1991:0819-840}}, pages 0819--840, 1991.

\bibitem{Ooguri:1992eb}
H.~Ooguri.
\newblock {Topological lattice models in four-dimensions}.
\newblock {\em Mod. Phys. Lett.}, A7:2799--2810, 1992, hep-th/9205090.

\bibitem{ORaifeartaigh:1997dvq}
L.~O'Raifeartaigh.
\newblock {\em {The dawning of gauge theory}}.
\newblock Princeton University Press, 1997.

\bibitem{Oriti:2005mb}
D.~Oriti.
\newblock {Quantum gravity as a quantum field theory of simplicial geometry}.
\newblock In J.~T. B.~Fauser and E.~Zeidler, editors, {\em Mathematical and
  Physical Aspects of Quantum Gravity}. Birkhaeuser, 2006, gr-qc/0512103.

\bibitem{Oriti:2006se}
D.~Oriti.
\newblock {The Group field theory approach to quantum gravity}.
\newblock In D.~Oriti, editor, {\em Approaches to Quantum Gravity - toward a
  new understanding of space, time, and matter}. Cambridge University Press,
  2006, gr-qc/0607032.

\bibitem{Oriti:2007qd}
D.~Oriti.
\newblock {Group field theory as the microscopic description of the quantum
  spacetime fluid: A New perspective on the continuum in quantum gravity}.
\newblock {\em PoS}, QG-PH:030, 2007, 0710.3276.

\bibitem{Oriti:2011jm}
D.~Oriti.
\newblock {The microscopic dynamics of quantum space as a group field theory}.
\newblock In {\em {Proceedings, Foundations of Space and Time: Reflections on
  Quantum Gravity: Cape Town, South Africa}}, pages 257--320, 2011, 1110.5606.

\bibitem{Oriti:2013jga}
D.~Oriti.
\newblock {Disappearance and emergence of space and time in quantum gravity}.
\newblock {\em Stud. Hist. Phil. Sci.}, B46:186--199, 2014, 1302.2849.

\bibitem{Oriti:2013}
D.~Oriti.
\newblock {Group field theory as the 2nd quantization of Loop Quantum Gravity}.
\newblock {\em Class. Quant. Grav.}, 33(8):085005, 2016, 1310.7786.

\bibitem{Oriti:2016acw}
D.~Oriti.
\newblock {The universe as a quantum gravity condensate}.
\newblock {\em Comptes Rendus Physique}, 18(235-245), 2016, 1612.09521.

\bibitem{Oriti:2015qva}
D.~Oriti, D.~Pranzetti, J.~P. Ryan, and L.~Sindoni.
\newblock {Generalized quantum gravity condensates for homogeneous geometries
  and cosmology}.
\newblock {\em Class. Quant. Grav.}, 32(23):235016, 2015, 1501.00936.

\bibitem{Oriti:2014yla}
D.~Oriti, J.~P. Ryan, and J.~Th{\"u}rigen.
\newblock {Group field theories for all loop quantum gravity}.
\newblock {\em New J. Phys.}, 17(2):023042, 2015, 1409.3150.

\bibitem{Oriti:2010}
D.~Oriti and L.~Sindoni.
\newblock {Towards classical geometrodynamics from Group Field Theory
  hydrodynamics}.
\newblock {\em New J. Phys.}, 13:025006, 2011, 1010.5149.

\bibitem{Oriti:2016qtz}
D.~Oriti, L.~Sindoni, and E.~Wilson-Ewing.
\newblock {Emergent Friedmann dynamics with a quantum bounce from quantum
  gravity condensates}.
\newblock {\em Class. Quant. Grav.}, 33(22):224001, 2016, 1602.05881.

\bibitem{Oriti:2016ueo}
D.~Oriti, L.~Sindoni, and E.~Wilson-Ewing.
\newblock {Bouncing cosmologies from quantum gravity condensates}.
\newblock {\em Class. Quant. Grav.}, 34(4):04LT01, 2017, 1602.08271.

\bibitem{Padilla:2013jza}
A.~Padilla, D.~Stefanyszyn, and M.~Tsoukalas.
\newblock {Generalised Scale Invariant Theories}.
\newblock {\em Phys. Rev.}, D89(6):065009, 2014, 1312.0975.

\bibitem{Page:1985}
D.~N. Page.
\newblock {Will Entropy Decrease if the Universe Recollapses?}
\newblock {\em Phys. Rev.}, D32:2496, 1985.

\bibitem{Palais:1979rca}
R.~S. Palais.
\newblock {The principle of symmetric criticality}.
\newblock {\em Commun. Math. Phys.}, 69(1):19--30, 1979.

\bibitem{Parker:2009uva}
L.~E. Parker and D.~Toms.
\newblock {\em {Quantum Field Theory in Curved Spacetime}}.
\newblock Cambridge Monographs on Mathematical Physics. Cambridge University
  Press, 2009.

\bibitem{Passian}
A.~Passian, H.~Simpson, S.~Kouchekian, and S.~Yakubovich.
\newblock On the orthogonality of the macdonald's functions.
\newblock {\em Journal of Mathematical Analysis and Applications}, 360(2):380
  -- 390, 2009.

\bibitem{Peloso:2002rx}
M.~Peloso and L.~Sorbo.
\newblock {Moduli from cosmic strings}.
\newblock {\em Nucl. Phys.}, B649:88--100, 2003, hep-ph/0205063.

\bibitem{Penrose:1964wq}
R.~Penrose.
\newblock {Gravitational collapse and space-time singularities}.
\newblock {\em Phys. Rev. Lett.}, 14:57--59, 1965.

\bibitem{Penrose:RoadReality}
R.~Penrose.
\newblock {\em {The Road to Reality: a complete guide to the laws of the
  Universe}}.
\newblock J. Cape, London, 2004.

\bibitem{Perez:2012wv}
A.~Perez.
\newblock {The Spin Foam Approach to Quantum Gravity}.
\newblock {\em Living Rev. Rel.}, 16:3, 2013, 1205.2019.

\bibitem{Peskin:1995ev}
M.~E. Peskin and D.~V. Schroeder.
\newblock {\em {An Introduction to quantum field theory}}.
\newblock Addison-Wesley, 1995.

\bibitem{Pithis:2016cxg}
A.~G.~A. Pithis and M.~Sakellariadou.
\newblock {Relational evolution of effectively interacting group field theory
  quantum gravity condensates}.
\newblock {\em Phys. Rev.}, D95(6):064004, 2017, 1612.02456.

\bibitem{Pithis:2016wzf}
A.~G.~A. Pithis, M.~Sakellariadou, and P.~Tomov.
\newblock {Impact of nonlinear effective interactions on group field theory
  quantum gravity condensates}.
\newblock {\em Phys. Rev.}, D94(6):064056, 2016, 1607.06662.

\bibitem{Postma:2014vaa}
M.~Postma and M.~Volponi.
\newblock {Equivalence of the Einstein and Jordan frames}.
\newblock {\em Phys. Rev.}, D90(10):103516, 2014, 1407.6874.

\bibitem{regge1961general}
T.~Regge.
\newblock General relativity without coordinates.
\newblock {\em Il Nuovo Cimento (1955-1965)}, 19(3):558--571, 1961.

\bibitem{Reisenberger:2000fy}
M.~Reisenberger and C.~Rovelli.
\newblock {Spin foams as Feynman diagrams}.
\newblock In {\em {2001: A relativistic spacetime odyssey. Proceedings, 25th
  Johns Hopkins Workshop on Problems in particle theory, Florence, Italy,
  September 3-5, 2001}}, pages 431--448, 2000, gr-qc/0002083.

\bibitem{Reisenberger:1996pu}
M.~P. Reisenberger and C.~Rovelli.
\newblock {`Sum over surfaces' form of loop quantum gravity}.
\newblock {\em Phys. Rev.}, D56:3490--3508, 1997, gr-qc/9612035.

\bibitem{Reisenberger:2000zc}
M.~P. Reisenberger and C.~Rovelli.
\newblock {Space-time as a Feynman diagram: The Connection formulation}.
\newblock {\em Class. Quant. Grav.}, 18:121--140, 2001, gr-qc/0002095.

\bibitem{Robertson:1935zz}
H.~P. Robertson.
\newblock {Kinematics and World-Structure}.
\newblock {\em Astrophys. J.}, 82:284--301, 1935.

\bibitem{Robertson:1936zza}
H.~P. Robertson.
\newblock {Kinematics and World-Structure. 2}.
\newblock {\em Astrophys. J.}, 83:187--201, 1935.

\bibitem{Robertson:1936zz}
H.~P. Robertson.
\newblock {Kinematics and World-Structure. 3}.
\newblock {\em Astrophys. J.}, 83:257--271, 1936.

\bibitem{Romero:2012hs}
C.~Romero, J.~B. Fonseca-Neto, and M.~L. Pucheu.
\newblock {General Relativity and Weyl Geometry}.
\newblock {\em Class. Quant. Grav.}, 29:155015, 2012, 1201.1469.

\bibitem{Rosen:2006bga}
J.~Rosen, J.~H. Jung, and G.~Khanna.
\newblock {Instabilities in numerical loop quantum cosmology}.
\newblock {\em Class. Quant. Grav.}, 23:7075--7084, 2006, gr-qc/0607044.

\bibitem{Rovelli:1990ph}
C.~Rovelli.
\newblock {What Is Observable in Classical and Quantum Gravity?}
\newblock {\em Class. Quant. Grav.}, 8:297--316, 1991.

\bibitem{Rovelli:2000aw}
C.~Rovelli.
\newblock {Notes for a brief history of quantum gravity}.
\newblock In {\em {Recent developments in theoretical and experimental general
  relativity, gravitation and relativistic field theories. Proceedings, 9th
  Marcel Grossmann Meeting, MG'9, Rome, Italy, July 2-8, 2000. Pts. A-C}},
  pages 742--768, 2000, gr-qc/0006061.

\bibitem{Rovelli:2001bz}
C.~Rovelli.
\newblock {Partial observables}.
\newblock {\em Phys. Rev.}, D65:124013, 2002, gr-qc/0110035.

\bibitem{Rovelli:2004tv}
C.~Rovelli.
\newblock {\em {Quantum gravity}}.
\newblock Cambridge Monographs on Mathematical Physics. Cambridge University
  Press, 2004.

\bibitem{Rovelli:2011eq}
C.~Rovelli.
\newblock {Zakopane lectures on loop gravity}.
\newblock {\em PoS}, QGQGS2011:003, 2011, 1102.3660.

\bibitem{Rovelli:2015}
C.~Rovelli.
\newblock {The strange equation of quantum gravity}.
\newblock {\em Class.Quant.Grav.}, 32:124005, 2015, 1506.00927.

\bibitem{Rovelli:1994ge}
C.~Rovelli and L.~Smolin.
\newblock {Discreteness of area and volume in quantum gravity}.
\newblock {\em Nucl. Phys.}, B442:593--622, 1995, gr-qc/9411005.
\newblock [Erratum: Nucl. Phys.B456,753(1995)].

\bibitem{Rovelli:2013zaa}
C.~Rovelli and E.~Wilson-Ewing.
\newblock {Why are the effective equations of loop quantum cosmology so
  accurate?}
\newblock {\em Phys. Rev.}, D90(2):023538, 2014, 1310.8654.

\bibitem{Rubakov:2014jja}
V.~A. Rubakov.
\newblock {The Null Energy Condition and its violation}.
\newblock {\em Phys. Usp.}, 57:128--142, 2014, 1401.4024.
\newblock [Usp. Fiz. Nauk184,no.2,137(2014)].

\bibitem{Sakellariadou:2016dfl}
M.~Sakellariadou and A.~Watcharangkool.
\newblock {Linear stability of noncommutative spectral geometry}.
\newblock {\em Phys. Rev.}, D93(6):064034, 2016, 1601.06397.

\bibitem{Salim:1996ei}
J.~M. Salim and S.~L. Sautu.
\newblock {Gravitational theory in Weyl integrable space-time}.
\newblock {\em Class. Quant. Grav.}, 13:353--360, 1996.

\bibitem{Sandvik:2001rv}
H.~B. Sandvik, J.~D. Barrow, and J.~Magueijo.
\newblock {A simple cosmology with a varying fine structure constant}.
\newblock {\em Phys. Rev. Lett.}, 88:031302, 2002, astro-ph/0107512.

\bibitem{Sasakura:1990fs}
N.~Sasakura.
\newblock {Tensor model for gravity and orientability of manifold}.
\newblock {\em Mod. Phys. Lett.}, A6:2613--2624, 1991.

\bibitem{Schutz:1970aa}
B.~F. Schutz.
\newblock Perfect fluids in general relativity: Velocity potentials and a
  variational principle.
\newblock {\em Physical Review D}, 2(12):2762--2773, 1970.

\bibitem{Schwabl:1997gf}
F.~Schwabl.
\newblock {\em {Advanced quantum mechanics}}.
\newblock Springer, 1997.

\bibitem{Seiberg:1999vs}
N.~Seiberg and E.~Witten.
\newblock {String theory and noncommutative geometry}.
\newblock {\em JHEP}, 09:032, 1999, hep-th/9908142.

\bibitem{Shaposhnikov:2008xb}
M.~Shaposhnikov and D.~Zenhausern.
\newblock {Scale invariance, unimodular gravity and dark energy}.
\newblock {\em Phys. Lett.}, B671:187--192, 2009, 0809.3395.

\bibitem{Sindoni:2014wya}
L.~Sindoni.
\newblock {Effective equations for GFT condensates from fidelity}.
\newblock 2014, 1408.3095.

\bibitem{Sindoni:2009fc}
L.~Sindoni, F.~Girelli, and S.~Liberati.
\newblock {Emergent gravitational dynamics in Bose-Einstein condensates}.
\newblock In {\em {Proceedings, 25th Max Born Symposium: The Planck Scale:
  Wroclaw, Poland, June 29-July 3}}, 2009, 0909.5391.
\newblock [AIP Conf. Proc.1196,258(2009)].

\bibitem{Singh:2005km}
P.~Singh.
\newblock {Effective state metamorphosis in semi-classical loop quantum
  cosmology}.
\newblock {\em Class. Quant. Grav.}, 22:4203--4216, 2005, gr-qc/0502086.

\bibitem{Smolin:1979uz}
L.~Smolin.
\newblock {Towards a Theory of Space-Time Structure at Very Short Distances}.
\newblock {\em Nucl. Phys.}, B160:253--268, 1979.

\bibitem{Smolin:2005mq}
L.~Smolin.
\newblock {The Case for background independence}.
\newblock In T.~Rickles, S.~French, and J.~T. Saatsi, editors, {\em The
  Structural Foundations of Quantum Gravity}, 2005, hep-th/0507235.

\bibitem{Smolin:2015fqa}
L.~Smolin.
\newblock {Dynamics of the cosmological and Newton's constant}.
\newblock {\em Class. Quant. Grav.}, 33(2):025011, 2016, 1507.01229.

\bibitem{Sorkin:2003bx}
R.~D. Sorkin.
\newblock {Causal sets: Discrete gravity}.
\newblock In {\em {Lectures on quantum gravity. Proceedings, School of Quantum
  Gravity, Valdivia, Chile, January 4-14, 2002}}, pages 305--327, 2003,
  gr-qc/0309009.

\bibitem{Sotiriou:2006hs}
T.~P. Sotiriou.
\newblock {f(R) gravity and scalar-tensor theory}.
\newblock {\em Class. Quant. Grav.}, 23:5117--5128, 2006, gr-qc/0604028.

\bibitem{Sotiriou:2007yd}
T.~P. Sotiriou.
\newblock {\em {Modified Actions for Gravity: Theory and Phenomenology}}.
\newblock PhD thesis, SISSA, Trieste, 2007, 0710.4438.

\bibitem{Sotiriou:2008rp}
T.~P. Sotiriou and V.~Faraoni.
\newblock {f(R) Theories Of Gravity}.
\newblock {\em Rev. Mod. Phys.}, 82:451--497, 2010, 0805.1726.

\bibitem{Sotiriou:2006qn}
T.~P. Sotiriou and S.~Liberati.
\newblock {Metric-affine f(R) theories of gravity}.
\newblock {\em Annals Phys.}, 322:935--966, 2007, gr-qc/0604006.

\bibitem{Speziale:2016axj}
S.~Speziale.
\newblock {Boosting Wigner's nj-symbols}.
\newblock {\em J. Math. Phys.}, 58(3):032501, 2017, 1609.01632.

\bibitem{Sundermeyer:1982gv}
K.~Sundermeyer.
\newblock {Constrained dynamics with applications to Yang-Mills theory, General
  Relativity, classical spin, dual string model}.
\newblock {\em Lect. Notes Phys.}, 169:1--318, 1982.

\bibitem{Svrcek:2006yi}
P.~Svrcek and E.~Witten.
\newblock {Axions In String Theory}.
\newblock {\em JHEP}, 06:051, 2006, hep-th/0605206.

\bibitem{Szmytkowski}
R.~Szmytkowski and S.~Bielski.
\newblock {Comment on the orthogonality of the MacDonald functions of imaginary
  order}.
\newblock {\em Journal of Mathematical Analysis and Applications}, 365(1):195
  -- 197, 2010.

\bibitem{Hooft:2010ac}
G.~'t~Hooft.
\newblock {Probing the small distance structure of canonical quantum gravity
  using the conformal group}.
\newblock 2010, 1009.0669.

\bibitem{Hooft:2014daa}
G.~'t~Hooft.
\newblock {Local Conformal Symmetry: the Missing Symmetry Component for Space
  and Time}.
\newblock {\em Int. J. Mod. Phys.}, D24(12):1543001, 2015, 1410.6675.

\bibitem{Tambornino:2011vg}
J.~Tambornino.
\newblock {Relational Observables in Gravity: a Review}.
\newblock {\em SIGMA}, 8:017, 2012, 1109.0740.

\bibitem{Taveras:2008ke}
V.~Taveras.
\newblock {Corrections to the Friedmann Equations from LQG for a Universe with
  a Free Scalar Field}.
\newblock {\em Phys. Rev.}, D78:064072, 2008, 0807.3325.

\bibitem{taylorlectures:lie00}
M.~Taylor.
\newblock Lectures on lie groups.
\newblock Available
  \href{http://www.unc.edu/math/Faculty/met/lieg.html}{online}.

\bibitem{Teitelboim:1981ua}
C.~Teitelboim.
\newblock {Quantum Mechanics of the Gravitational Field}.
\newblock {\em Phys. Rev.}, D25:3159, 1982.

\bibitem{1984qtg..book..327T}
C.~{Teitelboim}.
\newblock {The Hamiltonian Structure of Two-Dimensional Space-Time and its
  Relation with the Conformal Anomaly}.
\newblock In S.~M. {Christensen}, editor, {\em Quantum Theory of Gravity.
  Essays in honor of the 60th birthday of Bryce S. DeWitt.}, pages 327--344.
  Adam Hilger, 1984.

\bibitem{Thiemann:1996ay}
T.~Thiemann.
\newblock {Anomaly - free formulation of nonperturbative, four-dimensional
  Lorentzian quantum gravity}.
\newblock {\em Phys. Lett.}, B380:257--264, 1996, gr-qc/9606088.

\bibitem{Thiemann:1997rv}
T.~Thiemann.
\newblock {QSD 3: Quantum constraint algebra and physical scalar product in
  quantum general relativity}.
\newblock {\em Class. Quant. Grav.}, 15:1207--1247, 1998, gr-qc/9705017.

\bibitem{Thiemann:1996aw}
T.~Thiemann.
\newblock {Quantum spin dynamics (QSD)}.
\newblock {\em Class. Quant. Grav.}, 15:839--873, 1998, gr-qc/9606089.

\bibitem{Thiemann:2003zv}
T.~Thiemann.
\newblock {The Phoenix project: Master constraint program for loop quantum
  gravity}.
\newblock {\em Class. Quant. Grav.}, 23:2211--2248, 2006, gr-qc/0305080.

\bibitem{Thiemann:2007zz}
T.~Thiemann.
\newblock {\em {Modern canonical quantum general relativity}}.
\newblock Cambridge University Press, 2008, gr-qc/0110034.

\bibitem{Tyutin:1975qk}
I.~V. Tyutin.
\newblock {Gauge Invariance in Field Theory and Statistical Physics in Operator
  Formalism}.
\newblock Lebedev Institute preprint 39, 1975, 0812.0580.

\bibitem{Uzan:2010pm}
J.~P. Uzan.
\newblock {Varying Constants, Gravitation and Cosmology}.
\newblock {\em Living Rev. Rel.}, 14:2, 2011, 1009.5514.

\bibitem{Varadarajan:1999it}
M.~Varadarajan.
\newblock {Fock representations from U(1) holonomy algebras}.
\newblock {\em Phys. Rev.}, D61:104001, 2000, gr-qc/0001050.

\bibitem{Varadarajan:2001nm}
M.~Varadarajan.
\newblock {Photons from quantized electric flux representations}.
\newblock {\em Phys. Rev.}, D64:104003, 2001, gr-qc/0104051.

\bibitem{Veneziano:1986zf}
G.~Veneziano.
\newblock {A Stringy Nature Needs Just Two Constants}.
\newblock {\em Europhys. Lett.}, 2:199, 1986.

\bibitem{Vilenkin:1986cy}
A.~Vilenkin.
\newblock {Boundary Conditions in Quantum Cosmology}.
\newblock {\em Phys. Rev.}, D33:3560, 1986.

\bibitem{Vilenkin:1987}
A.~Vilenkin.
\newblock {Quantum Cosmology and the Initial State of the Universe}.
\newblock {\em Phys. Rev.}, D37:888, 1988.

\bibitem{Visser:1999de}
M.~Visser and C.~Barcelo.
\newblock {Energy conditions and their cosmological implications}.
\newblock In {\em {Proceedings, 3rd International Conference on Particle
  Physics and the Early Universe (COSMO 1999): Trieste, Italy, September
  27-October 3, 1999}}, pages 98--112, 2000, gr-qc/0001099.

\bibitem{Wald:1984rg}
R.~M. Wald.
\newblock {\em {General Relativity}}.
\newblock University of Chicago Press, 1984.

\bibitem{Walker:1937}
A.~Walker.
\newblock {On Milne's Theory of World-Structure}.
\newblock {\em Proc. Lond. Math. Soc.}, s2-42(1):90--127, 1937.

\bibitem{Ward:1990vs}
R.~S. Ward and R.~O. Wells.
\newblock {\em {Twistor geometry and field theory}}.
\newblock Cambridge University Press, 1991.

\bibitem{Weinberg:1988cp}
S.~Weinberg.
\newblock {The Cosmological Constant Problem}.
\newblock {\em Rev. Mod. Phys.}, 61:1--23, 1989.

\bibitem{Weinberg:2008zzc}
S.~Weinberg.
\newblock {\em {Cosmology}}.
\newblock Oxford University Press, 2008.

\bibitem{Weinberg:1983ph}
S.~Weinberg, H.~B. Nielsen, and J.~G. Taylor.
\newblock Overview of theoretical prospects for understanding the values of
  fundamental constants [and discussion].
\newblock {\em Philosophical Transactions of the Royal Society of London.
  Series A, Mathematical and Physical Sciences}, 310(1512):249--252, 1983.

\bibitem{weyl:1918}
H.~Weyl.
\newblock Gravitation und elektrizit{\"a}t.
\newblock {\em Sitzungsberichte der K{\"o}niglich Preu{\ss}ischen Akademie der
  Wissenschaften zu Berlin}, pages 465--480, 1918.

\bibitem{Wheeler:1988zr}
J.~A. Wheeler.
\newblock {Superspace and the nature of quantum geometrodynamics}.
\newblock In L.~Z. Fang and R.~Ruffini, editors, {\em Quantum Cosmology,
  Advanced Series in Astrophysics and Cosmology}, volume~3. World Scientific,
  1987.

\bibitem{Williams:2004qba}
J.~G. Williams, S.~G. Turyshev, and D.~H. Boggs.
\newblock {Progress in lunar laser ranging tests of relativistic gravity}.
\newblock {\em Phys. Rev. Lett.}, 93:261101, 2004, gr-qc/0411113.

\bibitem{Williams:1991cd}
R.~M. Williams and P.~A. Tuckey.
\newblock {Regge calculus: A Bibliography and brief review}.
\newblock {\em Class. Quant. Grav.}, 9:1409--1422, 1992.

\bibitem{WilsonEwing:2010rh}
E.~Wilson-Ewing.
\newblock {Loop quantum cosmology of Bianchi type IX models}.
\newblock {\em Phys. Rev.}, D82:043508, 2010, 1005.5565.

\bibitem{WilsonEwing:2012}
E.~Wilson-Ewing.
\newblock {Lattice loop quantum cosmology: scalar perturbations}.
\newblock {\em Class. Quant. Grav.}, 29:215013, 2012, 1205.3370.

\bibitem{WilsonEwing:2012pu}
E.~Wilson-Ewing.
\newblock {The Matter Bounce Scenario in Loop Quantum Cosmology}.
\newblock {\em JCAP}, 1303:026, 2013, 1211.6269.

\bibitem{Wipf:1993xg}
A.~W. Wipf.
\newblock {Hamilton's formalism for systems with constraints}.
\newblock {\em Lect. Notes Phys.}, 434:22, 1994, hep-th/9312078.

\bibitem{Witten:1996md}
E.~Witten.
\newblock {On flux quantization in M theory and the effective action}.
\newblock {\em J. Geom. Phys.}, 22:1--13, 1997, hep-th/9609122.

\bibitem{Witten:1998qj}
E.~Witten.
\newblock {Anti-de Sitter space and holography}.
\newblock {\em Adv. Theor. Math. Phys.}, 2:253--291, 1998, hep-th/9802150.

\bibitem{Woodard:2009ns}
R.~P. Woodard.
\newblock {How Far Are We from the Quantum Theory of Gravity?}
\newblock {\em Rept. Prog. Phys.}, 72:126002, 2009, 0907.4238.

\bibitem{Yadav:2005vv}
J.~Yadav, S.~Bharadwaj, B.~Pandey, and T.~R. Seshadri.
\newblock {Testing homogeneity on large scales in the Sloan Digital Sky Survey
  Data Release One}.
\newblock {\em Mon. Not. Roy. Astron. Soc.}, 364:601--606, 2005,
  astro-ph/0504315.

\bibitem{Yakubovich}
S.~Yakubovich.
\newblock {A distribution associated with the Kontorovich-Lebedev transform}.
\newblock {\em Opuscula Mathematica}, 26(1):161--172, 2006.

\bibitem{Yasue1977}
K.~Yasue.
\newblock {A note on the derivation of the Schr{\"o}dinger-Langevin equation}.
\newblock {\em Journal of Statistical Physics}, 16(1):113--116, 1977.

\bibitem{zeldovich1975structure}
I.~B. Zeldovich and I.~Novikov.
\newblock {Structure and Evolution of the Universe}.
\newblock {\em Moscow, Izdatel'stvo Nauka, 1975. 736 p. In Russian.}, 1975.

\bibitem{Zhang}
H.-H. Zhang, K.-X. Feng, S.-W. Qiu, A.~Zhao, and X.-S. Li.
\newblock {On analytic formulas of Feynman propagators in position space}.
\newblock {\em Chin. Phys.}, C34:1576--1582, 2010, 0811.1261.

\end{thebibliography}



\end{spacing}

\part{Appendices}
\begin{appendices} 

\chapter{Energy Conditions}\label{Appendix:ECs}
In this appendix we review briefly the formulation of some energy conditions in relation to their significance in Cosmology. The energy conditions represent different mathematical realisations of our physical intuition concerning positivity of the energy of matter fields, as measured by local observers. They play an important role in General Relativity, since suitable energy conditions must be assumed in order to prove singularity theorems.

The \emph{strong energy condition} (SEC) states that
\be\label{Eq:AppendixSEC:sec}
\left(T_{\mu\nu}-\frac{1}{2}Tg_{\mu\nu}\right)W^{\mu}W^{\nu}\geq0~,
\ee
for any time-like four-vector $W^{\mu}$. By continuity, this inequality must hold also for all null four-vectors. 
The particular form of SEC is motivated by a convergence condition imposed on timelike congruences of geodesics (which satisfy the Raychaudhuri equation), assuming that the Einstein equations hold (see Ref.~\cite{Hawking:1973uf}). SEC is one of the underlying assumptions of the singularity theorems formulated by Hawking and Penrose. As such, it has an important role in General Relativity. Other formulations of the singularity theorems only require a weaker inequality, namely the null energy condition, which we will discuss later.

We want to find a more convenient way to re-express this condition in the case of a perfect fluid with stress-energy tensor given in Eq.~(\ref{Eq:StressEnergyPefectFluid})
\be\label{Eq:AppendixSEC:fluid}
T_{\mu\nu}=(\rho+p)U_{\mu}U_{\nu}+p~g_{\mu\nu}~.
\ee
$U_{\mu}$ is the four-velocity of the fluid and satisfies $U_{\mu}U^{\mu}=-1$. A generic time-like four-vector $W^{\mu}$ can be expressed as follows
\be
W^{\mu}=\alpha U^{\mu}+\beta V^{\mu}~,
\ee
where $\alpha$, $\beta$ are two real parameters and $V^{\mu}$ is a normalised spatial four-vector, thus satisfying $V_{\mu}V^{\mu}=1$. By construction we have
\be\label{Eq:AppendixSEC:ConditionsDecomposition}
U^{\mu}V_{\mu}=0,\hspace{1em} |\alpha|>|\beta|~.
\ee
Given the isotropy of the fluid, see Eq.~(\ref{Eq:AppendixSEC:fluid}), $V^{\mu}$ is an eigenstate of the stress-energy tensor with eigenvalues given by the fluid pressure
\be
T_{\mu\nu}V^{\nu}=p~V_{\mu}~.
\ee
Explicit evaluation of the l.h.s. of the inequality (\ref{Eq:AppendixSEC:sec}) yields
\be\label{Eq:AppendixSEC:QuadraticForm}
\frac{1}{2}(\rho+3 p)\alpha^2+\frac{1}{2}(\rho-p)\beta^2\geq0~,
\ee
which is true if and only if
\be\label{Eq:AppendixSEC:SetInequalitiesI}
\rho+3 p\geq0~, \hspace{1em} (\rho-p)\geq0~.
\ee
Hence, the last two inequalities are equivalent to SEC. They give strong bounds on the pressure. For a fluid satisfying the equation of state
\be
p=w~\rho~,
\ee
one gets the following range for the equation of state parameter $w$
\be
-\frac{1}{3}\leq w\leq 1~.
\ee
This is satisfied for instance by dust and radiation but is clearly violated in other cases, such as \emph{e.g.} vacuum energy ($w=-1$), ekpyrotic terms ($w\gg1$) and scalar fields (\emph{e.g.} quintessence or the inflaton).

The inequality (\ref{Eq:AppendixSEC:QuadraticForm}) holds inside the light-cone, where $|\alpha|>|\beta|$. By continuity, it also holds on the boundary of the light-cone, \emph{i.e.} for null four-vectors. Hence, (\ref{Eq:AppendixSEC:QuadraticForm}) is true as long as the non-sharp inequality $|\alpha|\geq|\beta|$ is satisfied. This can be used to find the upper bound of the l.h.s. of the inequality (\ref{Eq:AppendixSEC:QuadraticForm}). In fact, we have on the light-cone (\emph{i.e.} for $\alpha=\beta$)
\be
(\rho+p)\alpha^2\geq0~,
\ee
which implies
\be\label{Eq:AppendixSEC:InequalityDensityPlusPressure}
\rho+p\geq 0~.
\ee
The last inequality can also be obtained by trivial algebraic manipulations of the two inequalities (\ref{Eq:AppendixSEC:SetInequalitiesI}).
Moreover, we note that (\ref{Eq:AppendixSEC:InequalityDensityPlusPressure}) and the second of (\ref{Eq:AppendixSEC:SetInequalitiesI}) together imply positivity of the energy density
\be
\rho\geq0~.
\ee

The importance of SEC for Cosmology can be seen by looking at the second Friedmann equation, Eq.~(\ref{Eq:AccelerationSecondFriedmann})
\be
\frac{\ddot{a}}{a}=-\frac{4\pi G}{3}(\rho+3 p)~.
\ee
From the second inequality (\ref{Eq:AppendixSEC:SetInequalitiesI}) we see that the acceleration in the expansion of the Universe is always negative as long as SEC is satisfied. This is indeed the case for ordinary types of matter, including dust and radiation. However, SEC does not hold in general, \emph{i.e.} for any type of matter. In fact, it is fairly easy to violate it by considering a matter field with self-interactions or non-minimal coupling to gravity, see \emph{e.g.} \cite{Hawking:1973uf,Visser:1999de}. Violations of SEC are crucial in Cosmology, since they
provide a natural way to accommodate for accelerated expansion within the framework of classical General Relativity, as in inflationary models.

At this point, we would like to discuss how SEC can be violated using a scalar field. The energy density and pressure of a scalar field $\phi$ are given by Eqs.~(\ref{Eq:ScalarFieldEnergyDensity}) and (\ref{Eq:ScalarFieldPressure}), respectively
\begin{align}
\rho_\phi&=\left(\frac{1}{2}\dot{\phi}^2+V(\phi)\right)~,\\
p_\phi&=\left(\frac{1}{2}\dot{\phi}^2-V(\phi)\right)~.
\end{align}
Even if the potential is positive, which ensures $\rho_\phi\geq0$, it is possible to accommodate for a negative pressure which is large enough so as to violate the first inequality in (\ref{Eq:AppendixSEC:SetInequalitiesI}). A sufficient condition for SEC violation is
\be
V(\phi)>\dot{\phi}^2~.
\ee
This is indeed the case in slow-roll inflation, where the potential of the scalar field is assumed to be dominant over its kinetic energy.

A weaker condition than SEC is given by the null energy condition (NEC), which states that the inequality (\ref{Eq:AppendixSEC:sec}) only holds on the light-cone. Hence, we have
\be
T_{\mu\nu}K^{\mu}K^{\nu}\geq0~,
\ee
for any null four-vector $K_\mu$. In the case of a perfect fluid, this is equivalent to the following inequality
\be
\rho+p\geq0~.
\ee
NEC is one of the hypothesis in Penrose's singularity theorem \cite{Penrose:1964wq} and also plays a role in Cosmology. More specifically, this is the relevant condition which needs to be violated in GR for cosmological singularity resolution, see Ref.~\cite{Rubakov:2014jja}. This is best seen by looking at the evolution equation for the Hubble rate $H=\frac{\dot{a}}{a}$, which can be obtained combining the two Friedmann equations
\be
\dot{H}=-4\pi G (\rho+p)+\frac{K}{a^2}~.
\ee
If NEC is violated, the Hubble rate can increase\footnote{The Hubble rate is decreasing in the open (K=-1) and flat (K=0) case, for matter satisfying NEC. The case of a closed Universe (K=1) is more delicate and deserves separate discussion, see Ref.~\cite{Rubakov:2014jja}.}. This is accompanied by an increase in the energy density, according to the continuity equation
\be
\frac{\de \rho}{\de t}=-(\rho+p)H~.
\ee


\chapter{Dirac's formalism for constrained Hamiltonian systems}\label{Appendix:Dirac}
Many interesting physical systems which admit a Lagrangian description are characterised by the existence of functional relations between the canonical variables, called \emph{constraints}. In fact, this is typical of all fundamental interactions, including gravity. In these cases the Legendre transformation is not invertible, since the Hessian of the Lagrangian with respect to the velocities is degenerate. Therefore, it is not possible to construct a Hamiltonian function following the usual procedure. It is nevertheless possible to provide a canonical formulation of the dynamics by appropriately taking into account the constraints. This can be done by following Dirac's algorithm, which we will briefly outline in this appendix following Refs.~\cite{dirac1964lectures,Henneaux:1992ig} closely\footnote{The reader is referred to those references for technical details which we will omit for brevity.}. For a more complete treatment we refer the reader to the many excellent reviews on the subject, \emph{e.g.} Refs.~\cite{Henneaux:1992ig,Sundermeyer:1982gv,Wipf:1993xg,hanson1976constrained}. For simplicity, we will illustrate the procedure in the case of a system with a finite number of degrees of freedom, although it is possible to generalise it to the case of an infinite (continuous) number of degrees of freedom (fields). This is actually the case which is relevant for most applications in theoretical physics. We will consider some physically relevant examples that show how the procedure works in the case of generally covariant systems.

\section*{Canonical formulation of a system with constraints}
Let us consider a physical theory with Lagrangian $L(q^i,p_i)$ and canonical momenta\footnote{We remark that the `time' used in the definition of the velocity and that also appears in the action functional $S=\int\de t\; L$ is not necessarily a coordinate time. Even though this is a possible choice even in a relativistic theory, it is by no means the only one. In fact, we will consider in this appendix several examples in which the time parameter is devoid of any physical meaning.} 
\be
p_i=\frac{\pa L}{\pa \dot{q}^i}~.
\ee
We assume that there is a set of functionally independent constraints\footnote{These are subject to some regularity conditions, see Ref.~\cite{Henneaux:1992ig}.}
\be\label{Eq:Dirac:PrimaryConstraints}
\phi_m(q,p)=0~, \hspace{1em} m=1,\dots,M.
\ee
These are called \emph{primary constraints}. The \emph{canonical Hamiltonian} is defined as
\be
H_c=p_i\dot{q}^i-L.
\ee
Its variation depends only on the variation of the position and that of the momenta, and is independent from the variation of the velocities. This implies that $H_c=H_c(q^i,p_i)$ does not depend on the velocities. However, this is not enough for a constrained system. In fact, the canonical Hamiltonian does not take into account the fact that some of the momenta can be expressed in terms of the remaining ones by means of the primary constraints\footnote{As it turns out, the canonical Hamiltonian is identically vanishing for systems that are reparametrisation invariant. See the examples below in this appendix.} (\ref{Eq:Dirac:PrimaryConstraints}). We are thus led to introduce a new Hamiltonian, which includes a linear combination of the primary constraints
\be\label{Eq:Dirac:PrymaryH}
H^*=H_c+u_m\phi_m~.
\ee
The coefficients $u_m$ are to be treated as Lagrange multipliers and are arbitrary functions of time (as well as of $q$ and $p$). The inclusion of the primary constraints in the Hamiltonian makes the Legendre transformation invertible. The Hamiltonian equations of motion obtained from Eq.~(\ref{Eq:Dirac:PrymaryH}) read as
\begin{align}\label{Eq:Dirac:EOM}
\dot{q}^i&=\frac{\pa H_c}{\pa p_i}+u_m\frac{\pa\phi_m}{\pa p_i}~,\\
\dot{p}_i&=-\frac{\pa H_c}{\pa q^i}-u_m\frac{\pa\phi_m}{\pa q^i}~.
\end{align}

The time derivative of a function $f$ defined on phase-space is given by its Poisson bracket with the Hamiltonian $H^*$ in Eq.~(\ref{Eq:Dirac:PrymaryH})
\be
\{f,H^*\}=\{f,H_c\}+u_m\{f,\phi_m\}~.
\ee
We remark that constraints must be imposed only \emph{after} Poisson brackets are computed. In particular, this applies to the derivation of the equations of motion (\ref{Eq:Dirac:EOM}). In this sense, the constraint equations are understood with the sign of \emph{weak equality}
\be
\phi_m(q,p)\approx0~.
\ee
The constraints define a hypersurface in phase space, known as the \emph{constraint hypersurface}. Such hypersurface must be stable under time evolution. This leads us to impose the requirement that the time derivative of primary constraints must (weakly) vanish
\be
0\approx\dot{\phi}_m=\{ \phi_m,H_c\}+u_n\{ \phi_m, \phi_n\}~.
\ee
Such consistency conditions can either lead to relations that are independent of the $u$'s (\emph{i.e.} a new constraint), or impose a restriction on the $u$'s (see Refs.~\cite{Henneaux:1992ig,dirac1964lectures} for details). The new constraints generated in this way are called \emph{secondary constraints} and will in turn lead to new consistency conditions. The procedure outlined above must be iterated until new secondary constraints or restrictions on the $u$'s can no longer be generated. The set of secondary constraints will be denoted by
\be
\phi_k\approx0~,\hspace{1em} k=1,\dots,K~.
\ee
Although they arise in different ways from a Lagrangian point of view, the distinction between primary and secondary constraints is not relevant for our purposes and they will be treated on the same footing in the final form of the theory. Hence we will denote them collectively using a uniform notation
\be\label{Eq:Dirac:FullSetConstraints}
\phi_j\approx0~,\hspace{1em} j=1,\dots, M+K\equiv J~.
\ee

Assuming that Eq.~(\ref{Eq:Dirac:FullSetConstraints}) specifies a complete set of constraints, \emph{i.e.} it does not lead to any further secondary constraints, the consistency conditions between them lead to restrictions on the Lagrange multipliers $u_n$. We have the inhomogeneous linear system
\be\label{Eq:Dirac:System}
\{ \phi_m,H_c\}+u_n\{ \phi_m, \phi_n\}\approx0~,
\ee
where the $u_n$ must be regarded as the unknowns.
Provided the system is compatible (otherwise the dynamics would be inconsistent), the solution is given by
\be
u_m=U_m+V_m~,
\ee
where $U_m$ is a particular solution of the inhomogeneous system and $V_m$ represents the general solution to the associated homogeneous system
\be
V_n\{ \phi_m, \phi_n\}\approx0~.
\ee
This is expressed as a linear combination of linearly independent solutions $V_m=v_a V_{am}$, with $a=1,\dots,A=J-r$, where $r$ is the rank of the homogeneous system\footnote{This does not vary moving along the constraint hypersurface if we assume that the rank of $\{ \phi_m, \phi_n\}$ is a constant there.}. Thus, we have
\be\label{Eq:Dirac:SolutionU}
u_m\approx U_m+v_a V_{am}~.
\ee

We can now substitute the solution (\ref{Eq:Dirac:SolutionU}) into the Hamiltonian (\ref{Eq:Dirac:PrymaryH}). This will give us the \emph{total Hamiltonian}, which includes the primary constraints \emph{and} the consistency conditions 
\be\label{Eq:Dirac:TotalHamiltonian}
H_{T}=H^{\prime}+v_a\phi_a~.
\ee
The two terms in Eq.~(\ref{Eq:Dirac:TotalHamiltonian}) respectively include the contributions to $u_m$ coming from the consistency conditions and those that instead remain arbitrary
\be
H^{\prime}=H_c+U_m\phi_m~,
\ee
\be
\phi_a=V_{a\,m}\phi_m
\ee
The equations of motion obtained from the total Hamiltonian (\ref{Eq:Dirac:TotalHamiltonian}) are by construction equivalent to those obtained from the variational principle in the Lagrangian formulation.

There is another classification of constraints, that is physically more important than the one in primary and secondary constraints. We define a function $R$ (not necessarily a constraint) to be \emph{first class} if it has (at least weakly) vanishing Poisson brackets with all of the other constraints
\be\label{Eq:Dirac:FirstClass}
\{R,\phi_j\}\approx0~,\hspace{1em} j=1,\dots, J~.
\ee
The Poisson bracket of two first class constraints is also first class. Hence, it is \emph{strongly} equal to a linear combination of first class constraints
\be\label{Eq:Dirac:1stClassConstraintsAlgebra}
\{\phi_i,\phi_j\}=f^k_{ij}\phi_k~.
\ee
Equation~(\ref{Eq:Dirac:1stClassConstraintsAlgebra}) shows that first class constraints close an algebra. Notice that the coefficients $f^k_{ij}$ are \emph{a priori} functions of $q$ and $p$.
If the condition (\ref{Eq:Dirac:FirstClass}) is not satisfied, \emph{i.e.} there is at least one constraint such that Eq.~(\ref{Eq:Dirac:FirstClass}) is not satisfied, the constraint is called \emph{second class}. 

It is possible to show that $H^{\prime}$ is first class. The primary constraints $\phi_a$ are also first class. It is worth pointing out that the splitting of $H_T$ (\ref{Eq:Dirac:TotalHamiltonian}) into $H^{\prime}$ and a linear combination of primary constraints with arbitrary coefficients $v_a\phi_a$ is also arbitrary. In fact, the quantities $U_m$ can be any solution of the inhomogeneous system (\ref{Eq:Dirac:System}), \emph{i.e.} they are defined up to a solution of the associated homogeneous system.

The importance of first class constraints lies in the fact that \emph{first class primary constraints} can be identified with generators of gauge transformations, \emph{i.e.} they do not change the physical state. In order to show this, let us consider a dynamical variable $F$. Considering the evolution generated by the total Hamiltonian from $t$ to $t+\de t$ and a variation $\delta v_a$ of the coefficients, one has
\be
\delta F=\delta v_a\{F,\phi_a\}~.
\ee
Since the coefficients $v_a$ are entirely arbitrary, states related by such transformation correspond to the same physical state. Dirac conjectured that \emph{all} first class constraints (\emph{i.e.} including secondary ones) are generators of gauge transformations\footnote{It is actually possible to construct counterexamples to such conjecture, see Ref.~\cite{Henneaux:1992ig}. However, the conjecture holds for all physically relevant systems that have been studied so far.}. In the quantisation of constrained systems all first class constraints are treated on the same footing, \emph{i.e.} Dirac's conjecture is assumed to hold.

It is possible to define the \emph{extended Hamiltonian}, that also accounts for the gauge degrees of freedom corresponding to the secondary first class constraints
\be
H_E=H^{\prime}+ u_a \gamma_a~,
\ee
with the index $a$ running over a complete set of first class constraints, here denoted by $\gamma_a$. It is worth observing that, strictly speaking, only the total Hamiltonian $H_T$ follows directly from the Lagrangian. In fact, the extended Hamiltonian $H_E$ introduces more arbitrary functions of time. The introduction of $H_E$ is nevertheless natural from the canonical point of view, since it allows to treat all of the gauge degrees of freedom on the same footing. The dynamics generated by the three Hamiltonian function $H^{\prime}$, $H_T$ and $H_E$ are of course equivalent, \emph{i.e.} they are the same up to gauge transformations.

Second class constraints deserve a separate discussion and require the introduction of a new mathematical object, known as the Dirac bracket. Since the systems we will be dealing with only have first class constraints, we will not discuss this topic and refer the interested reader to the references given above.

\section*{Examples}
In this section we wish to give two basic examples, drawn from Ref.~\cite{Kiefer:2007ria}, illustrating the canonical formulation of familiar systems obtained using the Dirac algorithm. Both examples share one of the most important features of the dynamical theory of the gravitational field, as given by GR, which is that of being reparametrisation invariant. An important consequence of this is that the Hamiltonian theory turns out to be \emph{fully constrained}.

The simplest example of a constrained system is the relativistic particle, with action
\be\label{Eq:Dirac:FreeParticle}
S=-m\int\de t\; \sqrt{-\dot{x}_\mu\dot{x}^\mu}~.
\ee
The integration is over the world-line of the particle.
The parameter $t$ is entirely arbitrary and shall not be confused with a coordinate time. In fact, the action (\ref{Eq:Dirac:FreeParticle}) is invariant under time reparametrisations $t\to f(t)$. The canonical momenta are
\be
p_\mu=m\frac{\dot{x}_\mu}{\sqrt{-\dot{x}_\mu\dot{x}^\mu}}~.
\ee
The Lagrangian is a homogenous function of the velocities of degree one, which implies that the canonical Hamiltonian $H_c=p_\mu\dot{x}^\mu-L$
is identically vanishing.
Following the above discussion, this must be understood in the sense of a weak equality in phase-space
\be\label{Eq:Dirac:FreeParticle}
H_c=p_\mu p^\mu+m^2\approx0~.
\ee
Thus, the total Hamiltonian is
\be\label{Eq:Dirac:TotalHamiltonianFreeParticle}
H_T=NH_c~.
\ee
Equation~(\ref{Eq:Dirac:TotalHamiltonianFreeParticle}) shows that $H_T$ is a primary first class constraint. Since $H_T$ is the generator of time evolution, we conclude that evolution in the parameter $t$ is actually a gauge transformation, \emph{i.e.} it does not change the physical state. Since $H^{\prime}=0$, the system is said to be \emph{fully constrained}. This is typical of reparametrisation invariant systems.

Another example is offered by the bosonic string, \emph{i.e.} a one-dimensional object moving in a higher-dimensional (flat) spacetime. Its dynamics is given by the Nambu-Goto action
\be
S=-\frac{1}{2\pi\alpha^{\prime}}\int\de^2\sigma\;\sqrt{\left|\det G_{\alpha\beta}\right|}~.
\ee
Here $\alpha^\prime$ is the Regge slope and $G_{\alpha\beta}$ the induced metric on the world-sheet\footnote{The world-sheet is the two-dimensional submanifold spanned by the string in its motion.}, parametrised by $\sigma^\alpha$
\be
G_{\alpha\beta}=\eta_{\mu\nu}\frac{\pa X^{\mu}}{\pa\sigma^\alpha}\frac{\pa X^{\nu}}{\pa\sigma^\beta}~.
\ee
The Nambu-Goto action is invariant under world-sheet reparametrisations $\sigma^\alpha\to f^\alpha(\sigma^\beta)$. The \emph{Hamiltonian density} is given by
\be
\mathcal{H}=N\mathcal{H}_\perp+N^1\mathcal{H}_1~.
\ee
The two functions $N$, $N_1$ are Lagrange multipliers and one has the following first class constraints\footnote{$\sigma^1$ denotes the space-like coordinate on the world-sheet.}
\begin{align}
\mathcal{H}_\perp&=\frac{1}{2}\left(P^2+\frac{(\pa_{\sigma^1}X)^2}{4\pi^2(\alpha^\prime)^2}\right)\approx0~,\\
\mathcal{H}_1&=P_\mu \pa_{\sigma^1}X^{\mu}\approx 0~.
\end{align}
We can construct the smeared generators of deformations
\begin{align}
H_\perp[N]&=\int \de\sigma^1\; N \mathcal{H}_\perp~,\\
H_1[N^1]&=\int \de\sigma^1\; N^1 \mathcal{H}_1~.
\end{align}
They generate $\sigma^0$ and $\sigma^1$ reparametrisations, respectively. They close the following algebra of world-sheet deformations (cf. Refs.~\cite{Marnelius:1982bz,Niemi:1986kc,Govaerts:1988ch,1984qtg..book..327T})
\begin{align}\label{Eq:Dirac:DeformationAlgebra}
\{H_\perp[N],H_\perp[\tilde{N}]\}&=H_1[N\partial_{\sigma^1}\tilde{N}-\tilde{N}\partial_{\sigma^1}N]~,\\
\{H_\perp[N],H_1[N^1]\}&=H_\perp[N\partial_{\sigma^1}N^1-N^1\partial_{\sigma^1}N]~,\\
\{H_1[N^1],H_1[\tilde{N_1}]\}&=H_1[N^1\partial_{\sigma^1}\tilde{N^1}-\tilde{N^1}\partial_{\sigma^1}N^1]~.
\end{align}
This is a realisation of the Lie algebra of the diffeomorphism group in 1+1 dimensions.

\section*{Quantisation}
We will discuss the Dirac quantisation scheme for a system with first class constraints and assume that there are no second class constraints\footnote{Second class constraints play nonetheless an important role in some physical situtations. In fact, they naturally arise when a gauge fixing is introduced \cite{Henneaux:1992ig}. Second class constraints also occur in the case of the superstring with spacetime supersymmetry \cite{Green:1987sp} and in the case of gravitational theories including torsion \cite{Esposito:1992xz}.}.
 For the more general case, the reader is referred to Refs.~\cite{dirac1964lectures,Henneaux:1992ig}.
 
 One starts off with a kinematical Hilbert space $H_{\rm kin}$. The first class constraints are then imposed as restrictions on the physically allowed wave functions \cite{Kiefer:2007ria}. That is, considering a set of classical (first class) constraints $\phi_a(q,p)\approx0$ and a quantisation map, we require 
 \be\label{Eq:Dirac:Quantization}
 \hat{\phi}_a\psi=0~.
 \ee
 The set of solutions of the constraint equations (\ref{Eq:Dirac:Quantization}) defines the physical Hilbert space $H_{\rm phys}$. Notice that consistency of the quantisation scheme demands that \emph{all} of the first class constraints (\emph{i.e.} both primary and secondary) must be treated on the same footing \cite{Henneaux:1992ig}.
 
 Applying the Dirac quantisation procedure to the free particle (see above), and using the standard quantisation rule $p_\mu\to -i \pa_\mu$ (setting $\hbar=1$) we obtain
 \be
 0=\hat{H}_T \psi=(- \pa_\mu\pa^\mu+m^2)\psi~,
 \ee
 which is the Klein-Gordon equation. The quantisation of the bosonic string is more delicate due to the presence of anomalies \cite{1984qtg..book..327T}.


\chapter{Canonical formalism for classical GR: the ADM action}\label{Appendix:ADM}
In this Appendix we briefly review the ADM formulation of GR, following Ref.~\cite{Kiefer:2007ria}. The ADM formalism made its first appearance in the seminal paper~\cite{Arnowitt:1962hi}\footnote{Ref.~\cite{Arnowitt:1962hi} is an unretouched republication.}.
In order to construct a canonical formulation for a dynamical theory of the gravitational field, such as GR, we need to introduce a foliation of spacetime. Thus, we assume that the spacetime ($\mathcal{M}$, $g$) is globally hyperbolic\footnote{For the definition of global hyperbolicity of a spacetime see \emph{e.g.} Ref.~\cite{Hawking:1973uf}.}; therefore, it has topology $\mathcal{M}\simeq\mathbb{R}\times\Sigma$, where $\Sigma$ is a Cauchy surface. The foliation corresponds to the existence of a global time function $t$, with associated vector field $t^\mu$ satisfying $t^\mu\nabla_\mu t=1$. The induced metric on the space-like sheets $\Sigma_t$ of the foliation is
\be
h_{\mu\nu}=g_{\mu\nu}+n_\mu n_\nu~,
\ee
where $n_\mu$ is a hypersurface orthogonal vector, normalised to $n^\mu n_\mu=-1$. 

We introduce the following decomposition of $t^{\mu}$ into its normal and tangential components to $\Sigma_t$
\be
t^{\mu}=N n^\mu+N^\mu~.
\ee
$N$ is called the lapse function, whereas $N^\mu$ is the shift vector and it is tangent to $\Sigma_t$. Hence, the metric reads as
\be
\de s^2=-(N^2-N_i N^i)\de t^2+2 N_i \de t \de x^i+h_{ij}\de x^i\de x^j~.
\ee
By means of such 3+1 decomposition and performing a Dirac analysis of the constraints, one finds that the Einstein-Hilbert action of GR leads to the Hamiltonian
\be\label{Eq:ADM:ADMHamiltonian}
H=\int\de^3x\;\mathcal{H}=\int\de^3x\; \left(N\mathcal{H}_\perp+N^i\mathcal{H}_i\right)~.
\ee
The constraints have the following expressions
\begin{align}
\mathcal{H}_\perp&=16\pi G~ G_{ijhk}p^{ij}p^{hk}-\frac{\sqrt{h}}{16\pi G}\left(\hspace{-6pt}\ph^{(3)}R-2\Lambda\right)\approx0\label{Eq:ADM:HamiltonianConstraint}~,\\
\mathcal{H}_i&=-2D_jp_i^j\approx0 \label{Eq:ADM:DiffeomorphismConstraint}~.
\end{align}
$\mathcal{H}_\perp$ and $\mathcal{H}_i$ are known as the Hamiltonian (or scalar) and diffeomorphism constraints, respectively.
$\hspace{-6pt}\ph^{(3)}R$ is the curvature of the three-dimensional spatial sheet and $D_i$ denotes the connection that is compatible with $h_{ij}$. We also have the DeWitt supermetric
\be\label{Eq:ADM:SuperMetric}
G_{ijhk}=\frac{1}{2\sqrt{h}}\left(h_{ih}h_{jk}+h_{ik}h_{jh}-h_{ij}h_{hk}\right)~.
\ee
This is a metric on a six-dimensional space, known as superspace. The DeWitt supermetric $G_{ijhk}$ has signature $(-+++++)$, which makes the geometry of superspace hyperbolic. Note that the signature of $G_{ijhk}$ is not related to that of the spacetime metric. In fact, it has the same expression in the Lorentzian and in the Euclidean case, given by Eq.~(\ref{Eq:ADM:SuperMetric}). However, the signature of spacetime makes its appearance in the Hamiltonian constraint (\ref{Eq:ADM:HamiltonianConstraint}) through the sign in front of the second term (`potential term').
 
The canonical momenta have the following expression in terms of the extrinsic curvature of $\Sigma_t$
\be
p^{ij}=\frac{1}{16\pi G}G^{ijhk}K_{hk}=\frac{\sqrt{h}}{16\pi G}\left(K^{ij}-Kh^{ij}\right)~.
\ee
The action can be written in canonical form as
\be\label{Eq:Dirac:ADMAction}
S=\frac{1}{16\pi G}\int\de t\de^3 x\; \left(p^{ij}\dot{h}_{ij}-N\mathcal{H}_\perp+N^i\mathcal{H}_i\right)~.
\ee
Eq.~(\ref{Eq:Dirac:ADMAction}) gives the canonical ADM action.

In complete analogy with the free relativistic particle and with the bosonic string, considered in Appendix~\ref{Appendix:Dirac}, GR is also a fully constrained system, \emph{i.e.} its Hamiltonian is a linear combination of first class constraints\footnote{We remark that, in constrast to the full theory, in linearised General Relativity the Hamiltonian is no longer a linear combination of constraints  \cite{Arnowitt:1959lin,Hinterbichler:2011tt,deCesare:2017xhv}. In fact, the system also has a `true' (\emph{i.e.} non-vanishing) Hamiltonian term, which is obtained from the ADM Hamiltonian after solving the constraints. The existence of a true Hamiltonian originates from the loss of background independence, which is due to the expansion of the metric around a fixed background (\emph{e.g.} Minkowski spacetime). The relation between the canonical formulation of the full theory and that of the linearised one is clarified in Ref.~\cite{Arnowitt:1959ah}.}
. This is the counterpart in the canonical formalism of diffeomorphism invariance of the Einstein-Hilbert action. In fact, the constraints close an algebra known as the \emph{hypersurface deformation algebra}. Let us consider the smeared version of the algebra generators
\begin{align}
H[N]&=\int_{\Sigma} \de^3x\; N \mathcal{H}_\perp~,\\
H[N^i]&=\int_{\Sigma} \de^3x\; N^i \mathcal{H}_i~.
\end{align}
The Poisson brackets of the generators are
\begin{align}
\{H[N],H[M]\}&=H[K^i]~,\hspace{1em} K^i=-h^{ij}(NM_{,i}-N_{,i}M)~,\label{Eq:ADM:DeformationAlgebra1}\\
\{H[N^i],H[N]\}&=H[M]~,\hspace{1em} M=N^iN_{,i}=\mathcal{L}_{\bf N}N~,\label{Eq:ADM:DeformationAlgebra2}\\
\{H[N^i],H[M^i]\}&=H[K^i]~,\hspace{1em} {\bf K}=[{\bf N},{\bf M}]=\mathcal{L}_{\bf N}M~.\label{Eq:ADM:DeformationAlgebra3}
\end{align}
The two-dimensional case is equivalent to the algebra of first constraints of the bosonic string \cite{Kiefer:2007ria,1984qtg..book..327T}. We observe that, with the exception of this case, the above algebra is not a Lie algebra. In fact, the Poisson brackets involve structure functions rather than structure constants as it can be seen from Eq.~(\ref{Eq:ADM:DeformationAlgebra1}), where the inverse metric $h^{ij}$ explictly appears in the definition of $K^i$. Such algebraic structures have been studied by Bergmann and Komar \cite{Bergmann:1981fc,komar1983generalized,bergmann1972coordinate} (see also Ref.~\cite{Thiemann:2007zz} and references therein), and the group they generate is known as the Bergmann-Komar group BK($\mathcal{M}$). This is to be contrasted with the fact that the spacetime diffeomorphism group $\mbox{Diff}(\mathcal{M})$ is a Lie algebra. In fact, the transformations generated by the algebra (\ref{Eq:ADM:DeformationAlgebra1})--(\ref{Eq:ADM:DeformationAlgebra3}) generate infinitesimal diffeomorphism only when the equations of motion hold, \emph{i.e.} on-shell \cite{Thiemann:2007zz}. It is worth pointing out that the difference between the two groups is a physical one and, since it manifests itself off-shell, is important for the quantum theory. Actually, $\mbox{Diff}(\mathcal{M})$ can be regarded as a kinematical symmetry, since it is shared by any generally covariant theory. On the other hand, the structure of BK($\mathcal{M}$) depends on the particular physical theory one considers, \emph{i.e.} it is a dynamical symmetry group \cite{Thiemann:2007zz}. Ref.~\cite{Kiefer:2007ria} gives a clear geometric interpretation of the algebra of BK($\mathcal{M}$) as generating infinitesimal deformations of a hypersurface embedded in spacetime. From this point of view, it is clear that such deformations form a larger class of transformations than $\mbox{Diff}(\mathcal{M})$. The way in which GR can be recovered from the BK($\mathcal{M}$) by means of the `principle of path-independence' for embedded hypersurfaces represents the so-called `seventh route to geometrodynamics', for which we refer the reader to Refs.~\cite{Kiefer:2007ria,Hojman:1976vp}.


\chapter{Loop Quantum Cosmology}\label{Appendix:LQC}
Loop Quantum Cosmology (LQC) is a symmetry reduced version of Loop Quantum Gravity (LQG), which is an alternative canonical approach based on different variables from the ones used in geometrodynamics\footnote{It would be beyond our purposes to provide the reader with an exhaustive review of the formulation of LQG and its recent developments. The interested reader is referred to Ref.~\cite{Dona:2010hm} for an introduction and to Ref.~\cite{Thiemann:2007zz} for a comprehensive and very detailed review of the subject.} that allow to reformulate GR using a language similar to that of Yang-Mills theories. We give below a very brief overview of the LQG programme, before turning our attention to LQC.

\section{LQG in a nutshell}
The basic variables in the phase space of LQG are a non-Abelian gauge connection
 $A^\alpha_i$ and a densitized `electric field' $E^i_\alpha$  which satisfy the algebra ($\{~,~\}$ denotes the Poisson bracket)
\be\label{Eq:LQC:PBalgebra}
\{A^\alpha_i,E^j_\beta\}=8\pi G\beta \delta^\alpha_\beta \delta({\bf x}, {\bf y})~,\hspace{1em} \{A^\alpha_i,A^\beta_j\}=\{E^i_\alpha,E^j_\beta\}=0~,
\ee
where $\beta$ is known as the Barbero-Immirzi parameter\footnote{The \emph{new variables} for QG where introduced by Ashtekar in Ref.~\cite{Ashtekar:1986yd}.  Originally the value of the Barbero-Immirzi parameter was chosen as $\beta=\pm i$ so that the Ashtekar connection $A^\alpha_i$ coincides with the pull-back on spatial slices of a spacetime spin-connection. However, when $\beta$ is complex, reality conditions must be imposed, which are difficult to implement at the quantum level \cite{Thiemann:2007zz}. This is the reason for preferring a real-valued $\beta$ in the formulation of the quantum theory.}.
The generators carry internal indices (here denoted by Greek letters) labelling components in an internal $\rm su(2)$ algebra. In fact, a local $\mbox{SU}(2)$ symmetry is introduced in the formalism, corresponding to the invariance under local rotations of the tetrad. As a consequence of this extra symmetry, a new constraint arises in the canonical formulation of the theory \emph{\`a la} Dirac. This is known as the Gauss constraint $G^\alpha\approx0$, which is also familiar from ordinary Yang-Mills theories. Together with the Hamiltonian and diffeomorphism constraints, it spans the algebra of first class constraints of the theory. A regular Poisson algebra can be obtained from (\ref{Eq:LQC:PBalgebra}) by smearing the generators, \emph{i.e.} by integrating them over paths and surfaces embedded in three-space, respectively. The algebra thus obtained is known as the \emph{holonomy-flux algebra}.

The Dirac quantisation procedure can then be applied. The kinematical Hilbert space of the theory is the space of \emph{cylindrical functions}. A cylindrical function(al) $\psi_{(\Gamma,f)}[A]$ is defined by introducing the pair $(\Gamma,f)$, where $\Gamma$ is a graph with $L$ links and $f:\mbox{SU}(2)^{ L} \to \mathbb{C}$ \cite{Dona:2010hm}. The space of cylindrical functions on a fixed graph $\Gamma$ is equipped with the inner product
\be
\langle \psi_{(\Gamma,f)}[A],\psi_{(\Gamma,f^\prime)}[A]\rangle=\int\prod_e \de h_e\; \overline{f(h_{e_1},\dots,h_{e_L})}f^\prime(h_{e_1},\dots,h_{e_L})~,
\ee
which turns it into a Hilbert space $\mathcal{H}_{\Gamma}$.
Here the variables $h_{e_i}$ denote holonomies of the connection over the corresponding $i$-th link and $\prod_e \de h_e$ is the Haar measure on $\mbox{SU}(2)^{ L}$. The kinematical Hilbert space is then defined as the direct sum of such Hilbert spaces, over all possible graphs $\Gamma$
\be
\mathcal{H}_{\rm kin}=\bigoplus_\Gamma \mathcal{H}_{\Gamma}~.
\ee
The inner product defined above has a natural extension to $\mathcal{H}_{\Gamma}$. The Gauss constraint $\hat{G}^{\alpha}=0$ is then imposed on $\mathcal{H}_{\rm kin}$. As a result, cylindrical functions satisfying the Gauss constraint must be gauge-invariant (with respect to the internal $\mbox{SU}(2)$) at each node. Such states are called \emph{spin-networks}. The subsequent imposition of the diffeomorphism constraint $\hat{H}_a=0$ makes the spin-networks independent from the embedding in a continuum space. The dynamics of spin-networks obtained by imposing the Hamiltonian constraint is a much more delicate issue and will not be discussed here. We refer the reader to Ref.~\cite{Thiemann:2007zz} for details.

\section{A bird's eye view on (homogeneous) LQC}
The formulation, and in particular the quantisation procedure of LQC are inspired from that of LQG. However, LQC is not, strictly speaking, derived from the full theory. In fact, the relation between the two is currently under investigation\footnote{\label{Footnote:QRLG}Interesting results come from the approach known as Quantum Reduced Loop Gravity (QRLG). In this approach, the cosmological sector is recovered from the full theory by implementing gauge fixing conditions that restrict the metric and the triad to a diagonal form \cite{Alesci:2012md,Alesci:2013xya,Alesci:2015nja,Alesci:2016gub}. The kinematical Hilbert space of LQG is thus truncated to spin networks based on a cubic lattice. The framework can accommodate anisotropic as well as well as inhomogeneous field configurations. It also provides an explanation for the different regularisation schemes adopted in LQC, such as the $\mu_0$ scheme and the (improved) $\bar{\mu}$ scheme (see below), which are obtained in the QRLG framework by summing over graphs \cite{Alesci:2016rmn}.}. 
As in the full theory, also in LQC the basic phase-space variables are represented by an $\mbox{SU}(2)$ gauge connection $A^\alpha_i$ and its conjugate momentum, \emph{i.e.} the densitized triad $E^i_\alpha$. In such symmetry reduced setting, one starts by solving the Gauss and the diffeomorphism constraints. 
A natural question to ask is then how unique is the solution. In order to give an answer to this question, let us start by considering the group of spatial isometries $\mathcal{S}$ corresponding to a given homogeneous foliation of spacetime. It is convenient to introduce the following definition: a pair $(A^\alpha_i,E^i_\alpha)$ is said to be \emph{symmetric} if and only if all isometries $s\in\mathcal{S}$ act on it as gauge transformations \cite{Ashtekar:2003hd}, \emph{i.e.} there exists a group element $g\in\mbox{SU}(2)$ such that\footnote{With a standard notation, $s^*$ denotes the pull-back of $\mathcal{S}$.}
\be
(s^{*}A,~s^{*}E)=(g^{-1}A g+g^{-1}\de g,~g^{-1}E g)~.
\ee
On the gauge orbit of each symmetric pair satisfying the Gauss and diffeomorphism constraints, there is one and only one representative with the following form \cite{Ashtekar:2003hd}
\be\label{Eq:LQC:SymmetricPair}
A=\tilde{c}~\pha{o}\omega^\alpha\tau_\alpha~,\hspace{1em} E=\tilde{p}\sqrt{\pha{o} q}~\pha{o}e_\alpha\tau^\alpha~.
\ee
In Eq.~(\ref{Eq:LQC:SymmetricPair}) we introduced the notation $\pha{o}e_i$, $\pha{o}\omega^i$ to denote the \emph{fiducial} triad and co-triad, respectively. They correspond to a fiducial metric $\pha{o}q_{ab}$. The $\tau_i$'s are the generators of the $\mbox{su}(2)$ algebra
\be
[\tau_\alpha,\tau_\alpha]=\epsilon_{\alpha\beta\gamma}\tau^\gamma~,\hspace{1em} \tau_\alpha=-\frac{i}{2}\sigma_\alpha,
\ee
$\sigma_\alpha$ being the Pauli matrices. In the parametrisation given by Eq.~(\ref{Eq:LQC:SymmetricPair}), all the information about the gravitational field configuration is encoded in the two parameters $\tilde{c}$ and $\tilde{p}$. The symplectic form on the \emph{reduced phase space} is given by
\be
\Omega= \frac{3V_0}{8\pi \beta G}\de\tilde{c}\wedge\de\tilde{p}~.
\ee
$V_0$ denotes the volume of a \emph{fiducial cell}\footnote{In the case of a spatially compact Universe such restriction is superfluous. $V_0$ can then be taken as the comoving volume of the Universe, computed using the fiducial metric.} $\mathcal{V}$ adapted to the fiducial metric $\pha{o}q_{ab}$. The volume of the fiducial cell can be rescaled at will, implying a corresponding rescaling of the variables $\tilde{c}$ and $\tilde{p}$. Physics is not altered by such rescalings\footnote{This situation is entirely analogous to the one encountered in the standard formulation of classical FLRW models and their geometrodynamics formulation. In fact, the freedom of rescaling comoving coordinates implies that the scale factor is defined only up to an arbitrary numerical factor.}. For this reason, it is convenient to introduce physical variables $c$ and $p$ that are unaffected by rescalings of the fiducial metric
\be\label{Eq:LQC:RescaleVariables}
c=V_0^{1/3}\tilde{c}~,\hspace{1em}p=V_0^{2/3}\tilde{p}~.
\ee
 Thus, the sympletic form has the following expression in terms of the new variables.
\be
\Omega= \frac{3}{8\pi \beta G}\de c\wedge\de p~.
\ee
The Hamiltonian constraint of the pure gravity theory is $H=N\int\de^3 x\; \mathcal{C}_{\rm grav}$, with 
\be\label{Eq:LQC:Constraint}
\mathcal{C}_{\rm grav}=-\frac{1}{\beta^2}\epsilon_{\alpha\beta\gamma}\frac{F^{\alpha}_{ij}E^{i\beta}E^{j\gamma}}{\sqrt{\det|E|}}~.
\ee

The Hilbert space of the quantum theory is $L^2(\overline{\mathbb{R}}_{\rm Bohr},\de \mu)$, with $\overline{\mathbb{R}}_{\rm Bohr}$ being the Bohr compactification of the real line and $\de \mu$ its Haar measure \cite{Ashtekar:2003hd}. Notice that this space is not separable. In fact, it is the Cauchy completion of the space of \emph{almost periodic} functions of $c$ with respect to the following inner product \cite{Ashtekar:2003hd}
\be
\langle \e^{i \frac{\mu_1 c}{2}} |\e^{i \frac{\mu_2 c}{2}} \rangle=\delta_{\mu_1,\mu_2}~.
\ee
The construction of the Hilbert space parallels the one of the full theory. In fact, almost periodic functions naturally arise when computing holonomies, which is done by smearing the connection over the edges of the fiducial cell. The Hilbert space representation of LQC is strikingly different from the one used in ordinary QM. In fact, it encodes the existence of an underlying discreteness of spacetime (polymer quantisation). In this representation, $\hat{c}$ is not a well-defined operator, whereas its exponentiated version (the holonomy) $\widehat{\e^{i\mu c/2}}$ is. Its conjugate momentum $p$ is represented in the quantum theory as
\be
\hat{p}=-i\frac{8\pi\beta \ell_{\rm Pl}^2}{3}\frac{\de}{\de c}~.
\ee
Introducing the Dirac bra-ket notation, we can express the basis states of $L^2(\overline{\mathbb{R}}_{\rm Bohr},\de \mu)$ as
\be\label{Eq:LQC:MuStatesDef}
\langle c|\mu\rangle=\e^{i \frac{\mu c}{2}}~.
\ee
The states in Eq.~(\ref{Eq:LQC:MuStatesDef}) are eigenstates of  $\hat{p}$
\be
\hat{p}|\mu\rangle=p_\mu |\mu\rangle~,\hspace{1em} \mbox{with}\hspace{0.5em} p_\mu=\frac{4\pi\beta \ell_{\rm Pl}^2}{3}~.
\ee
On the other hand we find
\be\label{Eq:LQC:HolonomiesAsShifts}
\widehat{\e^{i\mu^{\prime} c/2}}|\mu\rangle=|\mu+\mu^{\prime}\rangle~.
\ee
That is, the holonomies act as discrete shift operators. To complete the discussion of the kinematical framework of the quantum theory we need to identify a complete set of observables. In the homogeneous case, we only need one observable. A particularly convenient choice corresponds to the \emph{volume operator}, defined as
\be
\hat{V}=\widehat{|p|}^{3/2}~.
\ee
When we include a massless scalar field $\phi$ as a matter degree of freedom, a complete set of Dirac observables is given by the pair $\hat{V}|_{\phi}$, $\hat{p}_{(\phi)}$ \cite{Ashtekar:2007}. The operator $\hat{V}|_{\phi}$ represents the volume at any `instant' $\phi$ of the internal time, whereas $\hat{p}_{(\phi)}$ is the momentum of the scalar field $\phi$.

The dynamics needs a regularisation scheme for the constraint (\ref{Eq:LQC:Constraint}). In fact, the field strength is not defined in the quantum theory. Thus, one replaces the field strength $F^{\alpha}_{ij}$ which appears in Eq.~(\ref{Eq:LQC:Constraint}) by the following expression
 (see \emph{e.g.} the review \cite{Banerjee:2011qu})
\be\label{Eq:LQC:RegularizedFieldStrength}
-2\mbox{Tr}\left( \frac{h^{\bar{\mu}}_{\square_{\beta\gamma}}-\delta_{\beta\gamma} }{\bar{\mu}^2V_0^{2/3}} \tau^\alpha    \right)  \pha{o}e^\beta_j~ \pha{o}e^\gamma_i ~  ~,
\ee
where $h^{\bar{\mu}}_{\square_{\alpha\beta}}$ is the holonomy of the connection $A$ in Eq.~(\ref{Eq:LQC:SymmetricPair}) computed along the edges of a face of the fiducial cell\footnote{This regularisation of the field strength, which parallels the one in the full theory, is entirely analogous to what is done in lattice QCD, see \emph{e.g.} Ref.~\cite{Gupta:1997nd}.}. An additional ingredient is then `imported' from the full theory: it is assumed that loops cannot be shrunk to a point, due to the existence of an \emph{area gap}, \emph{i.e.} a minimum non-zero eigenvalue $\Delta$ of the area operator as in LQG \cite{Rovelli:1994ge,Ashtekar:1996eg}. This is a well-known property of the area operator in LQG, which has important consequences for the LQC dynamics. The quantity $\bar{\mu}$ is in principle a function of the flux, \emph{i.e.}
\be\label{Eq:LQC:MuBarGeneral}
\bar{\mu}=\bar{\mu}(p)~.
\ee
Different \emph{ans\"atze} can lead to radically different dynamics.

For a generic choice of $\bar{\mu}$ one can define the holonomies as $\widehat{\e^{i\bar{\mu}c/2}}$. Therefore, it is convenient to introduce a further pair of conjugate variables, leading to quantum states with simple transformation properties under the action of the `improved' holonomy operators \cite{Banerjee:2011qu}. One defines
\be
b=\frac{\hbar \bar{\mu} c}{2}~,\hspace{1em}\upsilon(p)=(2\pi\beta\ell_{Pl}^2\sqrt{\Delta})^{-1}\mbox{sgn}|p|^{3/2}~,
\ee
such that
\be
\{b,\upsilon\}=1~.
\ee
Therefore, at the quantum level, defining eigenstates of $\hat{\upsilon}$ as
\be
\hat{\upsilon}|\upsilon\rangle=\upsilon|\upsilon\rangle~,
\ee
one has, with the above definitions and choice of units
\be\label{Eq:LQC:ImprovedDynamics}
 \widehat{\e^{i \frac{\bar{\mu} c}{2}}}|\upsilon\rangle=|\upsilon+1\rangle~.
\ee

In the original formulation of LQC it was assumed that $\bar{\mu}=\mu_0$, a constant parameter. However, such a choice leads to a dynamics that is incompatible with the predictions of GR. In particular, quantum effects can be large even for values of the curvature that fall well below the Planck scale. In order to fix these issues, a new regularisation scheme was introduced, known as \emph{improved dynamics} \cite{Ashtekar:2006improved}. Thus, the following \emph{ansatz} is made
\be\label{Eq:LQC:ImprovedDynamics}
\bar{\mu}^{-1}=\sqrt{\frac{|p|}{\Delta}}~.
\ee
It is apparent that for increasingly large fluxes Eq.~(\ref{Eq:LQC:ImprovedDynamics}) implies that the expression (\ref{Eq:LQC:RegularizedFieldStrength}) approaches the field strength, thus recovering the correct classical limit. However, for small values of the flux, the dynamics becomes sensitive to the underlying discreteness of quantum geometry and departures from GR can take place.

The promotion of the Hamiltonian constraint (\ref{Eq:LQC:Constraint}) to an operator also requires further manipulations to be properly regularized, including a definition of a non-singular inverse volume operator \cite{Banerjee:2011qu}. Even then, different factor ordering prescriptions are possible. However, it is true in general that the Hamiltonian constraint leads to a finite difference equation, see \emph{e.g.} Refs.~\cite{Bojowald:2012we,Ashtekar:2003hd,Banerjee:2011qu}. The Hamiltonian constraint corresponding to a (not necessarily flat) FLRW Universe and pure gravity\footnote{When matter is coupled to the gravitational field, the Hamiltonian constraint is modified by the inclusion of additional terms.} acts on states as follows (cf. Refs.~\cite{Ashtekar:2006,Calcagni:2012vb})
\be\label{Eq:LQC:SuperSelection}
-B(\nu)\Theta\,\psi(\nu)\equiv
A(\nu)\psi(\nu+\nu_0)+C(\nu)\psi(\nu)+D(\nu)\psi(\nu-\nu_0)=0~
\ee
where the functions $A$, $B$, $C$, $D$ depend on the model (\emph{e.g.} the particular factor ordering and regularisation prescription one chooses). The quantity $\nu_0$ is an elementary shift, whose exact value and units depend on the model and conventions adopted (cf. Refs.~\cite{Ashtekar:2006,Calcagni:2012vb,Banerjee:2011qu}, see also the footnotes \ref{Footnote:LQCdifference1}, \ref{Footnote:LQCdifference2}). The finite difference equation (\ref{Eq:LQC:SuperSelection}) shows the occurrence of superselection sectors \cite{Banerjee:2011qu}. In the limit\footnote{Limit which is of course mathematically well-defined, although fictitious from a physical point of view. In fact, in this limit the underlying discreteness introduced by quantum geometry disappears.} $\nu_0\to0$ one recovers a Wheeler-DeWitt equation with a precise factor-ordering, of which (\ref{Eq:LQC:SuperSelection}) represents a finite difference version. 

Particularly useful from the point of view of physical applications of LQC is the tool of \emph{effective equations} \cite{Bojowald:2012xy}, which can be obtained by computing the expectation value of the Hamiltonian constraint operator on semiclassical states $|\Psi_{\rm sc}\rangle$ that are peaked on some point $(b,v)$ in phase space \cite{Banerjee:2011qu}
\be
\langle \Psi_{\rm sc}| \hat{\mathcal{C}}|\Psi_{\rm sc} \rangle \simeq -\frac{6}{\beta^2}\alpha\sqrt{p}\frac{\sin^2(\bar{\mu}c)}{\bar{\mu}^2}+\langle \Psi_{\rm sc}|\hat{\mathcal{C}}_{\rm matter}|\Psi_{\rm sc} \rangle~,
\ee
where the operator $\hat{\mathcal{C}}_{\rm matter}$ is the matter contribution to the Hamiltonian constraint. $\alpha$ is a parameter whose dependence on $v$ is fixed by the model. One then finds the modified Friedmann equation (obtained in Ref.~\cite{Taveras:2008ke} for a scalar field)
\be
H^2=\frac{8\pi G}{3}\rho\left(1-\frac{\rho}{\rho_{\rm crit}}\right)~,
\ee
with the critical density defined as
\be\label{Eq:LQC:CriticalDensity}
\rho_{\rm crit}=\frac{3}{8\pi G\beta^2\Delta}~.
\ee
Note that the finiteness of the critical density depends on the existence of the area gap. When the density of matter approaches $\rho_{\rm crit}$, gravity becomes repulsive as a result of the extra term in Eq.~(\ref{Eq:LQC:CriticalDensity}). This in agreement with the singularity resolution (bounce) observed both numerically and analitically in the solutions of the finite difference equation \cite{Ashtekar:2006rx,Ashtekar:2006D73}. The accuracy of the effective dynamics crucially depends on the smallness of quantum back-reaction effects \cite{Rovelli:2013zaa}. See also Ref.~\cite{Bojowald2011} for general aspects of effective descriptions of quantum systems.

The LQC framework can accommodate for the anisotropic case\footnote{See Refs.~\cite{Ashtekar:2009um,Ashtekar:2009vc,MartinBenito:2008wx,WilsonEwing:2010rh} for the LQC formulation of Bianchi models.}, but in the general case it is not suited to study inhomogeneous field configurations. An exception is represented by the Gowdy model 
whose loop quantisation can be carried out exactly\footnote{The canonical quantisation of the Gowdy model in \emph{complex} Ashtekar variables was performed in Refs.~\cite{HUSAIN1989205,MenaMarugan:1997us}.}. In the general case, a hybrid scheme is adopted \cite{Garay:2010sk,MartinBenito:2008ej}; the dynamics of the background is the same as in homogeneous LQC, while a Fock space quantisation is carried out for the cosmological perturbations. However, consistency of the approach requires some additional input coming from the full theory in determining the properties of states considered in LQC. In particular, the physical predictions are very sensitive to the particular regularisation scheme adopted.

As mentioned above, the so-called $\mu_0$ scheme leads to serious consistency problems with GR in the low curvature regime, as well as dynamical instabilities (similar to numerical ones) for the finite difference equations \cite{Cartin:2005an,Rosen:2006bga,Date:2005ik,Bojowald:2007ra}. From a heuristic point of view, this can be interpreted as due to the fact that the spacing of the underlying lattice is kept fixed. Denoting the lattice spacing (\emph{i.e.} the size of the fiducial cell) by $\ell_0$, this is stretched with the expansion of the Universe according to (see \emph{e.g.} Ref.~\cite{Banerjee:2011qu})
\be\label{Eq:LQC:Refinement}
L=a~\ell_0=\left(\frac{V}{\mathcal{N}}\right)^{1/3}~,
\ee
where we assumed a cubic lattice\footnote{In more general, anisotropic configurations, a different number of vertices must be introduced for each direction $x^i$, so that the total number of vertices is given by $\mathcal{N}=\prod_i \mathcal{N}_{x^i}$. The relation  (\ref{Eq:LQC:Refinement}) is modified accordingly \cite{Bojowald:2007ra}.} of $\mathcal{N}$ cells. The relation between the total size of the coordinate box and the lattice spacing is (assuming isotropy) \cite{Bojowald:2007ra}
\be\label{Eq:LQC:RescaleComovingLattice}
\ell_0=\left(\frac{V_0}{\mathcal{N}}\right)^{1/3}
\ee
 $V$ and $L$ are the \emph{physical} (as opposed to comoving) \emph{total} volume of the lattice and the lattice spacing, respectively. Since holonomies are computed along lattice edges by integrating the connection (\ref{Eq:LQC:SymmetricPair}), we have\footnote{Assuming as usual a lattice adapted to the fiducial metric, which then takes a diagonal form.} for the holonomy in the $x^3$ direction
 \be
 h_e=\exp \int_0^{\ell_0}\tilde{c}\tau_3=\exp \left(V_0^{-1/3}\ell_0c\tau_3\right)=\exp\left(\mathcal{N}^{-1/3}c\tau_3\right)~,
 \ee
 where we used Eqs.~(\ref{Eq:LQC:RescaleVariables}),~(\ref{Eq:LQC:RescaleComovingLattice}).
 The quantity $\mathcal{N}$ is a priori a function of the phase space of LQC, whose particular form (which must be \emph{assumed} as an external input, not determined by the theory) can lead to very different dynamics. In fact, it can be interpreted as a power of the general function $\bar{\mu}(p)$, introduced in Eq.~(\ref{Eq:LQC:MuBarGeneral})
 \be
 \bar{\mu}(p)=\mathcal{N}^{-1/3}~.
 \ee
 In particular, when $\mathcal{N}$ is a constant it leads to the (old) $\mu_0$ regularisation scheme. On the other hand, by assuming Eq.~(\ref{Eq:LQC:ImprovedDynamics}), one recovers the improved dynamics scheme.

Lattice refinement provides a physical interpretation of the $\bar{\mu}$ scheme, and generalizes it. It leads in general to finite difference equations with a variable step, for which suitable numerical techniques have been developed \cite{Nelson:2008bx}.
The interpretation of the quantity $\mathcal{N}$ as a phase space function, means that new lattice sites are created by the dynamics. This deviates considerably from the naive picture of a homogeneous and isotropic Universe whose geometry is entirely determined from that of a single, fiducial cell. Instead, it requires a description which is that of a system with a variable number of degrees of freedom. This is agreement with the fact that in general the Hamiltonian constraint acts on spin-network states by changing the graph, \emph{i.e.} creating new edges and nodes leading to a finer graph \cite{Bojowald2006}. The underlying lattice structure seems to suggest that the inclusion of inhomogeneities, as excitations of individual lattice sites, is unavoidable \cite{Banerjee:2011qu}.


\chapter{WKB approximation and asymptotics of the wave packets}\label{WKB solution}

Since Eq.~(\ref{EQUATION FOR ALPHA}) has the form of a time-independent
Sch\"rodinger equation, it is possible to construct approximate
solutions using the WKB method.
For a given $k$, we divide the real line in three regions, with a
neighborhood of the classical turning point in the middle. The
classical turning point is defined as the point where the potential is
equal to the energy
\be
\alpha_{k}=\frac{1}{2}\log{k}~.
\ee
The WKB solution to first order in the classically allowed region
$]-\infty, \alpha_{k}-\epsilon]$ (with $\epsilon$ an appropriately
    chosen real number, see below) is 
\be\label{allowed region}
C_{k}^{I}=\frac{2}{(k^2-e^{4\alpha})^{1/4}}\cos\left(\frac{1}{2}\sqrt{k^2-e^{4\alpha}}-\frac{k}{2}\mbox{arccoth}\left(\frac{k}{\sqrt{k^2-e^{4\alpha}}}\right)+\frac{\pi}{4}\right)~,
\ee
and reduces to a plane wave in the allowed region for
$\alpha\ll\alpha_{k}$. The presence of the barrier fixes the amplitude
and the phase relation of the incoming and the reflected wave through
the matching conditions.

The solution in the classically forbidden region $[
  \alpha_{k}+\epsilon,+\infty[ $ is instead exponentially decreasing
    as it penetrates the potential barrier and reads
\be\label{forbidden region}
C_{k}^{III}=\frac{1}{(e^{4\alpha}-k^2)^{1/4}}\exp\left(-\frac{1}{2}\sqrt{e^{4\alpha}-k^2}+\frac{k}{2}\arctan\left(\frac{k}{\sqrt{e^{4\alpha}-k^2}}\right)\right)~.
\ee
Finally, in the intermediate region
$]\alpha_{k}-\epsilon,\alpha_{k}+\epsilon[$ any semiclassical method
    would break down, hence the Schr\"odinger equation must be solved
    exactly using the linearised potential
\be
V(\alpha)\simeq V(\alpha_{k})+V^{\prime}(\alpha_{k})(\alpha-\alpha_{k})=4e^{4\alpha_{k}}(\alpha-\alpha_{k})=4k^2(\alpha-\alpha_{k})~.
\ee
In this intermediate regime, Eq.~(\ref{EQUATION FOR ALPHA}) can be
rewritten as the well-known Airy equation
\be\label{Airy equation}
\frac{\mbox{d}^2 C_{k}^{II} }{\mbox{d}t^2}-t C_{k}^{II}=0~,
\ee
where the variable $t$ is defined as
\be
t=(4k^2)^{1\over 3}(\alpha-\alpha_{k})~.
\ee
Of the two independent solutions of Eq.~(\ref{Airy equation}),
only the Airy function $\mbox{Ai}(t)$ satisfies the boundary
condition, and hence we conclude that
\be
C_{k}^{II}=c \; Ai(t)~,
\ee
where $c$ is a constant that has to be determined by matching the
asymptotics of $\mbox{Ai}(t)$ with the WKB approximations on both
sides of the turning point. Hence we get that, for $t\ll0$
\be
\mbox{Ai}(t)\approx\frac{\cos\left(\frac{2}{3}|t|^{3/2}-\frac{\pi}{4}\right)}{\sqrt{\pi}|t|^{1/4}}~,
\ee while for $t\gg0$ 
\be
\mbox{Ai}(t)\approx\frac{e^{-\frac{2}{3}t^{3/2}}}{2\sqrt{\pi}t^{1/4}}~.
\ee 
We thus fix $c=2\sqrt{\pi}$.
Note that the arbitrariness in the choice of $\epsilon$ can be solved,
\emph{e.g.} by requiring the point $\alpha_{k}-\epsilon$ to coincide
with the first zero of the Airy function.

  \begin{figure}\label{tre curve WKB}
  \centering
 \includegraphics[width=\textwidth]{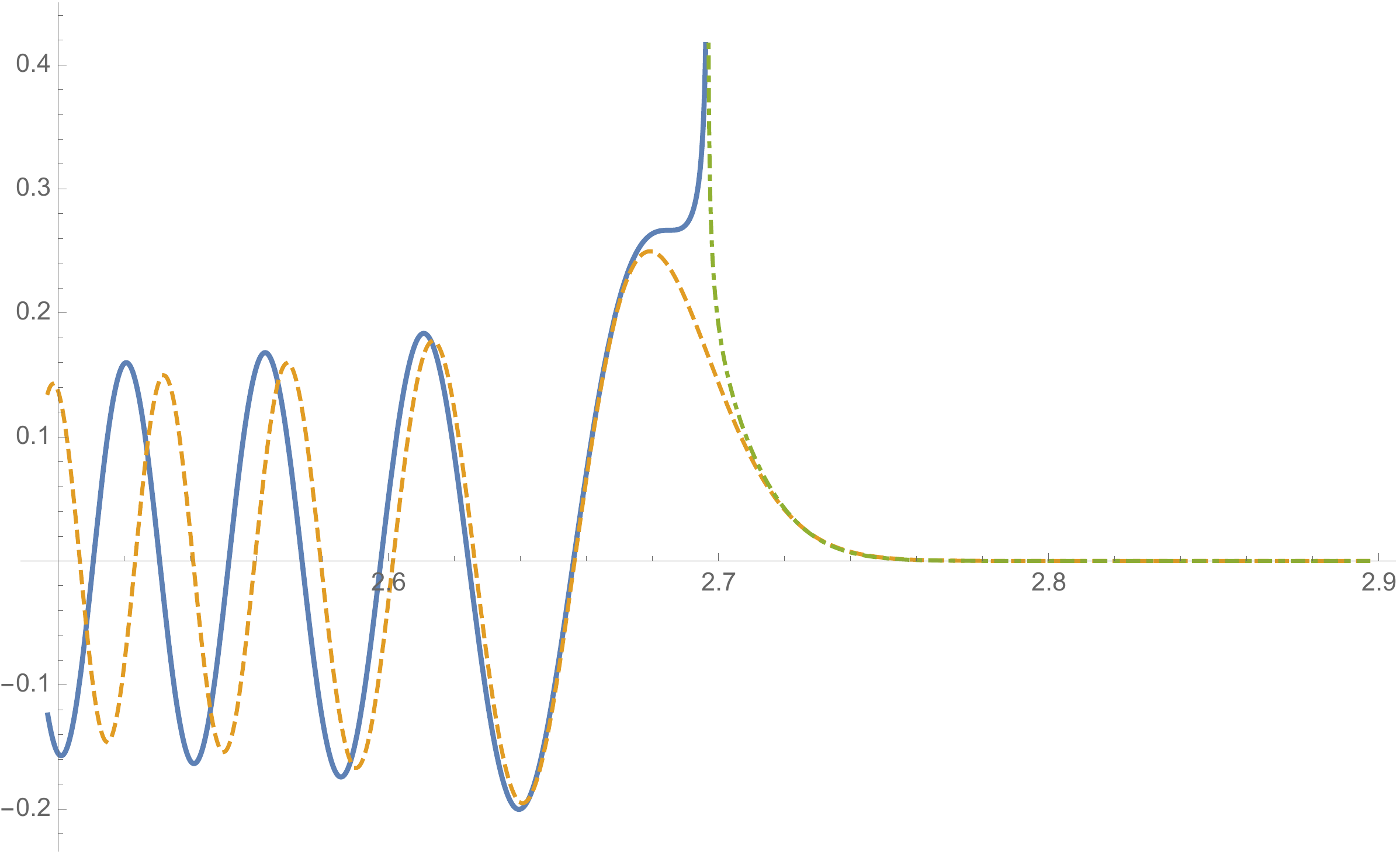}
\caption[WKB approximation to the WDW equation]{The thick (blue) and the dot-dashed (green) curve represent the
  WKB approximation of the wave function in the classically allowed
  and in the forbidden region, respectively. They both diverge at the classical
  turning point. The Airy function is represented by the dashed curve. The value
  $k=220$ was chosen.}
    \end{figure}
The WKB approximation improves at large values of $k$, as one should
expect from a method that is semiclassical in spirit. Yet it allows
one to capture some effects that are genuinely quantum, such as the
barrier penetration and the tunneling effect.

From the approximate solution we have just found, we can construct
wave packets as in Ref.~\cite{Kiefer:1988}.
We then restrict our attention to the classically allowed region and
the corresponding approximate solutions, \emph{i.e.}
$C_{k}^{I}(\alpha)$ for $\alpha\ll\alpha_{k}=\frac{1}{2}\log{k}$ and
compute the integral in Eq.~(\ref{Wave Packet}). If the Gaussian
representing the amplitudes of the monochromatic modes is narrow
peaked, \emph{i.e.} its variance $b^2$ is small enough, we can
approximate the amplitude in Eq.~(\ref{allowed region}) with that
corresponding to the mode with the mean frequency
$\overline{k}$. Thus, introducing the constant
\be
c_{\overline{k}}=\frac{2}{(\overline{k}^2-e^{4\alpha})^{1/4}}\frac{1}{\pi^{1/4}\sqrt{b}}~,
\ee
we have
\be\label{stima}
\psi(\alpha,\phi)\simeq c_{\overline{k}}\int_{-\infty}^{\infty}\mbox{d}k\; e^{-\frac{(k-\overline{k})^2}{2b^2}}\cos\left(\frac{1}{2}\sqrt{k^2-e^{4\alpha}}-\frac{k}{2}\mbox{arccoth}\left(\frac{k}{\sqrt{k^2-e^{4\alpha}}}\right)+\frac{\pi}{4}\right)e^{ik\phi}~.
\ee
Moreover, we have
\be\label{dominant term}
k\,\mbox{arccoth}\left(\frac{k}{\sqrt{k^2-e^{4\alpha}}}\right)=k\arccosh\frac{k}{e^{2\alpha}}\simeq k\arccosh\frac{\overline{k}}{e^{2\alpha}}~.
\ee
The last approximation in the equation above holds since the derivative of
the inverse hyperbolic function turns out to be much smaller than unity
in the allowed region. Furthermore, the term approximated in
Eq.~(\ref{dominant term}) dominates over the square root in the
argument of the cosine in the integrand in the r.h.s. of
Eq.~(\ref{stima}), so we can consider the latter as a constant. Hence,
we can write
\be\label{psi}
\psi(\alpha,\phi)\simeq c_{\overline{k}}\int_{-\infty}^{\infty}\mbox{d}k\; e^{-\frac{(k-\overline{k})^2}{2b^2}}\cos\left(\Lambda k+\delta\right)e^{ik\phi}~,
\ee
where we have introduced the notation
\begin{align}
\Lambda&\equiv\frac{1}{2}\arccosh\frac{\overline{k}}{e^{2\alpha}}~,\\
\delta&\equiv\frac{1}{2}\sqrt{\overline{k}^2-e^{4\alpha}}-\frac{\pi}{4}~.
\end{align}
for convenience.
Using Euler's formula we can express the cosine in the integrand in the r.h.s. of 
Eq.~(\ref{psi}) in terms of complex exponentials and evaluate $\psi(\alpha,\phi)$.
In fact, defining
\be
P_{\overline{k}}\equiv e^{i\left(\overline{k}(\Lambda+\phi)+\delta\right)}~,~\hspace{1em}Q_{\overline{k}}\equiv e^{-i\left(\overline{k}(\Lambda-\phi)+\delta\right)}~,
\ee
we have
\be
\psi(\alpha,\phi)\simeq \frac{c_{\overline{k}}}{2}\int_{-\infty}^{\infty}\mbox{d}k\; e^{-\frac{(k-\overline{k})^2}{2b^2}}\left(e^{i(k-\overline{k})(\Lambda+\phi)}P_{\overline{k}}+e^{-i(k-\overline{k})(\Lambda-\phi)}Q_{\overline{k}}\right)~.
\ee
Shifting variables and performing the Gaussian integrations we get the
result as in Ref.~\cite{Kiefer:1988} 
\be \psi(\alpha,\phi)\simeq
c_{\overline{k}}\sqrt{\frac{\pi}{2}}b\left(e^{-\frac{b^2}{2}(\Lambda+\phi)^2}P_{\overline{k}}+e^{-\frac{b^2}{2}(\Lambda-\phi)^2}Q_{\overline{k}}\right)~.
\ee


\chapter{Orthogonality of the MacDonald functions}
It was proved in Refs.~\cite{Szmytkowski, Yakubovich, Passian} that
\be
\int_{0}^{\infty}\frac{\mbox{d}x}{x}\;K_{i\nu}(kx)K_{i\nu^{\prime}}(kx)=\frac{\pi^2\delta(\nu-\nu^{\prime})}{2\nu\sinh(\pi\nu)}~,
\ee
which expresses the orthogonality of the MacDonald functions of
imaginary order. Performing a change of variables, the above formula
can be recast in the form
 \be
\int_{-\infty}^{\infty}\mbox{d}\alpha\;
K_{i\frac{h}{2}}\left(\frac{e^{2\alpha}}{2}\right)K_{i\frac{h^{\prime}}{2}}\left(\frac{e^{2\alpha}}{2}\right)=\frac{\pi^2\delta(h-h^{\prime})}{h\sinh(\pi\frac{h}{2})}~,
\ee
which is convenient for the applications considered in this work.
For large values of $k$ it is equivalent to the normalisation used
in Section~\ref{section 2}
\be \int_{-\infty}^{\infty}\mbox{d}\alpha
\;\psi^{*}_{k^{\prime}}(\alpha,\phi)\psi_{k}(\alpha,\phi)=\frac{\pi^2}{2}\delta(k-k^{\prime})~.
\ee
%


\chapter{Harmonic Analysis on $\mbox{SU(2)}$}\label{Appendix:HarmonicAnalysis}
Harmonic analysis is a generalisation of Fourier analysis to topological groups.
In this appendix we review the fundamentals of harmonic analysis on the Lie group $\mbox{SU(2)}$. We obtain the invariant volume element (Haar measure) and the eigenvectors of the Laplacian operator. The Peter-Weyl formula is then given as a generalisation of the Fourier series expansion.

There is a well known natural homeomorphism between the group manifold $\mbox{SU(2)}$ and the three-sphere $S^3$.
The three dimensional sphere $S^3$ has a natural embedding in Euclidean $\mathbb{R}^4$ as the set of points with Cartesian coordinates $(x_1,x_2,x_3,x_4)$ satisfying the equation
\be\label{eq:constraint}
x_1^2+x_2^2+x_3^2+x_4^2=\eta^{-1},
\ee
where $\eta^{-1/2}$ is the radius of the sphere. The parameter $\eta$ determines the group volume. Its value can be fixed by means of an appropriate normalisation, as we show below.
A possible parametrisation in terms of three angles $(\alpha_1,\alpha_2,\alpha_3)$ is given by the map
\begin{align}
x_1&=\eta^{-1/2}\sin\alpha_3\sin\alpha_2\cos\alpha_1\nonumber\\
x_2&=\eta^{-1/2}\sin\alpha_3\sin\alpha_2\sin\alpha_1\nonumber\\
x_3&=\eta^{-1/2}\sin\alpha_3\cos\alpha_2\nonumber\\
x_4&=\eta^{-1/2}\cos\alpha_3\label{eq:embedding}
\end{align}
Here $0\leq\alpha_1<2\pi$ and $0\leq\alpha_2,\alpha_3<\pi$. 
The generic $\mbox{SU(2)}$ group element can then be written as
\be
\begin{split}
g&=\eta^{1/2}\begin{pmatrix} x_4+ix_3 & ix_1+x_2\\ ix_1-x_2 &x_4-ix_3\end{pmatrix}\\
&=\begin{pmatrix} \cos\alpha_3+i\sin\alpha_3\cos\alpha_2 & i\sin\alpha_3\sin\alpha_2\e^{-i\alpha_1}\\i\sin\alpha_3\sin\alpha_2\e^{i\alpha_1} & \cos\alpha_3-i\sin\alpha_3\cos\alpha_2\end{pmatrix}~.
\end{split}
\ee
The induced metric $h_{\rho\sigma}$ on $\mbox{SU(2)}$ from the Euclidean metric $\delta_{ij}$ on $\mathbb{R}^4$ via the embedding~(\ref{eq:embedding}) is
\be
\de l^2=h_{\rho\sigma}\de\alpha^\rho\de\alpha^\sigma=\eta^{-1}\Big(\de\alpha_3^2+(\sin\alpha_3)^2\big(\de\alpha_2^2+(\sin\alpha_2)^2\de\alpha_1^2\big)\Big)~.
\ee
This is the unique (up to a factor) bi-invariant metric on SU(2) (see Footnote~\ref{Footnote:LieGroupBiInvMetric}).

A more convenient parametrisation of $\mbox{SU(2)}$ elements is given in terms of Euler angles. In fact, every element can be written as
\be
g=\e^{-i\psi \frac{\sigma_3}{2}}\e^{-i\theta \frac{\sigma_2}{2}}\e^{-i\phi \frac{\sigma_3}{2}}=\begin{pmatrix}e^{-\frac{i \phi }{2}-\frac{i \psi }{2}} \cos \left(\frac{\theta }{2}\right) & -e^{\frac{i
   \phi }{2}-\frac{i \psi }{2}} \sin \left(\frac{\theta }{2}\right) \\
 e^{\frac{i \psi }{2}-\frac{i \phi }{2}} \sin \left(\frac{\theta }{2}\right) & e^{\frac{i
   \phi }{2}+\frac{i \psi }{2}} \cos \left(\frac{\theta }{2}\right)\end{pmatrix}.
\ee
The Euler angles have range $0\leq\theta<\pi$, $0\leq\psi<2\pi$, $0\leq\phi<4\pi$. Cartesian coordinates on the sphere are thus expressed in terms of Euler angles as follows
\begin{align}
x_1&=-\eta^{-1/2}\sin \left(\frac{\theta }{2}\right) \sin \left(\frac{\phi }{2}-\frac{\psi }{2}\right)\nonumber\\
x_2&=-\eta^{-1/2}\sin \left(\frac{\theta }{2}\right) \cos \left(\frac{\phi }{2}-\frac{\psi
   }{2}\right)\nonumber\\
x_3&=-\eta^{-1/2}\cos \left(\frac{\theta }{2}\right) \sin \left(\frac{\psi }{2}+\frac{\phi }{2}\right)\nonumber\\
x_4&=\eta^{-1/2}\cos \left(\frac{\theta }{2}\right) \cos \left(\frac{\psi }{2}+\frac{\phi }{2}\right)\label{eq:EulerEmbedding}.
\end{align}

The metric on $S^3$ is obtained from the Euclidean metric on $\mathbb{R}^4$ after imposing the constraint Eq.~(\ref{eq:constraint}). In the chart given by the Euler angles it takes the form
\be\label{eq:MetricEuler}
\de l^2=\frac{\eta^{-1}}{4}\big(\de\theta^2+\de\psi^2+2\cos\theta~\de\psi\,\de\phi+\de\phi^2\big).
\ee
The invariant volume element is
\be\label{eq:VolumeElement}
\de\mu_{\rm Haar}=\sqrt{h}\;\de\theta\,\de\psi\,\de\phi=\frac{\sin\theta}{8}\eta^{-3/2}\de\theta\,\de\psi\,\de\phi,
\ee
where $h$ is the determinant of the metric on SU(2) in the chart $(\theta,\psi,\phi)$, given in Eq.~(\ref{eq:MetricEuler}).
This volume element defines the \emph{Haar measure} as the unique measure (up to rescalings) which is invariant under right and left action of the group onto itself.
It is convenient to rescale the volume element by the volume of the group. Integrating Eq.~(\ref{eq:VolumeElement}) over the whole group and normalising the Haar measure to one we have
\be
1=\int_{\rm SU(2)}\sqrt{h}\;\de\theta\,\de\psi\,\de\phi=\frac{1}{8}\eta^{-3/2}\int_0^{2\pi}\de\psi\int_0^{4\pi}\de\phi\int_0^\pi\de\theta\sin\theta=2\pi^2 \eta^{-3/2}.
\ee
Thus, $\eta$ is determined as
\be\label{eq:EtaValue}
\eta=(2\pi^2)^{2/3}.
\ee

We define the Laplacian corresponding to the metric in Eq.~(\ref{eq:MetricEuler}) via its action on smooth functions $f$ on SU(2) as
\be
\triangle_\eta f=\frac{1}{\sqrt{h}}\,\partial_\rho \Big(\sqrt{h}\,h^{\rho\sigma}\partial_\sigma f\Big)~.
\ee
For $\eta=1$ we have the standard Laplacian on the three-sphere with unit radius, which we denote by $\triangle$.
We observe that
\be
\triangle_\eta =\eta~\triangle .
\ee

Eigenfunctions of the standard Laplacian $\triangle$ are given by the Wigner matrices $D^j_{mn}(\psi,\theta,\phi)$. They are solutions of the eigenvalue equation
\be
-\triangle D^j_{mn}=E_j D^j_{mn},
\ee
with eigenvalues
\be
E_j=4j(j+1).
\ee
The indices $m, n$ are degenerate for fixed $j$. Thus, we have for the eigenvalues of the rescaled Laplacian $\triangle_\eta$
\be
E_{j}^{\eta}=\eta~4j(j+1)=\big(2\pi^2\big)^{2/3}\;4j(j+1).
\ee

The Wigner matrices form a complete basis of the Hilbert space $L^2(\mbox{SU(2)},\de\mu_{\textrm Haar})$. They satisfy the following orthogonality relations
\be\label{eq:Orthogonality}
\int \de\mu_{\textrm Haar}\; \overline{D^{j_1}_{m_1n_1}}D^{j_2}_{m_2 n_2}=\frac{1}{d_j}\delta_{j_1j_2}\delta_{m_1m_2}\delta_{n_1n_2},
\ee
where $d_j=2j+1$ is the dimension of the irreducible representation of $\mbox{SU(2)}$ with spin $j$. Under complex conjugation
\be
\overline{D^j_{mn}}=(-1)^{2j+m+n}D^j_{-m\,-n}.
\ee
Since they are the coefficients of representations of $SU(2)$, for any two group elements $g_1$, $g_2$ one has
\be
D^j_{mn}(g_1g_2)=\sum_{k=-j}^j D^j_{mk}(g_1)D^j_{kn}(g_2).
\ee

The Peter-Weyl formula gives the decomposition of a square integrable functions on $\mbox{SU(2)}$ in the basis given by the Wigner matrices
\be\label{eq:PeterWeyl}
f(g)=\sum_{j,m,n}f^{j}_{m,n}\sqrt{d_{j}}\;D^{j}_{m,n}(g).
\ee
In Section~\ref{Sec:GFT:MeanField} we used a straightforward generalisation of Eq.~(\ref{eq:PeterWeyl}) obtained by considering functions defined on the direct product of four copies of $\mbox{SU(2)}$. The scalar product can be re-expressed using Eqs.~(\ref{eq:Orthogonality}),~(\ref{eq:PeterWeyl}) as
\be
\int \de\mu_{\textrm Haar}\; \overline{f}g=\sum_{j,m,n}\overline{f^{j}_{m,n}}\,g^{j}_{m,n}.
\ee
For $f=g$, this is the analogue of the Plancherel theorem
\be
||f||^2=\int \de\mu_{\textrm Haar}\; |f|^2=\sum_{j,m,n}\left| f^{j}_{m,n}\right|^2.
\ee


\chapter{Intertwiner Space of a four-valent vertex}\label{sec:Intertwiner}
The Peter-Weyl decomposition of the GFT field can be used to make the degrees of freedom of the theory more transparent. In fact, as observed in Section~\ref{Sec:GFT:KinematicalOperators}, the right-invariance property of the GFT field implies that certain coefficients in the series expansion of the wave function in Eq.~(\ref{Eq:GFT:GenericState}) vanish. In order to have a non-zero coefficient, the spin labels of the different $\mbox{SU(2)}$ copies must satisfy certain algebraic conditions. In this Appendix, we will first review the constuction of the intertwiner space of a four-valent vertex and then clarify its relation with the kinematics of GFT.

In Eq.~(\ref{eq:GFTfield}) the series coefficients are labelled by four spins $j_\nu$. Each of them identifies a finite dimensional Hilbert space $\mathcal{H}_{j_i}$ which is also an irreducible representation of the Lie group $\mbox{SU(2)}$.
The intertwiner space $\accentset{\circ}{\mathcal{H}}_{j_\nu}$ is defined as the subspace of the tensor product $\mathcal{H}_{j_1}\otimes\mathcal{H}_{j_2}\otimes\mathcal{H}_{j_4}\otimes\mathcal{H}_{j_4}$ whose elements are invariant under the diagonal action of $\mbox{SU(2)}$, \emph{i.e.} we define it as the space of invariant tensors \cite{Haggard:2011qvx}
\be
\accentset{\circ}{\mathcal{H}}_{j_\nu}=\mbox{Inv}_{\mbox{\scriptsize SU(2)}}\left[\mathcal{H}_{j_1}\otimes\mathcal{H}_{j_2}\otimes\mathcal{H}_{j_3}\otimes\mathcal{H}_{j_4}\right]~.
\ee
Thus, $\accentset{\circ}{\mathcal{H}}_{j_\nu}$ is the space of singlets that can be constructed out of four spins. It can be interpreted as the Hilbert space of a quantum tetrahedron. In fact, we can give a geometric interpretation of the construction of the intertwiner space. Let us consider four links, each carrying a spin $j_i$ and meeting at a vertex. A tetrahedron is constructed by duality from the vertex; the spins $j_i$ are the quantum numbers of the areas of its faces. Global invariance of the vertex under $\mbox{SU(2)}$ amounts to the closure of the tetrahedron, see Fig.~\ref{Fig:Tetrahedron}.

A basis in $\accentset{\circ}{\mathcal{H}}_{j_\nu}$ can be found by first composing the four spins pairwise, then the two resultant spins together so as to form singlets. We partition the spins in two pairs $(j_1,j_2)$ and $(j_3,j_4)$, corresponding to the recoupling channel $\mathcal{H}_{j_1}\otimes\mathcal{H}_{j_2}$ \cite{Haggard:2011qvx}. The total spin of a pair is labelled by the quantum number $J$, which is the same for each of the two pairs since they sum to give a singlet. Basis vectors can thus be expressed in terms of the tensor product basis as
\be\label{eq:Singlets}
|j_\nu ;J\rangle=\sum_{m_{\nu}}\alpha^{j_\nu,J}_{m_\nu}|j_1,m_1\rangle|j_2,m_3\rangle|j_3,m_3\rangle|j_4,m_4\rangle~.
\ee
The coefficients $\alpha^{j_\nu,J}_{m_\nu}$ are the elements of a unitary matrix, which implements the change of basis from the tensor product basis $\{|j_1,m_1\rangle|j_2,m_3\rangle|j_3,m_3\rangle|j_4,m_4\rangle\}$ to $\{|j_\nu;J\rangle\}$ in the space of singlets $\accentset{\circ}{\mathcal{H}}_{j_\nu}$~\footnote{Notice that it is not a unitary matrix over the whole Hilbert space $\mathcal{H}_{j_1}\otimes\mathcal{H}_{j_2}\otimes\mathcal{H}_{j_4}\otimes\mathcal{H}_{j_4}$, since $\alpha^{j_\nu,J}_{m_\nu}$ vanishes when $m_\nu$ fails to satisfy Eq.~(\ref{eq:SpinsSumToZero})}.
$\alpha^{j_\nu,J}_{m_\nu}$ is an invariant tensor, \emph{i.e.} all of its components are invariant under $\mbox{SU(2)}$.

The quantum number $J$ satisfies the inequalities
\be\label{eq:InequalityJ}
\mbox{max}\left\{|j_1-j_2|,|j_3-j_4|\right\}\leq J\leq \mbox{min}\left\{j_1+j_2,j_3+j_4\right\}.
\ee
Moreover, in order to get a singlet one must have
\be\label{eq:SpinsSumToZero}
m_1+m_2+m_3+m_4=0~.
\ee
When Eqs.~(\ref{eq:InequalityJ}),~(\ref{eq:SpinsSumToZero}) are not satisfied for certain values of $J$ and $m_\nu$, $\alpha^{j_\nu,J}_{m_\nu}$ vanishes and the corresponding term gives a vanishing contribution to Eq.~(\ref{eq:Singlets}).
The coefficients of the decomposition in Eq.~(\ref{eq:Singlets}) can be expressed in terms of Clebsch-Gordan coefficients as
\be\label{eq:SingletSpinJ}
\alpha^{j_\nu,J}_{m_\nu}=\gamma~\frac{(-1)^{J-M}}{\sqrt{d_J}}C^{j_1j_2J}_{m_1m_2\,M}C^{j_3j_4J}_{m_3m_4\,-M}~,
\ee
where we defined $M=m_1+m_2=-(m_3+m_4)$ and $\gamma$ is a phase factor. The latter can depend on $J$ as well as on the fixed values of the four spins $j_\nu$. The functional dependence is omitted to avoid confusion with tensor indices. Clearly, the value of $\gamma$ does not affect the unitarity relation satisfied by the coefficients defined in Eq.~(\ref{eq:SingletSpinJ})
\be
\sum_{m_\nu}\overline{\alpha}^{j_\nu,J}_{m_\nu}\alpha^{j_\nu,J^{\prime}}_{m_\nu}=\delta^{J J^{\prime}}~.
\ee
We choose the value of the phase $\eta$ so as to have
\be\label{eq:DefAlphaIntertwiner}
\alpha^{j_\nu,J}_{m_\nu}=(-1)^{J-M}\sqrt{d_J}\left(\begin{array}{ccc} j_1 & j_2 & J\\ m_1 & m_2 & -M\end{array}\right)\left(\begin{array}{ccc} J & j_4 & j_3\\ M & m_4 & m_3\end{array}\right)~.
\ee
The convenience of this particular choice of conventions lies in the fact that the contraction of five four valent intertwiners thus defined coincides with the definition of the 15j symbol, see Ref.~\cite{Finocchiaro}.

At this stage, we can clarify the relation with the kinematics of GFT. In fact, since $\alpha^{j_\nu,J}_{m_\nu}$ is an invariant tensor it satisfies the following equation
\be
\sum_{n_\nu}\alpha^{j_\nu,J}_{n_\nu}\prod_{i=1}^4D^{j_{i}}_{k_{i},n_{i}}(h)=\alpha^{j_\nu,J}_{k_\nu}.
\ee
This is the same as Eq.~(\ref{Eq:GFT:Algebraic constraint}), which in Section~\ref{sec:1} was shown to be a consequences of right-invariance of the GFT field. Hence, right-invariance implies that we can associate a quantum tetrahedron to the harmonic components of the GFT field.  
 
All intertwiners, \emph{i.e.} elements of $\accentset{\circ}{\mathcal{H}}_{j_\nu}$ can be expressed as linear combinations of the $\alpha^{j_\nu,J}_{n_\nu}$ coefficients given above
\be\label{eq:LinearCombinationIotaAlpha}
\mathcal{I}^{j_\nu,\iota}_{n_\nu}=\sum_{J} c^{J\iota}\,\alpha^{j_\nu,J}_{n_\nu}~.
\ee
 Hence,  we can attach the label $\iota$ to any linear subspace in the intertwiner space $\accentset{\circ}{\mathcal{H}}_{j_\nu}$. Different choices correspond to different physical properties of the quanta of geometry.
  In Appendix~\ref{sec:Volume} we construct explicitly the intertwiners of volume eigenstates in a simple example.

\chapter{Volume Operator}\label{sec:Volume}
There are several different definitions of the volume operator in LQG \cite{Ashtekar:1997fb,Rovelli:1994ge,Bianchi:2010gc}. However, they all agree in the case of a four-valent vertex \cite{Haggard:2011qvx} and match the operator introduced in Ref.~\cite{Barbieri:1997ks}. In this Appendix we will largely follow Ref.~\cite{Brunnemann:2004xi} for the definition of the volume operator and the derivation of its spectrum. The volume operator acting on a spin network vertex (embedded in a differentiable manifold) is defined as
\be\label{eq:VolumeDefinition}
\hat{V}=\sqrt{\left|\sum_{I<J<K}\epsilon(e_I,e_J,e_K)\epsilon_{ijk}J^i_I J^j_J J^k_K\right|}=\sqrt{\left|\frac{i}{4}\sum_{I<J<K}\epsilon(e_I,e_J,e_K)\hat{q}_{IJK}\right|}~.
\ee
In the formula above $(e_I,e_J,e_K)$ is a triple of edges adjacent to the vertex and $\epsilon(e_I,e_J,e_K)$ is their orientation, given by the triple product of the vectors tangent to the edges. A copy of the angular momentum algebra is attached to each edge, \emph{i.e.} there is one spin degree of freedom per edge. Angular momentum operators corresponding to distinct edges commute
\be
\left[J^i_I,J^j_J\right]=i \delta_{IJ}\epsilon^{ijk}J^k_I~.
\ee
In the last step of Eq.~(\ref{eq:VolumeDefinition}) we introduced the operator
\be
\hat{q}_{IJK}=\left(\frac{2}{i}\right)^3\epsilon_{ijk}J^i_I J^j_J J^j_K~.
\ee

Spin network vertices are gauge-invariant, \emph{i.e.} the angular momenta carried by the edges entering a vertex satisfy a closure condition. In the case of a four-valent vertex the closure condition reads as
\be\label{eq:Closure}
\mathbf{J}_1+\mathbf{J}_2+\mathbf{J}_3+\mathbf{J}_4=0~.
\ee
Hence, the Hilbert space of the vertex is that of Eq.~(\ref{eq:HilbertSpace4Vertex}).
Eq.~(\ref{eq:Closure}) leads to the following simplification in the evaluation of the sum in Eq.~(\ref{eq:VolumeDefinition})
\be
\sum_{I<J<K}\epsilon(e_I,e_J,e_K)\hat{q}_{IJK}=2\,\hat{q}_{123}~.
\ee
Therefore, the squared volume operator can be rewritten as
\be\label{eq:VolumeExpressionQ}
\hat{V}^2=\left|\frac{i}{2}\hat{q}_{123}\right|~.
\ee
Note that, while the definition (\ref{eq:VolumeDefinition}) makes explicit reference to the embedding map, the final expression (\ref{eq:VolumeExpressionQ}) clearly does not depend on it.

Using the recoupling channel $\mathcal{H}_{j_1}\otimes\mathcal{H}_{j_2}$ as in Appendix~\ref{sec:Intertwiner} and labelling with $J$ the eigenvalue of $(\mathbf{J}_1+\mathbf{J}_2)^2$, we find that the non-vanishing matrix elements in the recoupling basis are \cite{Brunnemann:2004xi}
\be\label{eq:VolumeSpectrum}
\begin{split}
\langle J | \hat{q}_{123} |J-1\rangle&=\frac{1}{\sqrt{4J^2-1}}\Big[(j_1+j_2+J+1)(-j_1+j_2+J)(j_1-j_2+J)(j_1+j_2-J+1)\Big.\\
&\phantom{=\frac{1}{\sqrt{4J^2-1}} [}  \Big.(j_3+j_4+J+1)(-j_3+j_4+J)(j_3-j_4+J)(j_3+j_4-J+1)\Big]^{\frac{1}{2}}\\ &=-\langle J -1 | \hat{q}_{123} |J\rangle~.
\end{split}
\ee
The eigenvalues of $\hat{q}_{123}$ are non-degenerate. Moreover, if $\hat{q}_{123}$ has a non-vanishing eigenvalue $a$, also $-a$ is an eigenvalue. The sign corresponds to the orientation of the vertex.
If the dimension of the intertwiner space is odd, $\hat{q}_{123}$ has a non-degenerate zero eigenvalue.
\subsection*{Monochromatic Vertex}
If the four spins are all identical ($j_1=j_2=j_3=j_4=j$) the vertex is called monochromatic. In this case Eq.~(\ref{eq:VolumeSpectrum}) simplifies to
\be\label{eq:SpectrumMonochromatic}
\langle J | \hat{q}_{123} |J-1\rangle=\frac{1}{\sqrt{4J^2 -1}} J^2 (d_j^2 - J^2 )~,
\ee
where $d_j=2j+1$ is the dimension of the irreducible representation with spin $j$ \cite{Brunnemann:2004xi}.

For the applications of Sections~\ref{Sec:GFT:Background},~\ref{sec:Perturbations}, it is particularly interesting to consider the fundamental representation $j=\frac{1}{2}$. In this case the intertwiner space is two-dimensional, with a basis given by $\{|0\rangle,|1\rangle\}$, \emph{i.e.} the four-valent gauge-invariant vertex is constructed using two singlets and two triplets, respectively. Using Eq.~(\ref{eq:SpectrumMonochromatic}), we have that the squared volume operator is written in this basis as
\be
\hat{V}^2=\frac{\sqrt{3}}{2}\left|\begin{pmatrix}0 & -i \\ i & 0 \end{pmatrix}\right|=\left|\hat{Q}\right|~,
\ee
where we introduced a new matrix $\hat{Q}=\frac{\sqrt{3}}{2}\sigma_2$, which is equal to $\hat{V}^2$ up to a sign. The sign of the eigenvalues of $\hat{Q}$ gives the orientation of the vertex.

The normalised eigenvectors of $\hat{Q}$ are
\be
|+\rangle=\frac{1}{\sqrt{2}}\begin{pmatrix}1\\i\end{pmatrix}, \hspace{1em} |-\rangle=\frac{1}{\sqrt{2}}\begin{pmatrix}1\\-i\end{pmatrix}~,
\ee
with eigenvalues $\pm\frac{\sqrt{3}}{2}$. The volume eigenstates $|\pm\rangle$ can be decomposed in the tensor product basis of $\mathcal{H}_{\frac{1}{2}}\otimes\mathcal{H}_{\frac{1}{2}}\otimes\mathcal{H}_{\frac{1}{2}}\otimes\mathcal{H}_{\frac{1}{2}}$ as follows
\be
|\pm\rangle=\sum_J c^{J\,\pm} |J\rangle=\sum_{m_{\nu}J}c^{J\,\pm}\alpha^{\frac{1}{2}\,J}_{m_\nu}|{\scriptstyle\frac{1}{2}},m_1\rangle|{\scriptstyle\frac{1}{2}},m_2\rangle|{\scriptstyle\frac{1}{2}},m_3\rangle|{\scriptstyle\frac{1}{2}},m_4\rangle~.
\ee
We define the intertwiners corresponding to the volume eigenstates $|\pm\rangle$ as
\be
\mathcal{I}^{\frac{1}{2}\,\pm}_{m_\nu}=\sum_{J}c^{J \pm}\alpha^{\frac{1}{2}\,J}_{m_\nu}=\frac{1}{\sqrt{2}}\left(\alpha^{\frac{1}{2}\,0}_{m_\nu}\pm i\, \alpha^{\frac{1}{2}\,1}_{m_\nu}\right)~.
\ee
Hence, we can write
\be
|\pm\rangle=\sum_{m_{\nu}}\mathcal{I}^{\frac{1}{2}\,\pm}_{m_\nu}|{\scriptstyle\frac{1}{2}},m_1\rangle|{\scriptstyle\frac{1}{2}},m_2\rangle|{\scriptstyle\frac{1}{2}},m_3\rangle|{\scriptstyle\frac{1}{2}},m_4\rangle~.
\ee

\chapter{Effective Friedmann equations}\label{Appendix:EffectiveFriedmann}
Here we review how the effective equations that give the relational
evolution of the volume of the universe,
Eqs.~(\ref{eq:EmergingFriedmannI}), (\ref{eq:EmergingFriedmannII}) can
be recast in a form that is closer to that of ordinary FLRW models. In
fact, in the GFT framework, spacetime is an
emergent concept. Since there is no natural notion of proper time, the
evolution of physical observables is more appropriately defined using a matter clock.
In the models that we considered, this is represented by a massless scalar which
is minimally coupled to the gravitational field.
Below we show in more detail the connection between the relational
description of the effective dynamics and the more conventional formulation
of the Friedmann equations.

In the regime of validity of the mean field approximation to the GFT dynamics,
we can assume the classical relation between the velocity of the
scalar field $\phi$ and its canonically conjugate momentum
$\pi_{\phi}$ \be\label{eq:Momentum} \pi_{\phi}=V\dot{\phi}~. \ee
The scale factor can be defined so as to satisfy the standard relation with the proper volume
\be\label{eq:VolumeScaleFactor} V\propto a^3~. \ee
It is important to stress that Eq.~(\ref{eq:Momentum}) has not been derived from GFT so far.
Thus, it can be regarded as a \emph{definition} of proper time, which is sufficient for our purposes.

We can
express the Hubble expansion rate as
\be\label{eq:AppendixFirstDerVolume}
H=\frac{\dot{a}}{a}=\frac{1}{3}\frac{\dot{V}}{V}=\frac{1}{3}\pi_{\phi}\frac{\partial_{\phi}V}{V^2}~.
\ee The last equation is the same as Eq.~(\ref{eq:HubbleRate}), which
leads to the modified Friedmann equation of
Eq.~(\ref{eq:FriedmannHubbleRate}).
In a similar fashion, it is also
possible to get the Raychaudhuri equation for the acceleration. In
fact, from Eq.~(\ref{eq:VolumeScaleFactor}) one obtains 
 \be\label{eq:AppendixSecondDerScale}
\frac{\ddot{a}}{a}=\frac{1}{3}\left[\frac{\ddot{V}}{V}-\frac{2}{3}\left(\frac{\dot{V}}{V}\right)^2\right].
\ee Furthermore, using Eq.~(\ref{eq:Momentum}), one finds \be
\dot{V}=(\partial_{\phi}V) \dot{\phi}=(\partial_{\phi}V)
\frac{\pi_{\phi}}{V} \ee and \be\label{eq:AppendixSecondDerVolume}
\ddot{V}=\left(\frac{\pi_{\phi}}{V}\right)^2 \left[\partial^2_{\phi}V
  - \frac{\left(\partial_{\phi}V\right)^2}{V} \right].  \ee From
Eqs.~(\ref{eq:AppendixFirstDerVolume}),~(\ref{eq:AppendixSecondDerScale}),~(\ref{eq:AppendixSecondDerVolume})
one has \be
\frac{\ddot{a}}{a}=\frac{1}{3}\left(\frac{\pi_{\phi}}{V}\right)^2\left[\frac{\partial^2_{\phi}V}{V}-\frac{5}{3}\left(\frac{\partial_{\phi}V}{V}\right)^2\right].
\ee The last equation justifies the definition of the acceleration $\mathfrak{a}$
given in Eq.~(\ref{eq:DefinitionAcceleration}) as
\be
\mathfrak{a}=\frac{\partial^2_{\phi}V}{V}-\frac{5}{3}\left(\frac{\partial_{\phi}V}{V}\right)^2~.
\ee
%


\chapter{Stochastic Processes and Stochastic Differential Equations}\label{Appendix:Stochastic}
A stochastic process\footnote{Most of this Appendix is based on Ref.~\cite{oksendal2003stochastic}. The notation has been adapted to that of Chapter \ref{Chapter:VariableG} and generality has been kept down to the minimum necessary.} is a collection of random variables depending on a parameter $t$ (which can be either discrete or continuous)
\be
\{X_t\}_{t\in T}
\ee
defined on a probability space $(\Omega,\mathcal{F},P)$ and taking values in $\mathbb{R}^n$ \cite{oksendal2003stochastic}. In physical applications $t$ is usually interpreted as time. The probability space is defined as follows. $\mathcal{F}$ is a $\sigma$-algebra on the set $\Omega$ and $P$ a probability distribution function satisfying a suitable set of axioms (see Ref.~\cite{oksendal2003stochastic}). Measurable subsets of $\Omega$ are called \emph{events}. We will restrict our attention to the case of one-dimensional stochastic processes, which is the one relevant for our applications of Chapter \ref{Chapter:VariableG}.

According to the definition given above, for a given $t\in T$ we have that
\be
\omega\rightarrow X_t(\omega), \hspace{1em} \omega\in\Omega
\ee
is a random variable. On the other hand, fixing the subset $\omega$ defines a \emph{path}
\be
t\rightarrow X_t(\omega),\hspace{1em} t\in T~.
\ee
The reason for the term \emph{path} is clear if $X_t$ represents a diffusion process, \emph{e.g.} a random walk (see below). In fact, in this case a given event $\omega$ represents a sequence of outcomes which determines the physical trajectory of a particle.

A particularly relevant stochastic process is the \emph{Wiener process}, which is used to model Brownian motion. The Wiener process $W_t$, defined for $t\in[0,\infty)$, satisfies the following properties:
\begin{enumerate}[label=$\roman*$.]
\item $W_t$ has a.s.\footnote{That is, almost surely. In probability theory this means that the event occurs with probability one.} the origin as a starting point\footnote{The generalisation to a Wiener process starting from a different point $x$ is immediate, see Ref.~\cite{oksendal2003stochastic}. However, in that case the formulae given here must also be suitably generalised.}, \emph{i.e.} $P(B_0=0)=1$.
\item $W_t$ is a Gaussian process, \emph{i.e.} for any ordered time-sequence $0\leq t_1\leq \dots \leq t_k$, the random variable $Z=(W_{t_1},\dots,W_{t_k})\in\mathbb{R}^k$ has a multi-normal distribution. This means that there is a vector $M\in\mathbb{R}^k$ and a positive-definite matrix $C=(c_{jk})\in \mathbb{R}^{2k}$ such that the characteristic function is given by
\be
E\left[\exp\left(i~u_k Z_k\right)\right]=\exp\left(-\frac{1}{2}c_{jk}u_j u_k+i~u_kM_k\right)~,
\ee
where $E$ denotes the expectation value with respect to the probability measure $P$. Moreover, one has that $W_t$ is normally distributed with  mean zero and variance $t$ 
\be
E[W_t]=0~, \hspace{1em} E[(W_t)^2]=t~.
\ee
For $t\geq s\geq 0$ one has that $W_t-W_s$ is also normally distributed with mean  zero and variance equal to the time difference
\be
E[(W_t-W_s)^2]=t-s~.
\ee
\item $W_t$ has independent and normally distributed increments, \emph{i.e.} for any non-overlapping time intervals $[t_1,t_2]$, $[t_3,t_4]$ one has
\be
E[(W_{t_2}-W_{t_1})(W_{t_4}-W_{t_3})]=0~.
\ee
\item Finally, there is a precise sense in which any path $W_t$ can be regarded as continuous in $t$ a.s., see Ref.~\cite{oksendal2003stochastic}. However, paths are a.s.~not differentiable anywhere.
\end{enumerate}

Since it defines a probability measure on the space of continuous functions on the real half-line $[0,\infty)$, the Wiener process $W_t$ can be used to define an integral. More precisely, given $0\leq S\leq T$ and a partition $0=t_0\leq t_1\leq\dots\leq t_k$, one can consider the Riemann-Stieltjes sum
\be
\sum_{j=0}^{k-1} f(t^*_j,\omega) (W_{t_{j+1}}-W_{t_j})(\omega)~,
\ee
where $t^*_j$ is a point in the interval $[t_j,t_{j+1})$.
However, unlike the Riemann-Stieltjes case, the $k\to\infty$ limit of such sums (provided it exists) depends on the point $t^*_j$ we choose. This is due to the fact that paths of the Wiener process $W_t$ are a.s. not differentiable\footnote{There is a remarkable similarity with the problem of defining the path integral in quantum mechanics, which requires a discretisation of the action of a point particle. In that case, different prescriptions for the choice of the points $t^*_j$ give rise to quantisation ambiguities \cite{Feynman:1948ur}. The mid-point prescription corresponds to Weyl quantisation, or symmetric factor-ordering.}. There are two main choices that are commonly encountered in the literature:
\begin{enumerate}[label=$\roman*$.]
\item The \^{I}to integral corresponds to the choice $t^*_j=t_j$ (left-end point) and is denoted by
\be
\int_{S}^{T}f(t,\omega) \de W_{t}(\omega)~;
\ee
The \^Ito integral is evaluated differently from ordinary integrals. For instance, one finds
\be\label{Eq:Stochastic:ItoIntegralExample}
\int_0^t W_s\de W_s=\frac{1}{2}W_t^2-\frac{1}{2}t~.
\ee
\item The Stratonovich integral corresponds to the choice $t^*_j=(t_j+t_{j+1})/2$ (mid-point) and is denoted by
\be
\int_{S}^{T}f(t,\omega) \circ\de W_{t}(\omega)~.
\ee
The Stratonovich integral satisfies
\be\label{Eq:Stochastic:StratonovichIntegralExample}
\int_0^t W_s\circ\de W_s=\frac{1}{2}W_s^2~,
\ee
which is formally analogous to the result one gets by computing an ordinary integral. It is interesting to compare Eqs.~(\ref{Eq:Stochastic:ItoIntegralExample}),~(\ref{Eq:Stochastic:StratonovichIntegralExample}) to illustrate the difference between different prescriptions for the choice of $t^*_j$.

We introduce the concept of stochastic differential equations (SDEs). For our purposes, we will only deal with ordinary differential equations (ODEs), although generalisation to partial differential equations (PDEs) are possible (see Ref.~\cite{oksendal2003stochastic} and references therein). The prototype is the Langevin equation, which is used to study Brownian motion. The Langevin equation describes the dynamics of a dust particle subject to random forces due to the collision with molecules of a fluid
\be\label{Eq:Stochastic:Langevin}
m\ddot{x}=-\gamma\dot{x}+\sigma\xi(t)~.
\ee
$m$ is the mass of the Brownian particle, $\gamma$ is the damping coefficient in the friction term. The last term in Eq.~(\ref{Eq:Stochastic:Langevin}) is modelled as white noise with strength $\sigma$. That is
\begin{align}
\langle\xi(t)\rangle&=0~,\\
\langle\xi(t)\xi(t^{\prime})\rangle&=\delta(t-t^{\prime})~.
\end{align}
In order to avoid mathematical subtleties due to the distributional nature of the white noise process, we introduce a differential notation
\be\label{Eq:Stochastic:LangevinDifferentialNotation}
m \de v_t=-\gamma v_t \de t+\sigma \de W_t,
\ee
where $v_t$ is the (stochastic) velocity of the particle and $W_t$ is the Wiener process defined above. However, one must bear in mind that the differential considered in Ref.~(\ref{Eq:Stochastic:LangevinDifferentialNotation}) should be understood as a finite difference. The continuum limit is not uniquely defined but depends in general on the particular differential calculus one adopts, \emph{e.g.} \^Ito or Stratonovich. In order to better illustrate this point, let us consider a more general case, replacing the noise strength with a more general function $\sigma\to\sigma(t,W_t)$ in Eq.~(\ref{Eq:Stochastic:LangevinDifferentialNotation}), and rewrite Eq.~(\ref{Eq:Stochastic:LangevinDifferentialNotation}) in integral form over the time interval $[0,t]$
\be\label{Eq:Stochastic:LangevinIntegralForm}
v_t=v_0+\frac{1}{m}\left[-\gamma \int_0^t v_s\de s+ \mbox{``}\int_0^t \sigma(s,W_s) \de W_s~\mbox{''} \right]~.
\ee
Depending on the interpretation of the last term $\mbox{``}\int_0^t \sigma(s,W_s) \de W_s~\mbox{''}$ according to the rules of a given stochastic calculus, the integral equation (\ref{Eq:Stochastic:LangevinIntegralForm}) will have a different interpretation and lead to a different solution $v_t$. Therefore, in writing down SDEs using the differential notation as in Eq.~(\ref{Eq:Stochastic:LangevinDifferentialNotation}), one must be aware that a choice of stochastic calculus is also implied. Such a choice depends on the physical situation we are studying. In the particular case of a constant $\sigma(t,W_t)$, the example given above does not lead to any difference between different choices of stochastic calculus.

SDEs such as Eq.~(\ref{Eq:Stochastic:LangevinDifferentialNotation}) can be solved analytically by using similar methods to those known for standard ODEs, suitably generalised to account for the modification introduced stochastic calculus, see Ref.~\cite{oksendal2003stochastic}. Analogous generalisations also exist for numerical methods, which we employed for our applications of Chapter \ref{Chapter:VariableG}. For a review of such methods see Refs.~\cite{higham2001algorithmic,talay1994numerical}. 

\end{enumerate}

\chapter{Conformal Invariance of the Dirac Action}\label{Appendix:ConformalDirac}
In this Appendix we wish to add more details showing the motivation for considering the action (\ref{Eq:ActionDirac2}). Furthermore, we will prove that the two Dirac Lagrangians in Eqs.~(\ref{eq:ActionDirac}) and (\ref{Eq:ActionDirac2}) are the same.

The reason to look at the action (\ref{Eq:ActionDirac2}) in first place, is to have an action functional which is manifestly conformally invariant. 
We will say that a field $F$ has conformal weight $w$ if, under a local conformal transformation, it transforms as
\be\label{eq:ConformalWeight}
F\rightarrow\tilde{F}=\Omega^{w}F~.
\ee
For scalar fields and half-spin fermions the conformal weight is the opposite of their canonical mass-dimension, \emph{i.e.} $w=-1,-3/2$, respectively\footnote{One must be aware that this statement cannot be generalised to fields with arbitrary canonical mass-dimension. In fact, for a gauge vector field, one must have $w=0$, see the discussion in Section~\ref{Sec:CouplingSM} and Ref.~\cite{Wald:1984rg}.} . Therefore, we can construct a `gauge covariant derivative' for the conformal symmetry as
\be\label{eq:CovariantDerivative}
\mathcal{D}_\mu F=\pa_\mu F -\frac{w}{2}B_\mu F~.
\ee
It is straightforward to check, using Eqs.~(\ref{eq:TransformationLawWeylVector}),~(\ref{eq:ConformalWeight}),~(\ref{eq:CovariantDerivative}), that
\be
\mathcal{D}_\mu F\rightarrow\tilde{\mathcal{D}}_\mu \tilde{F}=\Omega^{w}\,\mathcal{D}_\mu F~,
\ee
which justifies calling $\mathcal{D}_\mu$ a gauge covariant derivative. This is all we need to build the kinetic term of a scalar field in Eq.~(\ref{eq:PhiSector}), but it is not enough for fermions. In fact, the spin connection must appear explicitly in the action of a spinor. In the metric-compatible case, the spin connection is given by (see Ref.~\cite{Wald:1984rg})
\be\label{eq:LCSpinConnection}
\omega^{\rm\scriptscriptstyle LC}_{\mu\, ab}=e_a^{\;\nu}\nabla^{\rm \scriptscriptstyle LC}_\mu e_{b\,\nu}~.
\ee
In order to be consistent with the principle of local conformal invariance, it is natural to replace this object with the one constructed out of the Weyl connection
\be\label{eq:WeylSpinConnection}
\omega^{\rm\scriptscriptstyle W}_{\mu\, ab}=e_a^{\;\nu}\nabla^{\rm \scriptscriptstyle W}_\mu e_{b\,\nu}~.
\ee
Under a conformal transformation
\be
\omega^{\rm\scriptscriptstyle W}_{\mu\, ab}\rightarrow\tilde{\omega}^{\rm\scriptscriptstyle W}_{\mu\, ab}=\omega^{\rm\scriptscriptstyle W}_{\mu\, ab}+\left(\Omega^{-1}\pa_{\mu}\Omega\right)\eta_{ab}~.
\ee
Notice that in the case of Weyl geometry, the spin connection fails to be antisymmetric in the internal indices $\omega^{\rm\scriptscriptstyle W}_{\mu\, ab}\neq-\omega^{\rm\scriptscriptstyle W}_{\mu\, ba}$, as it is instead the case in Riemannian geometry.
However, since in the Dirac action (\ref{Eq:ActionDirac2}) $\omega^{\rm\scriptscriptstyle W}_{\mu\, ab}$ is contracted with the generator of Lorentz transformations in spinor space (which is $\propto [\gamma^a,\gamma^b]$), only the antisymmetric part gives a non-vanishing contribution\footnote{Which is also the reason why it does not make a difference whether we define $\omega^{\rm\scriptscriptstyle W}_{\mu\, ab}$ as equal to $e_a^{\;\nu}\nabla^{\rm \scriptscriptstyle_\mu} e_{b\,\nu}$ or $e_{a\nu}\nabla^{\rm \scriptscriptstyle_\mu} e_{b}^{\;\nu}$, although they are not the same in the non metric-compatible case.}. Hence, the third term in the bracket in the action (\ref{Eq:ActionDirac2}) is conformally invariant.

We will now proceed to show that the Lagrangian in Eq.~(\ref{Eq:ActionDirac2}) is equal to the Dirac Lagrangian in the metric case, which appears in Eq.~(\ref{eq:ActionDirac}). In order to do this, we expand the spin connection in Eq.~(\ref{eq:WeylSpinConnection}) in terms of its counterpart in the metric case, given by Eq.~(\ref{eq:LCSpinConnection}), plus terms involving the Weyl vector
\be
\omega^{\rm\scriptscriptstyle W}_{\mu\, ab}=\omega^{\rm\scriptscriptstyle LC}_{\mu\, ab}+e_{[a}^{\; \nu}e_{b]\mu}B_{\nu}+\frac{1}{2}\eta_{ab}B_{\mu}~.
\ee
Hence, we have for the contribution to the action (\ref{Eq:ActionDirac2}) coming from the last term in the round bracket
\be
\frac{1}{8}\gamma^c e_c^{\;\mu}[\gamma^a,\gamma^b]\,\omega^{\rm\scriptscriptstyle W}_{\mu\, ab}=\sum_{c\neq a}\frac{1}{4}\gamma_c\gamma^a\gamma^c e_{a}^{\; \nu}B_\nu
=-\frac{3}{4}\gamma^a e_{a}^{\; \nu}B^{\nu}~,
\ee
which cancels exactly the contribution due to the gauge-covariant coupling to $B_{\mu}$.

\end{appendices}

\printthesisindex 

\end{document}